\documentclass[twoside,12pt,a4paper]{report}
\usepackage{graphicx}
\usepackage{pgf}
\usepackage{fancyhdr}
\usepackage{dcolumn}
\usepackage{multirow}
\usepackage{amssymb}
\usepackage{amsbsy}
\usepackage{amsmath}
\usepackage{epsfig,subfigure}
\usepackage{color}
\usepackage{bm}
\usepackage{verbatim} 
\usepackage{rotating} 
\usepackage{hhline}
\usepackage{xspace}
\usepackage{subfigure}
\usepackage{anysize}
\usepackage{fixmath}
\usepackage{color}
\usepackage{cite}
\usepackage[toc,page]{appendix}
\usepackage[utf8]{inputenc}
\usepackage[T1]{fontenc}
\usepackage{comment}
\usepackage[acronym]{glossaries}
\usepackage{lineno} 
\usepackage{titlesec}
\usepackage{titlesec}
\usepackage{braket}

\titleclass{\subsubsubsection}{straight}[\subsection]

\newcounter{subsubsubsection}[subsubsection]
\renewcommand\thesubsubsubsection{\thesubsubsection.\arabic{subsubsubsection}}

\titleformat{\subsubsubsection}
{\normalfont\normalsize\bfseries}{\thesubsubsubsection}{1em}{}
\titlespacing*{\subsubsubsection}{0pt}{3.25ex plus 1ex minus .2ex}{1.5ex plus .2ex}

\makeatletter
\renewcommand\paragraph{\@startsection{paragraph}{5}{\z@}%
	{3.25ex \@plus1ex \@minus.2ex}%
	{-1em}%
	{\normalfont\normalsize\bfseries}}
\renewcommand\subparagraph{\@startsection{subparagraph}{6}{\parindent}%
	{3.25ex \@plus1ex \@minus .2ex}%
	{-1em}%
	{\normalfont\normalsize\bfseries}}
\def\toclevel@subsubsubsection{4}
\def\toclevel@paragraph{5}
\def\toclevel@paragraph{6}
\def\l@subsubsubsection{\@dottedtocline{4}{7em}{4em}}
\def\l@paragraph{\@dottedtocline{5}{10em}{5em}}
\def\l@subparagraph{\@dottedtocline{6}{14em}{6em}}
\makeatother

\begin{document}
\begin{titlepage}
	
	\begin{center}
		{
			\large
			A DOCTORAL DISSERTATION\\
			PREPARED IN THE INSTITUTE OF PHYSICS\\
			OF THE JAGIELLONIAN UNIVERSITY,\\
			SUBMITTED TO THE FACULTY OF PHYSICS, ASTRONOMY
			AND APPLIED COMPUTER SCIENCE\\
			OF THE JAGIELLONIAN UNIVERSITY
		}
	\end{center}

	\begin{figure}[h]
		\begin{center}
			\includegraphics[width=4.0cm,height=5.0cm]{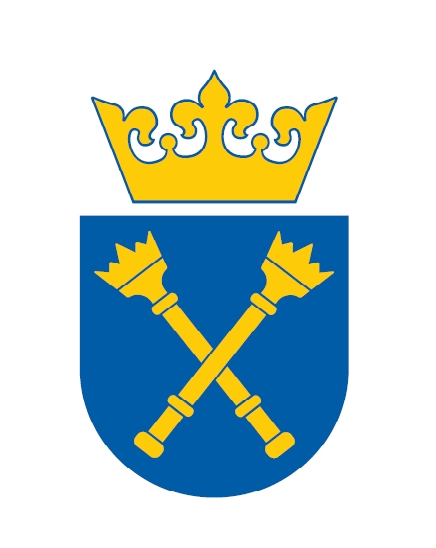}
		\end{center}
	\end{figure}

	\begin{center}
		\Huge \textbf{Search for $\eta$-mesic $^3He$ in non-mesonic final states\\}
	\end{center}

	\begin{center}
		\Large \textbf{OLEKSANDR RUNDEL}
	\end{center}
	
	\vspace{1.0cm}
	
	\begin{center}
		\normalsize{THESIS ADVISOR:\\ PROF. DR HAB. PAWEŁ MOSKAL}\\
		\vspace{0.5cm}
		\normalsize{CO-ADVISOR:\\ DR MAGDALENA SKURZOK}
		
		\vspace{2.0cm}
		
		Cracow, 2018
		
	\end{center}
	
\end{titlepage}
\newpage\thispagestyle{empty}\begin{center}\end{center}
\newpage\thispagestyle{empty}
\begin{center}
	{
		\large
		ROZPRAWA DOKTORSKA\\
		PRZYGOTOWANA W INSTYTUCIE FIZYKI\\
		UNIWERSYTETU JAGIELLOŃSKIEGO,\\
		ZŁOŻONA NA WYDZIALE FIZYKI, ASTRONOMII\\
		I INFORMATYKI STOSOWANEJ\\
		UNIWERSYTETU JAGIELLOŃSKIEGO
	}
\end{center}

\begin{figure}[h]
	\begin{center}
		\includegraphics[width=4.0cm,height=5.0cm]{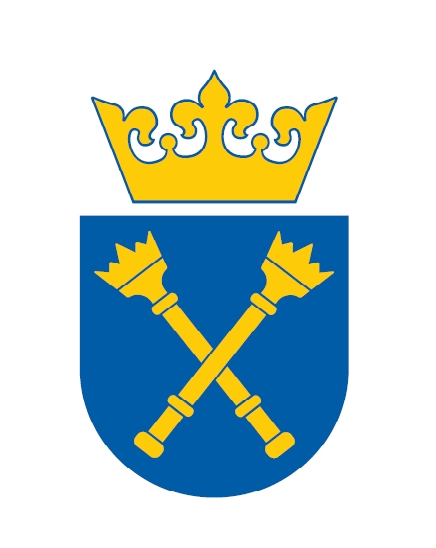}
	\end{center}
\end{figure}

\vspace{-0.5cm}

\begin{center}
	\Huge \textbf{Poszukiwanie $\eta$-mezonowego $^3$$He$ w niemezonowym stanie końcowym}
\end{center}

\begin{center}
	\Large \textbf{Oleksandr Rundel}
\end{center}

\vspace{0.7cm}

\begin{center}
	\normalsize{PROMOTOR:\\ PROF. DR HAB. PAWEŁ MOSKAL}\\
	\vspace{0.5cm}
	\normalsize{PROMOTOR POMOCNICZY:\\ DR MAGDALENA SKURZOK}
	
	\vspace{1.5cm}
	
	Kraków, 2018
	
\end{center}

\newpage\thispagestyle{empty}\begin{center}\end{center}
\newpage\thispagestyle{empty}
~\\
Wydział Fizyki, Astronomii i Informatyki Stosowanej\\
Uniwersytet Jagielloński
~\\~\\~\\~\\
\begin{center}\Large \textbf{Oświadczenie}\end{center}
~\\~\\
Ja ni\.{z}ej podpisany Oleksandr Rundel (nr indeksu: 1116401) doktorant Wydziału Fizyki, Astronomii i Informatyki Stosowanej Uniwersytetu Jagiello\'{n}skiego o\'{s}wiadczam, \.{z}e przed-
{\l}o\.{z}ona przeze mnie rozprawa doktorska pt. ,,Search for $\eta$-mesic $^3$$He$ in non-mesonic final states’’ jest oryginalna i przedstawia wyniki bada\'{n} wykonanych przeze mnie osobi\'{s}cie, pod kierunkiem prof. dr hab. Paw{\l}a Moskala. 
Pracę napisałem samodzielnie.\\
Oświadczam, że moja rozprawa doktorska została opracowana zgodnie z Ustawą o prawie autorskim i prawach pokrewnych z dnia 4 lutego 1994 r. (Dziennik Ustaw 1994 nr 24 poz. 83 wraz z późniejszymi zmianami).
Jestem świadom, że niezgodność niniejszego oświadczenia z prawdą ujawniona w dowolnym czasie, niezależnie od skutków prawnych wynikających z ww. ustawy, może spowodować unieważnienie stopnia nabytego na podstawie tej rozprawy.\\
~\\~\\~\\~\\
Kraków, dnia ……………….....~~~~~~~~~~~~~~~~~~~~~~~~~~~~~.............................................\\
\newpage\thispagestyle{empty}\begin{center}\end{center}
\newpage\thispagestyle{empty}
\begin{center}\large \textbf{Abstract}\end{center}

The $\eta$-mesic nucleus that is the bound state of a nucleus and $\eta$ meson has been postulated theoretically in 1986 but has not been observed in the experiments yet.
 
 In May-June 2014, the experiment devoted to the search for the bound state of the $\eta$ meson and $^3$$He$ nucleus has been performed at COSY accelerator in Research Center Jülich in Germany with WASA-at-COSY facility.
 The excitation functions for 
 $pd\rightarrow^3$$He2\gamma$, $pd\rightarrow^3$$He6\gamma$,
 $pd\rightarrow ppp\pi^{-}$, $pd\rightarrow ppn\pi^{0}$, 
 $pd\rightarrow dn\pi^{-}$, $pd\rightarrow dp\pi^{0}$,
 $pd\rightarrow pd$, and $pd\rightarrow ppn$
 reactions have been measured in the vicinity of the $\eta$ meson production threshold.
 The experiment was carried out using COSY ramped proton beam and deuterium pellet target.
 The beam momentum varied continuously in the range of $1.426-1.635~GeV/c$ corresponding to $^3$$He-\eta$ excess energy range from $-70$ to $+30~MeV$.

This dissertation is devoted to the search for $\eta$ mesic $^3$$He$ nucleus in the non-mesonic decay channels: 
$pd\rightarrow^3$$He2\gamma$ and $pd\rightarrow^3$$He6\gamma$ reactions.
The excitation functions for these processes were obtained after identification of all outgoing particles and selection of events with conditions based on the results of Monte Carlo simulations of direct decay of $\eta$ meson bound in the $^3$$He$ nucleus.
The integrated luminosity dependence on the excess energy, used for the excitation function normalization, was calculated based on $pd\rightarrow^3$$He\eta$ and $pd\rightarrow ppn_{spectator}$ reactions.

The analysis of the obtained excitation functions for the  $pd\rightarrow^3$$He\eta$ indicate slightly the signal from the bound state for $\Gamma>20~MeV$ and $B\in[0;15]~MeV$.  
However, the observed indication is in the range of the systematic error. 
Therefore the final conclusion of this thesis is that no narrow structure that could be interpreted as $\eta$-mesic nucleus was observed in both excitation curves.
Thus, the upper limit for the total cross section of the bound state formation was estimated assuming that the $\eta$ decay branching ratios in the bound state is the same as in free space.
The upper limit at the $90\%$ confidence level varies from $2$ to $15~nb$ 
within
the binding energy range from $0$ to $60~MeV$ and the width from $2.5$ to $40~MeV$.

\newpage\thispagestyle{empty}\begin{center}\end{center}
\newpage\thispagestyle{empty}
\begin{center}\large \textbf{Streszczenie}\end{center}

Jądra $\eta$-mezonowe, które są stanem związanym jądra i mezonu $\eta$, zostały zapostulowane teoretycznie w roku 1986, jednak do tej pory nie zostały zaobserwowane eksperymentalnie.

Na przełomie maja i czerwca 2014 roku przeprowadzono eksperyment na akceleratorze COSY w Centrum Badawczym w Jülich w Niemczech z wykorzystaniem detektora WASA, którego celem było poszukiwanie stanu związanego $^3$ $He$ z mezonem $\eta$.
Krzywe wzbudzenia dla reakcji
$pd\rightarrow^3$ $He2\gamma$, $pd\rightarrow^3$ $He6\gamma$,
$pd\rightarrow ppp\pi^{-}$, $pd\rightarrow ppn\pi^{0}$, 
$pd\rightarrow dn\pi^{-}$,  $pd\rightarrow dp\pi^{0}$,
$pd\rightarrow pd$ oraz $pd\rightarrow ppn$
zostały zmierzone wokół progu na produkcję mezonu $\eta$.
Pomiar przeprowadzono z wykorzystaniem wiązki protonów oraz tarczy pelletowej deuteru.
Pęd wiązki zmieniano w sposób ciągły w zakresie od
$1.426$ do $1.635~GeV/c$, co odpowiada zakresowi energii wzbudzenia $Q_{^3He\eta}$ od $-70$ do $+30~MeV$.

Tematem niniejszej pracy jest poszukiwanie jądra mezonowego  $^3$ $He$-$\eta$ w niemezonowych kanałach rozpadu: $pd\rightarrow^3$ $He2\gamma$ oraz $pd\rightarrow^3$ $He6\gamma$.
Krzywe wzbudzenia dla tych procesów otrzymano po identyfikacji wszystkich emitowanych cząstek i zastosowaniu odpowiednich warunków opartych na wynikach symulacji Monte Carlo dla rozpadów związanego mezonu $\eta$.
Zależność całkowalnej świetlności od $Q_{^3He\eta}$ potrzebna do normalizacji krzywych wzbudzenia została obliczona w oparciu o reakcje $pd\rightarrow^3$ $He\eta$ oraz $pd\rightarrow ppn_{spectator}$.

Wyniki analizy funkcji wzbudzenia wskazują iż możliwe jest istnienie stanu związanego o szerokości $\Gamma>20~MeV$ i energii wiązania $B\in[0;15]~MeV$, jednakże zaobserwowana struktura mieści się w zakresie niepewności systematycznej. Ostatecznym wynikiem analizy jest więc brak obserwacji sygnału, który mógłby zostać interpretowany jako sygnatura na istnienie jądra mezonowego. Dlatego też, wyznaczono górną granicę na przekrój czynny dla produkcji stanu związanego, zakładając że współczynniki rozgałęzienia dla rozpadów mezonu $\eta$ związanego w jądrze mezonowym pozostają takie same jak dla rzeczywistego mezonu $\eta$. Górna granica na poziomie ufności 90\%  zmienia się od $2$ do $15~nb$ w zakresie energii wiązania od $0$ do $60~MeV$ oraz szerokości od $2.5$ do $40~MeV$.

\clearpage
\newpage\thispagestyle{empty}
\begin{center}\end{center}
\pagestyle{fancy}
\renewcommand{\chaptermark}[1]{ \markboth{#1}{}}
\renewcommand{\sectionmark}[1]{ \markright{#1}{}}
\fancyhf{}
\fancyhead[LE,RO]{\thepage}
\fancyhead[RE]{\textbf{\nouppercase{\leftmark}}}
\fancyhead[LO]{\textbf{\nouppercase{\rightmark}}}
\tableofcontents
\clearpage
\chapter{Introduction}

The nature of the strong interaction mechanism is much more complex than in case of electromagnetic and weak interactions.
The main reason is that the perturbative approach that allows to build the theory of electro-weak interaction cannot be applied for strong forces at low energies.
According to Quantum Chromodynamics (QCD), the strong interaction, unlike electromagnetic forces, is caused by color charge that has more complicated $SU(3)$ symmetry.
Strongly interacting particles (hardons) consist of quarks that are interacting via exchanging gluons.
Hardons are neutral from the color charge point of view.
Most frequently observed hadrons consist of three quarks (baryons) or quark-antiquark pairs (mesons).
There are many known particles belonging to these two groups.
Most important of them are proton and neutron, two baryons that are the building blocks for all atomic nuclei.
However, there are another objects such as 
hypernuclei~\cite{hypernuclei}, 
tetraquarks~\cite{tetraquarks}, 
pentaquarks~\cite{pentaquarks}, or 
dybarions~\cite{dibaryons1,dibaryons2,dibaryons3} 
that are less stable and thus harder to be registered.
The experimental investigations connected with such exotic matter are very useful for testing the theories describing the strong interaction.

One of theoretically predicted kind of exotic nuclear matter that has not been observed yet is  mesic nucleus.
It consists of nucleus bound with neutral meson e.~g. $\eta$, $\eta'$, or $\omega$.
Neutral meson can be bound only due to the strong interaction, thus investigations in this field can improve the knowledge about the strong forces.
From theoretical point of view, the most promising are $\eta$ mesic nuclei
that have been postulated in 1986~\cite{mesic_nuclei_first_postulated1}.
This study is complex because of several reasons.
Due to the short $\eta$ lifetime, direct measurements of $\eta N\rightarrow\eta N$ scattering are not actually possible.
Coupled channel calculations taking into account $\pi N\rightarrow\eta N$ and several other reactions induced by $\pi$ mesons allow to obtain the $\eta$-nucleon scattering matrix~\cite{coupled_channel_calculations}.
Calculated without elastic data, the $\eta$-nucleon scattering length is known with huge uncertainty that causes that the predictions about $\eta$ mesic nuclei are inexact as well.
However, some theoreticians postulate the existing of $\eta$-mesic helium~\cite{bound_theory3,bound_theory4}
and even deuteron~\cite{bound_theory5}.

The question about changing the $\eta$ meson properties when it is bound by nuclei is one of especial interest.
There are theoretical models predicting $\eta$ mixing with $\eta'$ when embedded in nucleus~\cite{bound_theory8,bound_theory9}.
Therefore, the investigations of $\eta$ mesic nuclei is important for understanding $\eta$ and $\eta'$ structure.

The experimental method developed by WASA-at-COSY collaboration allows to search for $\eta$ mesic $^3$$He$ nucleus in proton-deuteron collisions.
The proposal for the experiment~\cite{our_proposal} was presented at the
meeting of the Program Advisory Committee in Research Center Jülich in Germany
 and accepted for the realization.
The measurements, in which the author o this thesis took part, were performed in May-June 2014 at COSY accelerator by means of WASA detection system.
Proton beam and deuterium pellet target system were used.
For reducing the systematic uncertainties in excitation curve measurement, the ramped beam technique was applied.
The beam momentum was changed constantly in the range of $1.426-1.635~GeV/c$ that corresponds to $^3$$He\eta$ excess energy range from $-70$ to $+30~MeV$.

If the $\eta$ mesic nucleus exists, it will influence the shape of excitation curves for 
 $pd\rightarrow^3$$He2\gamma$, $pd\rightarrow^3$$He6\gamma$,
$pd\rightarrow ppp\pi^{-}$, $pd\rightarrow ppn\pi^{0}$, 
$pd\rightarrow dn\pi^{-}$ $pd\rightarrow dp\pi^{0}$,
$pd\rightarrow pd$, and $pd\rightarrow ppn$
reactions.
This dissertation contains the description of experimental data analysis connected with the first two non-mesonic reactions that can indicate the direct decay of the $\eta$ meson bound by nucleus.
For the interpretation of obtained excitation curves, Monte Carlo simulation for these two reactions was performed.

The thesis is divided into seven chapters.
The second chapter contains the brief review of existing theories and experimental data connected with $\eta$ mesic nuclei.
The experiment description is given in the third chapter.
The part of data analysis related to the luminosity estimation is described in the fourth chapter.
The fifth chapter is devoted to the determination of excitation curves for $pd\rightarrow^3$$He2\gamma$ and $pd\rightarrow^3$$He6\gamma$ reactions.
The interpretation of the obtained results is given in the chapter number six
while the seventh chapter contains the conclusions.
\chapter{Theoretical and experimental background}

\section{Theoretical predictions for $\eta$-mesic nuclei}

The question about $\eta$-nucleus bound state is connected with the question about $\eta$-nucleon interaction.
The coupled channel calculations
performed based on the experimental data 
reviewed in Refs.~\cite{eta_nucleon_prev_exp_review1,eta_nucleon_prev_exp_review2,eta_nucleon_recent_exp_review1,eta_nucleon_recent_exp_review2}
result in the strong attractive interaction between $\eta$ meson and nucleon~\cite{Nstar_decay_channels_calc1,Nstar_decay_channels_calc2,Nstar_decay_channels_calc3,Nstar_decay_channels_calc4,Nstar_decay_channels_calc5,Nstar_decay_channels_calc6}.
However, these results contain uncertainty that yields in different predictions about $\eta$-nucleus bound state.

The standard theoretical approach to describe $\eta$-nucleus interaction is to build optical potential based on the knowledge about $\eta$-nucleon interaction. 
One of the ways to do that is "$T\rho$" approximation~\cite{bound_t_rho}:
\begin{equation}
	U_{opt} = V + iW = -\frac{2\pi}{\mu} T_{\eta N\rightarrow\eta N}A\rho(r),
\label{Trho_approx}
\end{equation}
where $T_{\eta N\rightarrow\eta N}$ is the transition matrix (known from $\eta-N$ interaction), $A$ denotes the nucleus mass number, $\rho(r)$ is nuclear density, and $\mu$ is the reduced $\eta$-nucleus mass.
The calculations using older $\eta-N$ interaction data%
 ~\cite{mesic_nuclei_first_postulated1,mesic_nuclei_first_postulated2,bound_theory1,bound_theory6,bound_theory7}
 result in prediction of $\eta$-nucleus bound state existence for $A\ge12$
while the calculations using wider range of possible $\eta-N$ scattering length
do not exclude the existence of $\eta$-mesic helium~\cite{bound_theory3,bound_theory4}
and even deuteron~\cite{bound_theory5}.

Another popular approach is QCD based quark-meson-coupling (QMC). 
It assumes that the $\eta$ is submerged into the nucleus, couples to quarks, and mixes with $\eta'$~\cite{bound_theory8,bound_theory9} ($\eta$ meson properties are explained in Appendix~\ref{eta_properties}).
Solving Klein-Gordon equation in frame of such model results in prediction of $\eta$ mesic $^6He$, $^{11}B$, and $^{26}Mg$ existence.

For lighter nuclei, the bound state can be manifested via poles of $\eta$-nucleus scattering matrix.
Solving few body equations results in the existing of  $d-\eta$, $^3He-\eta$, and $^4He-\eta$ bound states~\cite{bound_theory10,bound_theory11}.
Newer calculations for
$d\eta$, $^3He\eta$, and $^4He\eta$ interaction
taking into account the data
about $\eta$ production mechanism and the FSI (Final State Interaction)%
\cite{bound_theory12,bound_theory2}
show that the bound states can exist only for small values of $a_{\eta N}$ scattering length in the range that is actually postulated. The higher values correspond to resonances.

Theoretical investigations connected with $\eta$-mesic nuclei are reviewed in Refs.~\cite{bound_theory2,krzemien_phd,skurzok_phd}.

One of recent $\eta^3$$He$ interaction theoretical investigations~\cite{helium3_theory_potential} takes
into account the total cross sections and asymmetries for $pd\rightarrow^3$$He\eta$ reaction near threshold.
The optical potential is calculated from these data and the scattering amplitude is determined.
Due to these results, a bound state with the binding energy of $0.3~MeV$ and width of $3~MeV$ is expected.
The estimated cross section of the bound state forming and decay in $pd\rightarrow(^3$$He-\eta)_{bound}\rightarrow^3He3\pi^0$ is about $0.4~nb$~\cite{our_proposal}.

For the case of $\eta$-mesic $^4$$He$, theoretical investigations taking into account the data about $\eta$ production near threshold have been performed~\cite{helium4_theory_potential2}.
Phenomenological approach allowed to build an optical potential reproducing the experimental data quite well for broad range of the potential parameters.
The cross section for $dd\rightarrow(^4$$He\eta)_{bound}\rightarrow^3$$He~p~\pi^{-}$ bound state decay channel is estimated at $4.5~nb$~\cite{helium4_theory_potential1}.

\section{Previous experiments on search for $\eta$-mesic nuclei}

After being postulated~\cite{mesic_nuclei_first_postulated1}, 
$\eta$-mesic nuclei were searched for in experiments with 
pion~\cite{pion_beam_experiment2,pion_beam_experiment3}, 
photon~\cite{photoproduction1,photoproduction2,photoproduction3},
proton~\cite{proton_induced_almost1,proton_induced_almost2},
and deuteron~\cite{deuteron_induced_JINR,deuteron_induced_GSI}
beams.
More detailed review of previous experiments devoted to $\eta$-mesic nuclei search can be found in Refs.~\cite{krzemien_phd,skurzok_phd}.

The strongest claim about existing of such bound state was made by COSY-GEM Collaboration~\cite{proton_induced_almost1}.
The reaction $p+^{27}$$Al\rightarrow^3$$He+\eta^{25}Mg\rightarrow^3$$He+p+\pi^{-}+X$ was studied.
The excitation energy spectrum obtained in these measurements shows an enhancement at the energy about $13~MeV$ below the $\eta$ creation threshold that is in agreement with the theoretical prediction~\cite{bound_theory7}.

\begin{figure}[h!]
\begin{center}
\includegraphics[width=300pt]{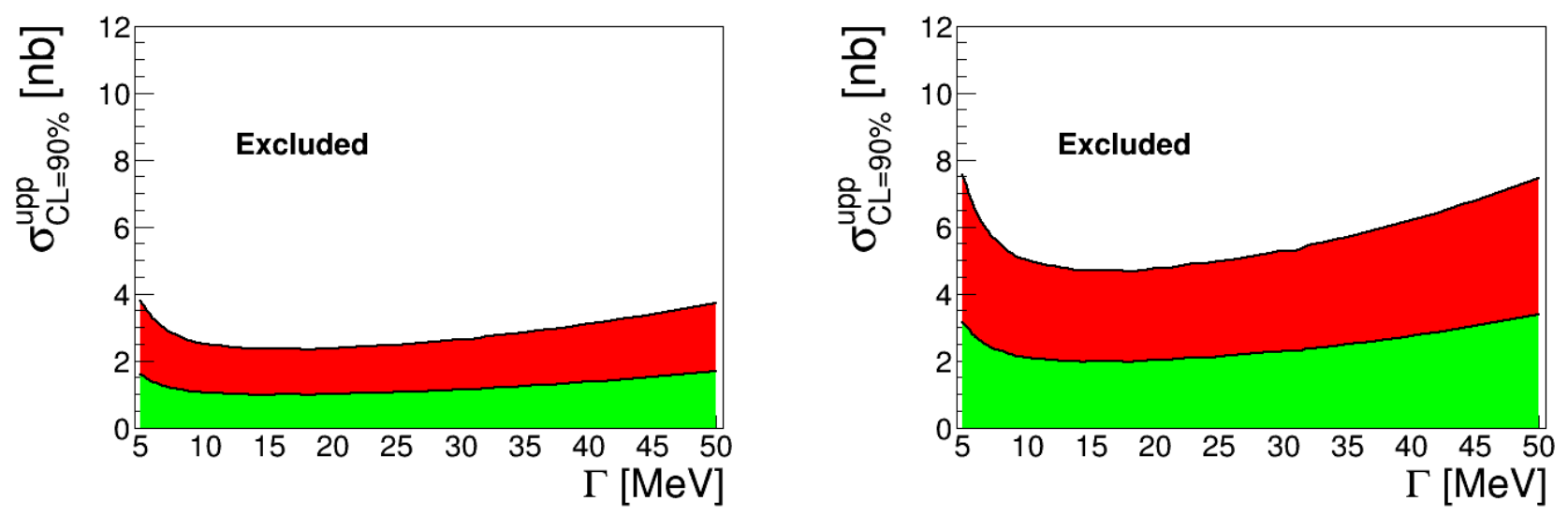}
\end{center}
\caption{
The upper limit for 
$dd\rightarrow(^4$$He\eta)_{bound}\rightarrow^3$$Hen\pi^0$ (left panel)
and $dd\rightarrow(^4$$He\eta)_{bound}\rightarrow^3$$Hep\pi^{-}$ (right panel)
reactions obtained in Ref.\cite{skurzok_phd}.
The assumed binding energy is equal to $30~MeV$. 
Red color shows the upper limit. Green color shows systematic uncertainty.
The picture was taken from Ref.~\cite{skurzok_phd}.
}
\label{fig_skurzok_phd}
\end{figure}

The recent search of $\eta$-mesic $^4$$He$ provided by WASA-at-COSY Collaboration~\cite{skurzok_phd,light_he4_one,light_he4_two} results in the upper limit of $3-6$ nanobarns for $dd\rightarrow^3$$He~n~\pi^0$ and $dd\rightarrow^3$$He~p~\pi^{-}$ bound state decay channels (Fig.~\ref{fig_skurzok_phd}).
This result was compared~\cite{light_he4_three_compare} with the theoretical estimations~\cite{helium4_theory_potential2}.
The experimental data allow to exclude a wide range of possible parameters for $\eta-^4$$He$ optical potential.
However, extremely narrow bound states with small binding energy within the model~\cite{helium4_theory_potential2} are not excluded~\cite{light_he4_three_compare}.

Previous indirect experimental studies of near threshold $\eta$ production in $pd\rightarrow^3$$He\eta$ reaction~\cite{light_region3_he3eta_pole,light_region4_he3eta_pole} show that probably there is a pole in the production amplitude at excitation energy of $Q_0 =[(-0.36\pm0.11\pm0.04)\pm i(0.19\pm0.28\pm0.06)]~MeV$ that is in agreement with data for $\gamma~^3$$He\rightarrow\eta~^3$$He$ reaction~\cite{light_region5_photo,light_region6_photo}.
The experimental search for $\eta$-mesic $^3$$He$ provided by COSY-11 collaboration~\cite{light_he3_old_upperlimit,old_review} resulted in the cross section upper limit of $70~nb$ for $pd\rightarrow(^3$$He\eta)_{bound}\rightarrow^3$$He\pi^0$ reaction.

The existing experimental data about $\eta$-mesic nuclei search are reviewed in details in Refs.~\cite{krzemien_phd,skurzok_phd,old_review,new_review1,new_review2}.

\section{Motivation}

Actual knowledge about $\eta$-nucleon interaction potential is not exact because experimental investigations of $\eta N\rightarrow\eta N$ scattering are actually not possible.
This uncertainty does not allow to determine the potential of $\eta$-nucleus interaction exactly and to definitely conclude if the $\eta$-mesic nuclei exist.

In case if the mesic nuclei are observed, it would become possible to investigate changing of $\eta$ meson properties when it is bound by nucleus. 
Such effects are postulated in some theories~\cite{bound_theory8,bound_theory9} and can provide us better understanding of strong interaction mechanism at low energies.

In case if the $\eta$-mesic nucleus is not observed in the experiment, a new more accurate upper limit value for the bound state formation cross section is determined.
This value can be useful for excluding a part of $\eta$-nucleus and $\eta$-nucleon interaction parameter range postulated in the theories but not realistic according to new experimental data.
Such comparison between experiment and theory was made in Ref.~\cite{light_he4_three_compare} after the new upper limit for $^4$$He\eta$ bound state formation cross section was obtained~\cite{skurzok_phd}.

Actually, no experiment has proven the existing of $\eta$-mesic $^3$$He$ nucleus.
However, the $pd\rightarrow^3$$He\eta$ cross section above the threshold is better described by FSI calculations assuming that the bound state exists~\cite{bound_theory2} and has the binding energy of several $MeV$.

The statistics gathered by WASA-at-COSY Collaboration in May-June 2014~\cite{our_proposal} for proton-deuteron collisions with beam momentum values close to $\eta$ creation threshold is the best one ever obtained for similar conditions.
The sensitivity of current experiment allows to measure the cross sections with better accuracy than currently measured upper limit for the bound state formation in proton-deuteron collisions~\cite{light_he3_old_upperlimit,old_review}.
\chapter{Experiment}

This chapter contains the description of experiment on searching for $\eta$-mesic $^3He$
that was carried out in Jülich (Germany) in May-June 2014.
The first section contains COSY accelerator complex brief description.
In the second section, the WASA-at-COSY detection system description is presented.
Brief data analysis software tools review is presented in the third section.
The conditions of current experiment are described in the fourth section.

\section{COSY accelerator}

\begin{figure}
	\begin{center}
\includegraphics[width=450pt]{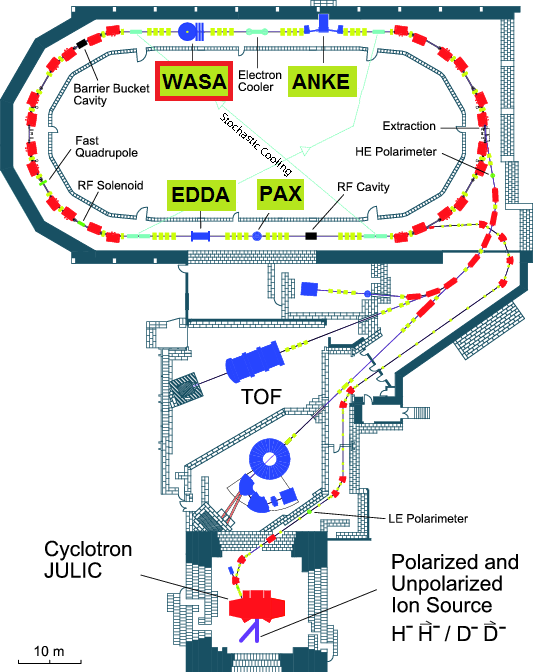}
	\end{center}
\caption{
	The scheme of COSY accelerator facility. The figure is taken from Ref.~\cite{skurzok_phd}.
	$TOF$, $PAX$, $EDDA$, and $ANKE$ names show other experiments implemented at COSY accelerator. 
	The label $WASA$ shows the detection system used for the experiment described in this thesis.
}
\label{Fig_COSY_ring}
\end{figure}

Cooler Synchrotron COSY  accelerator complex \cite{COSY_description1}
in Jülich Research Center consists of
184~m synchrotron ring, isochronous cyclotron JULIC (injector), and
internal and external
experimental targets (Fig.~\ref{Fig_COSY_ring}).
The accelerator works with proton and deuteron beams either polarized or unpolarized
that can be accelerated to the momenta of $0.3$ - $3.7~GeV/c$.
The ring can be filled with $10^{11}$ unpolarized particles.
Such intensity allows to reach the luminosity of $10^{31}~cm^{-2}s^{-1}$ for experiments with cluster target (ANKE, COSY11)~\cite{COSY_description2,COSY_description3}
and $10^{32}~cm^{-2}s^{-1}$ for pellet target applied at WASA~\cite{WASA_description1}.
Beam injection, accumulation, and acceleration with COSY facilities takes few seconds and its lifetime in case of using pellet target like in WASA is about several minutes.
Beams at injection are cooled by means of electron cooling, while stochastic cooling is used for high energies~\cite{COSY_description4}.
More detailed COSY accelerator description can be found in Ref.~\cite{COSY_description5}.
The review of experiments performed at this accelerator can be found in Ref.~\cite{COSY_description6}.

\section{WASA-at-COSY detector setup}

WASA (Wide Angle Shower Aparatus) detection
system~\cite{WASA_description1,WASA_description2,WASA_description3,WASA_description4} 
was installed at COSY since 2007 till 2014. 
Before 2005 it was operating at CELSIUS storage ring at Svedberg Laboratory in Uppsala, Sweden~\cite{WASA_description2}.
WASA detector has almost $4\pi$ geometry and consists of two parts: Central Detector and Forward Detector (Fig.~\ref{WASA-detector-scheme}).

\subsection{Pellet target}
\begin{figure}
	\begin{center}
		\includegraphics[width=450pt]{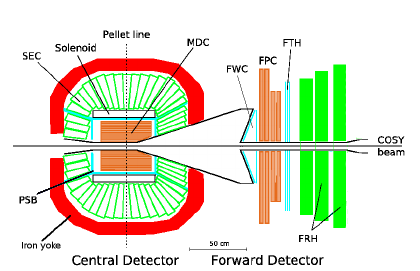}
	\end{center}
	\caption{
		The scheme of WASA detector vertical cross section in the configuration used in experiment described in this thesis.
		The reaction takes place in the center of the detector in the pellet line and COSY beam axis cross point.
		The Central Detector is designed for registering neutral and charged meson decay products.
		The scattered projectiles and recoil nuclei are registered in Forward Detector.
		The abbreviations used as detectors names are explained in the text (Sections~\ref{pellet_target}, \ref{central_detector} and \ref{forward_detector}).
	}
	\label{WASA-detector-scheme}
\end{figure}

The internal pellet type target~\cite{WASA_pellett_target} is installed in the Central Detector and it's position is marked by a vertical line in Fig.~\ref{WASA-detector-scheme}.
\label{pellet_target}
The target system provides frozen gas pellets (in current experiment it was deuterium) into the interaction point.
The production of pellets starts in the pellet generator that forms them of high purity liquid gas using a vibrating nozzle.
Vibration frequency of $70~kHz$ allows to produce pellets with the average diameter of $\approx 35~\mu m$.
After production, the pellets get into $7~cm$ vacuum-injection capillary where they are accelerated up to $60-80~m/s$.
The accelerated pellets are colimated and provided into the interaction region.
The average rate of pellets passing the interaction point is few thousands per second.

\subsection{Central Detector}
The Central Detector built around the interaction point
is designed mainly for measuring photons and charged particles originating from mesons decays.
\label{central_detector}
It consists of several sub-detectors playing different roles in particles registration and identification.

The part closest to the interaction point is the Mini Drift Chamber (\textbf{MDC}).
It consists of 1738 straw tubes arranged in 17 layers and covers the angular range from $24^o$ to $159^o$.
The straw diameter is $4~mm$ for the first 5 inner layers.
Next 6 middle layers consist of straw tubes with $6~mm$ diameter.
The 6 outer layers are formed by $8~mm$ diameter straw tubes.
The straws are made of $25~\mu m$ thin aluminized mylar foil 
and are filled with argon-ethane $50\%-50\%$ gas mixture.
Inside the straws, $20~\mu m$ diameter gold wire is used as an anode.
Nine inner layers are parallel with respect to the beam axis 
while the next layers are situated with $6^o-9^o$ skew angles.
The Mini Drift Chamber main purpose is particle momenta directions and reaction vertex position determination.

Plastic Scintillator Barrel (\textbf{PSB}) surrounds the Mini Drift Chamber and is used to identify charged particles.
It consists of cylindrical part (48 scintillator bars) and two endcaps (48 "cake-piece" shaped scintillators each one) covering almost full wide angular range.
This sub-detector can also be used for $\Delta E-E$ particle identification method together with SEC (Fig.~\ref{WASA-detector-scheme}) or $\Delta E-p$ method together with MDC.

The Superconducting \textbf{Solenoid}~\cite{WASA_solenoid} surrounds the Central Detector parts described above and provides the magnetic field used for charged particle identification.
This magnetic field is taken into account in momentum reconstruction based on information from MDC.
The Solenoid is cooled with liquid helium and produces the magnetic fields up to $1.3~T$.

The Scintillation Electromagnetic Calorimeter (\textbf{SEC}) is situated between Superconducting Solenoid and the iron yoke covering the whole Central Detector.
It is composed of 1012 sodium-doped CsI scintillating crystals.
The angular range covered by this sub-detector is from $20^o$ to $169^o$.
The crystals have truncated pyramid shape and are organized in 24 layers.
The energy resolution is about $3\%$ for stopped charged particles, and about $8\%$ for $0.1~GeV$ photons.
More detailed Calorimeter description is given in Ref.~\cite{WASA_calorimeter}.

\subsection{Forward Detector}

\label{forward_detector}
The Forward Detector covers angular range from $3^o$ to $18^o$ and is designed mainly for charged particles registration.

The first part of the Forward Detector along the beam direction is Forward Window Counter (\textbf{FWC}).
It contains two layers (FWC1, FWC2) of $5~mm$ thick plastic scintillators connected to the photomultipliers via lightguides.
The layers are mounted on paraboloidal stainless steel vacuum window.
The layers are shifted with respect to each other by a half of an element.
The FWC is used for the first level of the trigger logic and allows to identify charged particles originating from the reaction point and to reduce the background of particles scattered downstream the beam pipe.

The Forward Proportional Chamber (\textbf{FPC}) is located directly after the FWC.
This module contains four layers of straw tubes.
Each layer consists of 122 tubes.
The tubes are made of thin mylar foil and have $8~mm$ diameter.
They are filled with argon-ethane gas mixture and work as proportional drift detectors.
The layers have orientations respectively $-45^o$, $+45^o$, $0^o$ and $90^o$ with respect to $x$ direction.
These straw tube layers are used for measuring charged particle track angles.
The module provides $0.2^o$ angular resolution.

\label{removed_forward_layers}
The Forward Trigger Hodoscope (\textbf{FTH}) earlier consisted of three thin plastic scintillator layers.
It was used for angular information measuring but two of three layers were removed before current experiment.
Only one layer (FTH1) consisting of 48 radial elements was left.
Thus, in current experiment, it can be used only for charged particle identification using $\Delta E-E$ method together with FRH module.

The Forward Range Hodoscope (\textbf{FRH}) contained three layers of $11~cm$ thick plastic scintillators during the current experiment.
Earlier, it contained more layers but they were removed.
This thick scintillators are used mainly for measuring charged particles energy.

\subsection{Data Acquisition system (DAQ)}

\label{DAQ_section}
\begin{figure}[h!]
	\begin{center}
		\includegraphics[width=290pt]{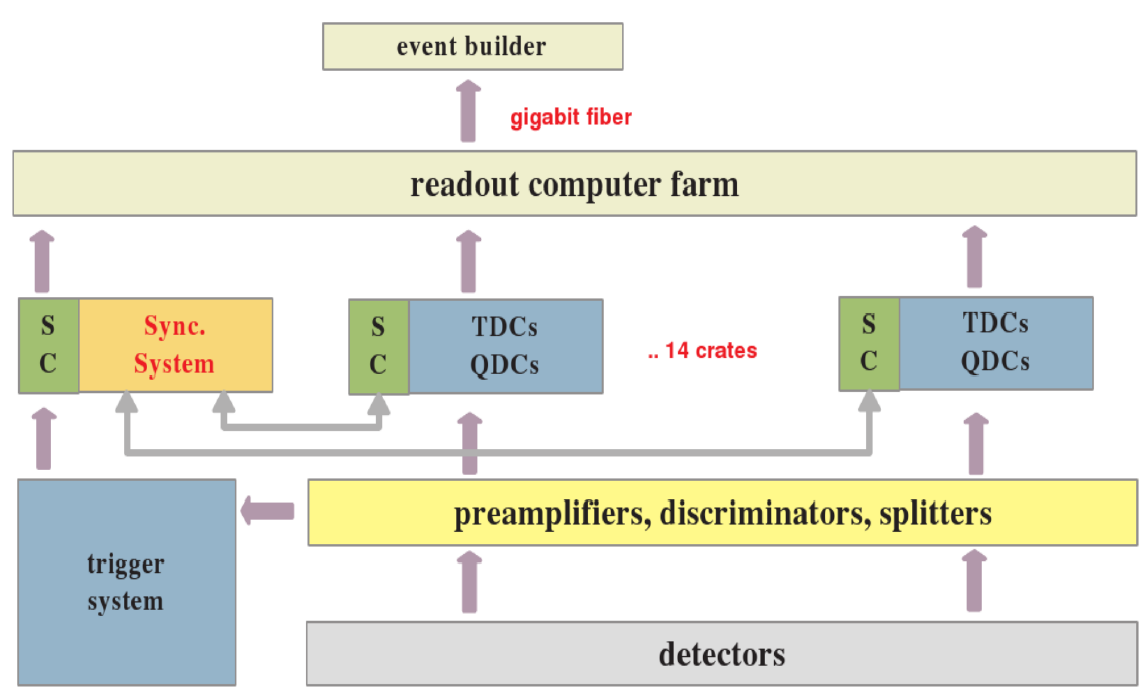}
	\end{center}
	\caption{
		The scheme of data acquisition system for WASA-at-COSY. 
		The figure is taken from Ref.~\cite{DAQ_link2_withscheme}.
	}
	\label{DAQ_scheme}
\end{figure}

The DAQ system stores the digitized signals from the detector modules to make them available for the analysis (Sec.~\ref{data_analysis_tools}).
In current experiment, the third generation of COSY DAQ system was used (Fig.~\ref{DAQ_scheme}).
It is optimized for experiments with high luminosities~\cite{DAQ_link1} and
allows to reach the event rate of $10^4~s^{-1}$ with at least $80\%$ lifetime~\cite{DAQ_link3}.

The analogue signals from the detectors are processed by FPGA based front-end electronics and then digitized by QDC (Charge-to-Digital Converter) and TDC (Time-to-Digital Converter) modules.

The digitized signals are marked with timestamps and put in FIFO queue ("First In First Out" queue).
Trigger system checks the conditions that are set up for a particular experiment and drives the process of events forming.
The synchronization system, called by trigger system, calculates the event number, and sends it together with the time stamp to all QDC and TDC modules.
Signals with matching timestamps are marked with this event number and pass to computer readout and to the event builder.
Finally, the events are stored to the discs.
More detailed DAQ system description is given in
Refs.~\cite{DAQ_link1,DAQ_link2_withscheme,DAQ_link3}.

\section{Data analysis tools}

\label{data_analysis_tools}
Monte Carlo simulations for 
$pd\rightarrow^3$$He\eta$ (Sec.~\ref{he3eta_simulation}), 
$pd\rightarrow ppn_{spec}$ (Sec.~\ref{ppn_simulation}), $pd\rightarrow(^3$$He\eta)_{bound}\rightarrow^3$$He2\gamma$,
and $pd\rightarrow(^3$$He\eta)_{bound}\rightarrow^3$$He6\gamma$
(Sec.~\ref{he3eta_bound_simulation})
reactions kinematics were done by software developed by the author of this thesis implementing the proper theoretical models.
Background reactions kinematics was simulated by PLUTO software.
The WASA detector response was simulated by WASA Monte Carlo (WMC) software that is based on GEANT software~\cite{GEANT_link}.
The analysis of both data and simulation results was performed by software developed by the author of this thesis
 based on RootSorter framework~\cite{RootSorter_link} 
 that is using data analysis software package ROOT~\cite{Root_link} developed at CERN.
Other calculations, fits, and preparing the histograms shown in this thesis were performed by the software developed by the author of this thesis.

\section{Current experiment conditions}

\subsection{Ramped beam}
\label{ramped_beam}
The current experiment was carried out using ramped beam technique.
The beam momentum was changed continuously in the range between $1.426$ and $1.635~GeV/c$
that corresponds to $pd\rightarrow^3$$He\eta$ reaction excess energy from $-70~MeV$ to $+30~MeV$.
The beam momentum is known as a function on time-in-cycle that is stored by DAQ system in each event header.
Though relative energy changes for the beam are known precisely, the absolute values depend on magnets settings.
Thus, the beam momentum contains unknown constant offset of the order of few $MeV/c$.
This offset was determined from kinematic conditions as it is described in Sec.~\ref{beam_momentum_offset_constant}.

\subsection{Trigger settings}
Trigger system is used to determine which events are written to discs by the DAQ system (Sec.~\ref{DAQ_section}).
\label{trigger_table}
The conditions provided by trigger system are simple enough to be checked in real time but allow to roughly reduce background events.

\begin{table}[h!]
	\begin{center}
		\begin{tabular}{|p{15pt}|p{150pt}|p{70pt}|}
			\hline
			& Condition & Scaling factor\\
			\hline
			0 & $fwca1$ & 20000\\
			1 & $fwcb1$& 20000\\
			3 & $fwHea1$& 200\\
			4 & $fwHeb1$& 200\\
			7 & $seln4$& 10\\
			\textbf{10} &$\textbf{fwHea1|fwHeb1|fHedwr1}$ & \textbf{1}\\
			13 & $fHedwr1|seln4$& 10\\
			\textbf{17} & $\textbf{psf1|psc1}$&\textbf{4000} \\
			21 & $frha1|psc1$ & 1\\
			26 & $Vfwc1|seln4s$&10 \\
			29 &$fhdwr2|seln2$ & 10\\
			30 &$fhdwr2|selc2$ & 10\\
			\hline
		\end{tabular}
	\end{center}
	\caption{
		Trigger settings used in current experiment.
		Active triggers are shown with bold font.
		The abbreviations used for trigger conditions are shown in Table~\ref{trigger_table_abbr}.
		Letter V before the abbreviation means veto condition.
		Triggers used in this thesis are shown with bold font.
	}
	\label{trigger_table_table}
\end{table}

\begin{table}[h!]
	\begin{center}
	\begin{small}
		\begin{tabular}{|p{70pt}|p{350pt}|}
			\hline
Abbreviation& Meaning \\
			\hline
$fwcaN$     & at least N modules above low threshold in FWC1\\
$fwcbN$     & at least N modules above low threshold in FWC2\\
$fwHeaN$  & at least N modules above high threshold in FWC1\\
$fwHebN$  & at least N modules above high threshold in FWC2\\
$selnN$      & at least N neutral groups in SEC (low threshold)\\
$selcN$      & at least N charged groups in SEC (low threshold)\\
$fhdwrN$   & at least N tracks matching between FWC, FTH, and FRH. Low threshold of FWC used\\
$fHedwrN$ & same as $fhdwrN$ but high threshold for FWC\\
$frhaN$      & at least N modules above threshold in FRH1\\
$psfN$        & at least N modules above threshold in PSB forward endcap \\
$pscN$       & at least N modules above threshold in PSB cylindrical part\\
			\hline
		\end{tabular}
	\end{small}
	\end{center}
	\caption{
		The abbreviations used for trigger conditions shown in the Table~\ref{trigger_table_table}.
	}
	\label{trigger_table_abbr}
\end{table}

Some trigger conditions occur with the event rate higher than the one that can be reached by DAQ system.
For such triggers, scaling factor is provided.
Scaling means that not every event matching the trigger conditions is written to disc but only 
those events for which the event number modulo scaling factor  is equal to zero.

The trigger conditions and corresponding scaling factors used in current experiment are shown in Table.~\ref{trigger_table_table} and the abbreviations used for the conditions are explained in Table.~\ref{trigger_table_abbr}.
 
 \subsection{Data preselection criteria}
 \label{preselection_conditions}
 The data stored by DAQ system using the trigger conditions still contains huge amount of background events that can be reduced in order to decrease time of main analysis.
This reduction is performed by preselection procedure.
A set of conditions that take very few time to check is applied to the whole amount of raw experimental data obtained in the measurements and only events fulfilling these conditions are stored into preselected data.

For $pd\rightarrow^3$$He\eta$, $pd\rightarrow^3$$He2\gamma$, and $pd\rightarrow^3$$He6\gamma$ reactions analysis (Section~\ref{luminosity_he3eta} and Chapter~\ref{analysis_chapter}), the events corresponding to trigger number 10 were taken into account and the condition of at least one charged particle track in Forward Detector that contains signal in FPC and corresponds to a particle stopping in FRH1 module is applied.
Particles stopping in FRH1 are required according to the simulation results showing that $^3$$He$ ions from these reactions stop in FRH1.
The "stopping" condition means that the signal in FWC1, FWC2, FTH1, and FRH1 is above the threshold and for FRH2 it is below the threshold. 
The threshold values used in this analysis are given in Table.~\ref{forward_layers_thresholds}.

\begin{table}[h!]
	\begin{center}
		\begin{tabular}{|p{50pt}|p{50pt}|}
			\hline
			Module & Threshold \\
			\hline
			FWC1 & $2.0~MeV$\\
			FWC2 & $2.0~MeV$\\
			FTH1 & $1.5~MeV$\\
			FRH1 & $1.0~MeV$\\
			FRH2 & $1.0~MeV$\\
			\hline
		\end{tabular}
	\end{center}
	\caption{Deposited energy thresholds applied for different Forward Detector modules.}
	\label{forward_layers_thresholds}
\end{table}

The proton-proton quasielastic scattering analysis performed for luminosity determination  	(Sec.~\ref{luminosity_ppn}) required another preselection conditions.
The events corresponding to trigger number 17 were selected and the condition that at least two charged particle tracks with the deposited energy above the threshold of $30~MeV$ registered in the Central Detector was applied.
\chapter{Luminosity determination}

\section{The method of luminosity determination}

\label{luminosity_chapter}
One of the most important issues in the experimental data analysis is normalization.
Luminosity is such normalization constant.
It can be obtained e.~g. by using a reaction that has known cross section and then it is used for measuring other reactions cross sections.
The luminosity is defined by the following formula:
\begin{equation}
	L 
	= \frac{dN_{true}}{dt} \frac{1}{\sigma}
	= \frac{dN_{registered}}{dt} \frac{1}{\epsilon~\sigma},
\end{equation}
where $\sigma$ is known reaction total cross section,
$\frac{dN_{true}}{dt}$ is the rate of this reaction events,
$N_{registered}$ is the number of registered  events.
It needs to be divided by the efficiency $\epsilon$ to obtain the true events count.

The efficiency is obtained from Monte Carlo simulation for each particular reaction.
First, the reaction products kinematics is simulated.
Then, WASA Monte Carlo software is used to simulate the detector response (Sec.~\ref{data_analysis_tools}). 
These simulated data are analyzed by the same algorithm that is used for the experimental data analysis
(after the preselection that is described in Sec.~\ref{preselection_conditions}).

In this experiment, the value of integrated luminosity is used
\begin{equation}
	\int L\,dt
	= \frac{N_{true}}{\sigma}  
	= \frac{N_{registered}}{\epsilon~\sigma} 
	= \frac{N_{data}~S_{trigger}~S_{MC}}{N_{MC}~\sigma},
	\label{luminosity_description}
\end{equation}
where 
$N_{registered}$ is the count of events registered by the detector,
$N_{data}$ is the events count obtained in raw data analysis, and
$N_{MC}$ is the events count obtained with the same algorithm analyzing Monte Carlo simulation results.
$S_{trigger}$ is the scaling factor for the trigger used in the experiment (Sec.~\ref{trigger_table}).
$S_{MC}$ is the total generated events count.
$N_{MC}$ divided by this value is equal to the efficiency.

\label{Q-bins}
The range of beam momentum used in current experiment corresponds to the following range of the excess energy:
\begin{equation}
	Q_{^3He\eta}~\in~[-70;+30]~MeV.
\end{equation}
This range is split into 40 bins with the width of $2.5~MeV$.
The integrated luminosity is obtained for each $Q$-bin separately.

\section{Luminosity determination based on $p+d\rightarrow^3$$He+\eta$ reaction}

\label{luminosity_he3eta}
The $pd\rightarrow^3$$He\eta$ reaction is visible only above the $\eta$ creation threshold however this channel analysis provides a lot of data useful for further analysis procedures.
The most important feature is obtaining the beam momentum correction constant.
Also for this reaction, the $^3$$He$ tracks and energy reconstruction algorithm was tested before it was used for analysis of the channels that may show the existing bound state.
In the beam momentum range where the $pd\rightarrow^3$$He\eta$ reaction is visible, it has the cross section about $300~nb$ \cite{3Heeta_cross_section,3Heeta_cross_section_older1}.

\subsection{Monte Carlo simulation of $pd\rightarrow^3$$He\eta$ reaction}

\label{he3eta_simulation}
Monte Carlo simulation was performed with the aim to obtain the detection and reconstruction efficiency.
Beam momentum values were generated uniformly distributed in the part of the range used in current experiment (sec.~\ref{ramped_beam})  corresponding to  $Q_{^3He\eta}>0$.
Assuming that target deuteron is at rest we have enough variables to obtain the total invariant mass.

The final $^3$$He$ nucleus and $\eta$ meson momenta are obtained in the center of mass frame.
They have opposite directions and equal magnitudes that are determined by the total invariant mass value.
The most complicated issue is the angular distribution.

\begin{figure}[h!]
	\begin{center}
	\includegraphics[width=145pt]{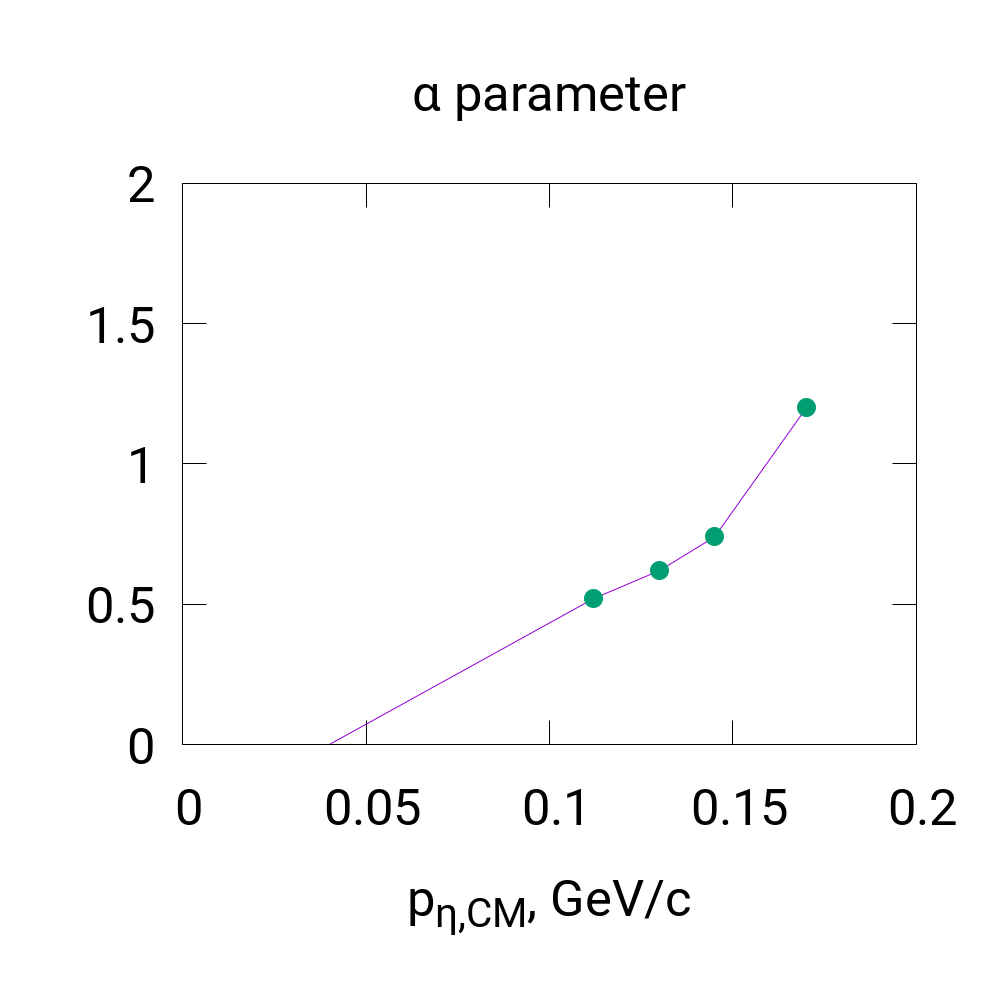}
	\includegraphics[width=145pt]{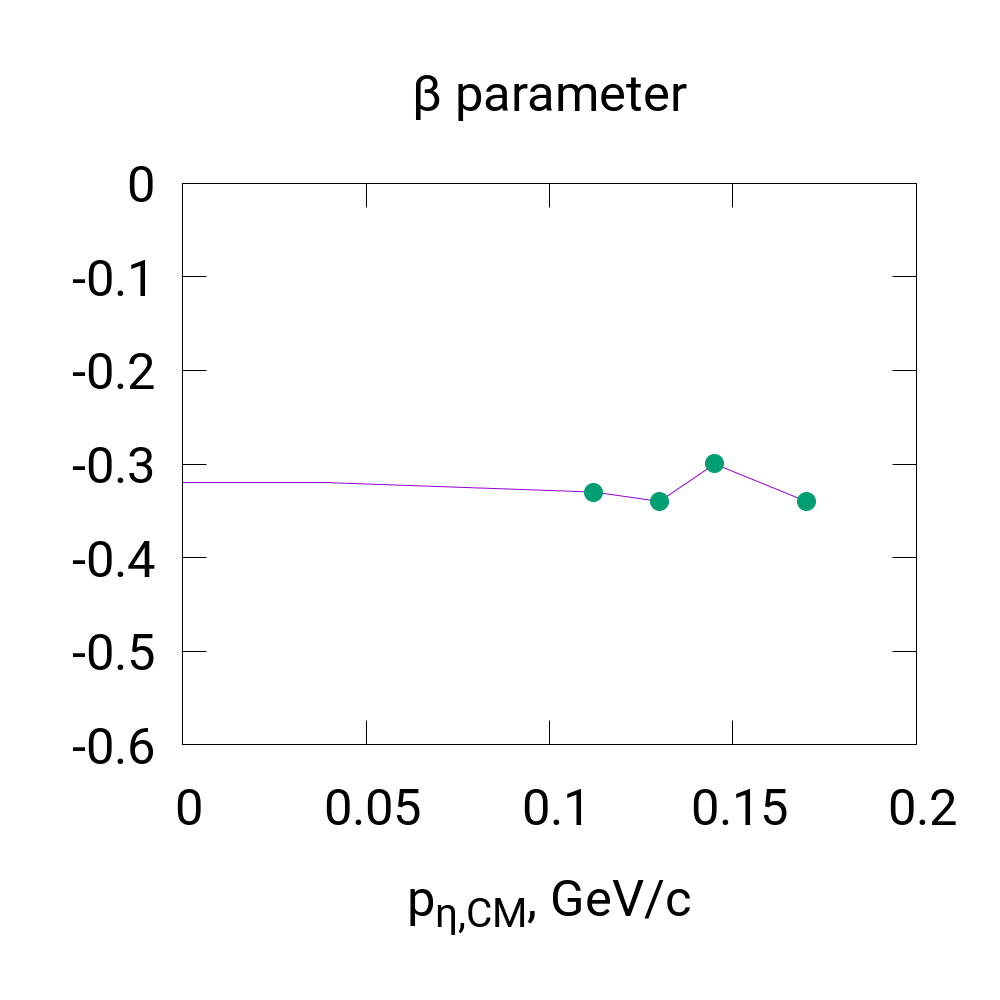}
	\includegraphics[width=145pt]{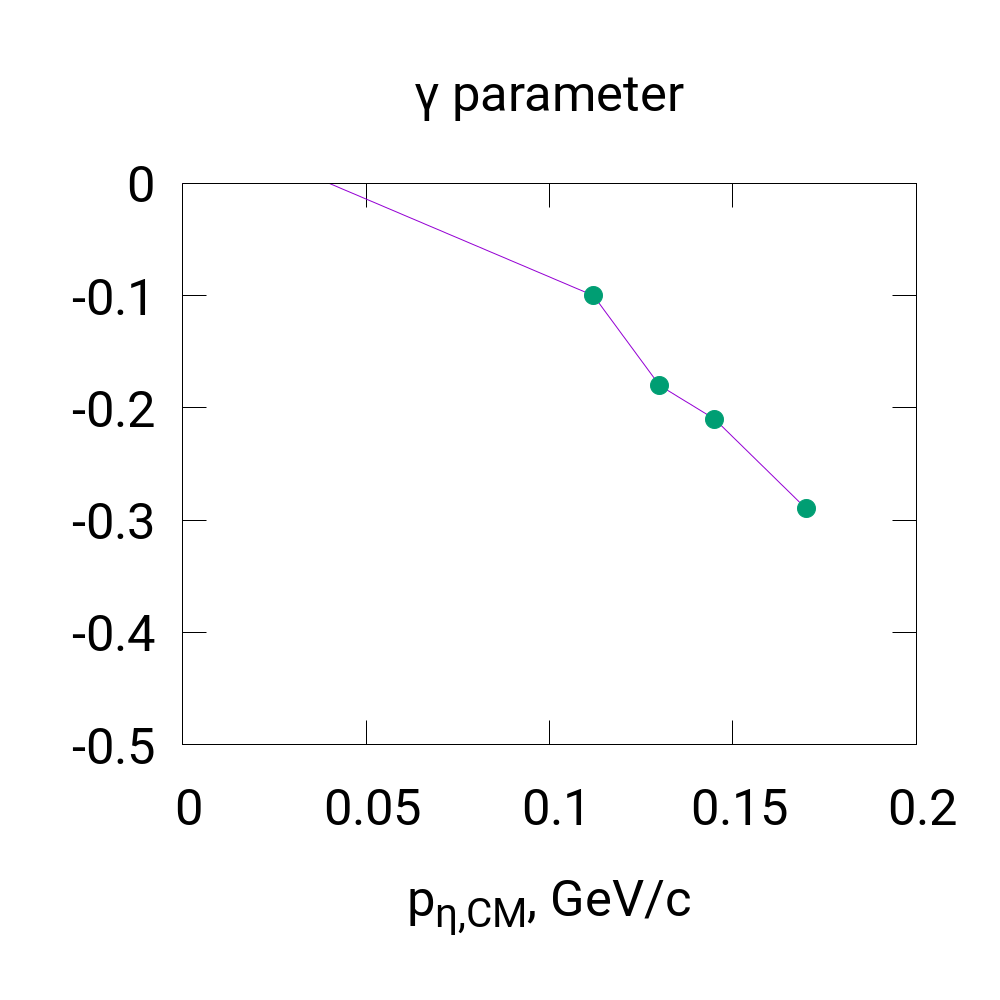}
	\end{center}
	\caption{
		The values of $\alpha$, $\beta$, and $\gamma$ parameters for $\theta_{\eta,CM}$ distribution (Eq.~\ref{angular_distribution_formula}) assumed in current work for 
		$pd\rightarrow^3$$He\eta$ reaction.
		Points show data from Ref.~\cite{3Heeta_cross_section}.
		Line shows the interpolation that was used for Monte Carlo simulation.		
	}
	\label{he3eta_angulardistr_parameters}
\end{figure}

Taking into account the most recent experimental data \cite{3Heeta_cross_section} for this reaction, the following approximation for the angular distribution in the center of mass system was applied:
\begin{equation}
\label{angular_distribution_formula}
f(cos(\theta_{\eta,CM})) = N_0(1+\alpha cos(\theta_{\eta,CM}) + \beta cos(\theta_{\eta,CM})^2 + \gamma cos(\theta_{\eta,CM})^3),
\end{equation}
where the $\alpha$, $\beta$ and $\gamma$ parameters depend on the $\eta$ meson momentum in the center of mass frame that can be obtained from beam momentum value. 
Ref.~\cite{3Heeta_cross_section} contains cross sections and angular distributions for the beam momentum values corresponding to $Q_{^3He\eta}$ values of $13.6$~MeV and higher. 
The earlier experiments \cite{3Heeta_cross_section_older1,3Heeta_cross_section_older2,3Heeta_cross_section_older3} allow to assume that the angular distribution becomes symmetric with $Q_{^3He\eta}$ going down to zero. 
Since the papers~\cite{3Heeta_cross_section_older1,3Heeta_cross_section_older2,3Heeta_cross_section_older3} contain only $\alpha$ parameter values, $\beta$ and $\gamma$ parameters extrapolation into $Q_{^3He\eta}$ region below $13.6$~MeV is needed (Fig.~\ref{he3eta_angulardistr_parameters}). 
The dependence of $\gamma$ that corresponds to "asymmetric" part of the distribution together with $\alpha$ parameter is extrapolated linearly as a function of $p_{\eta,CM}$.
The $\beta$ parameter, according to the data from Ref.~\cite{3Heeta_cross_section}, does not change so drastically with the beam momentum.
Thus,  it was extrapolated as remaining at the same level with $p_{\eta,CM}$ going down to zero.

$^3He$ and $\eta$ meson's momenta vectors are transferred from the center of mass frame into the laboratory frame.
$^3He$ ion and $\eta$ meson decay products are registered. 
As far as this part of analysis does not take $\eta$ decay products into account, the decay simulation is described later in sec.~\ref{eta_decay_simulation}.

\subsection{$^3He$ tracks reconstruction}

\label{3He_tracks_reconstruction}
In order to obtain the tracks corresponding to $^3He$ ions in Forward Detector the trigger number $10$ was used (sec.~\ref{trigger_table}).
This trigger implements the set of conditions providing selection of events containing at least one charged particle track in Forward Detector.
The routine for finding these tracks is implemented in {RootSorter} framework.

\label{forward_identification_of_he3}
The first condition applied is that there is a signal in FPC-layer.
The angle reconstruction algorithm implemented in RootSorter is applicable only in that case.
When the signal in FPC layer is present, the resolution of angular reconstruction is about $\approx0.2^o$ \cite{krzemien_phd}.

For being identified as $^3He$ the tracks have to fulfill several conditions.
Monte Carlo simulation has shown that $^3He$ ions that are products of $pd\rightarrow^3$$He\eta$ reaction mostly stop in FRH1 layer.
Thus the condition selecting tracks of charged particles stopped in this layer was provided. 

\begin{figure}[h!]
	\begin{center}
	\includegraphics[width=225pt]{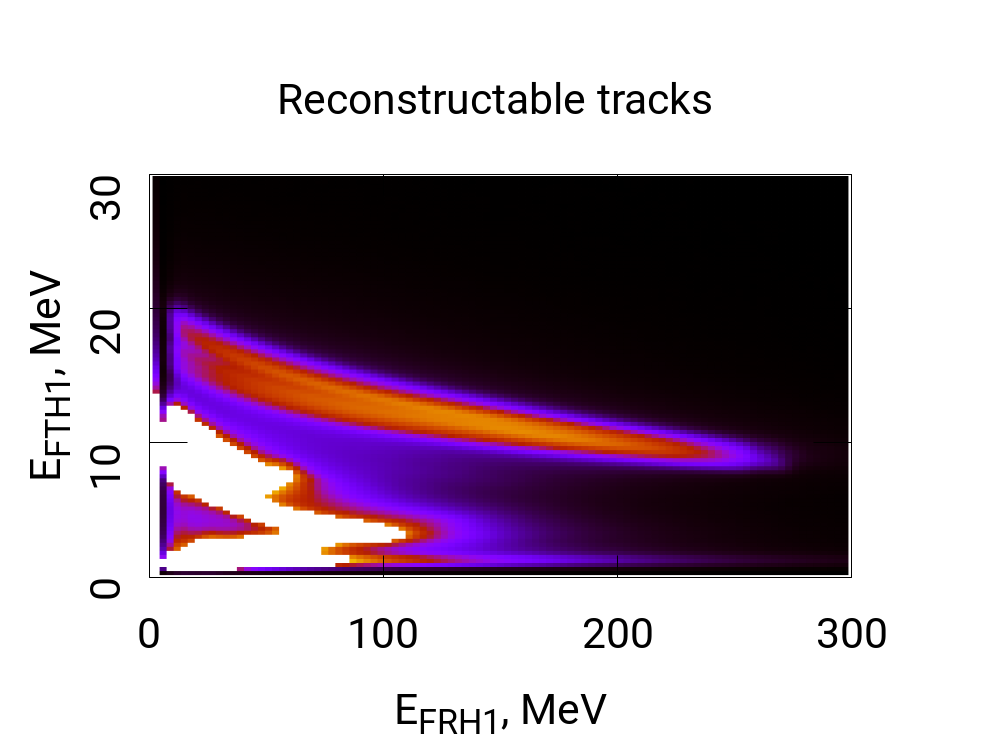}
	\includegraphics[width=225pt]{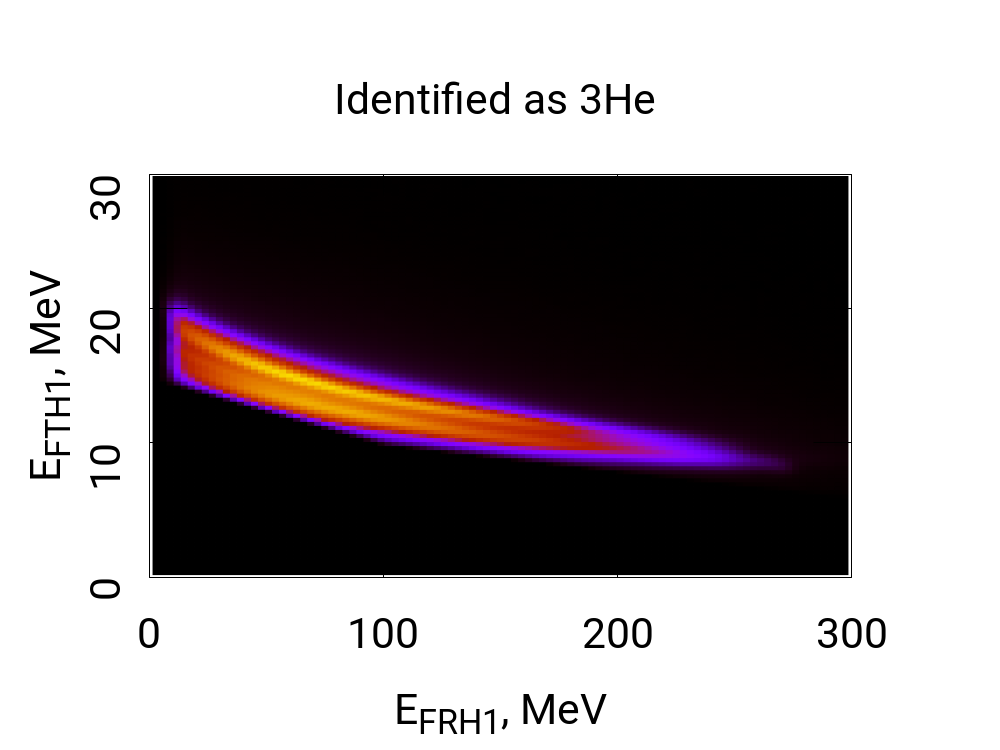}
	\end{center}
	\caption{
		2-D histograms of energies deposited in {FTH1} (vertical axis) and {FRH1} (horizontal axis)
		for all events with signal in {FPC} (left panel) and events that were identified as $^3He$ (right panel).
	}
	\label{he3_cut_data}
\end{figure}

For separating $^3He$ ions from lighter particles, a condition on the energy deposited in {FTH1} and {FRH1} layers was applied (Fig.~\ref{he3_cut_data}).
\begin{equation}
	E_{FTH1} > h+max( 0.05(100 - E_{FRH1}), -0.02(E_{FRH1}-100) ),
	\label{he3_cut_height}
\end{equation}
where all numeric constants added to energies are given in $MeV$ units.
{FTH1} and {FRH1} layers were selected as two last layers where $^3He$ ions deposit their energy.
The $h$ parameter equals $10~MeV$ and is used for systematic error estimation (Table~\ref{systematics_he3eta_lum_parameters}).

\label{forward_he3_reconstruction}
The reconstruction of kinetic energy required new algorithm because two layers were removed from Forward Detector before current experiment. 

The energy reconstruction was performed based on Monte Carlo simulations of  
$pd\rightarrow^3$$He\eta$, $pd\rightarrow^3$$He 2\pi^0$, and $pd\rightarrow^3$$He 3\pi^0$ reactions.
Using the simulation results, the kinetic energy was fitted by the following function: 
\begin{equation}
E_{kin} = f_0(\theta) + f_1(\theta)*E_{FRH1},
\end{equation}
where each function
\begin{equation}
f_i(\theta) = a_i + b_i \theta + c_i \theta^2
\end{equation}
is a polynomial with fitted $a$, $b$, and $c$ coefficients.
To avoid using of each single event as a unique point for fitting algorithm the $\theta$, $E_{FRH1}$, and $E_{kin}$ ranges were split into bins (Table.~\ref{energy_reconstruction_bins}).
After the binning, each cell was used as a point to fit and the weight of each point was the number of events in the cell.

\begin{table}[h!]
	\begin{center}
	\begin{tabular}{|p{70pt}|p{100pt}|p{100pt}|}
		\hline
		Variable & Range & Bin width \\
		\hline
		$\theta$ & $0.100~-~0.130~rad$ & $0.002~rad$\\
		$E_{FRH1}$ & $0~-~300~MeV$ & $5~MeV$\\
		$E_{true}$ & $200~-~500~MeV$ & $5~MeV$\\
		\hline
	\end{tabular}
	\end{center}
	\caption{Splitting of $\theta-E_{FRH1}-E_{true}$ space into bins for performing energy reconstruction of forward $^3He$ tracks.}
	\label{energy_reconstruction_bins}
\end{table}

\label{3he_theta_cut}
Also on this stage of $^3$$He$ tracks reconstruction, the cut on polar $\theta$ angle was provided.
The cut position is set to the value of
\begin{equation}
\theta_{min} = 4.5^o.
\label{he3_theta_cut_pos}
\end{equation}

This cut was applied for all analysis procedures that require $^3$$He$ track reconstruction.
It allows to separate $pd\rightarrow^3$$He\eta$ reaction from bound state decay processes due to different $^3$$He$ emission angular distributions.

 \subsection{The beam momentum calibration correction}

\label{beam_momentum_offset_constant}
Current experiment was performed using a ramped beam technique with beam momentum changing slowly and constantly during every accelerator cycle.
The calibration provided by COSY accelerator team
precisely describes the momentum changes but may contain few $MeV/c$ order of magnitude unknown constant offset $\Delta P$ (Sec.~\ref{ramped_beam}):

\begin{equation}
P_{final} = P_{rec}(t_{cycle}) + \Delta P.
\end{equation}

\begin{figure}[h!]
	\begin{center}
	\includegraphics[width=225pt]{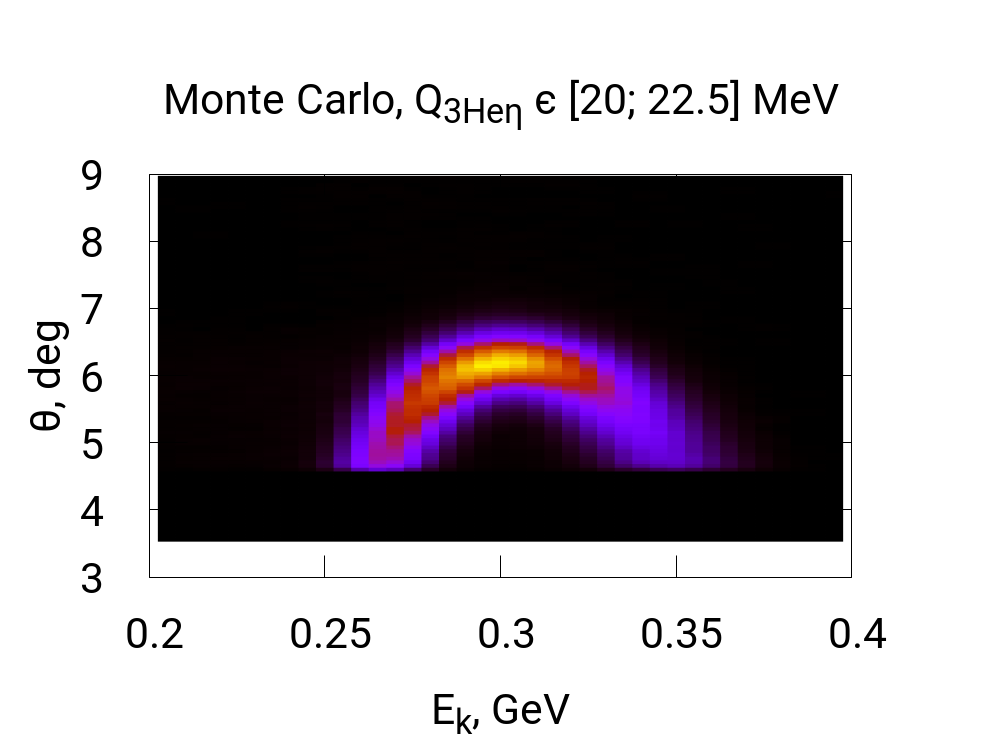}
	\includegraphics[width=225pt]{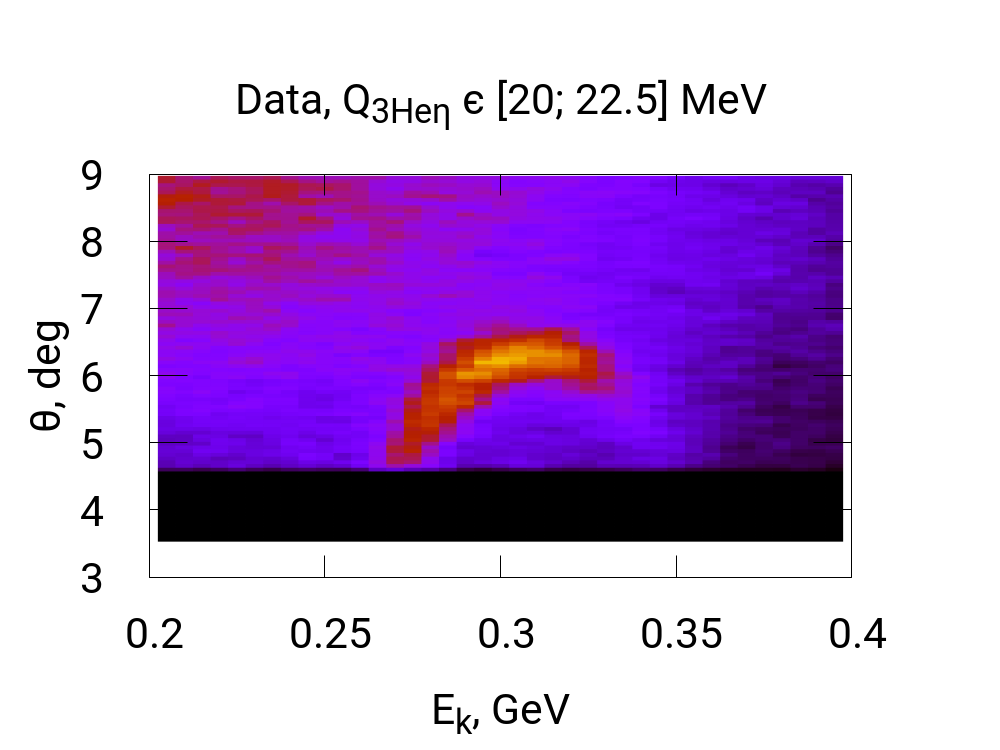}
	\end{center}
	\caption{
		Distribution of reconstructed $\theta$ vs. $E_{kin}$ values for $^3$$He$. Monte Carlo simulation (left panel), and the data (right panel).
		The values are shown for $Q_{~^3He\eta}$ bin $[20.0;22.5]~MeV$.
	}
	\label{he3_reconstruction_kinematic_hist}
\end{figure}

\begin{figure}[h!]
	\begin{center}
		\includegraphics[width=220pt]{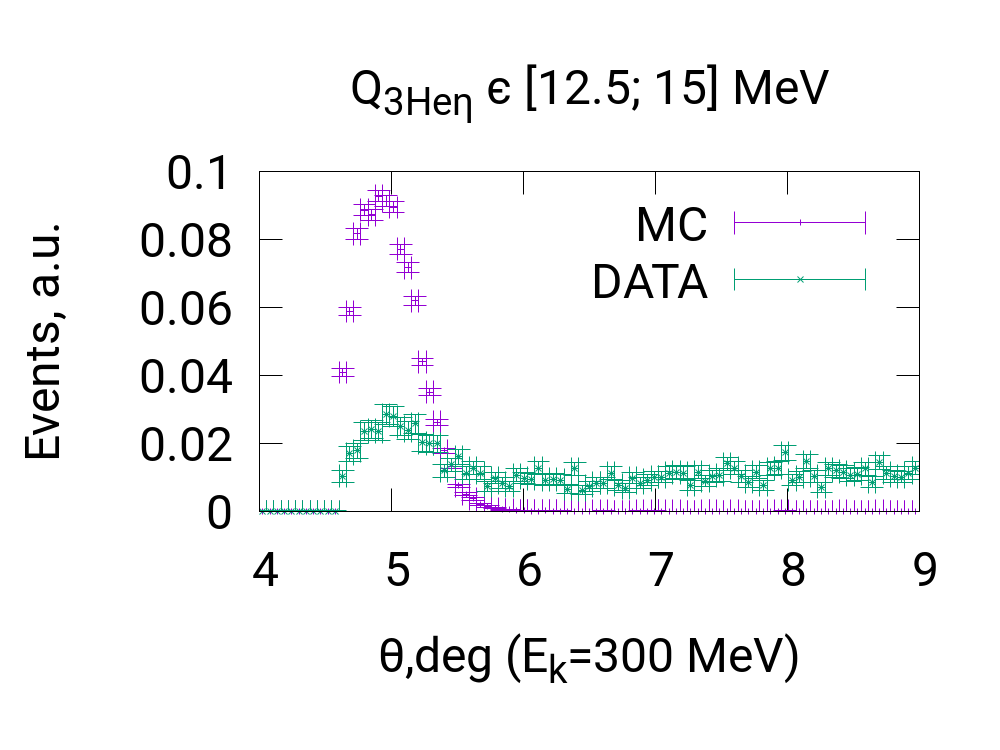}
		\includegraphics[width=220pt]{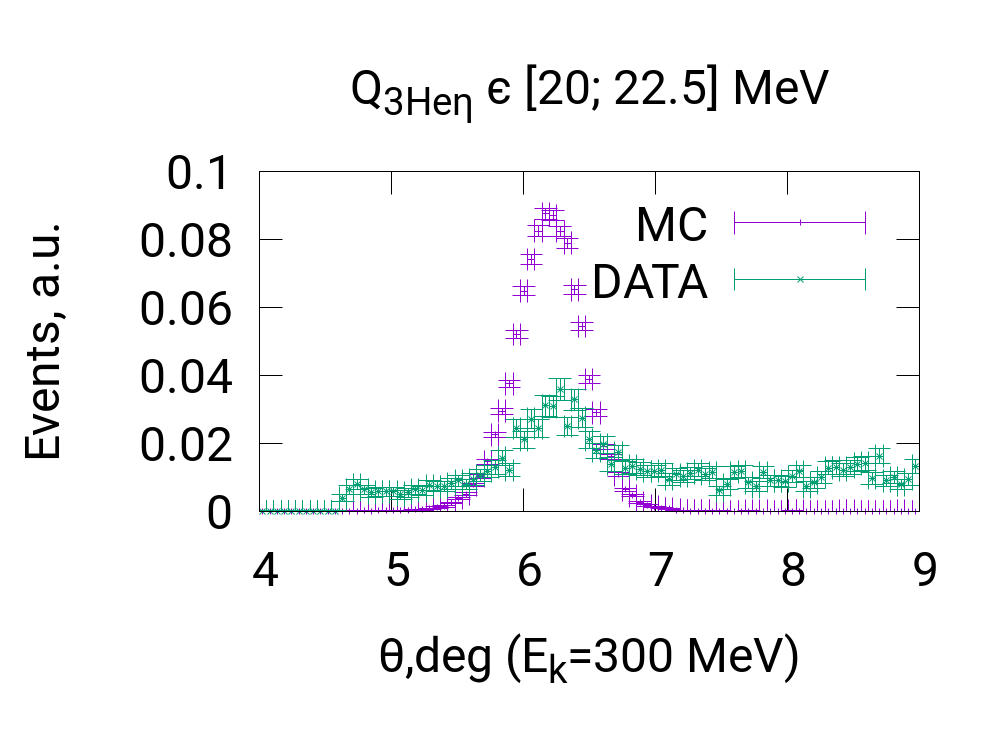}
	\end{center}
	\caption{
		The experimental $\theta$~angle distributions for the kinetic energy bin where the maximum $\theta$~value is observed. Each plot shows events for different $Q$-bins (see \ref{Q-bins}). The magenta curve shows the peak position obtained from simulation. The beam momentum correction (eq. \ref{beam_momentum_correction_constant}) is applied.
		Both curves are divided by corresponding total events counts to fit into the same scale.
	}
	\label{he3_reconstruction_kinematic_hists_theta}
\end{figure}

The distribution of $\theta$ angle and kinetic energy for $^3$$He$ tracks from $pd\rightarrow^3$$He\eta$ reaction (Fig.~\ref{he3_reconstruction_kinematic_hist}) allows to provide the correction.
The maximum $\theta$ angle in the distribution allows to obtain the beam momenta from the reaction kinematics.
The offset was set for the best agreement between Monte Carlo simulation of $pd \rightarrow ^3$$He\eta$ reaction and the corresponding kinematic histogram obtained from the data (Fig.~\ref{he3_reconstruction_kinematic_hist}, \ref{he3_reconstruction_kinematic_hists_theta}).
The offset value was found to be:
\begin{equation}
\Delta P = 4.0~MeV/c.
\label{beam_momentum_correction_constant}
\end{equation}

\subsection{Fitting $^3He$ missing mass distributions}

The $^3He$ missing mass spectra obtained in the experiment contain peak corresponding to $\eta$ meson mass.
For each $Q$-bin separately, the background around $\eta$ mass peak was fit by fourth power polynomial.

\begin{figure}[h!]
	\begin{center}
	\includegraphics[width=220pt]{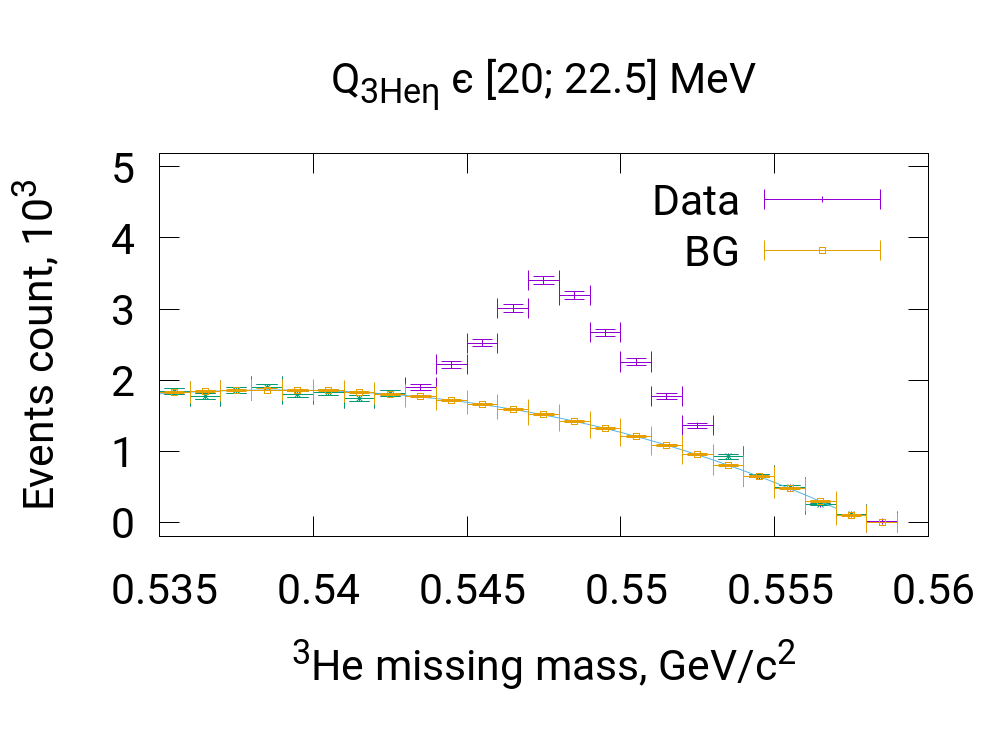}
	\includegraphics[width=220pt]{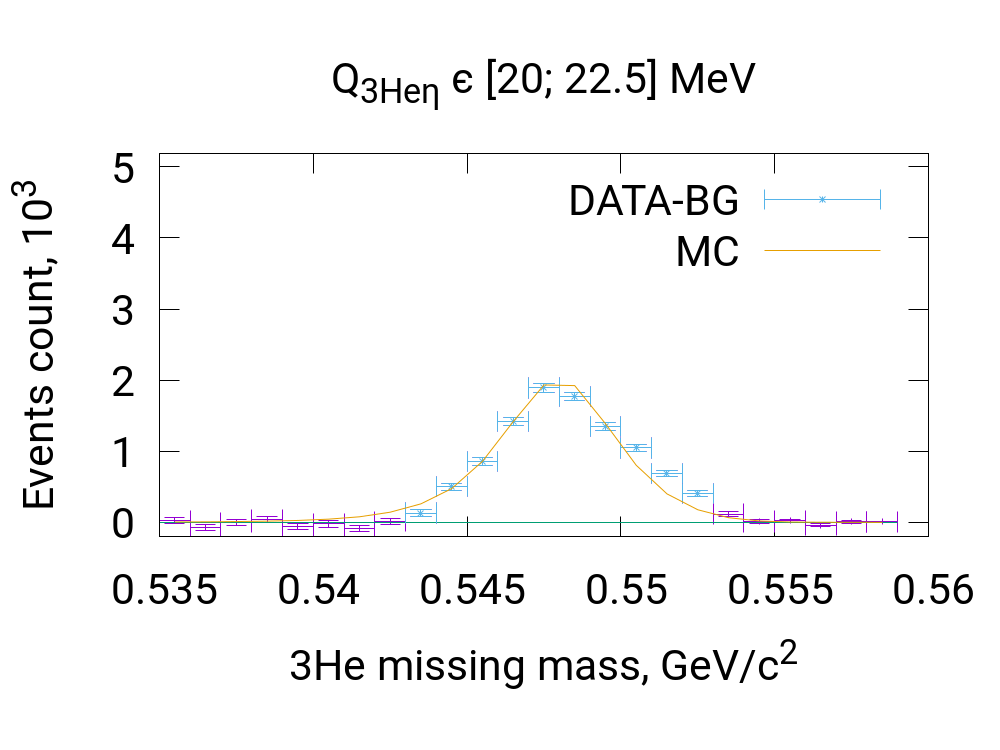}
	\end{center}
	\caption{\small
		Left:
		$^3He$ missing mass spectrum obtained from data for the range of $Q_{^3He\eta}~\in~[20.0;22.5]~MeV$. The part of the spectrum that is considered to be background is shown with green color and is fit with polynomial of fourth power.
		Right:
		The spectrum after background subtraction. The part of the spectrum that was taken into account for $\eta$ peak area calculation is shown in cyan color.
		Monte Carlo simulation for the peak shape is shown with orange line.
	}
	\label{he3_mm_spectra_fit}
\end{figure}

\begin{figure}[h!]
	\begin{center}
		\includegraphics[width=250pt]{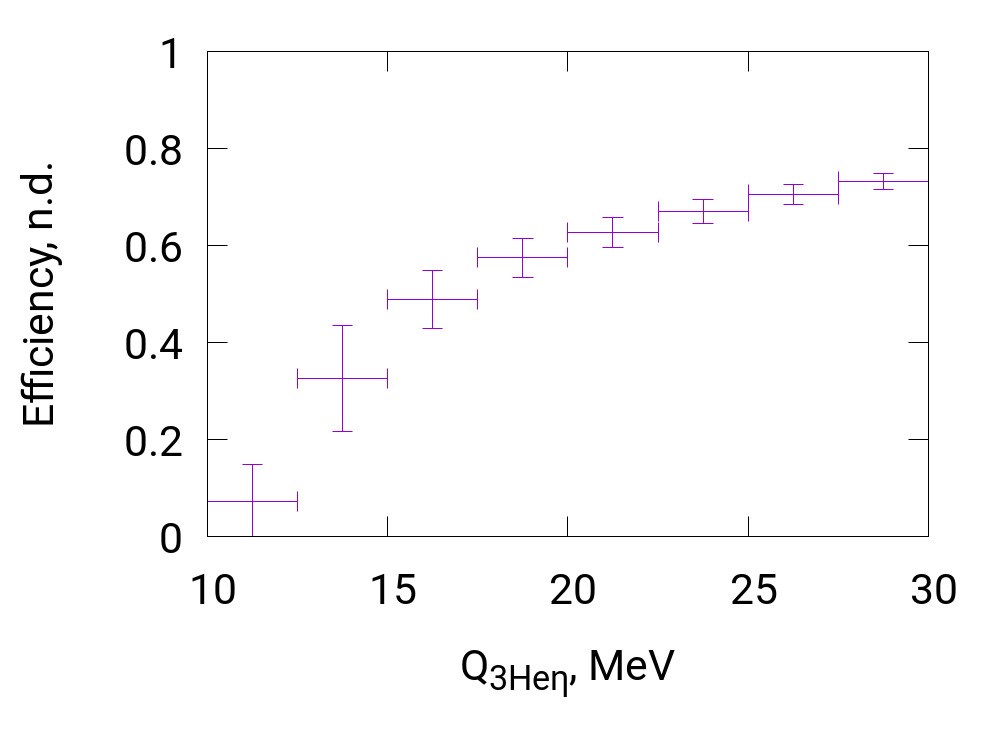}
	\end{center}
	\caption{
		The efficiency for the registration of $pd\rightarrow^3$$He\eta$ reaction.
		The vertical error bars show the systematic uncertainties (Sec.~\ref{luminosity_systematics}).
	}
	\label{he3_efficiency}
\end{figure}

The $pd\rightarrow^3He\eta$ events count was obtained as the $\eta$-peak area after background subtraction (Fig.~\ref{he3_mm_spectra_fit}).
The area of $\eta$-peak obtained from Monte Carlo simulation divided by total generated events count equals the efficiency for this reaction (Fig.~\ref{he3_efficiency}).

\begin{figure}[h!]
	\begin{center}
	\includegraphics[width=220pt]{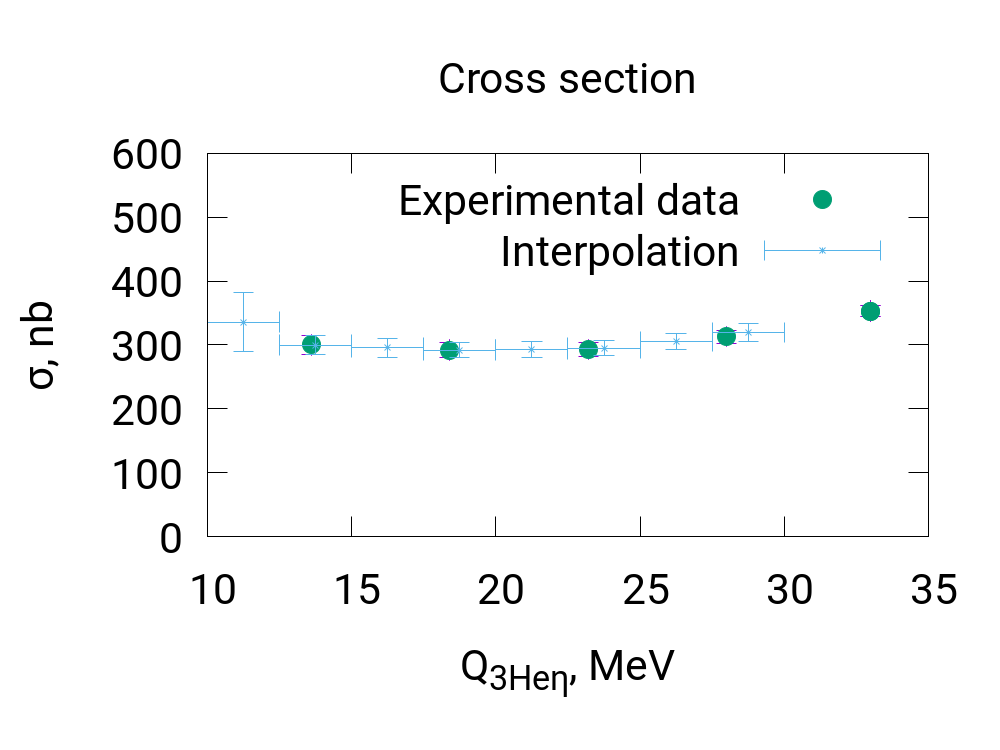}
	\includegraphics[width=220pt]{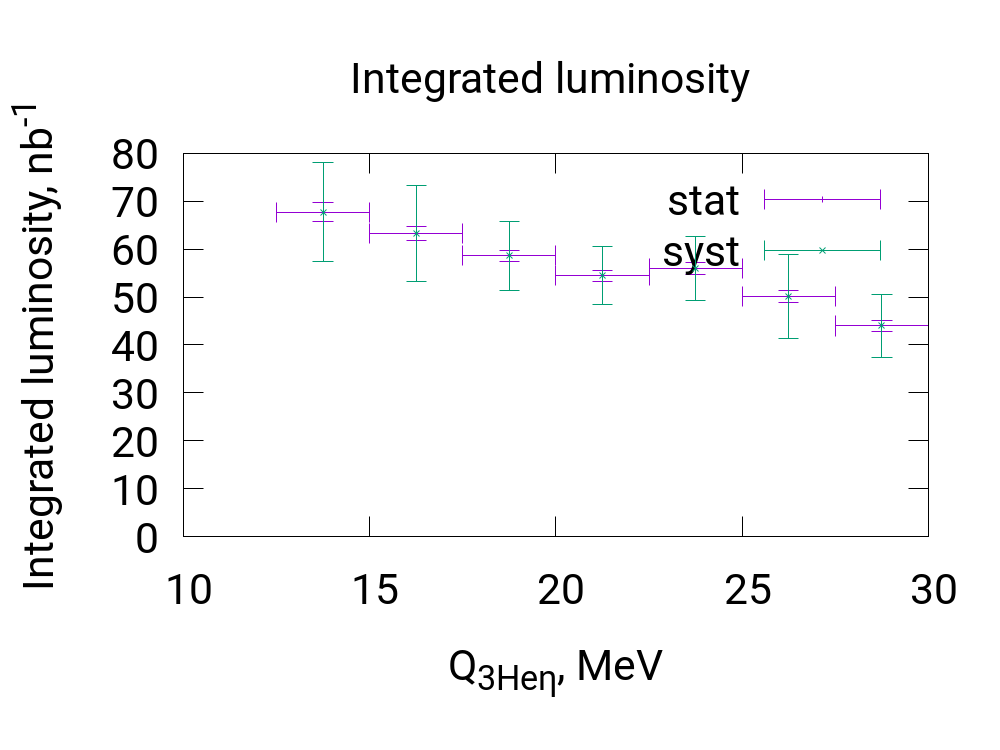}
	\end{center}
	\caption{\small
		Left:
		The estimation of the cross section values for $pd\rightarrow^3$$He\eta$ channel.
		Points show experimental data from Ref.~\cite{3Heeta_cross_section}.
		Error bars show linear interpolation used for luminosity estimation.
		Right: Integrated luminosity obtained using $pd\rightarrow^3$$He\eta$ channel.
		Blue error bars show statistical uncertainties while cyan error bars show systematic uncertainties.
	}
	\label{he3_cross_section}
	\label{he3_fwd_luminosity}
\end{figure}

To obtain luminosity in function of excess energy $Q_{^3He\eta}$, the number of experimental $pd\rightarrow^3$$He\eta$ events for each $Q$-bin was divided by the corresponding efficiency and the cross section (Eq.~\ref{luminosity_description}). 
The cross section values are taken from
\cite{3Heeta_cross_section,3Heeta_cross_section_older1}
and linearly interpolated in our excess energy range (Fig.~\ref{he3_cross_section}).

\section{Luminosity determination based on $pd\rightarrow ppn_{spec}$ reaction}

\label{luminosity_ppn}
Quasielastic proton-proton scattering is a good reaction for luminosity determination because it is visible in the whole beam momentum range used in this experiment, has large cross section about $30~mb$ 
(sec.~\ref{ppn_cross_section_integrating}), and is easy to identify.
This reaction produces two charged particles that have almost coplanar emission directions (just smeared due to deuteron nucleons Fermi motion) and they appear in wide range of $\theta$ angles.

\subsection{Monte Carlo simulation of $pd\rightarrow ppn_{spec}$ reaction}

\begin{figure}[h!]
	\begin{center}
	\includegraphics[width=220pt]{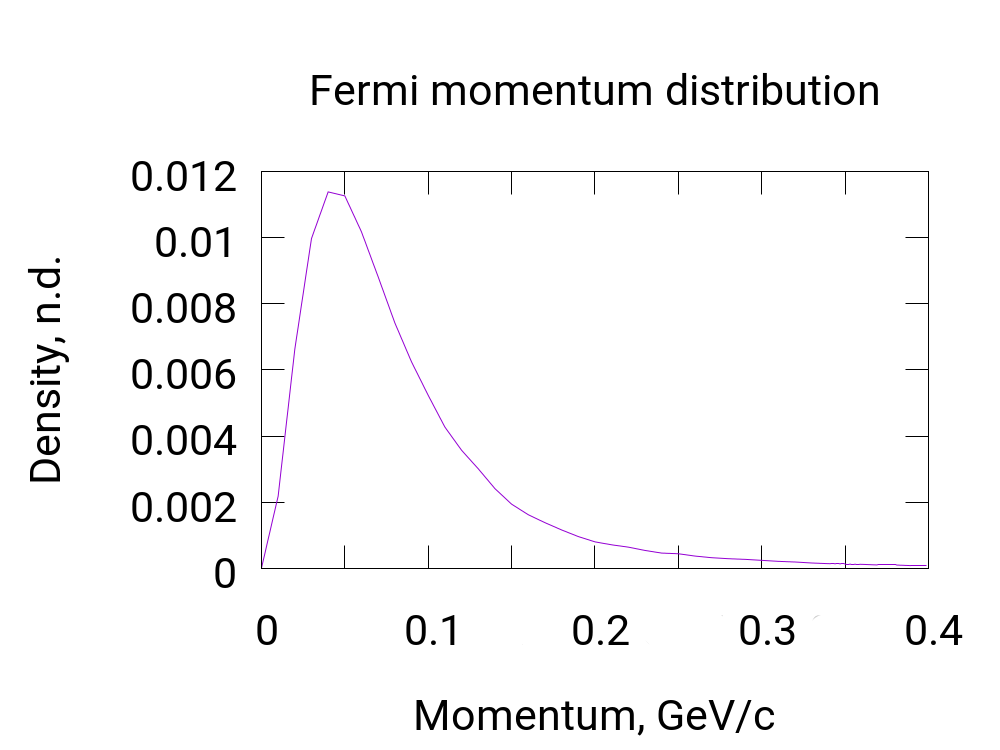}
	\includegraphics[width=220pt]{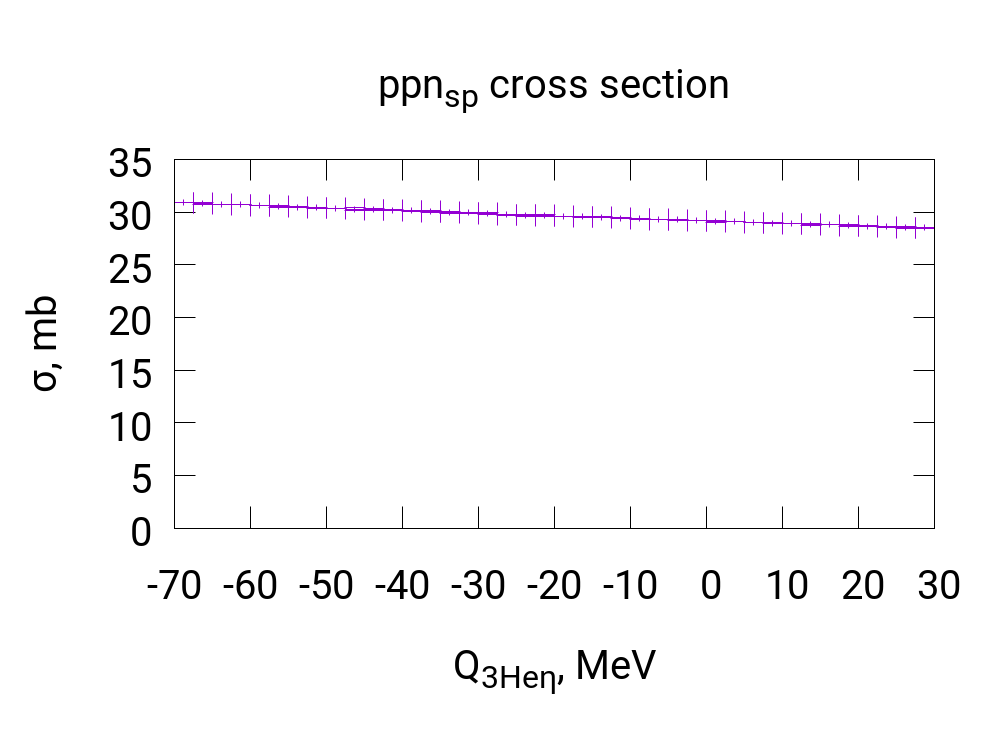}
	\end{center}
	\caption{
Left: Fermi momentum distribution for nucleons in target deuteron obtained from PARIS model \cite{PARIS_model}.
Right: The cross section of $pd\rightarrow ppn_{spectator}$ reaction calculated in this work for the luminosity determination.
}
\label{deuteron_fremi_figure}
\label{ppn_cs_figure}
\end{figure}

The simulation is performed in frame of the spectator model.
The beam momentum values are generated uniformly in the range that was used in the experiment (Sec.~\ref{Q-bins}).
\label{ppn_simulation}
The target deuteron is assumed to be at rest and its nucleons momenta values are generated according to 
Fermi momentum distribution that was calculated in frame of PARIS model \cite{PARIS_model} (left panel of Fig.~\ref{deuteron_fremi_figure}).

The spectator model assumes that effective coupled neutron mass is equal to free neutron mass.
The effective coupled proton mass is lesser than free proton mass due to deuteron binding energy.

The data from Ref.~\cite{PP_CS_link} about proton-proton scattering cross sections were used.
The differential cross section as a function depending on beam momentum in target proton frame and scattering angle in proton-proton center of mass frame is given there.
Thus the beam momentum 4-vector was transferred into target proton frame to obtain the products angular distributions and
the scattered protons momenta were generated in the proton-proton center of mass frame and then transferred to the laboratory frame.

\subsection{Total $pd\rightarrow ppn_{spec}$ cross section}

\label{ppn_cross_section_integrating}
To obtain quasielastic scattering cross section, proton-proton scattering cross section was integrated:

\begin{equation}
\label{integrate_pp_cs}
\sigma_{pp}=\frac{1}{2} \int_0^{2\pi}d\phi\int_0^{\pi}\frac{d\sigma_{pp}}{d\Omega}sin\theta~d\theta
= \int_0^{2\pi}d\phi\int_0^{\pi/2}\frac{d\sigma_{pp}}{d\Omega}sin\theta~d\theta.
\end{equation}

The formula contains coefficient $\frac{1}{2}$ because protons are undistinguishable \cite{PP_undistinguishable}. As far as the angular distribution in center of mass frame is symmetric, one can integrate in half of the angular range instead of multiplying by $\frac{1}{2}$.

After the proton-proton cross section is integrated, Monte Carlo simulation is performed in order to obtain quasielastic scattering cross section.
The deuteron nucleons momenta are generated according to the Fermi momentum distribution like in the simulation described in Sec.~\ref{ppn_simulation} \cite{PARIS_model}.
Then the beam momentum 4-vector is transferred into target proton frame and its absolute value is used to obtain proper proton-proton scattering cross section.
The resulting quasielastic scattering cross section is calculated as an average of proton-proton scattering cross sections obtained during the simulation. 
This value is multiplied by the factor of $0.96$ taking into account the shading effect~\cite{shading_effect} 
caused by neutron shading the scattered protons. 
The result is shown in the right panel of Fig.~\ref{ppn_cs_figure}.

\subsection{The algorithm of $pd\rightarrow ppn_{spec}$ events selection}

\begin{figure}[h!]
	\begin{center}
		\includegraphics[width=200pt]{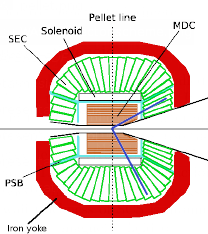}
	\end{center}
	\caption{A scheme explaining the condition of trigger 17 (sec.~\ref{trigger_table}).
		Blue lines show probable charged particles tracks directions.
	}
	\label{trigger_17_expalin}
\end{figure}

In this work, quasielastically scattered protons are searched in the Central Detector.
Such events correspond to trigger number 17 (sec.~\ref{trigger_table}) with condition of at least two charged particle tracks in CD: at least one track registered in forward part of PSB and at least one track in sidelong part of PSB (Fig.~\ref{trigger_17_expalin}).

\begin{figure}[h!]
\begin{center}
	\includegraphics[width=220pt]{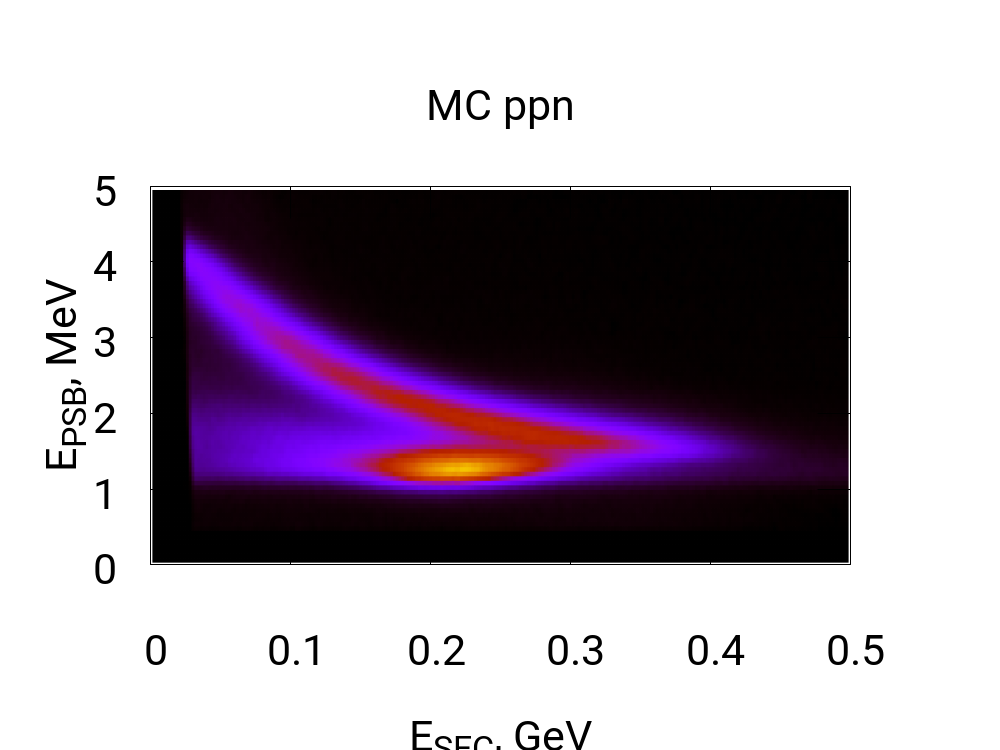}
	\includegraphics[width=220pt]{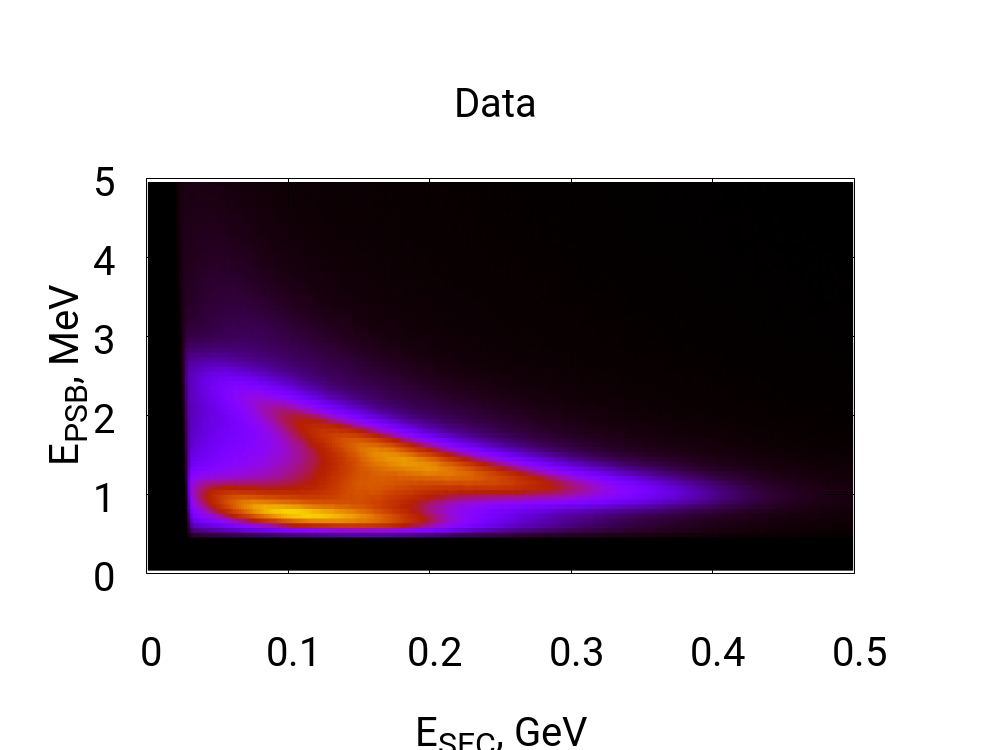}
\end{center}
\caption{
	The distribution of energy deposited in two parts of Central Detector by charged particles registered in pairs.
	Horizontal axis: energy deposited in SEC; vertical axis: in PSB (Sec.~\ref{central_detector}).
	Left side: Monte Carlo simulations for $pd\rightarrow$$ppn_{spec}$; right side: data analysis.
}
\label{ppn_trackid_figure}
\end{figure}

\begin{figure}[h!]
\begin{center}
	\includegraphics[width=220pt]{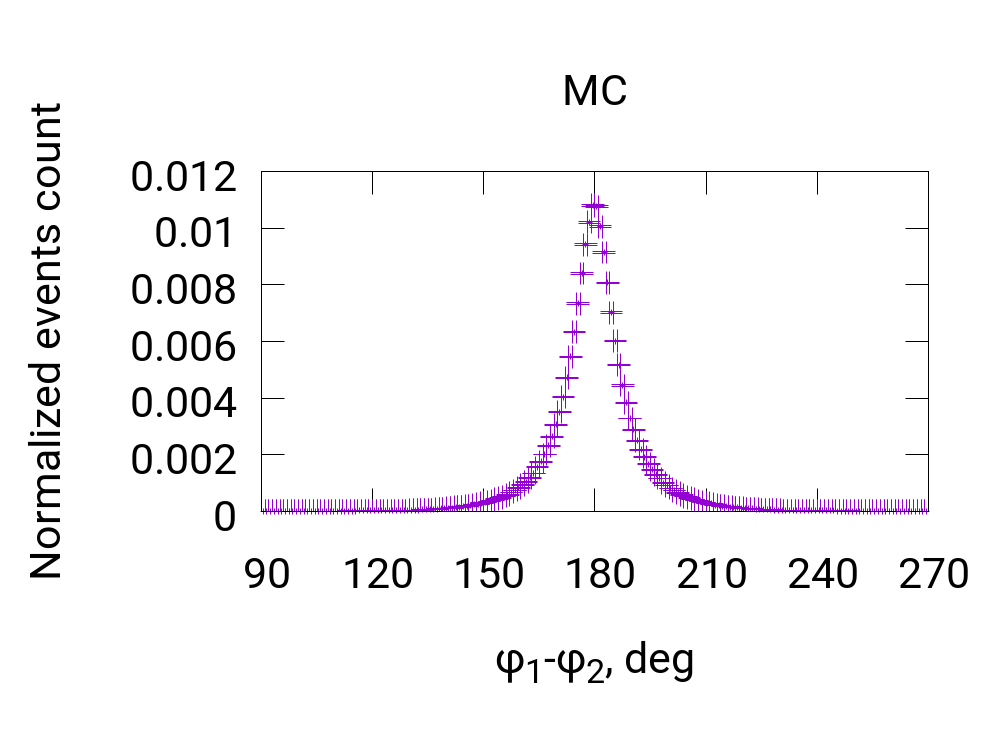}
	\includegraphics[width=220pt]{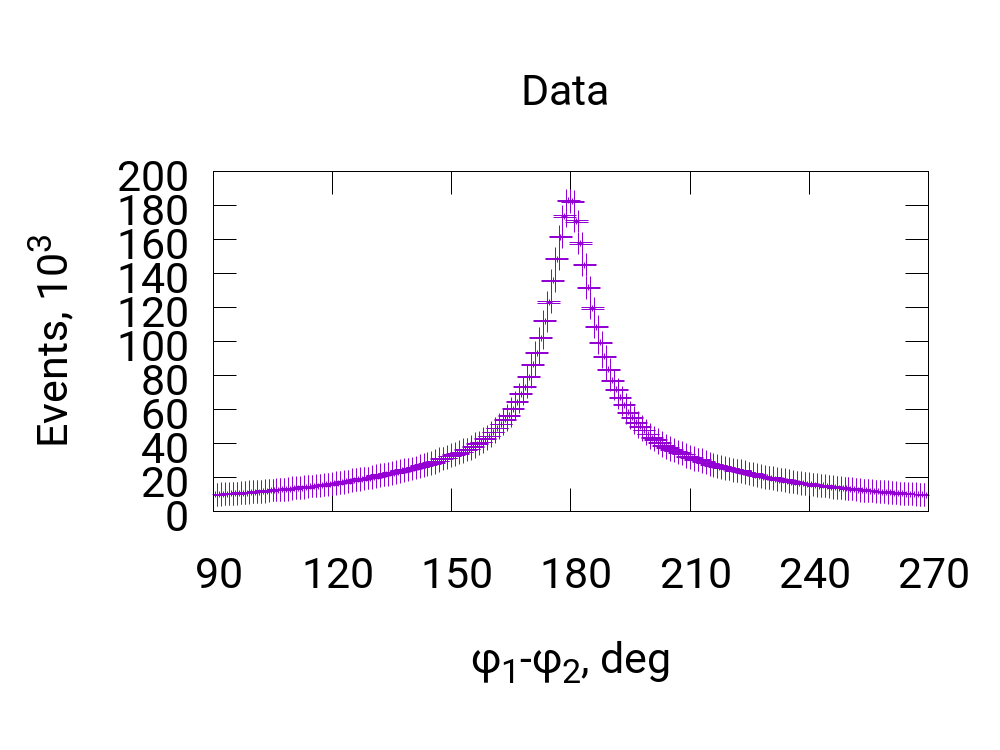}
\end{center}
\caption{\small
	The distribution of $\phi$ angle difference for charged particle pairs being the candidates for quasielastic proton-proton scattering products.
	Left side: Monte Carlo simulations for $pd\rightarrow$$ppn_{spec}$; right side: data analysis.
}
\label{ppn_coplanarity_figure}
\end{figure}

It is seen in $\Delta$$E-E$ spectra (Fig.~\ref{ppn_trackid_figure}) that the visible charged particles are protons and background formed the most probably by $\pi$ mesons.
Nevertheless, no $\Delta$$E-E$ particle identification was applied because there is a stronger condition like particles flight directions coplanarity (Fig.~\ref{ppn_coplanarity_figure}).
As far as no charged $\pi$ mesons appear as product of either binary or quasibinary reaction, this condition can be assumed to be strong enough to reduce possible $\pi^{+}$ and $\pi^{-}$ background.

One more important question is how to choose a pair of charged particle tracks if the event contains more than two tracks.
In that case the algorithm tests all possible charged track pairs and selects the one which has the asimuthal $\phi$ angle difference the closest to $180^o$.


\begin{figure}[h!]
\begin{center}
	\includegraphics[width=300pt]{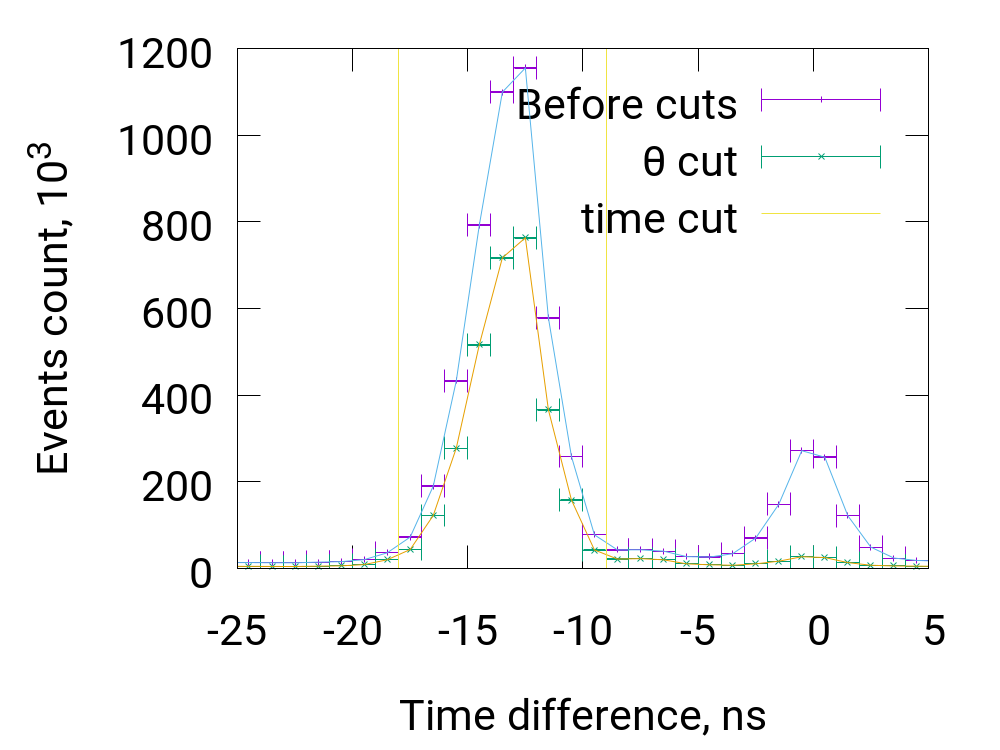}
\end{center}
\caption{\small
	Time difference distribution for the charged particle track pairs being candidates for quasielastic proton-proton scattering.
	The distributions before and after cut on $\theta$ angular distribution (Fig.~\ref{ppn_theta_cut_figure}) are shown.
}
\label{ppn_time_diff_cut_figure}
\end{figure}

\begin{figure}[h!]
	\begin{center}
		\includegraphics[width=220pt]{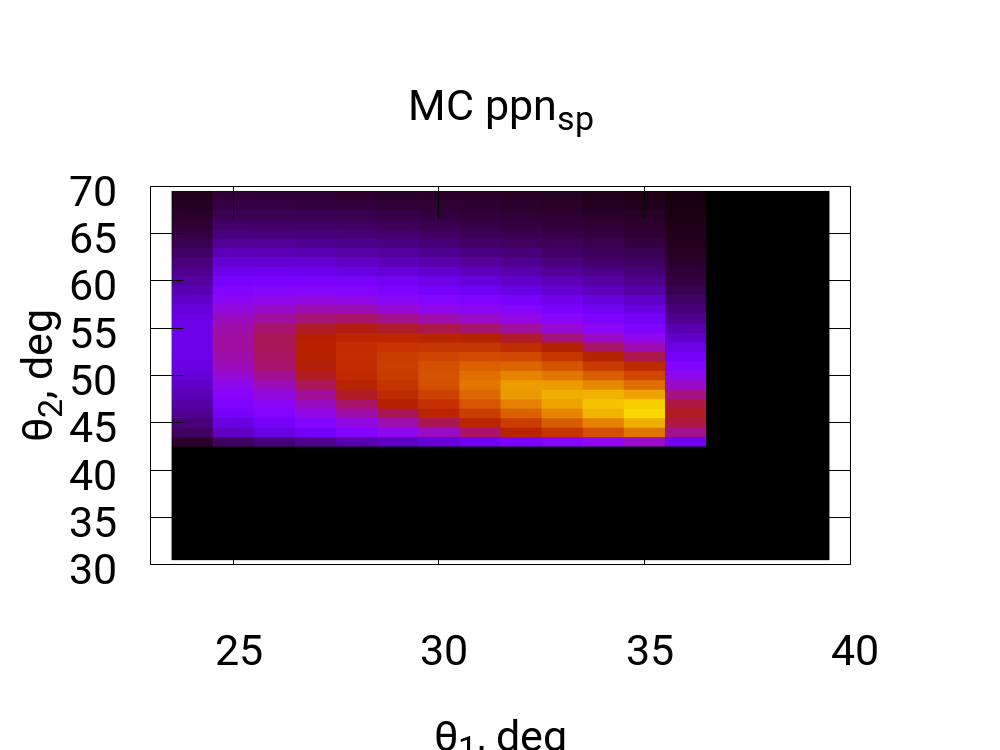}
		\includegraphics[width=220pt]{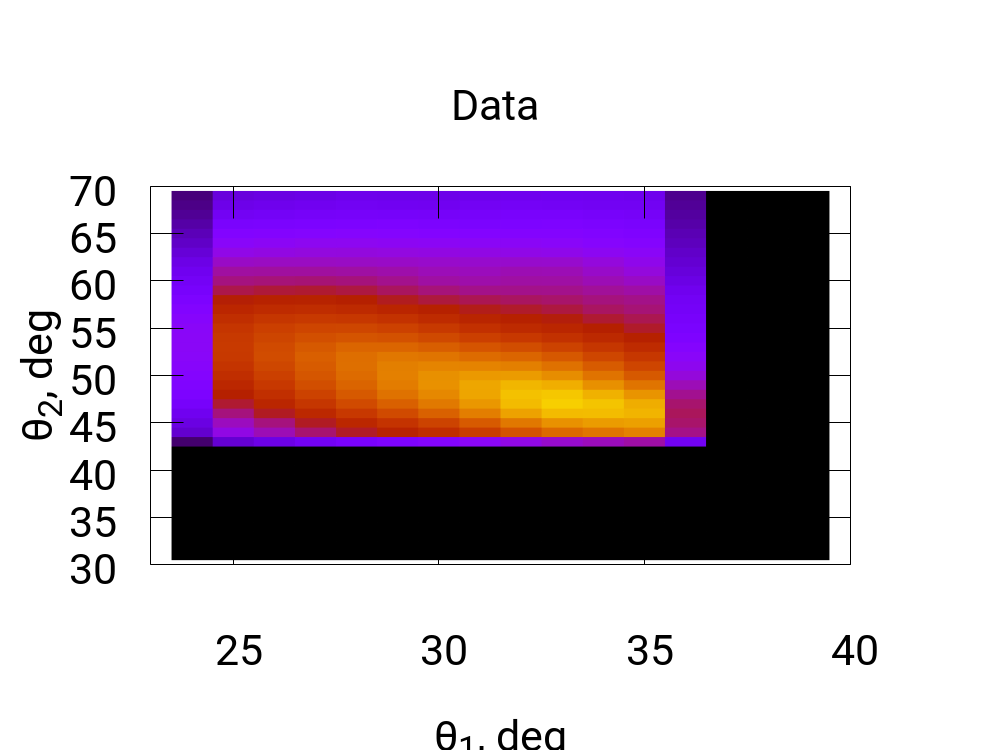}
	\end{center}
	\caption{\small
		The distribution of $\theta$ angles for charged particle track pairs being the candidates for quasielastic proton-proton scattering products.
		The cut on charged particle $\theta$ angles is applied in order to reduce second peak on time difference distribution (Fig.~\ref{ppn_time_diff_cut_figure}).
		Left side: Monte Carlo simulations for $pd\rightarrow$$ppn_{spec}$; right side: experimental data.
	}
	\label{ppn_theta_cut_figure}
\end{figure}

Another selection criterium was associated with the choice of the proper range of time difference for two protons as it is shown in Fig.~\ref{ppn_time_diff_cut_figure}.
Appearing of two peaks is caused by the fact that particles with different $\theta$ angles are registered by different parts of PSB (Sec.~\ref{central_detector}) and the time of registration is measured differently and the offsets are not corrected.
The lower peak on time difference spectrum is reduced by the $\theta$ cut (Fig.~\ref{ppn_theta_cut_figure}) applied due to the condition of trigger 17.
After the angular cut, the additional cut on time difference distribution is applied in order to reduce the remaining background (vertical lines in Fig.~\ref{ppn_time_diff_cut_figure}).


\begin{figure}[h!]
	\begin{center}
\includegraphics[width=220pt]{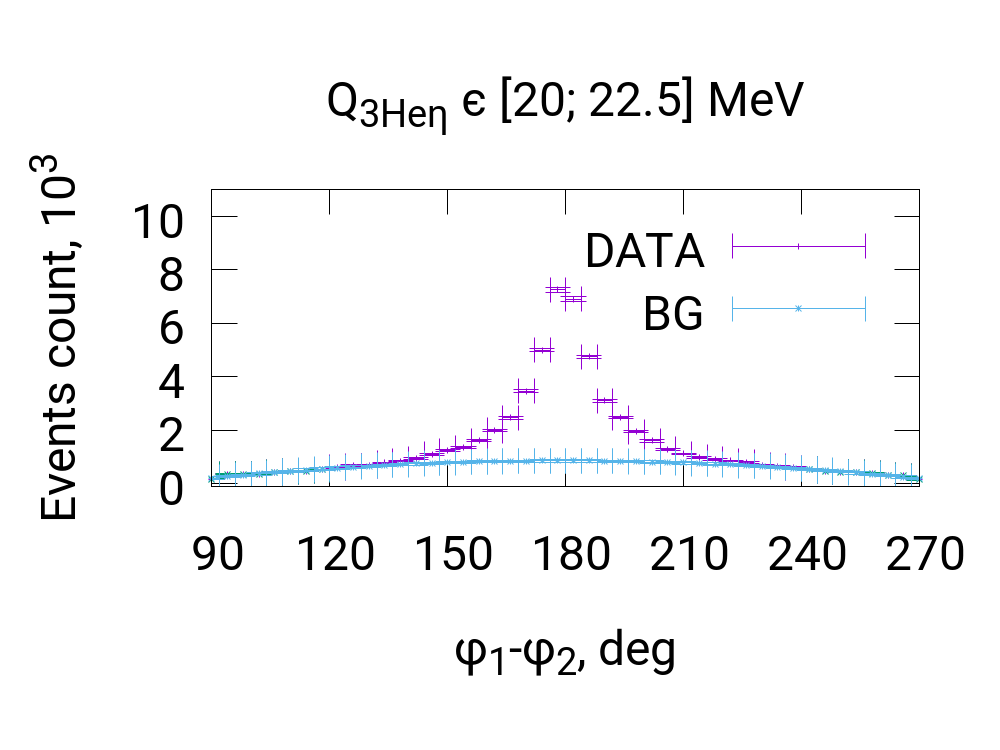}
\includegraphics[width=220pt]{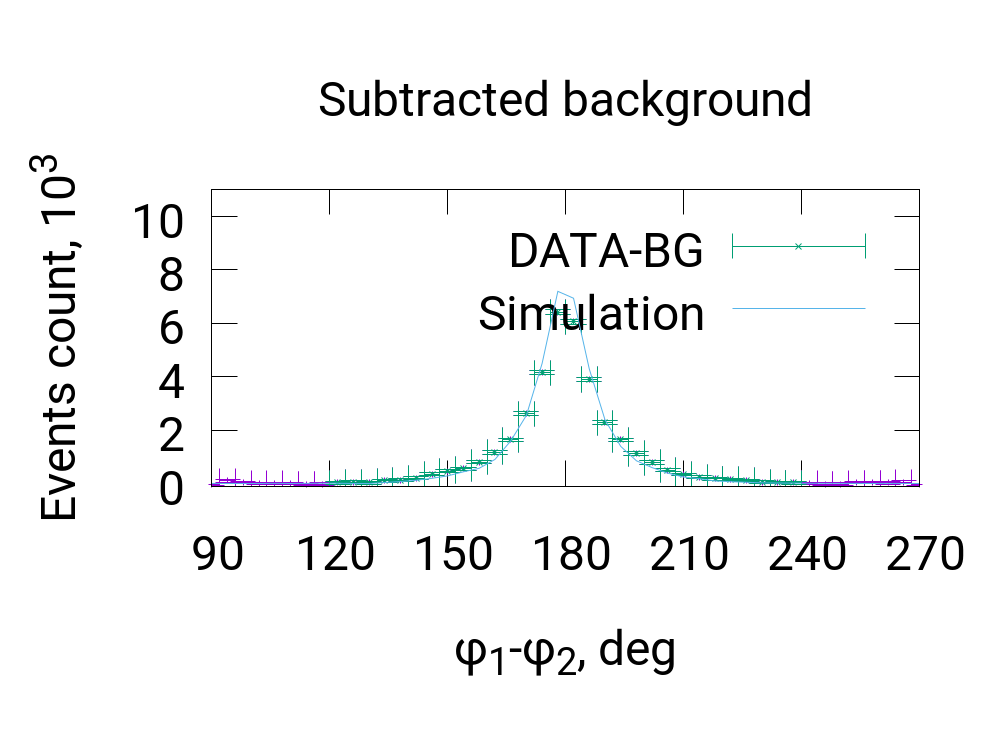}
	\end{center}
\caption{
	Left: the $\phi$ angle difference distribution obtained for $Q_{^3He\eta}~\in~[20.0;22.5]~MeV$ (Sec.~\ref{Q-bins}).
	The background around the peak is fitted by the second power polynomial (cyan line).
	Right: $\phi$ angle difference distribution after background subtraction. Data are shown by points while the line shown the simulation results.
}
\label{ppn_events_obtain_figure}
\end{figure}

After applying the conditions described above, the most essential part of background formed by other charged particles registered in Central Detector is reduced.
The rest of it can be easily seen in $\phi$ angle difference distributions (Fig.~\ref{ppn_events_obtain_figure}).
For each $Q$-bin separately (sec.~\ref{Q-bins}), the background around the peak is fitted by a second power polynomial.
This fit allows to subtract the background in the peak area and obtain the events count 
for the $pd\rightarrow ppn_{spec}$ reaction.

\begin{figure}[h!]
	\begin{center}
		\includegraphics[width=250pt]{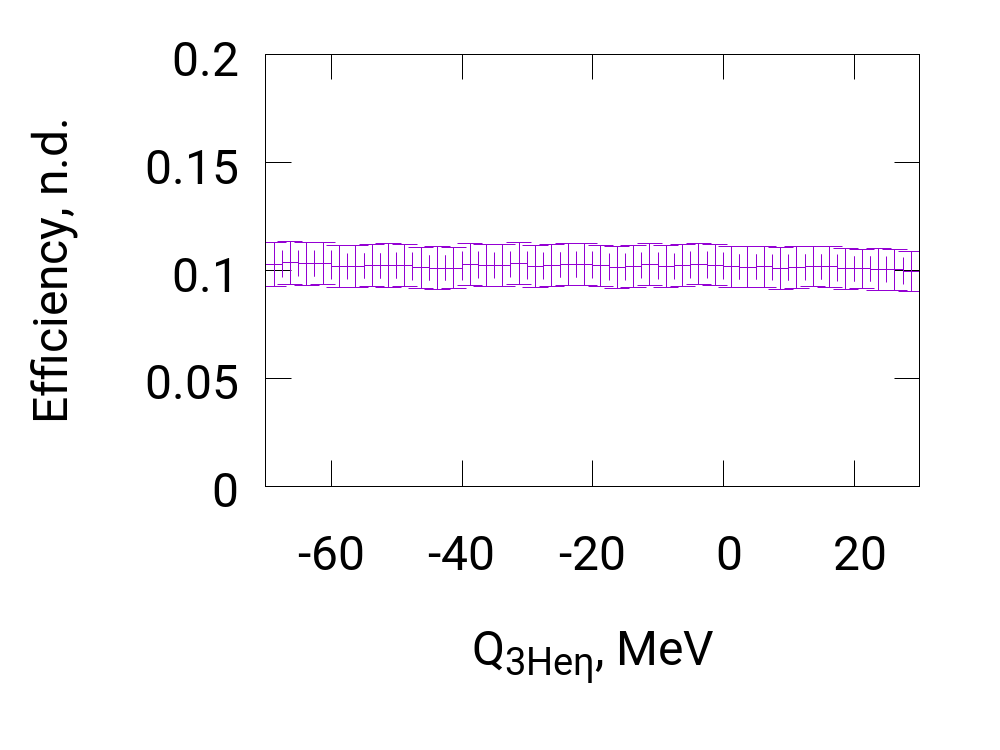}
	\end{center}
	\caption{
	The efficiency for $pd\rightarrow$$ppn_{spec}$ reaction obtained from Monte Carlo simulation.
	The error bars show the systematic uncertainties.
	}
	\label{ppn_efficiency}
\end{figure}

\begin{figure}[h!]
	\begin{center}
\includegraphics[width=300pt]{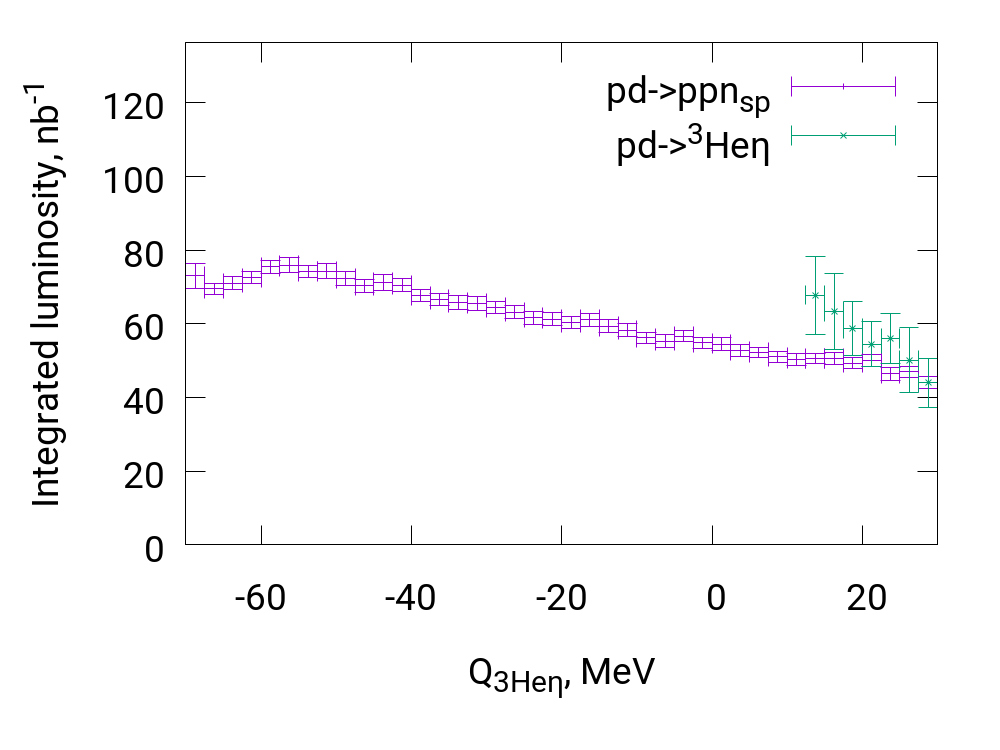}
	\end{center}
\caption{
Integrated luminosity calculated based on Eq.~\ref{luminosity_description}.
The results obtained using $pd\rightarrow$$^3He\eta$ (cyan) reaction and quasielastic proton-proton scattering (magenta) are shown.
The statistical and systematic uncertainties are taken into account.
}
\label{luminosity_final_plot}
\end{figure}

The efficiency obtained from Monte Carlo simulation does not essentially depend on beam momentum and is about $10\%$ (Fig.~\ref{ppn_efficiency}).
The integrated luminosity was calculated by the formula~\ref{luminosity_description} and the result is shown in Fig.~\ref{luminosity_final_plot}.

\section{The systematic uncertainties estimation}

\label{luminosity_systematics}
In case of $pd\rightarrow^3$$He\eta$ reaction analysis, the systematic error has the following sources (Table.~\ref{systematics_he3eta_lum_parameters}).
Due to smearing of kinematic distributions (Fig.~\ref{he3_reconstruction_kinematic_hist}, \ref{he3_reconstruction_kinematic_hists_theta}) that are used to define the beam momentum correction constant, the accuracy for this constant is assumed to be about $5\%$.
The accuracy for the positions of the cuts (Fig.~\ref{he3_cut_data}, \ref{he3_reconstruction_kinematic_hist}) that are used to identify $^3$$He$ tracks is also assumed to be about $5\%$.
The range used for background fit on $^3$$He$ missing mass distributions (Fig.~\ref{he3_mm_spectra_fit}) is defined with the accuracy of bin width used in the analysis that equals $1~MeV.$

The systematic error of $p-p$ quasielastic scattering process analysis originates from beam momentum correction constant inaccuracy, $\theta$ angular cut and time difference cut positions (Fig~\ref{ppn_time_diff_cut_figure}, \ref{ppn_theta_cut_figure}), and background fit range (Fig.~\ref{ppn_events_obtain_figure}) inaccuracies (Table.~\ref{systematics_ppn_lum_parameters}).

\begin{table}[h!]
	\begin{tabular}{|p{280pt}|p{50pt}|p{50pt}|}
		\hline
		Parameter description & Value & Parameter variation \\
		\hline
		Beam momentum correction constant (Eq.~\ref{beam_momentum_correction_constant})& $4.0~MeV$& $\pm0.2~MeV$\\
		\hline
		$\theta$ angular cut for forward tracks (Fig.~\ref{he3_reconstruction_kinematic_hist}, Eq.~\ref{he3_theta_cut_pos}) & $4.5^o$& $\pm0.2^o$\\
		\hline
		$^3$$He$ identification cut height (Fig.~\ref{he3_cut_data}, Eq.~\ref{he3_cut_height}) & $10~MeV$ & $\pm0.5~MeV$ \\
		\hline
		Background fit range (Fig.~\ref{he3_mm_spectra_fit}, left)& $543~\frac{MeV}{c^2}$ & $\pm1~\frac{MeV}{c^2}$\\
		\hline
		Background fit range (Fig.~\ref{he3_mm_spectra_fit}, right)& $553~\frac{MeV}{c^2}$ & $\pm1~\frac{MeV}{c^2}$\\
		\hline
	\end{tabular}
	\caption{The list of parameters contributing into systematic error for $pd\rightarrow^3$$He\eta$ reaction analysis.}
	\label{systematics_he3eta_lum_parameters}
\end{table}
\begin{table}[h!]
	\begin{tabular}{|p{280pt}|p{50pt}|p{50pt}|}
		\hline
		Parameter description & Value & Parameter variation \\
		\hline
		Beam momentum correction constant (Eq.~\ref{beam_momentum_correction_constant})& $4.0~MeV$& $\pm0.2~MeV$\\
		\hline
		Angular cut position (Fig.~\ref{ppn_theta_cut_figure}, horizontal axis)& $35^o$ & $\pm1^o$\\
		\hline
		Angular cut position (Fig.~\ref{ppn_theta_cut_figure}, vertical axis)& $42^o$ & $\pm1^o$\\
		\hline
		Time cut position (Fig.~\ref{ppn_time_diff_cut_figure}, left) & $-18~ns$ & $\pm1~ns$ \\
		\hline
		Time cut position (Fig.~\ref{ppn_time_diff_cut_figure}, right) & $-9~ns$ & $\pm1~ns$ \\
		\hline
		Background fit range (Fig.~\ref{ppn_events_obtain_figure}). Peak range is assumed from $180^o-x$ to $180^o+x$.& $60^o$ &$\pm5^o$ \\
		\hline
	\end{tabular}
	\caption{The list of parameters contributing into systematic error for $pd\rightarrow ppn_{spectator}$ reaction analysis.}
	\label{systematics_ppn_lum_parameters}
\end{table}

For each analysis, the total systematic uncertainty was calculated by the formula:
\begin{equation}
	\Delta L_{syst} = \sqrt{ \sum_i (\frac{|L^{P_i = P^{max}_i} - L^{final}|
			+ |L^{P_i = P^{min}_i}  - L^{final}|}{2})^2 },
	\label{luminosity_systematics_formulae}
\end{equation}
 where $L$ is integrated luminosity value obtained in the analysis, $i$ index denotes sum over all parameters given in Table.~\ref{systematics_he3eta_lum_parameters} or \ref{systematics_ppn_lum_parameters} respectively, $P_{i}$ in the index means which varied parameter value is used for each $L$ value in the formula,  and $P^{min}_{i}$ and $P^{max}_i$ are minimum and maximum values for the parameter $P_i$. Index $final$ denotes luminosity value that was actually used as the final result.

\section{Discussion of the results}

Two reactions were used to determine the integrated luminosity for this experiment.
$pd\rightarrow$$^3He\eta$ reaction has the cross section about $300~nb$ and is visible only above the $\eta$ creation threshold (Sec.~\ref{Q-bins}). 
Furthermore, there is strong efficiency dependence on $Q_{^3He\eta}$ that makes this analysis less accurate.
Quasielastic proton-proton scattering has drastically larger cross section, about $30~\mu$$b$, and is visible in the whole beam momentum range used in this experiment with the efficiency that does not essentially depend on the excess energy $Q_{^3He\eta}$.

The more precise and informative curve obtained for $pd\rightarrow$$ppn_{spec}$ reaction shows that the luminosity decreases with beam momentum increasing.
It is in agreement with the prediction that target overlapping by the beam is changing during the acceleration cycle.

For normalization of excitation functions for $pd\rightarrow^3$$He 2\gamma$ and $pd\rightarrow^3$$He 6\gamma$ reactions (Sec.~\ref{reactions_that_are_used}), the integrated luminosity curve obtained for the quasielastic proton-proton scattering is used. 
It is however, important to stress that above the threshold luminosity determined based on the $pd\rightarrow^3He\eta$ and $pd\rightarrow ppn_{spectator}$ are consistent (Table.~\ref{luminosity_table_compare})
giving more confidence to the obtained results. 
Moreover, 
$pd\rightarrow$$^3He\eta$ reaction analysis allowed to obtain very important beam momentum correction constant (Sec.~\ref{beam_momentum_offset_constant}) and to test the algorithms for $^3He$ tracks identification and reconstruction which are used for further analysis.

\begin{table}[h!]
	\begin{tabular}{|p{100pt}|p{160pt}|p{160pt}|}
		\hline
		Reaction & $Q_{3He\eta}$ range $[-70;+30]~MeV$ & $Q_{3He\eta}$ range $[+12.5;+30]~MeV$ \\
		\hline
		$pd\rightarrow^3$$He\eta$ & - & $399.7\pm3.6\pm53\pm18~nb^{-1}$ \\
		\hline
		$pd\rightarrow ppn_{spec}$ & $2446\pm3\pm66\pm4~nb^{-1}$ &
														 $337.8\pm1.3\pm10\pm0.7~nb^{-1}$ \\
		\hline
	\end{tabular}
	\caption{Luminosity values calculated based on $pd\rightarrow^3$$He\eta$ and 
		$pd\rightarrow ppn_{spec}$ reactions.
		The uncertainty values given in the table: statistical, systematic, normalization.
	}
	\label{luminosity_table_compare}
\end{table}
\chapter{The analysis of $pd\rightarrow^3$$He2\gamma$ and $pd\rightarrow^3$$He6\gamma$ reactions}
\label{analysis_chapter}

\section{Cross sections determination in current experiment}

For the measurement of the cross sections, events count, efficiency, and the luminosity are needed (Sec.~\ref{luminosity_chapter}).

The first analysis stage is Monte Carlo simulation of the analyzed reaction kinematics and processing it with WASA Monte Carlo software (Sec.~\ref{data_analysis_tools}) in order to simulate the detection system response.
Then, the events identification algorithm is developed and both simulation results and the data are processed by it.

After the events identification, the analyzed  reaction cross section $\sigma$ can be calculated by the following formula:
\begin{equation}
\sigma 
= \frac{N_{true}}{\int L\,dt}
= \frac{N_{signal} S_{trigger}}{\epsilon \int L\,dt}
= \frac{N_{signal} S_{trigger} S_{MC}}{N_{MC} \int L\,dt},
\end{equation}
where $\int L dt$ is the integrated luminosity from Eq.~\ref{luminosity_description},
$N_{signal}$ denotes the count of signal events,
and all other symbols have the same meanings like in this equation.

\section{What reactions are studied?}

\label{reactions_that_are_used}
This work is devoted to searching for $^3$$He\eta$ bound state using neutral channels of direct $\eta$ decay without being absorbed by any nucleons.
Assuming that the decay branching ratio for bound $\eta$ meson does not essentially differ from free $\eta$ decay branching ratio, $\eta\rightarrow\gamma\gamma$ and 
$\eta\rightarrow\pi^0\pi^0\pi^0\rightarrow 6\gamma$ decay channels are expected to be the most intensive~\cite{eta_decay_modes_new}.

For both of these reactions, the expected analysis result is significant events count for excess energy above zero (sec.~\ref{Q-bins}) because of $pd\rightarrow^3$$He\eta$ reaction.
Its cross section~\cite{3Heeta_cross_section, 3Heeta_cross_section_older1, 3Heeta_cross_section_older2, 3Heeta_cross_section_older3}  is about $300~nb$ in the $Q_{^3He\eta}$ range $[10;30]~MeV$.
Below the threshold, in case if the bound state is not observed, some count of background events is expected but essentially lesser than above the threshold.
In case if the bound state exists and is observed, a Breit-Wigner shaped peak is expected for $Q_{^3He\eta}~<~0$.

It is also possible that bound state creation and decay process can interfere with other channels that have the same particles in the final state but the analysis performed in this thesis does not take such possibility into account.

\section{Bound state theoretical model}
\label{bound_State_theoretical_model}
In current experiment, the $^3$$He\eta$ bound state is searched for in proton-deuteron collisions.
The mass of a bound state is a sum of $^3He$ and $\eta$ masses reduced by the binding energy:
\begin{equation}
		m_{bs} = m_{^3He} + m_{\eta} - B_s.
		\label{bound_state_mass}
\end{equation}
According to the model, the $\eta$ meson in the bound system is orbiting around the $^3$$He$ nucleus and decays into $2\gamma$ or $3\pi^0$.
The decaying $\eta$ is assumed to move due to Fermi motion.
For calculating this momentum distribution the following $^3He-\eta$ interaction potential is postulated~\cite{Bound_state_model_Hirenzaki}:
\begin{equation}
	V(r) = (V_0+iW_0)\frac{\rho(r)}{\rho_0},
	\label{potential_formula}
\end{equation}
where the assumed $V_0$ and $W_0$ parameters values are given in Table.~\ref{bound_state_potential_parameters}, 
and $\rho(r)$ is Hiyama's density distribution ($\rho_0>>0.17~fm^{-3}$) \cite{Bound_state_model_cite2,Bound_state_model_cite3,Bound_state_model_Hirenzaki}.

Then the Klein Gordon equation is considered:
\begin{equation}
	[-\nabla^2 + \mu^2 + 2\mu V(r)]\psi(\vec{r}) = E^2_{KG} \psi(\vec{r}),
	\label{klein_gordon_equ}
\end{equation}
where $\psi(\vec{r})$ is the bound system's wave function,
$\mu$ is $^3$$He-\eta$ reduced mass,
and $E_{KG}$ is Klein Gordon energy.
This equation was solved by S.~Hirenzaki and H.~Nagahiro \cite{Bound_state_model_Hirenzaki}
with potential parameters given in Table.~\ref{bound_state_potential_parameters} for the purpose of current analysis.

\begin{table}[h!]
	\begin{center}
		\begin{tabular}{|p{100pt}|p{100pt}|}
			\hline
			$(V_0,W_0)~[MeV]$ & $(B_s,\Gamma)~[MeV]$\\
			\hline
			$-(75,20)$ & $(-4.02,15.60)$\\
			$-(80,20)$ & $(-6.19,17.39)$\\
			$-(90,20)$ & $(-11.10,20.59)$\\
			\hline
		\end{tabular}
	\end{center}
	\caption{Different combinations of $^3$$He-\eta$ interaction potential assumed in relative motion momentum distribution calculation and corresponding binding energies and bound state widths obtained from Eq.~\ref{bound_state_parameters_formula}.}
	\label{bound_state_potential_parameters}
\end{table}

The binding energy and the bound state width can be obtained after Klein Gordon equation is solved and are equal to
\begin{equation}
    B_s = Re(E_{KG} - \mu),
    ~\Gamma = -2 Im(E_{KG}).
	\label{bound_state_parameters_formula}
\end{equation}

Then the radial wave function $\phi(r)$ defined by the following way:
\begin{equation}
	\psi({\vec{r}}) = \phi(r) Y_{lm}(\hat{r})
	\label{radial_wave_function_def}
\end{equation}
is obtained from Eq.~\ref{klein_gordon_equ} and then is transferred into momentum space:

\begin{equation}
	\tilde{\phi}(p) = \int e^{i\vec{p}\vec{r}} \phi(r) d\vec{r}
	\label{radial_function_transfer}
\end{equation}

\begin{figure}[h!]
	\begin{center}
		\includegraphics[width=300pt]{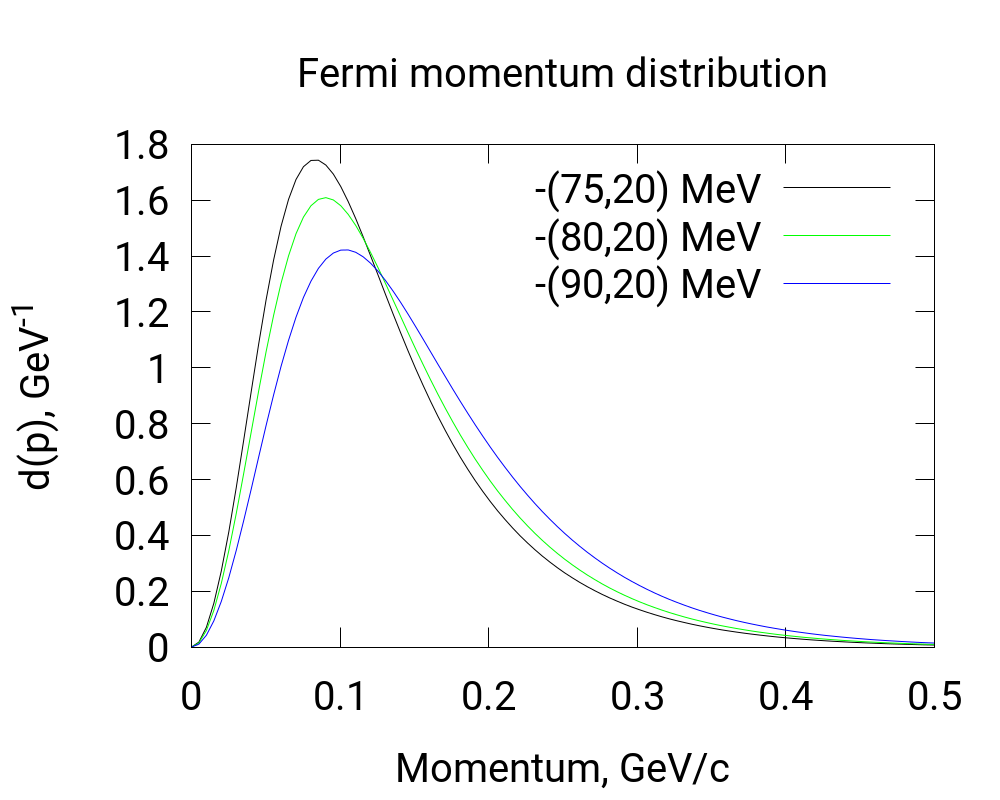}
	\end{center}
	\caption{
		The distribution of bound $^3$$He-\eta$ relative motion momentum calculated for different
		assumption of interaction potential (Eq.~\ref{bound_Fermi_density_formula}).
	}
	\label{bound_fermi_momentum_fig}
\end{figure}

The function $\tilde{\phi}(p)$ is used to obtain relative momentum distribution density (Fig.~\ref{bound_fermi_momentum_fig}):
\begin{equation}
    d(p) = |\tilde{\phi}(p)|^2 p^2,
	\label{bound_Fermi_density_formula}
\end{equation}
that is used in the $^3$$He\eta$ bound state production and decay Monte Carlo simulations.

\section{The bound state production and decay Monte Carlo simulations}

\label{he3eta_bound_simulation}
The $^3$$He\eta$ bound state is simulated in frame of the spectator model. 
The beam momentum values are generated uniformly in the range that fulfills the experimental conditions (Sec.~\ref{Q-bins}). 
The bound state invariant mass is obtained based on beam proton momentum.

The distribution of relative $^3He-\eta$ momentum is assumed to be isotropic and absolute value distribution was calculated by S.~Hirenzaki and H.~Nagahiro
\cite{Bound_state_model_Hirenzaki} (Fig.~\ref{bound_fermi_momentum_fig}, Table~\ref{bound_state_potential_parameters}).

The $^3He$ nucleus is assumed to be a spectator that means that its effective mass is assumed to be equal to free $^3He$ mass
while bound $\eta$ meson mass has to differ from free $\eta$ mass due to the total invariant mass conservation.
This difference has to be taken into account in $\eta$ decay simulation.

\label{eta_decay_simulation}
$\eta\rightarrow\gamma\gamma$ decay channel simulation requires only $\eta$ meson 4-momentum.
In the $\eta$ rest frame, two $\gamma$ quanta have equal momenta with opposite directions and their energy sum is equal to $\eta$ effective mass. 
In the $\eta$ rest frame, the decay is isotropic.
Then $\gamma$ quanta momenta are transferred to the laboratory frame.

$\eta\rightarrow\pi^0\pi^0\pi^0$ channel simulation requires more complicated calculations.
Let's consider the following invariants:
\begin{equation}
	s_{ij} = |\wp_{\pi^0_i}+\wp_{\pi^0_j}|^2,
\label{kin_invariant}
\end{equation}
where $\wp$ symbol means 4-momentum and $\pi^0_i$ or $\pi^0_j$ means one of three decay products ($i=1,2,3$, $j=1,2,3$).
These invariants fulfill the following conditions
\begin{equation}
s_{12}+s_{13}+s_{23} = m^2_\eta + 3m^2_{\pi^0},
\label{kin_invariant_condition1}
\end{equation}
\begin{equation}
(2m_{\pi^0})^2 < s_{ij} < (m_\eta - m_{\pi^0})^2.
\label{kin_invariant_condition2}
\end{equation}

These conditions arrange the part of phase space that can be populated.
This part is assumed to be populated uniformly.
The next stage is the energy and momentum calculation for each $\pi^0$:
\begin{equation}
E_1 = \frac{m^2_\eta+m^2_{\pi^0} - s_{23}}{2m_\eta};
E_2 = \frac{m^2_\eta+m^2_{\pi^0} - s_{13}}{2m_\eta};
E_3 = \frac{m^2_\eta+m^2_{\pi^0} - s_{12}}{2m_\eta}
\label{kin_pi0_energy}
\end{equation}
\begin{equation}
p_1 = \sqrt{E^2_1 - m^2_{\pi^0}};~
p_2 = \sqrt{E^2_2 - m^2_{\pi^0}};~
p_3 = \sqrt{E^2_3 - m^2_{\pi^0}}
\label{kin_pi0_momentum}
\end{equation}
\begin{equation}
\vec{p}_1 + \vec{p}_2 + \vec{p}_3 = \vec{0}.
\label{kin_pi0_momentum2}
\end{equation}

The vectors in Eq.~\ref{kin_pi0_momentum2} are required to lie in one plane.
Hereby, the planar vectors are obtained first and then the decay plane is rotated in order to attain isotropic distribution of its orientation.

Each $\pi^0$ decays into two $\gamma$ quanta.
Such decay is simulated similarly to $\eta\rightarrow\gamma\gamma$ decay.

\section{The analysis of $pd\rightarrow^3$$He2\gamma$ reaction}

\label{3He_2gamma_analysis}
In this analysis, for $pd\rightarrow^3$$He2\gamma$ events selection both $^3$$He$ track in Forward Detector and signals from two $\gamma$ quanta in Central Detector are required.
The $^3$$He$ track reconstruction algorithm is similar to the one used for $pd\rightarrow^3$$He\eta$ reaction analysis and described above in Sec.~\ref{3He_tracks_reconstruction}.

\begin{figure}
\begin{center}
\includegraphics[width=220pt]{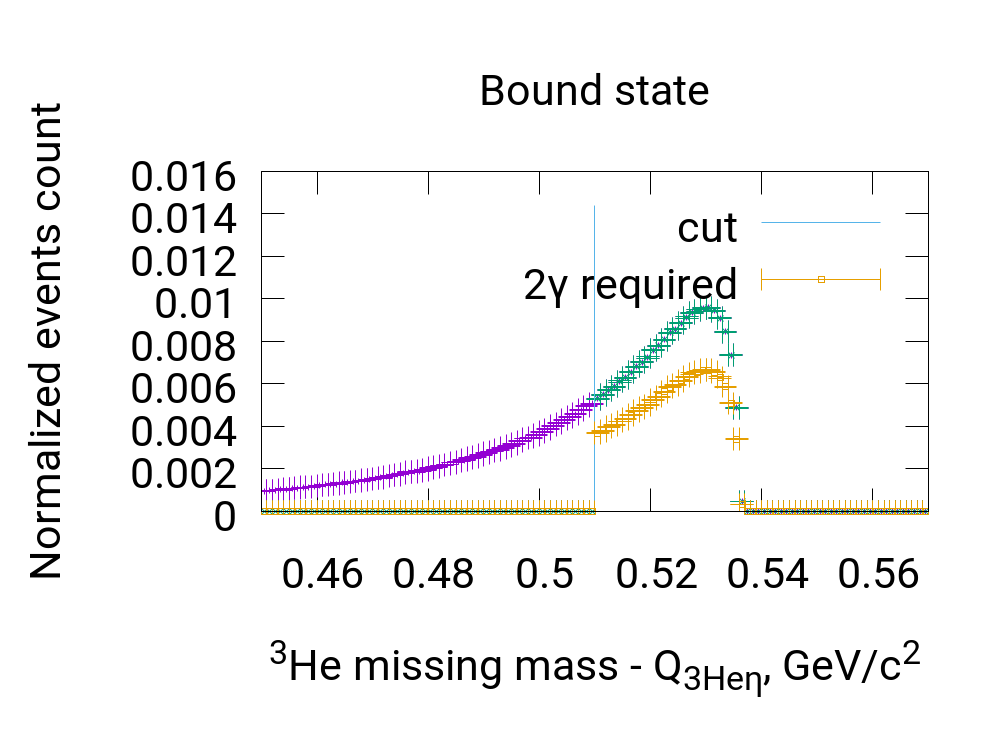}
\includegraphics[width=220pt]{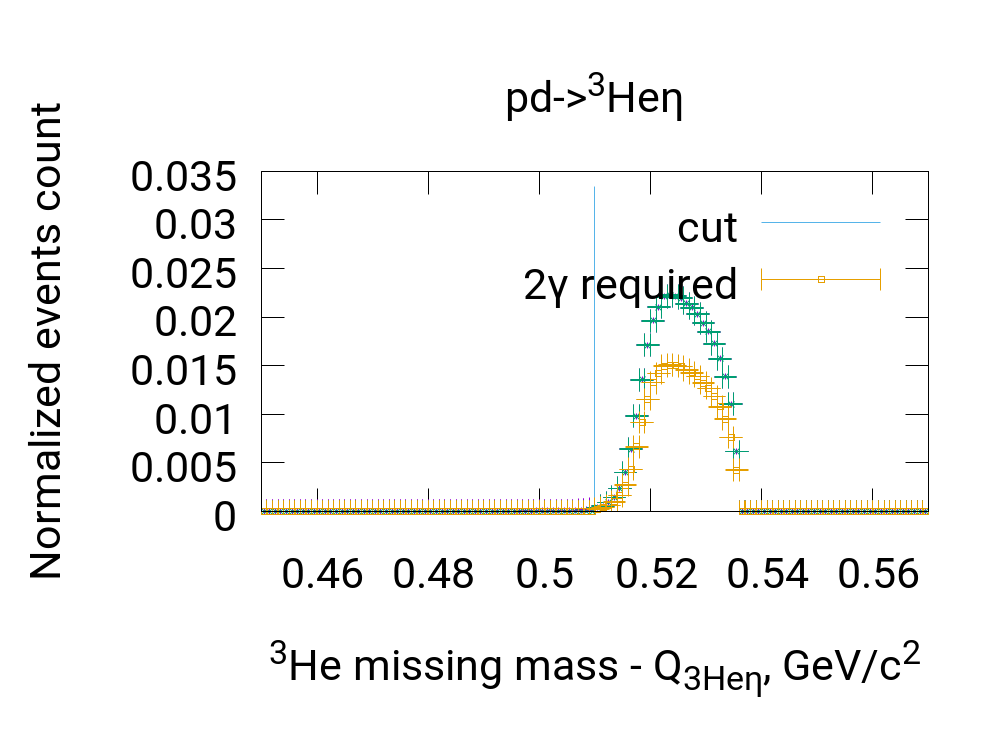}\\
~\\~\\
\includegraphics[width=220pt]{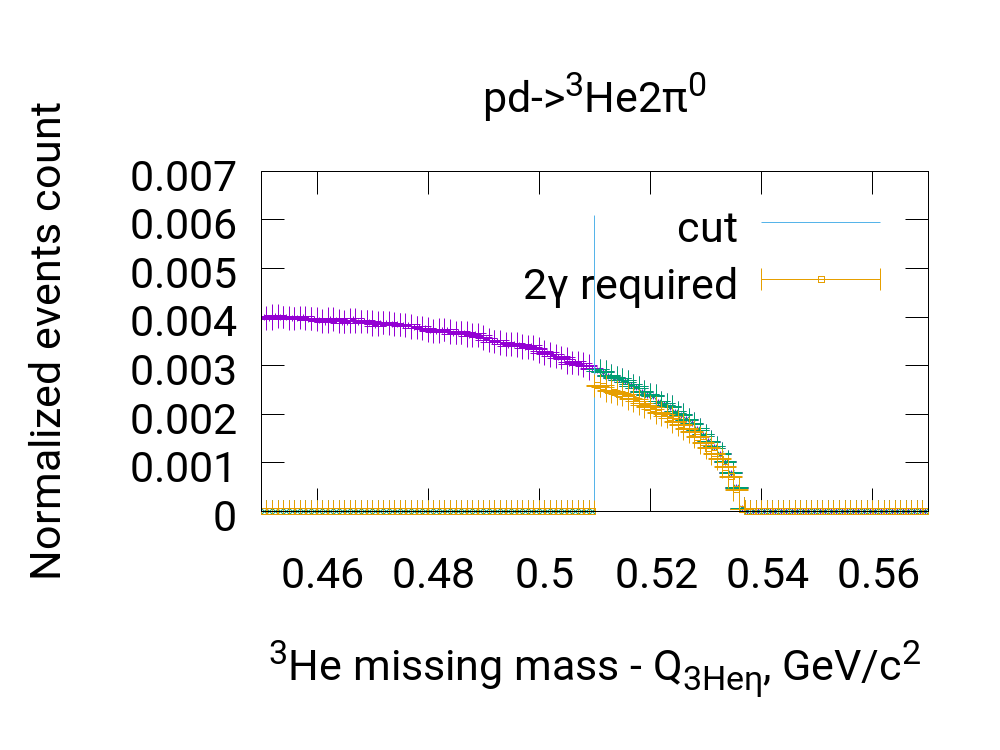}
\includegraphics[width=220pt]{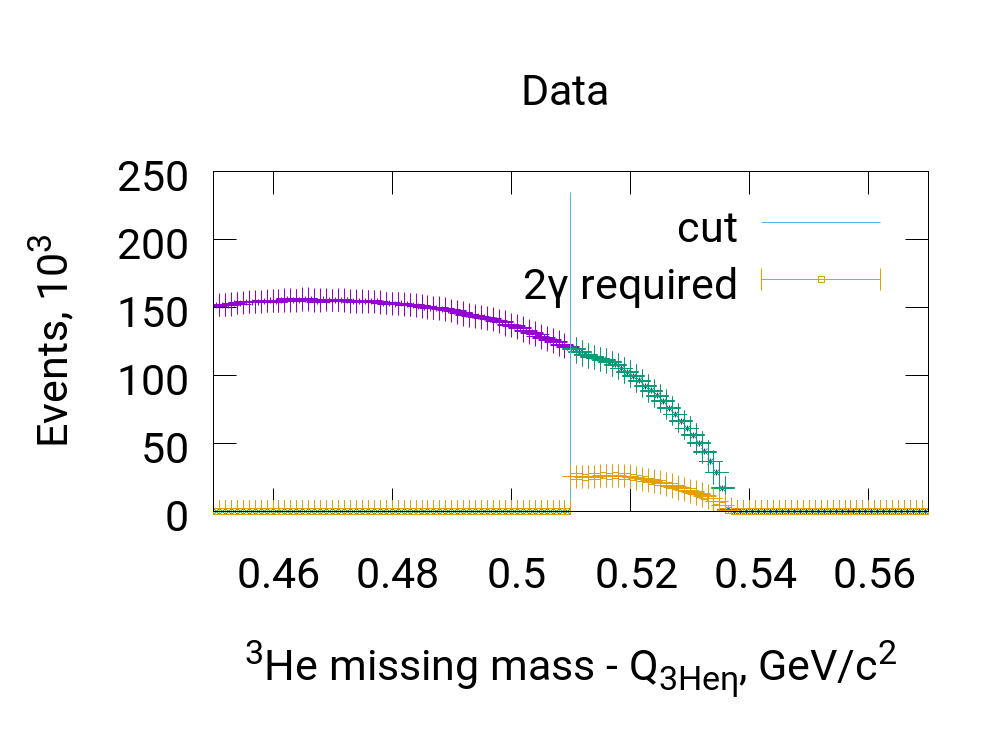}
\end{center}
\caption{
The distribution of $^3$$He$ missing mass corrected by $Q_{^3He\eta}$ obtained in $pd\rightarrow^3$$He2\gamma$
reaction analysis.
Vertical line shows the cut position.
Yellow points show the events count after the requirement of two $\gamma$ quanta in the Central Detector is provided.
As indicated in the legend above the pictures, the figure shows results of the analysis of data simulated for
$pd\rightarrow(^3$$He\eta)_{bound}\rightarrow^3$$He\gamma\gamma$, 
$pd\rightarrow^3$$He\eta\rightarrow^3$$He\gamma\gamma$, 
$pd\rightarrow^3$$He\pi^0\pi^0\rightarrow^3$$He\gamma\gamma\gamma\gamma$ 
and result of analysis of experimental data.
}
\label{two_gamma_3He_missing_mass_cut}
\end{figure}

\begin{figure}
	\begin{center}
		\includegraphics[width=210pt]{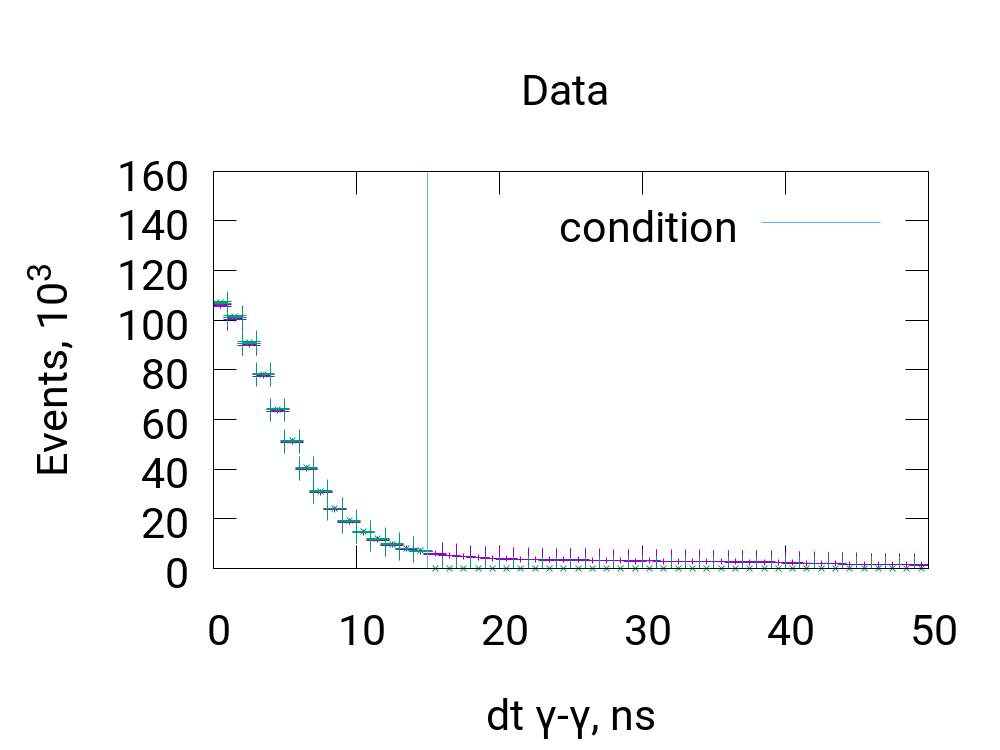}
		\includegraphics[width=220pt]{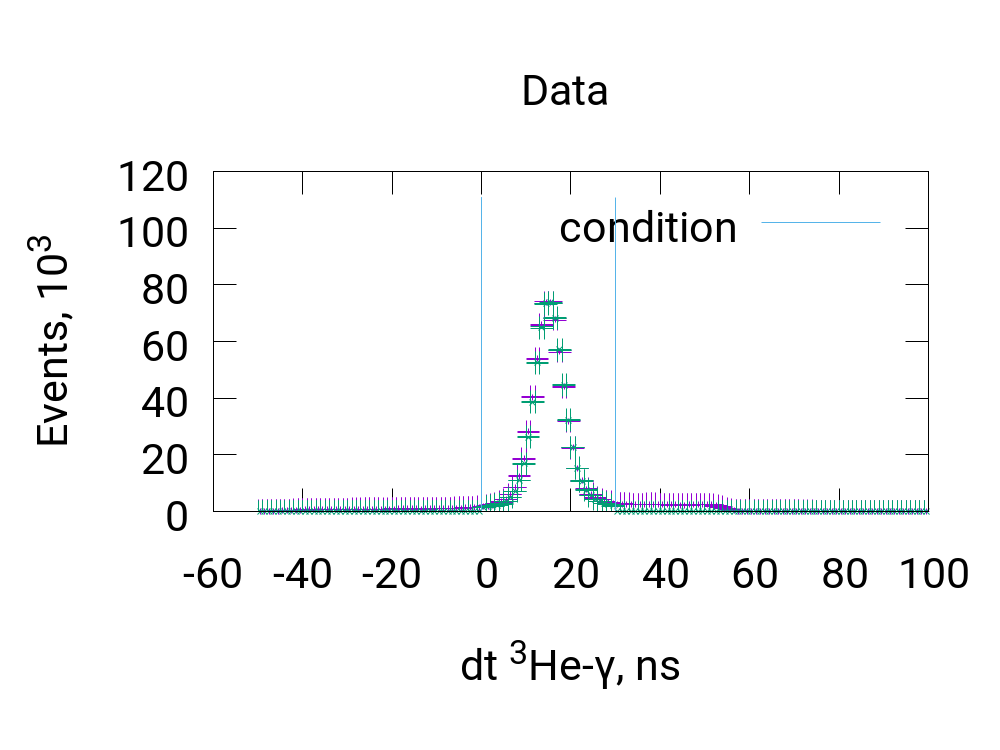}
	\end{center}
	\caption{
Time differences distribution plots showing the conditions applied for selecting proper $\gamma$ pair in Central Detector for $pd\rightarrow^3$$He2\gamma$ reaction analysis. 
Left: time difference between the $\gamma$ quanta;
right: time difference between $^3$$He$ track and the earliest signal from $\gamma$.
Vertical lines show the ranges used in the applied conditions.
Blue points show the distributions before applying the conditions and green points show the distributions after applying the conditions.
	}
	\label{two_gamma_time_condition}
\end{figure}

\begin{figure}
	\begin{center}
		\includegraphics[width=220pt]{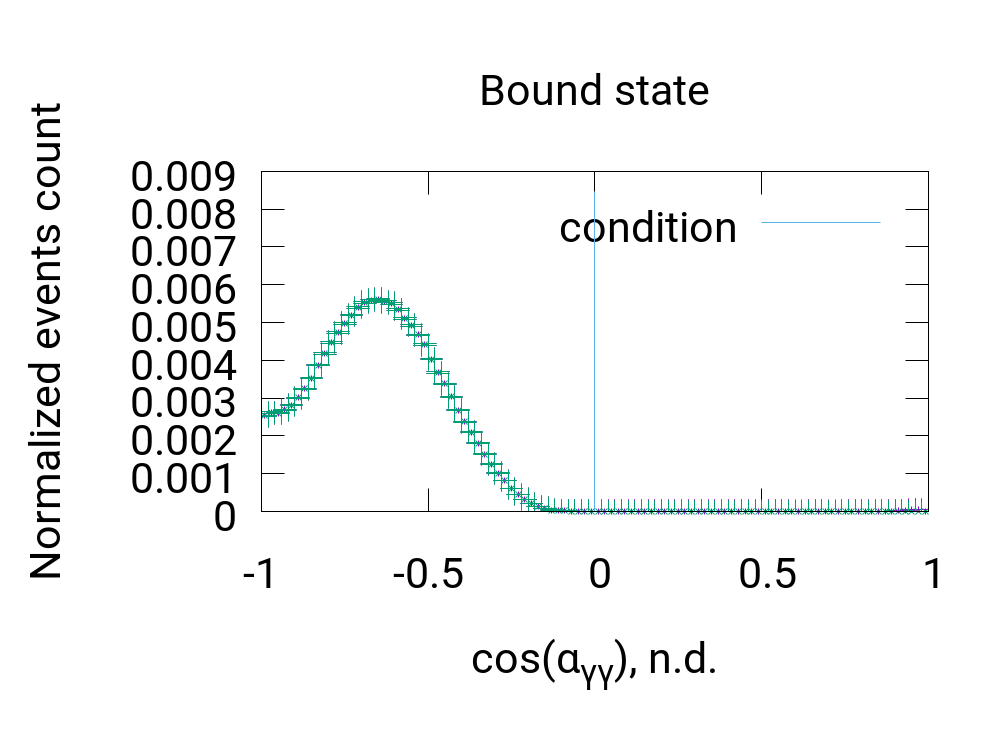}
		\includegraphics[width=220pt]{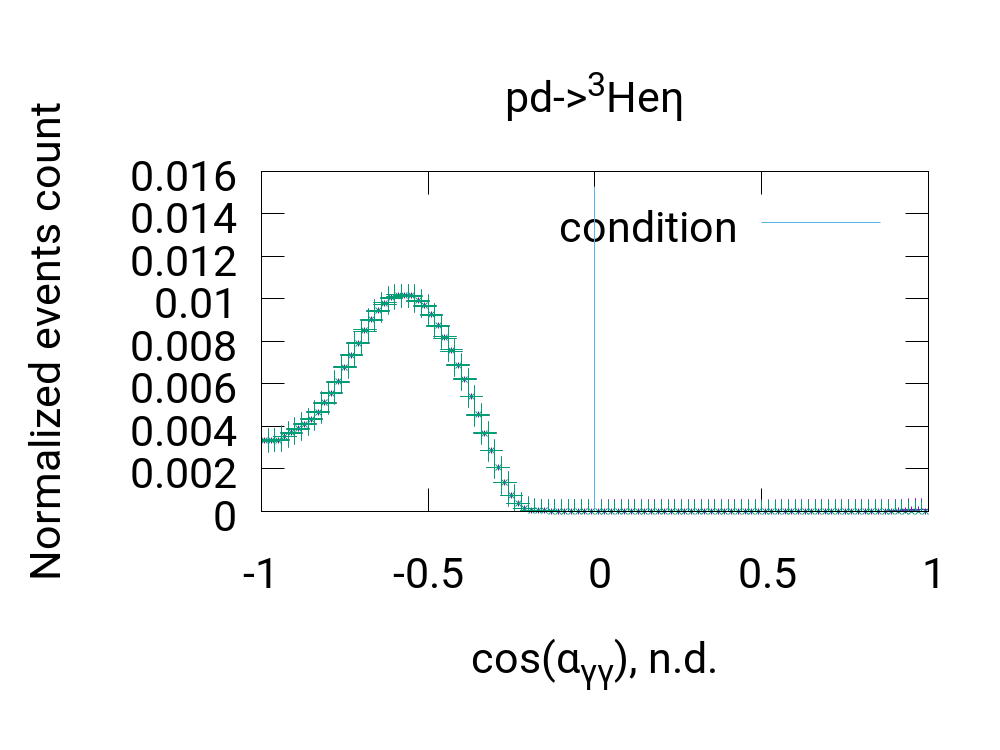}\\
		~\\~\\
		\includegraphics[width=220pt]{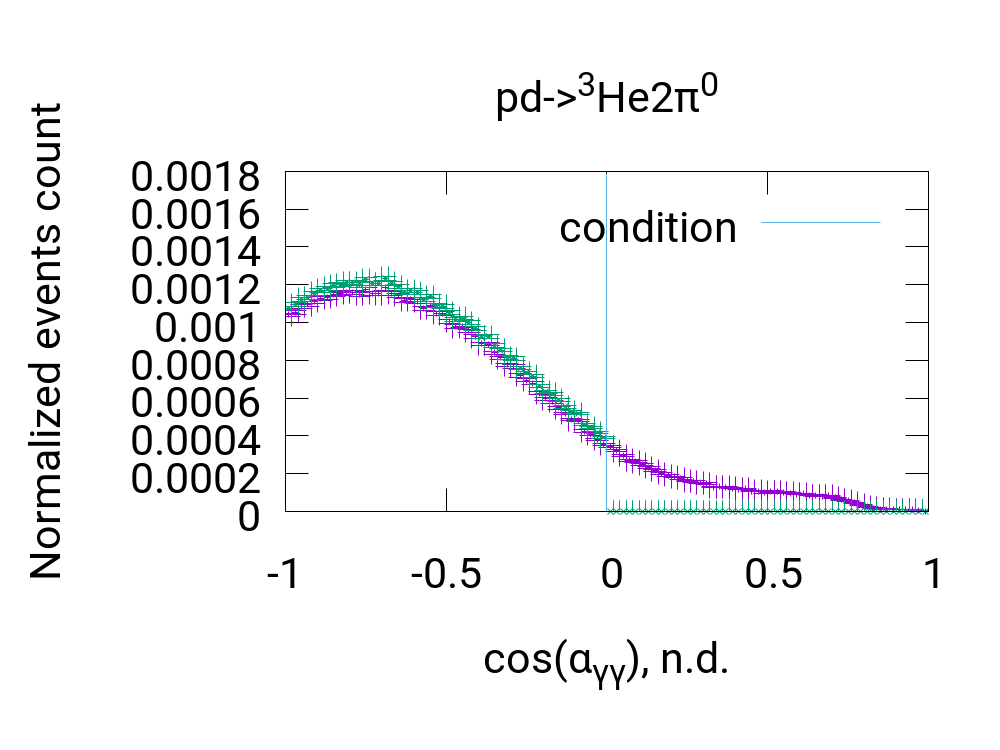}
		\includegraphics[width=220pt]{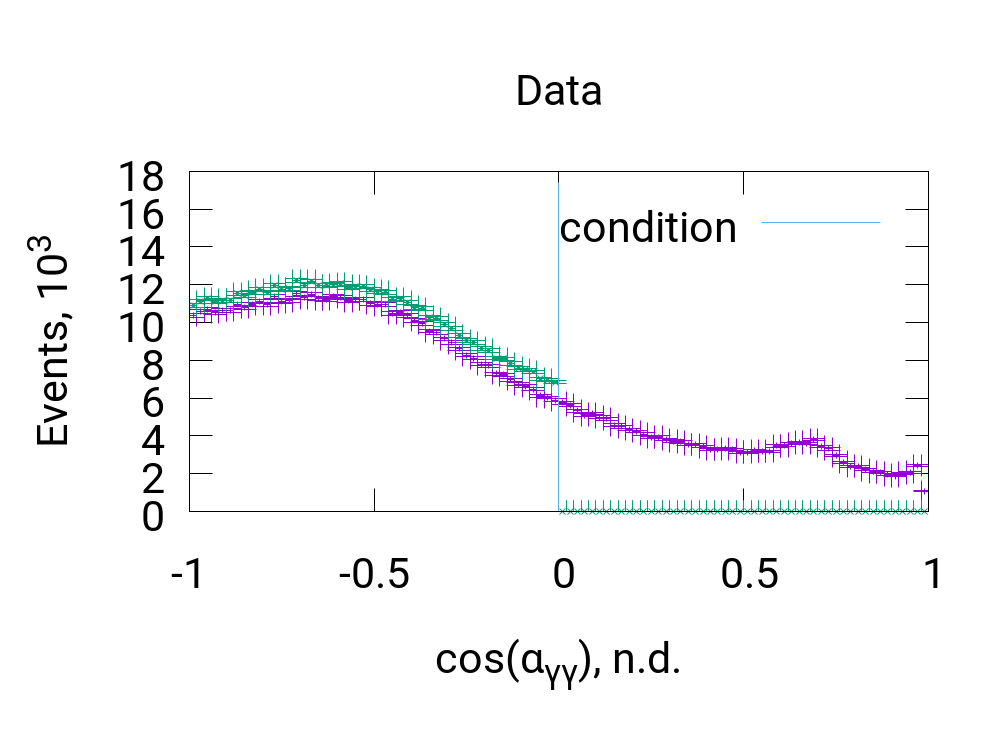}
	\end{center}
	\caption{
		$cos(\alpha_{\gamma\gamma})$ distribution obtained in
		 $pd\rightarrow^3$$He2\gamma$ reaction analysis.
		Vertical line shows the cut position.
		Monte Carlo simulations for
		$pd\rightarrow(^3$$He\eta)_{bound}\rightarrow^3$$He\gamma\gamma$, 
		$pd\rightarrow^3$$He\eta\rightarrow^3$$He\gamma\gamma$, 
		$pd\rightarrow^3$$He\pi^0\pi^0\rightarrow^3$$He\gamma\gamma\gamma\gamma$ 
		and data analysis.
Magenta points show the distributions without applying the conditions and green points show the distributions with applying the conditions.
	}
	\label{two_gamma_cos_alpha_condition}
\end{figure}

The first condition that is applied for the events acts on $^3$$He$ missing mass.
The momentum of the beam was varied in the range of the excess energies betwen $-70$ and $+30~MeV$.
To take into account 
differences in the total reaction energy $\sqrt{s}$, the 
$^3$$He$ missing mass is corrected by  $Q_{^3He\eta}$ (Fig.~\ref{two_gamma_3He_missing_mass_cut}).
Then $\gamma$ pair is searched among all $\gamma$ in the Central Detector.
It must fulfill the condition applied for time difference between $\gamma$ quanta (left panel of Fig.~\ref{two_gamma_time_condition}), time difference between $^3$$He$ and the quickest $\gamma$
(right panel of Fig.~\ref{two_gamma_time_condition}),
and condition applied for angle between $\gamma$ quanta directions $\alpha_{\gamma\gamma}$
(Fig.~\ref{two_gamma_cos_alpha_condition}).
If there are several possible combinations of $\gamma$ tracks fulfilling this condition,
the one having invariant mass closest to $(m_\eta + Q_{^3He\eta})$ value is chosen.
This value is equal to bound $\eta$ meson mass in frame of spectator model
(Sec.~\ref{he3eta_bound_simulation})
that is assumed for the bound state decay.

\begin{figure}
	\begin{center}
		\includegraphics[width=220pt]{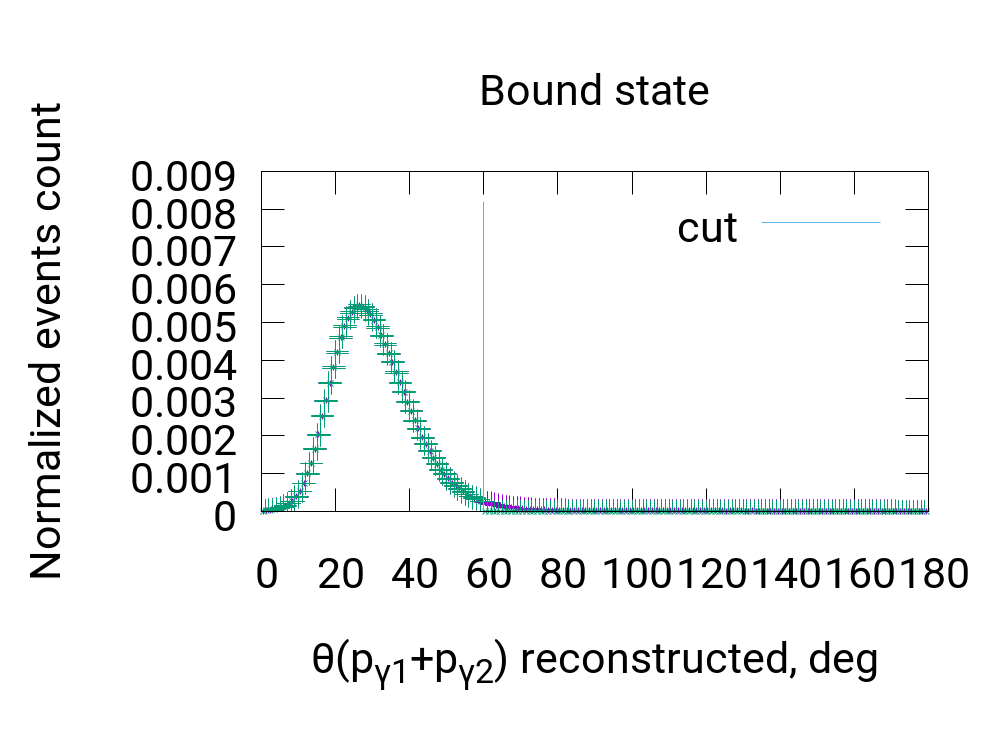}
		\includegraphics[width=220pt]{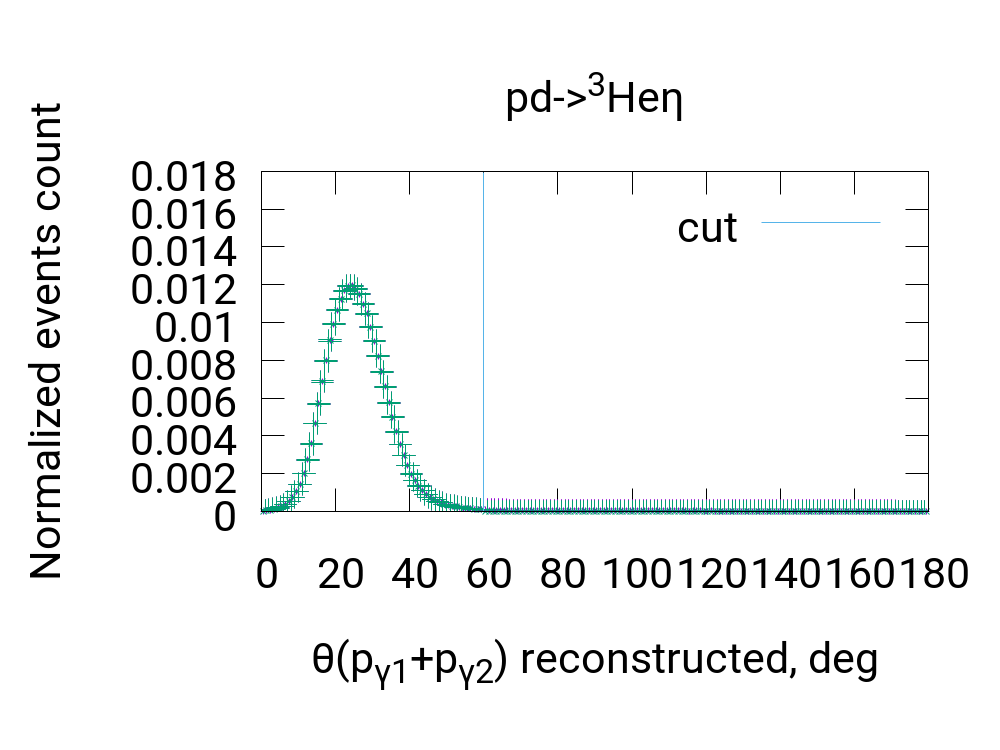}\\
		~\\~\\
		\includegraphics[width=220pt]{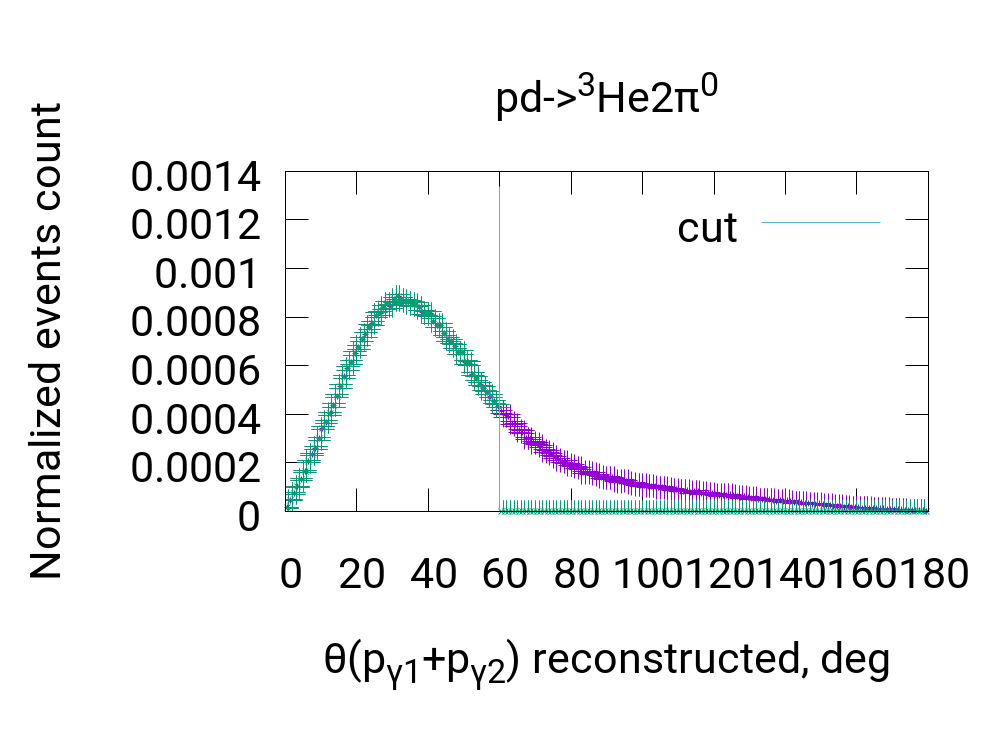}
		\includegraphics[width=220pt]{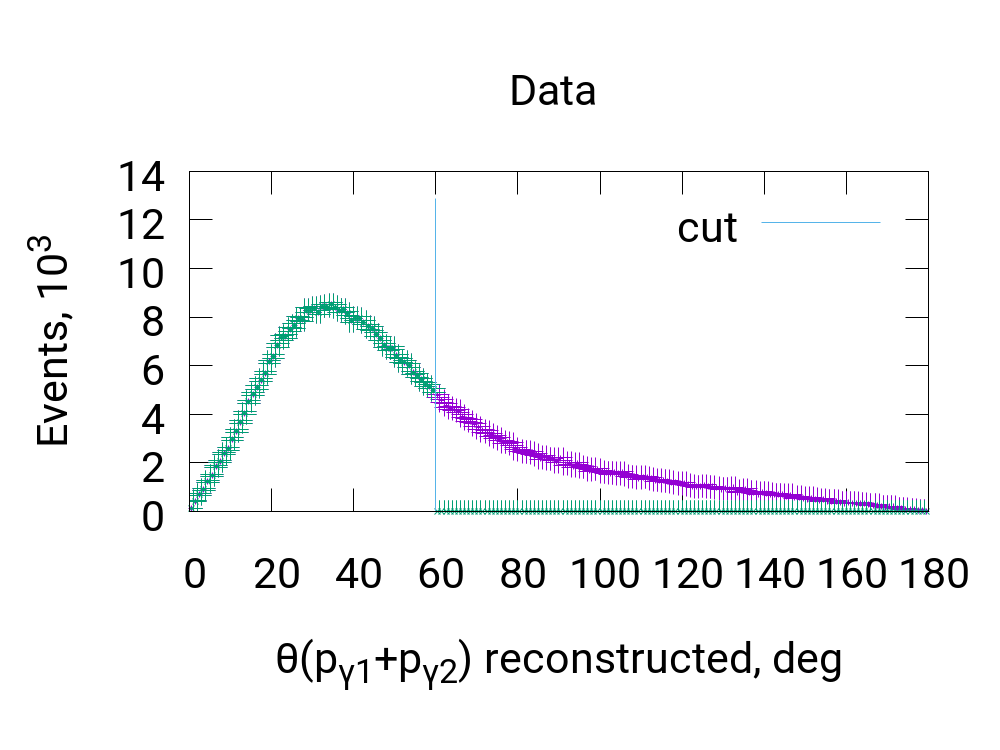}
	\end{center}
	\caption{
		$\theta(\vec{p}_{\gamma_1}+\vec{p}_{\gamma_2})$ distribution obtained in $pd\rightarrow^3$$He2\gamma$ reaction analysis.
		Vertical line shows the cut position.
As indicated in the legend above the pictures, the figure shows results of the analysis of data simulated for
$pd\rightarrow(^3$$He\eta)_{bound}\rightarrow^3$$He\gamma\gamma$, 
$pd\rightarrow^3$$He\eta\rightarrow^3$$He\gamma\gamma$, 
$pd\rightarrow^3$$He\pi^0\pi^0\rightarrow^3$$He\gamma\gamma\gamma\gamma$ 
and result of analysis of experimental data.
Magenta points show the distributions before applying the conditions and green points show the distributions after applying the conditions.
	}
	\label{two_gamma_eta_theta_cut}
\end{figure}

\begin{figure}
	\begin{center}
		\includegraphics[width=220pt]{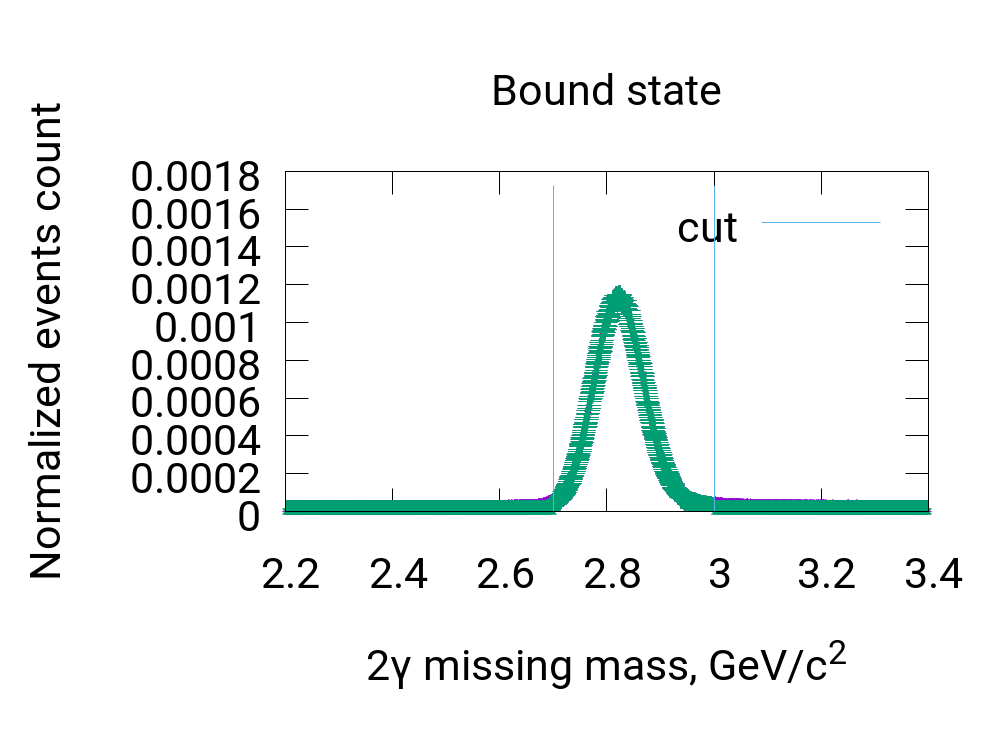}
		\includegraphics[width=220pt]{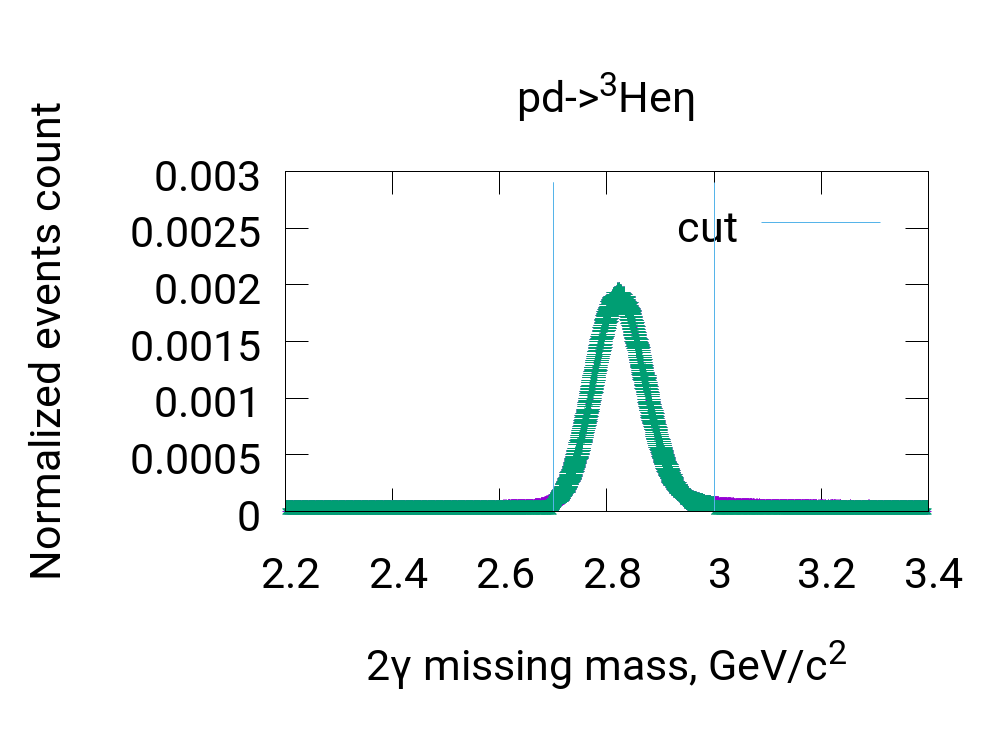}\\
		~\\~\\
		\includegraphics[width=220pt]{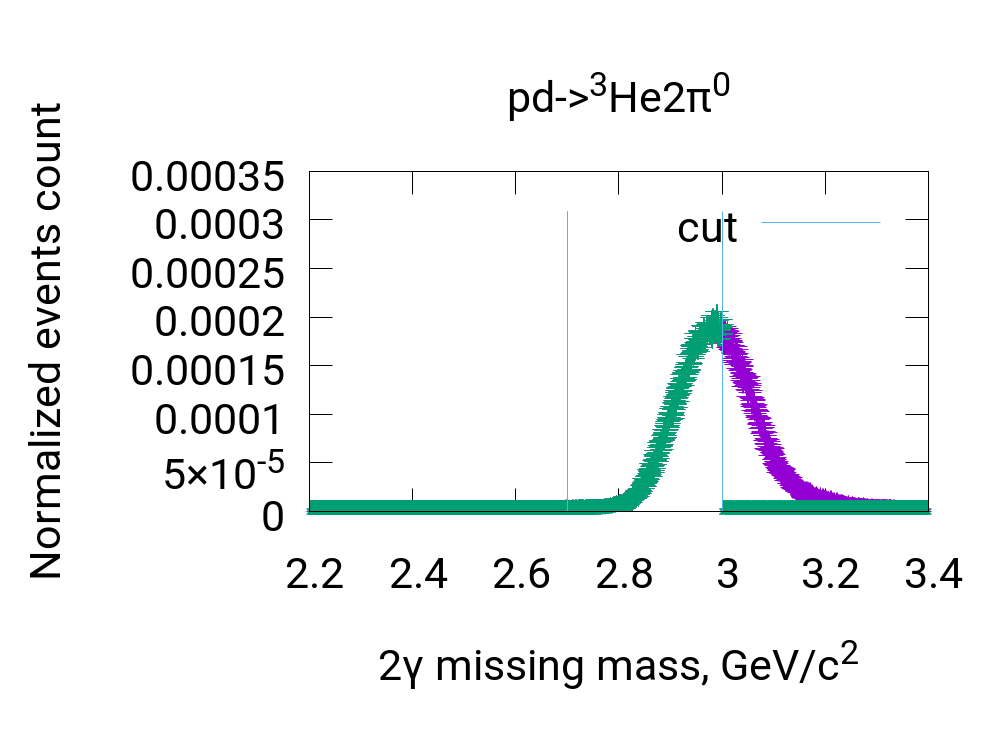}
		\includegraphics[width=220pt]{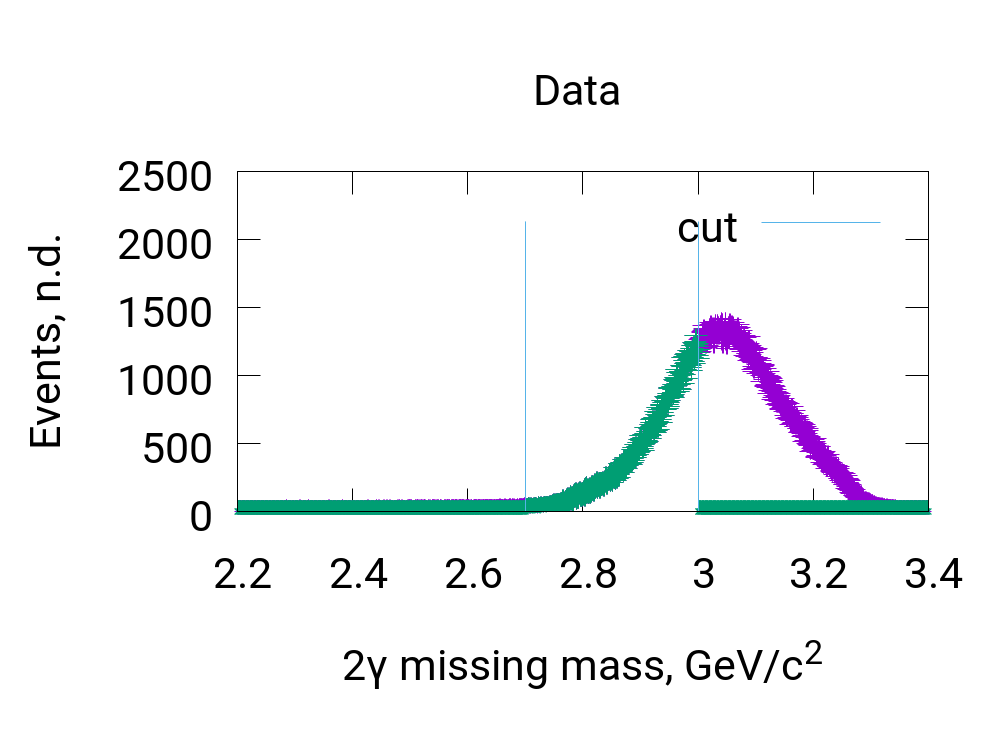}
	\end{center}
	\caption{
		$\gamma\gamma$ missing mass distribution obtained in
		 $pd\rightarrow^3$$He2\gamma$ reaction analysis.
		Vertical lines show the cuts positions.
As indicated in the legend above the pictures, the figure shows results of the analysis of data simulated for
$pd\rightarrow(^3$$He\eta)_{bound}\rightarrow^3$$He\gamma\gamma$, 
$pd\rightarrow^3$$He\eta\rightarrow^3$$He\gamma\gamma$, 
$pd\rightarrow^3$$He\pi^0\pi^0\rightarrow^3$$He\gamma\gamma\gamma\gamma$ 
and result of analysis of experimental data.
Magenta points show the distributions before applying the conditions and green points show the distributions after applying the conditions.
	}
	\label{two_gamma_ggmm_cut}
\end{figure}

\begin{figure}
	\begin{center}
		\includegraphics[width=220pt]{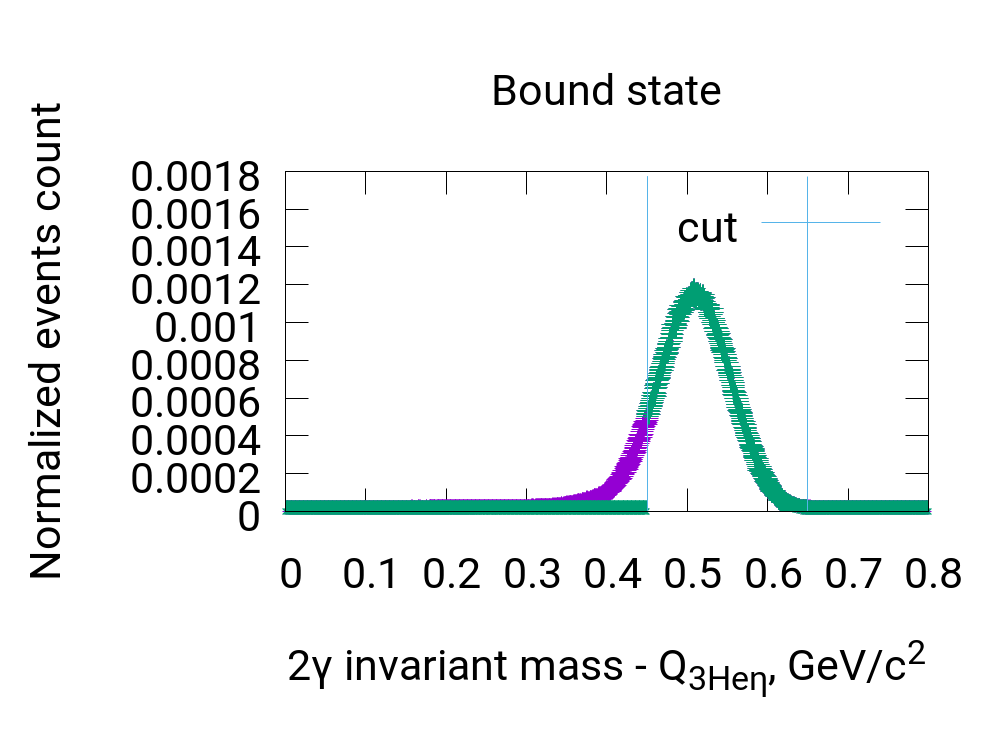}
		\includegraphics[width=220pt]{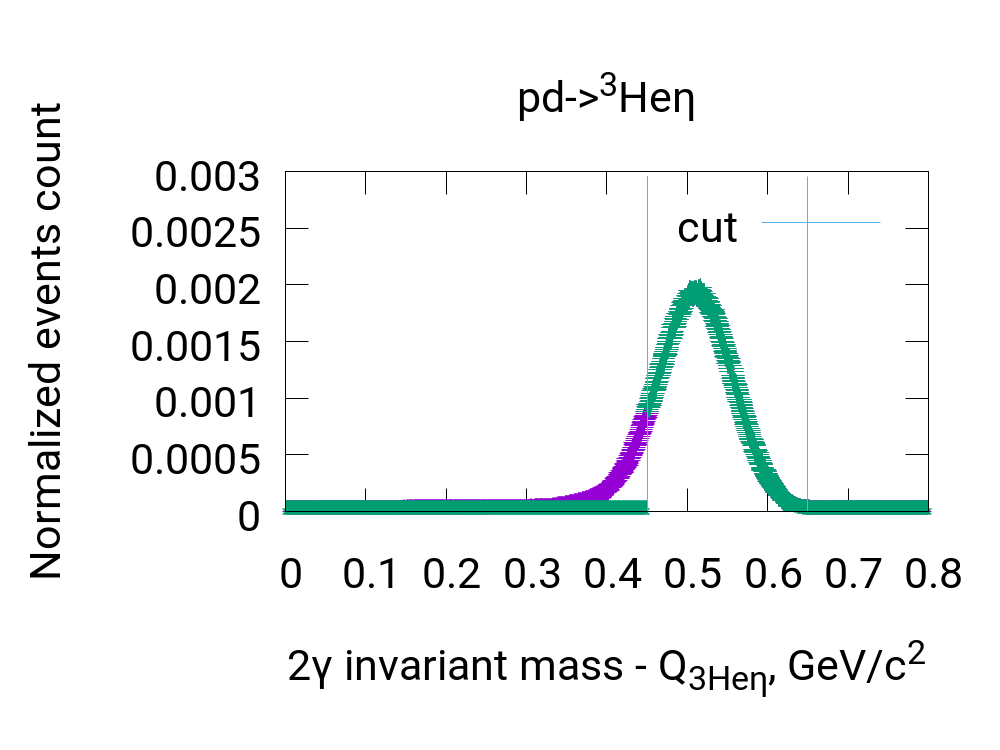}\\
		~\\~\\
		\includegraphics[width=220pt]{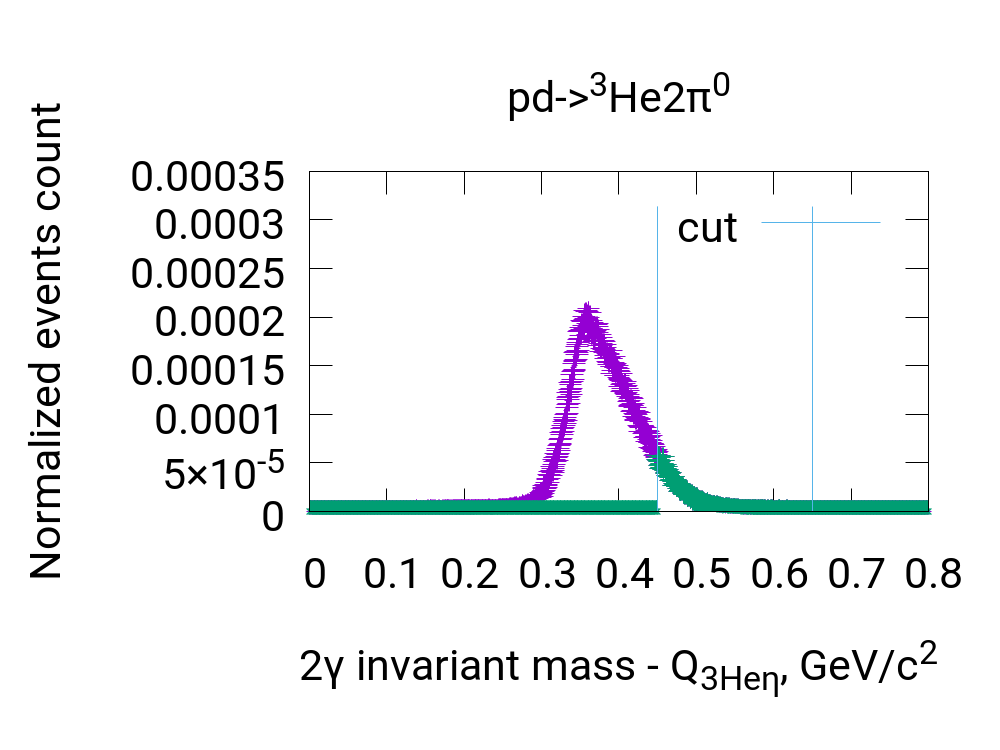}
		\includegraphics[width=220pt]{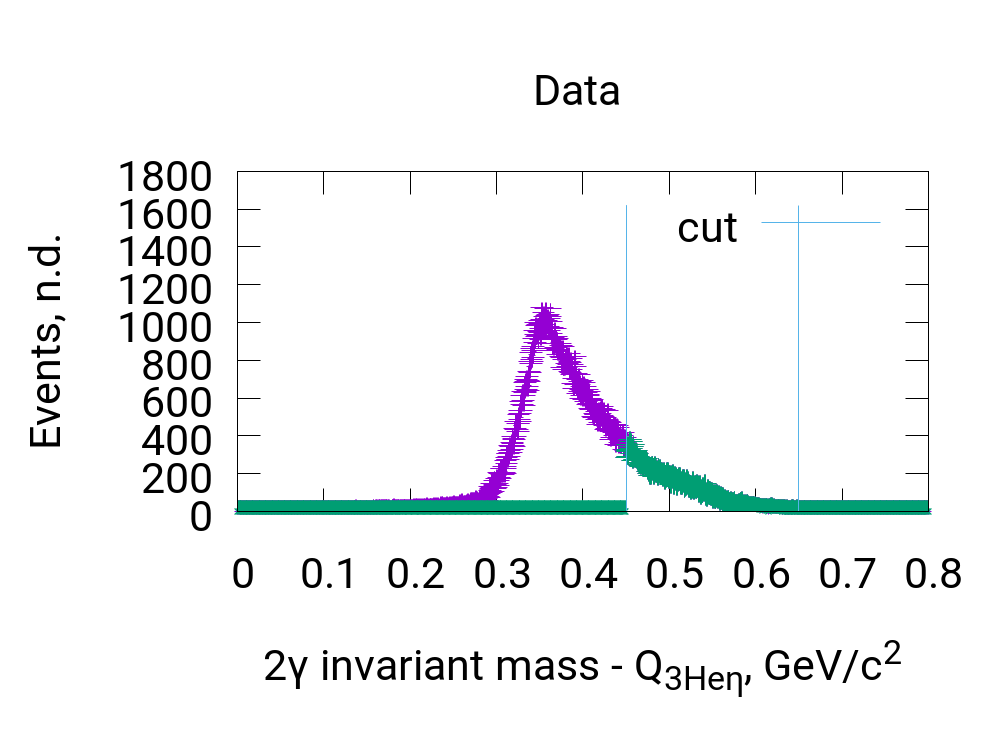}
	\end{center}
	\caption{
		The distribution of $\gamma\gamma$ invariant mass corrected by $Q_{^3He\eta}$ obtained in
		$pd\rightarrow^3$$He2\gamma$ reaction analysis.
		Vertical lines show the cuts positions.
As indicated in the legend above the pictures, the figure shows results of the analysis of data simulated for
$pd\rightarrow(^3$$He\eta)_{bound}\rightarrow^3$$He\gamma\gamma$, 
$pd\rightarrow^3$$He\eta\rightarrow^3$$He\gamma\gamma$, 
$pd\rightarrow^3$$He\pi^0\pi^0\rightarrow^3$$He\gamma\gamma\gamma\gamma$ 
and result of analysis of experimental data.
Magenta points show the distributions before applying the conditions and green points show the distributions after applying the conditions.
	}
	\label{two_gamma_ggim_cut}
\end{figure}

In order to suppress background, additional cuts are applied.
The first one is $\theta(\vec{p}_{\gamma_1}~+~\vec{p}_{\gamma_2})$ cut (Fig.~\ref{two_gamma_eta_theta_cut}).
$\theta$ angle means the angle between the beam axis and the direction of $\gamma$ quanta momenta sum vector.
Then cuts on $\gamma$ quanta missing mass (Fig.~\ref{two_gamma_ggmm_cut}) 
and invariant mass (Fig.~\ref{two_gamma_ggim_cut}) are applied.

As far as missing mass conditions are sensitive to possible beam momentum correction inaccuracy, they cannot be strict enough to suppress all background.
Hereby, the additional angular conditions independent on beam momentum correction have been provided.


\begin{figure}
	\begin{center}
		\includegraphics[width=220pt]{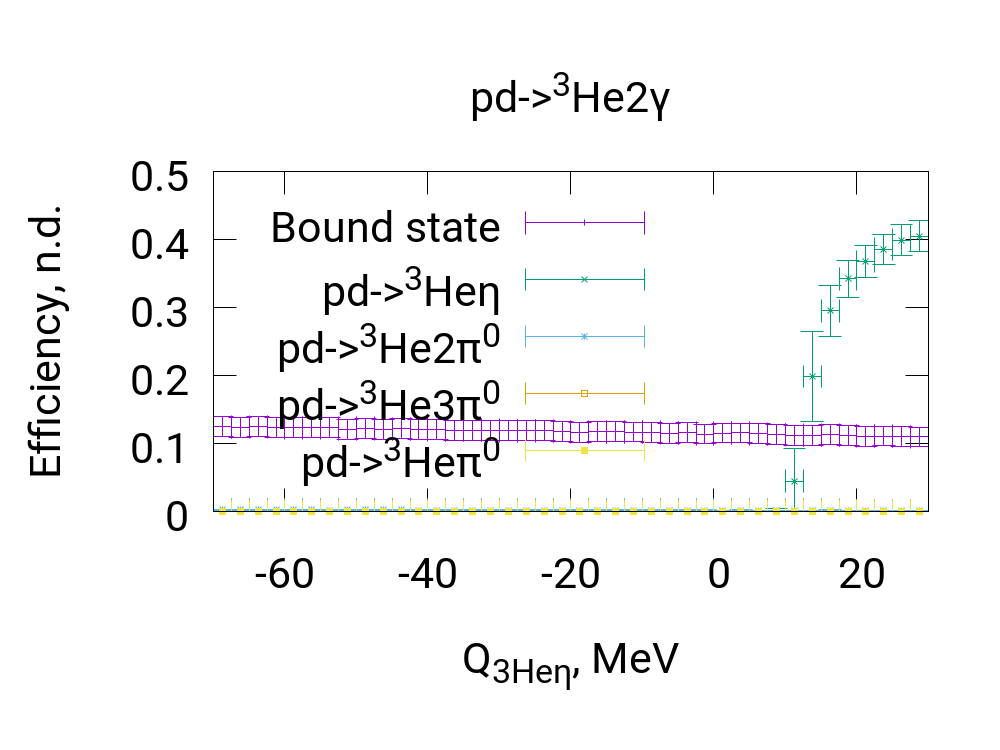}
		\includegraphics[width=220pt]{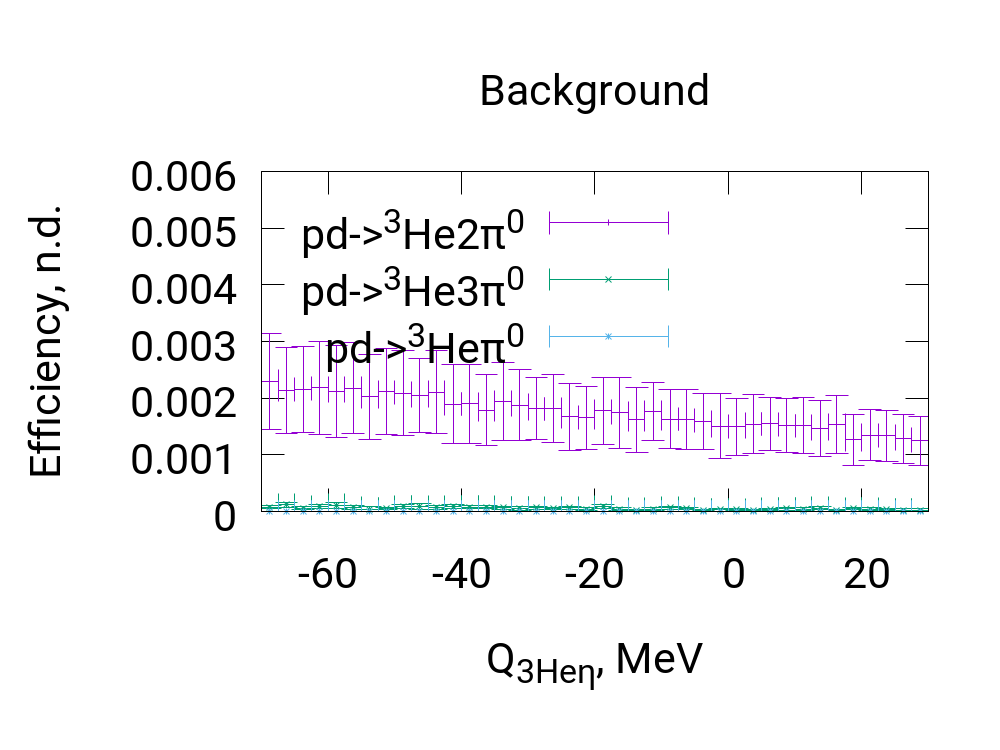}
	\end{center}
	\caption{
		The efficiency for different reactions
		when applying selection criteria defined for the
		$pd\rightarrow^3$$He2\gamma$ reaction analysis.
		Left: wider scale where the signal reactions are visible;
		Right: more narrow scale to see the background reactions.
		Systematic uncertainties are shown.
	}
	\label{two_gamma_acceptance}
\end{figure}

\begin{figure}
	\begin{center}
		\includegraphics[width=220pt]{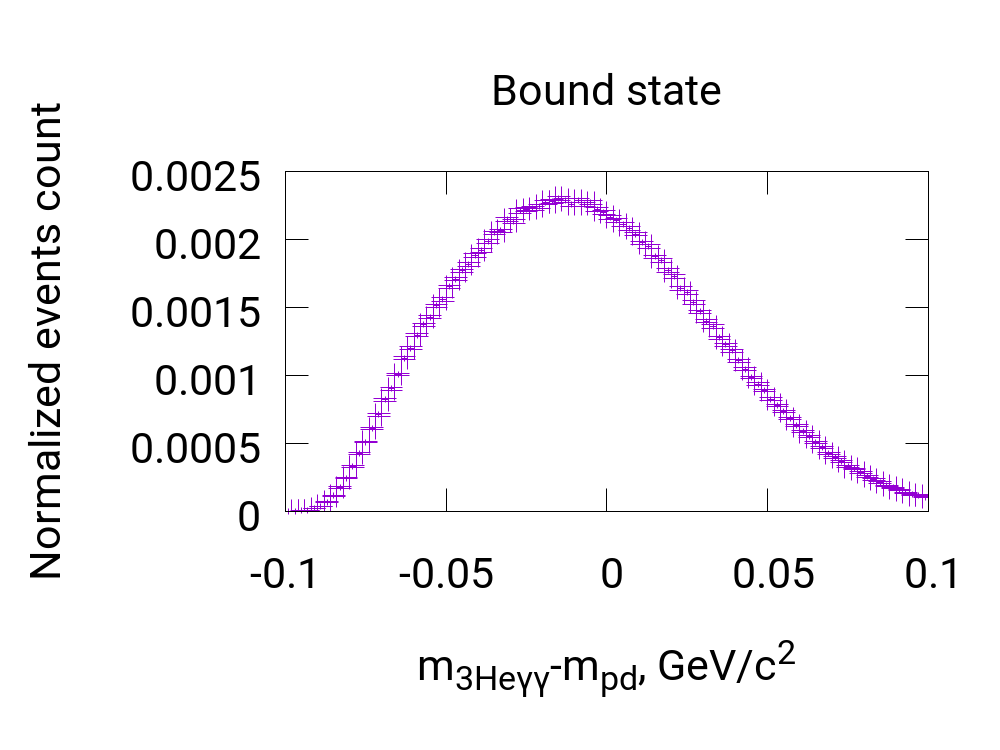}
		\includegraphics[width=220pt]{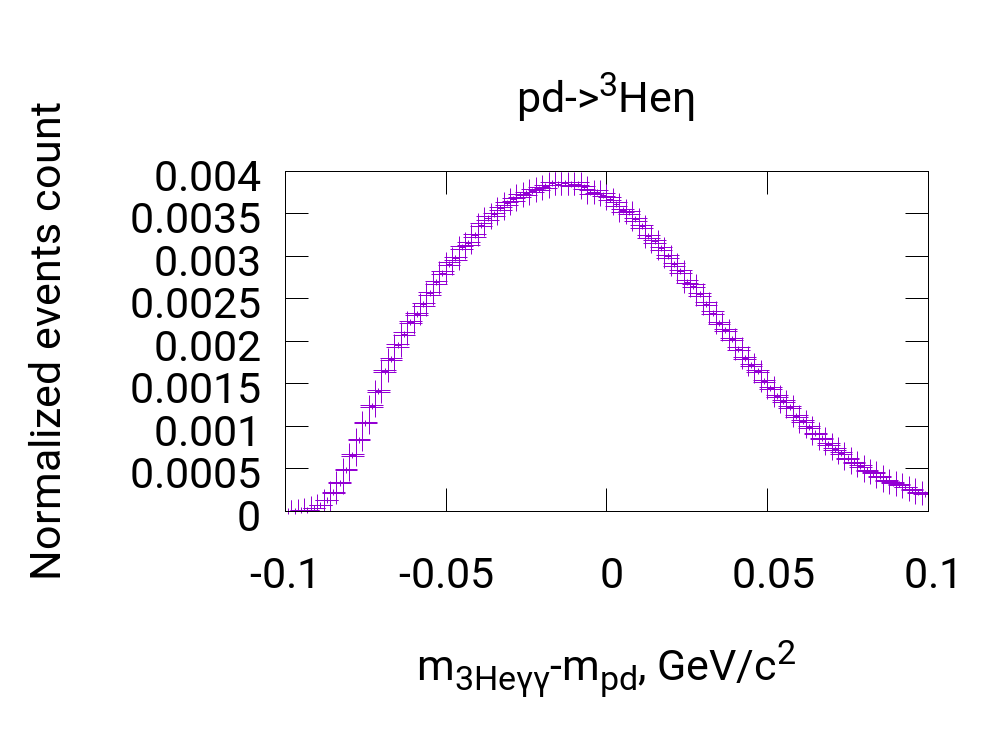}\\
		~\\~\\
		\includegraphics[width=220pt]{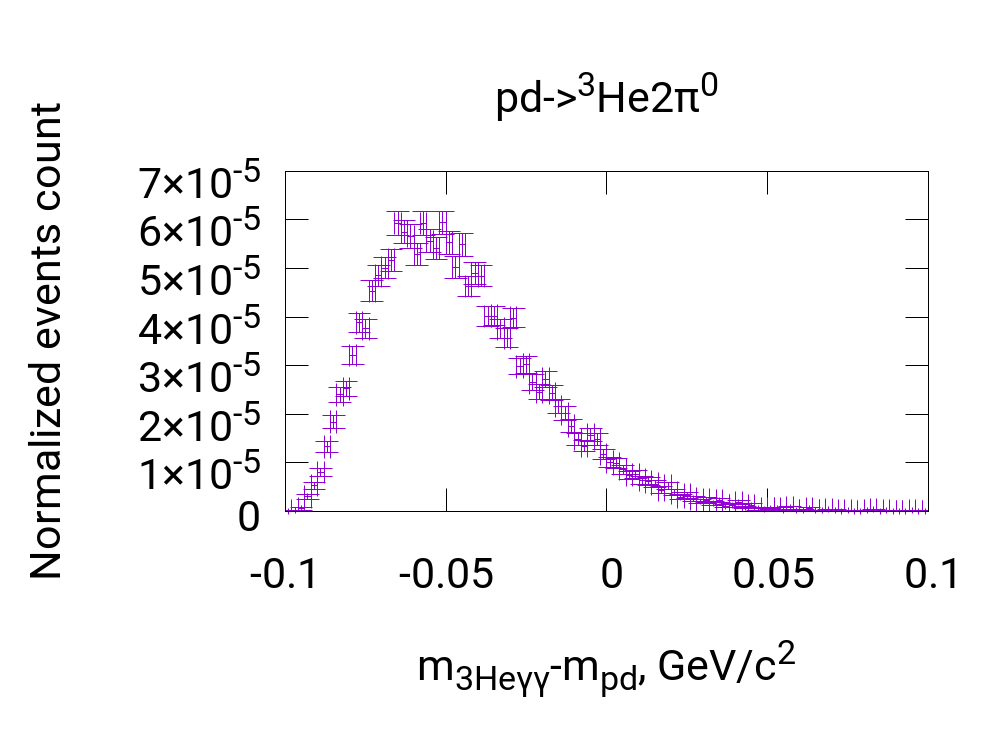}
		\includegraphics[width=220pt]{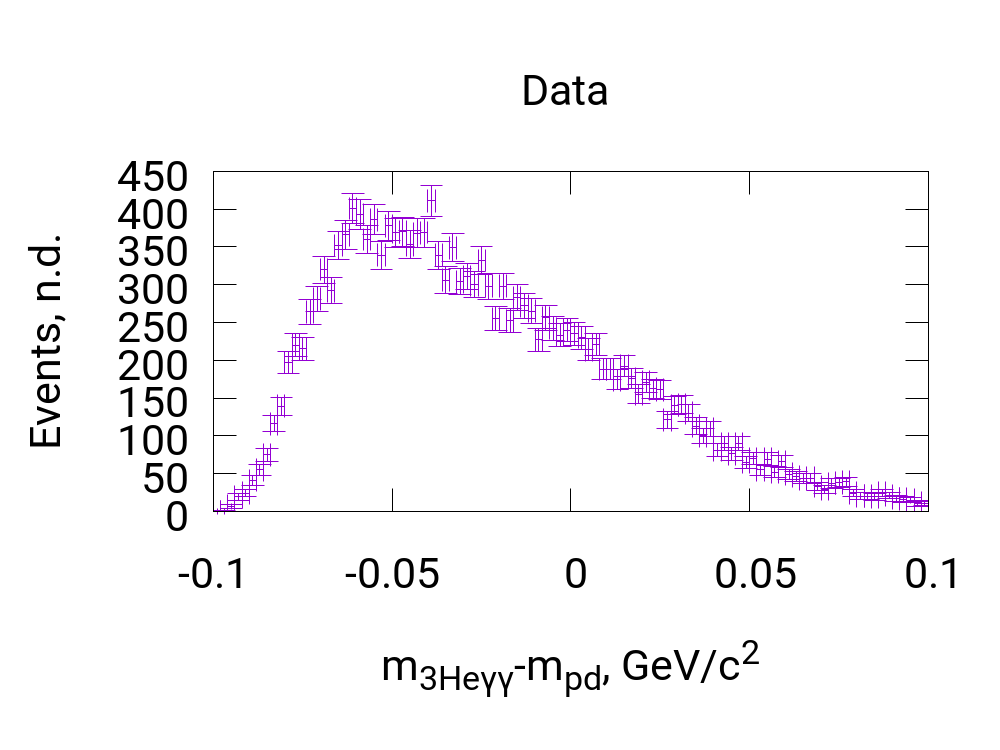}
	\end{center}
	\caption{
		Distribution of $m_{^3He\gamma\gamma}-m_{pd}$ obtained after all conditions in $pd\rightarrow^3$$He2\gamma$ reaction analysis are applied.
As indicated in the legend above the pictures, the figure shows results of the analysis of data simulated for
$pd\rightarrow(^3$$He\eta)_{bound}\rightarrow^3$$He\gamma\gamma$, 
$pd\rightarrow^3$$He\eta\rightarrow^3$$He\gamma\gamma$, 
$pd\rightarrow^3$$He\pi^0\pi^0\rightarrow^3$$He\gamma\gamma\gamma\gamma$ 
and result of analysis of experimental data.
	}
	\label{two_gamma_tim}
\end{figure}

These conditions reduce background reactions registration efficiency to the level below $0.5\%$ while the signal reaction efficiency is about $10\%$ (Fig.~\ref{two_gamma_acceptance}).
For $Q_{^3He\eta}~>~10~MeV$, $pd\rightarrow^3$$He\eta$ reaction is visible for the current analysis algorithm and the efficiency is increasing up to almost $40\%$.
The shape of total invariant mass difference $m_{^3He\gamma\gamma}-m_{pd}$ distribution plot is in agreement with assumption that mainly $pd\rightarrow^3$$He\pi^0\pi^0$ background reaction is observed (Fig.~\ref{two_gamma_tim}).

\begin{figure}
\begin{center}
		\includegraphics[width=300pt]{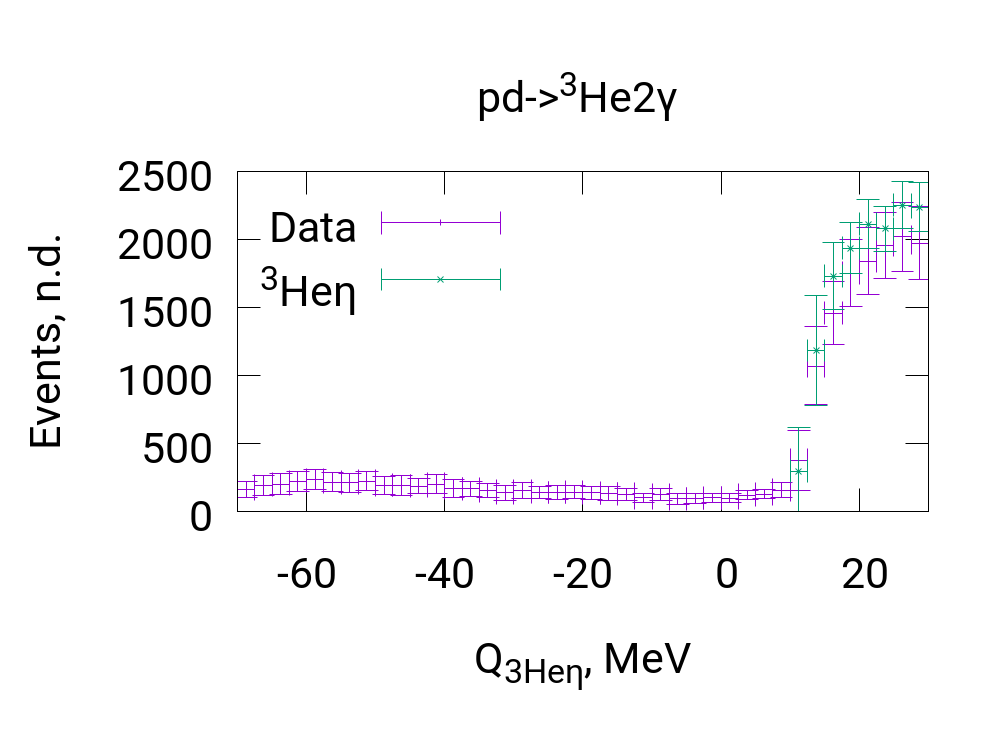}
\end{center}
\caption{
The dependence of events count on $Q_{^3He\eta}$ for 
$pd\rightarrow^3$$He2\gamma$ reaction.
Blue color: registered events count for full $Q_{^3He\eta}$ range; 
green color: expected $pd\rightarrow^3$$He\eta$ events count calculated by the Eq.~\ref{he3eta_contrib_estimation}.
Shown error bars include both statistical and systematic uncertainties (Sec.~\ref{reactions_systematics}).
}
	\label{two_gamma_events}
\end{figure}

The events count in excess energy region $Q_{^3He\eta}~>~10~MeV$ is compared with $pd\rightarrow^3$$He\eta$ reaction events count estimated by the formula
\begin{equation}
	N_{^3He\eta} = \frac{\int L dt ~\epsilon_{^3He\eta} ~\sigma_{^3He\eta}}{S_{trigger}},
	\label{he3eta_contrib_estimation}
\end{equation}
where all symbols have the same meaning like in Eq.~\ref{luminosity_description}.
Both curves are shown in Fig.~\ref{two_gamma_events} and are in agreement.
\newpage

\section{The analysis of $pd\rightarrow^3$$He6\gamma$ reaction}

\label{3He_6gamma_analysis}
\begin{figure}
	\begin{center}
		\includegraphics[width=220pt]{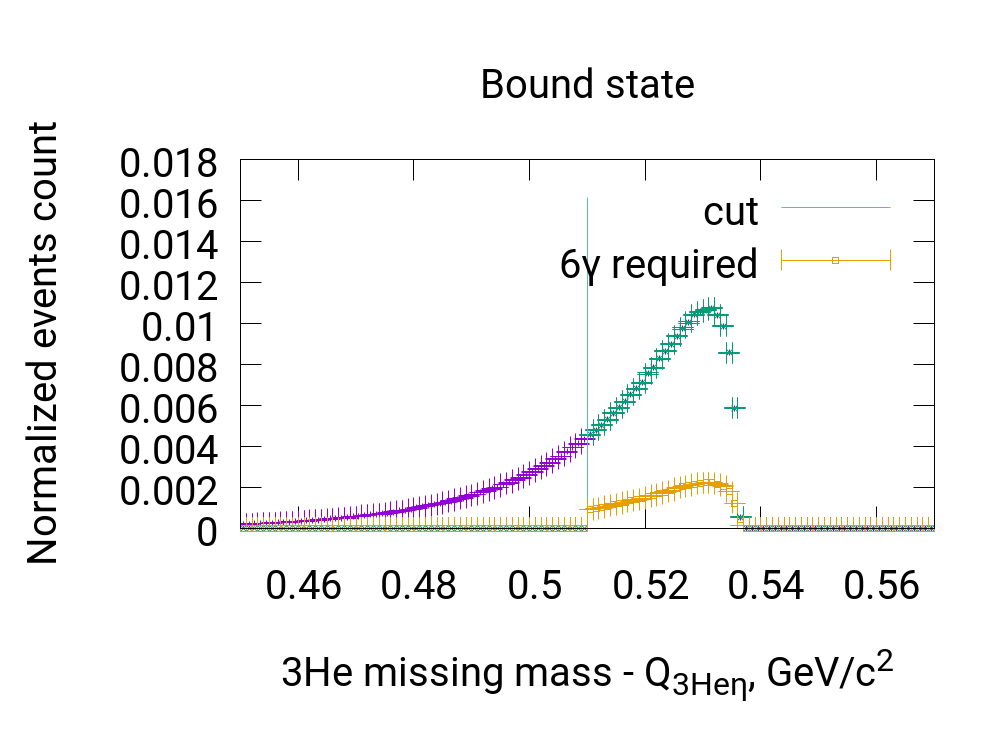}
		\includegraphics[width=220pt]{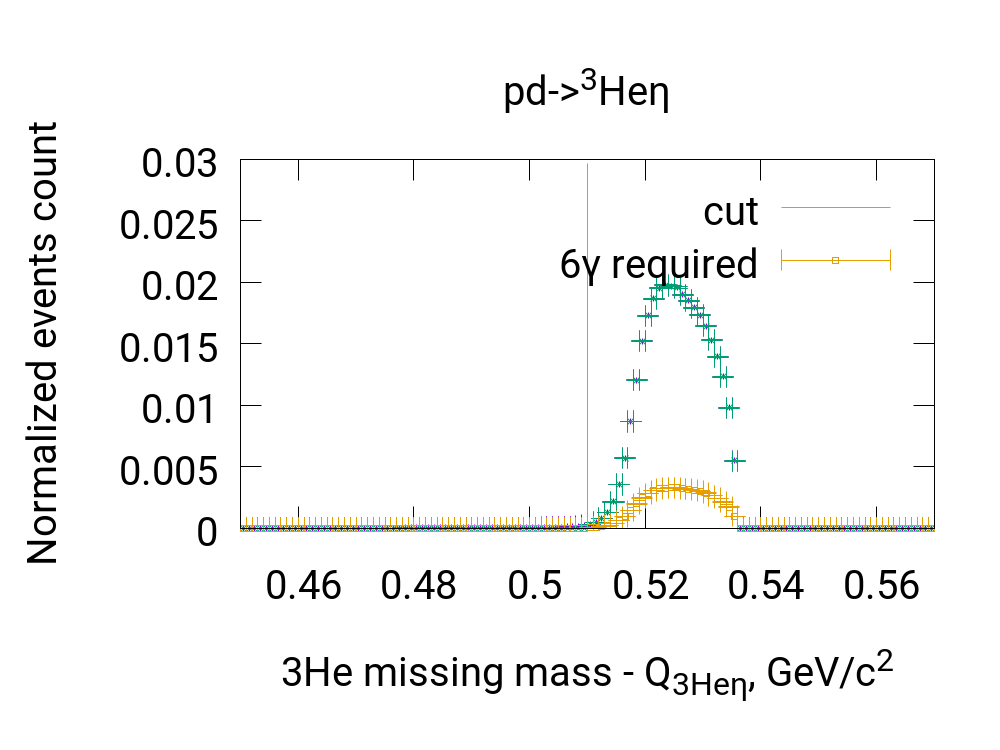}\\
		~\\
		\includegraphics[width=220pt]{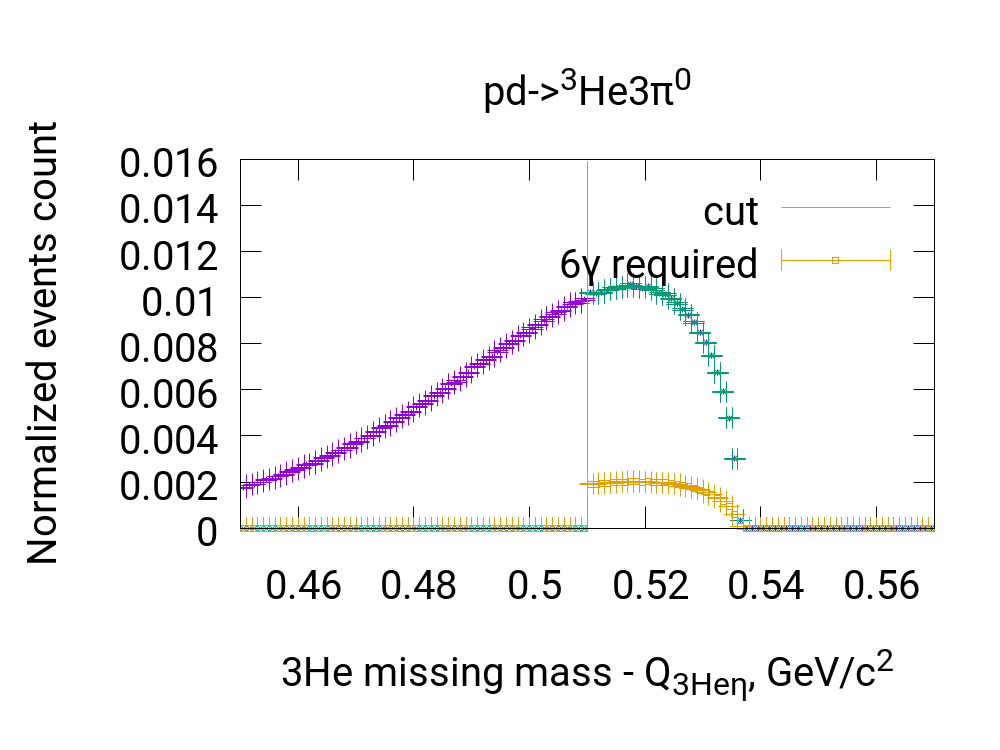}
		\includegraphics[width=220pt]{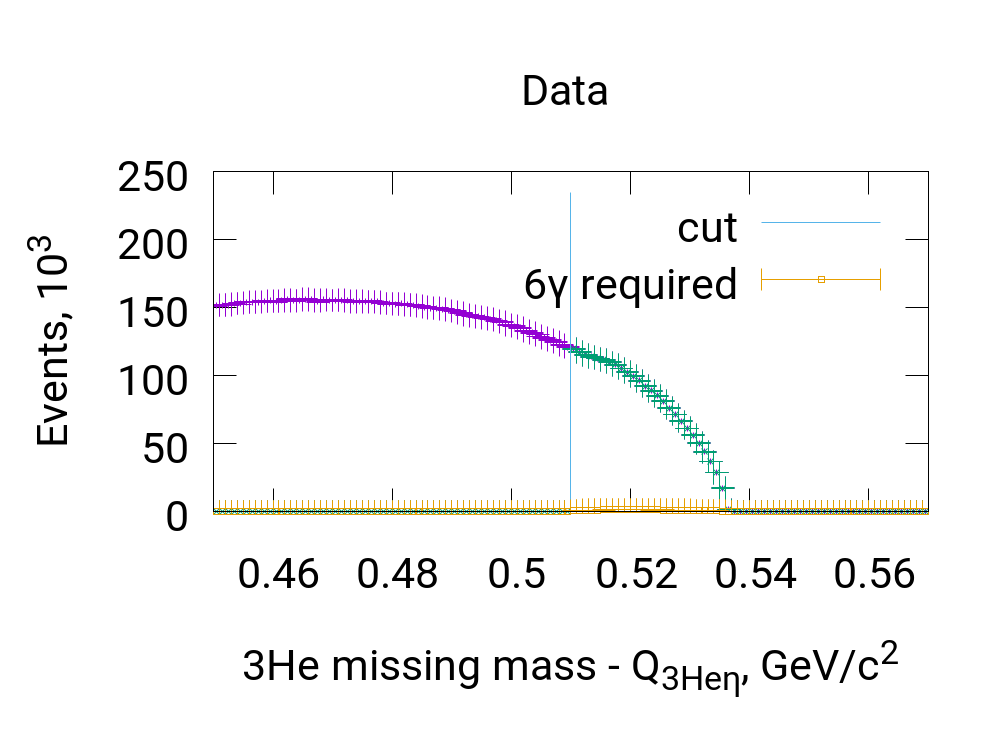}
	\end{center}
	\caption{
The distribution of $^3$$He$ missing mass corrected by $Q_{^3He\eta}$ obtained in
$pd\rightarrow^3$$He6\gamma$ reaction analysis.
Vertical line shows the cut position.
Yellow points show events count after six $\gamma$ quanta in Central Detector are requested.
As indicated in the legend above the pictures, the figure shows results of the analysis of data simulated for
$pd\rightarrow(^3$$He\eta)_{bound}\rightarrow^3$$He6\gamma$, 
$pd\rightarrow^3$$He\eta\rightarrow^3$$He6\gamma$, 
$pd\rightarrow^3$$He\pi^0\pi^0\pi^0\rightarrow^3$$He6\gamma$
and result of analysis of experimental data.
	}
	\label{six_gamma_3He_missing_mass_cut}
\end{figure}

Like in $pd\rightarrow^3$$He2\gamma$ reaction analysis (Sec.~\ref{3He_2gamma_analysis}), trigger number $10$ corresponding to the condition of at least one charged particle track in Forward Detector was used (Sec.~\ref{trigger_table}).
The same condition like in Sec.~\ref{3He_2gamma_analysis} on missing mass was used (Fig.~\ref{six_gamma_3He_missing_mass_cut}).

\begin{figure}
	\begin{center}
		\includegraphics[width=220pt]{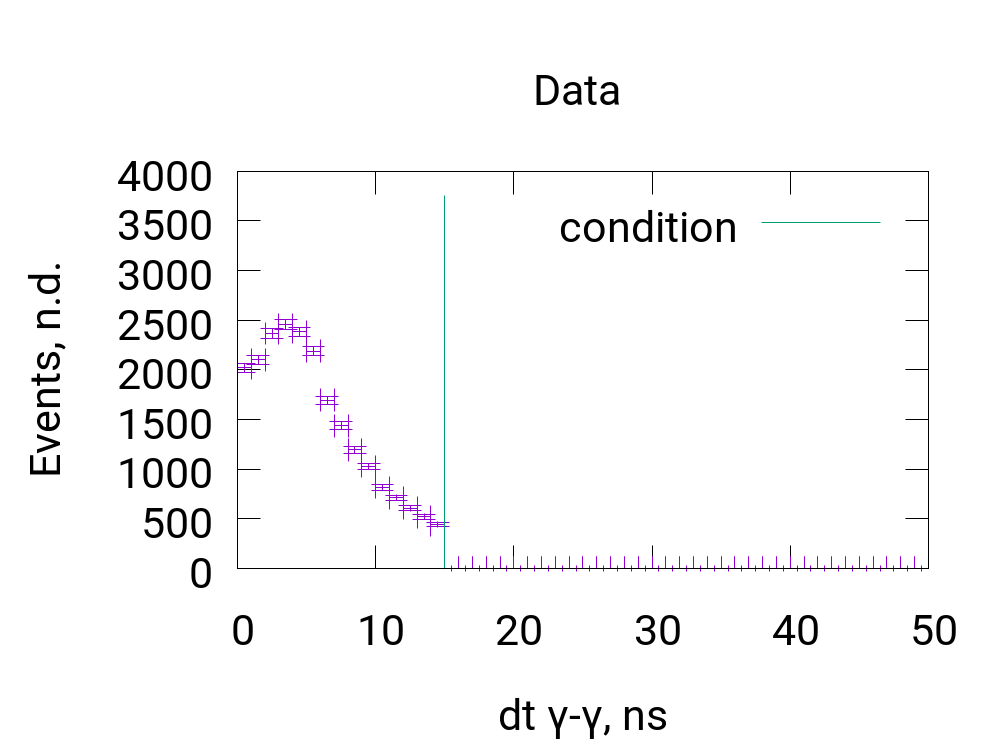}
		\includegraphics[width=220pt]{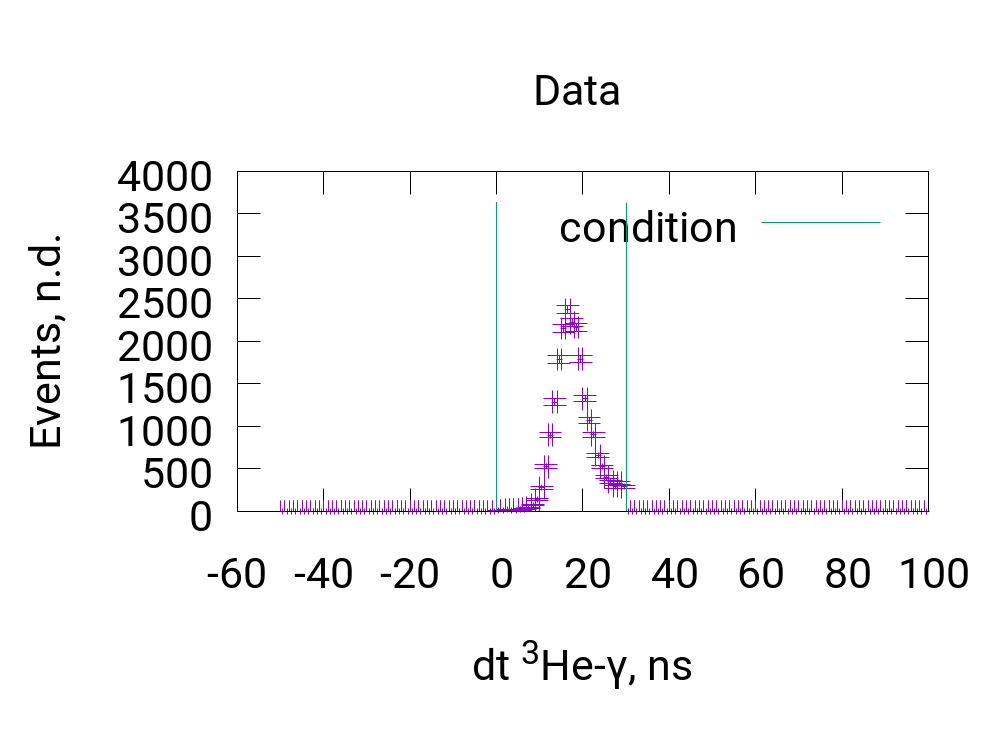}
	\end{center}
	\caption{
		Time differences distributions showing the conditions applied for selecting proper $\gamma$ track combination in the Central Detector for $pd\rightarrow^3$$He6\gamma$ reaction analysis. 
		Vertical lines show cuts positions.
		Left: time difference between the quickest and the slowest $\gamma$ quanta;
		right: time difference between $^3$$He$ track and the quickest $\gamma$ track.
	}
	\label{six_gamma_time_condition}
\end{figure}

\begin{figure}
	\begin{center}
		\includegraphics[width=220pt]{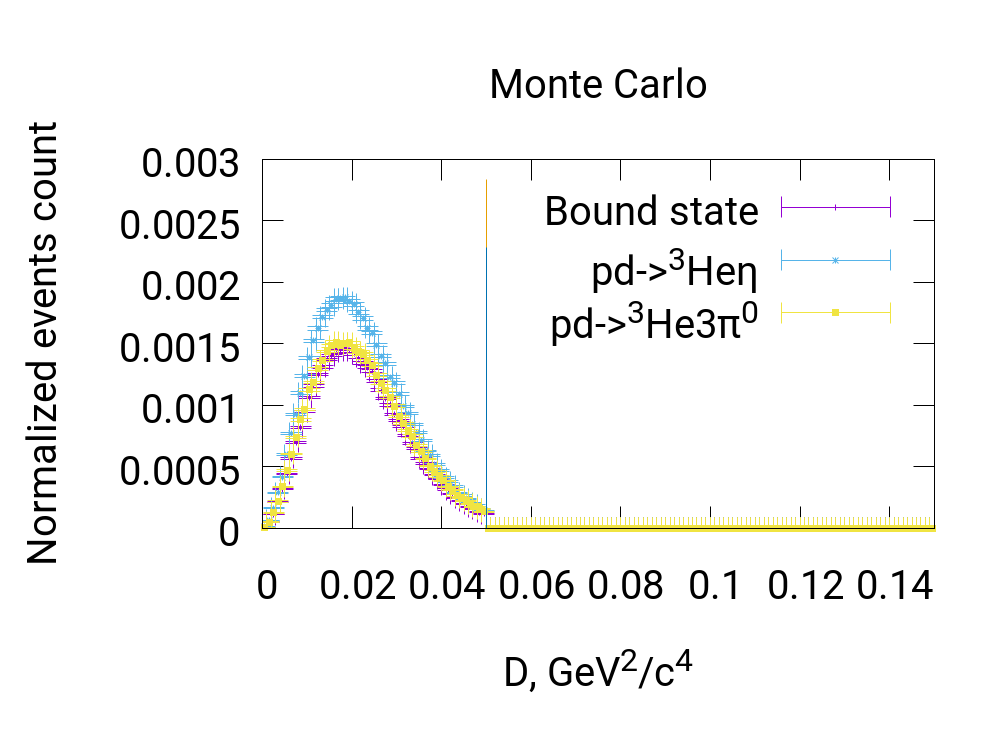}
		\includegraphics[width=220pt]{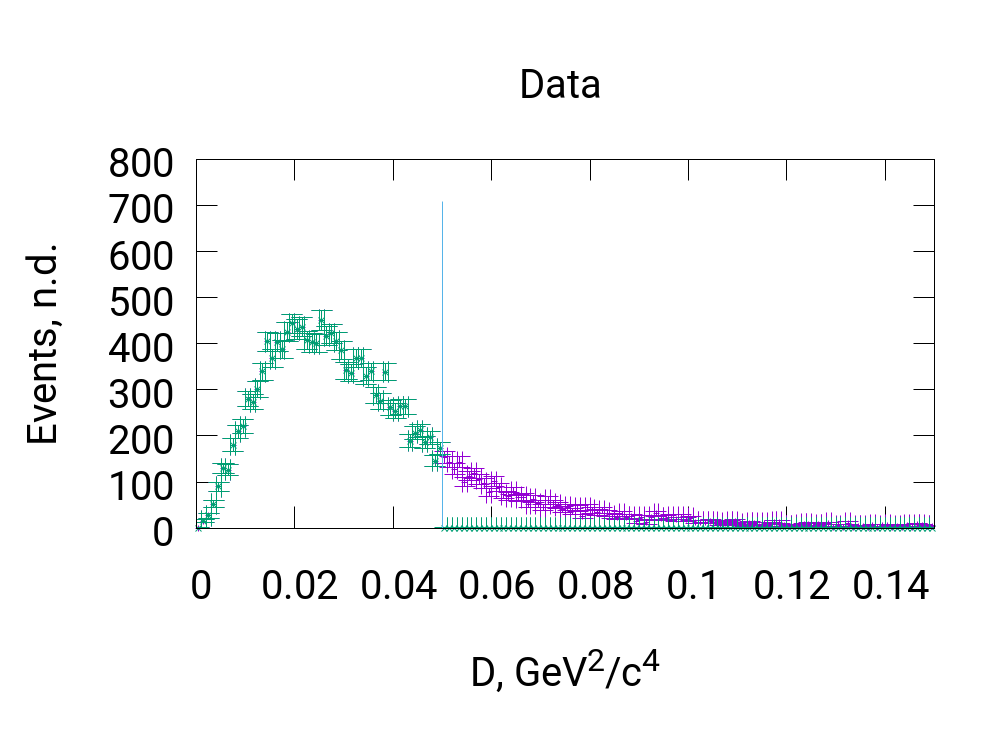}
	\end{center}
	\caption{
		The distribution of $D = \sum_{i=1}^{3}(m_{\gamma_{2i-1}\gamma_{2i}} - m_{\pi^0})^2$ 
		magnitude that is used to identify $\gamma$ quanta being the product of $3\pi^0$ decay (Eq.~\ref{6gamma_3pi0_identification}).
		On the right panel, blue points show the distributions before applying the conditions and cyan points show the distributions after applying the conditions.
	}
	\label{six_gamma_im_pi_diff_cut}
\end{figure}

Then  registration of at least six $\gamma$ quanta tracks in the Central Detector is required.
Among them, all combinations forming three pairs are tested.
They must fulfill the condition on time differences. 
Two time differences are considered: between $^3$$He$ and the quickest $\gamma$, and between the quickest $\gamma$ and the slowest one (Fig.~\ref{six_gamma_time_condition}).
As far as it makes the difference, how the six $\gamma$ quanta are divided into three pairs, this condition is not enough.
For each combinations forming three pairs, the following quantity is calculated:
\begin{equation}
	D= \sum_{i=1}^{3} (m_{\gamma_{(2i-1)}\gamma_{2i}} - m_{\pi^0})^2 ,
	\label{6gamma_3pi0_identification}
\end{equation}
where $m_{\gamma_{(2i-1)}\gamma_{2i}}$ is the $\gamma$ pair invariant mass and 
$m_{\pi^0}$ is $\pi^0$ mass.
The combination having this value the closest to zero is chosen. It is the most probable candidate to be product of $3\pi^0$ decay.
Then value from Eq.~\ref{6gamma_3pi0_identification} is required to be below the threshold of $0.05~GeV^2/c^4$ (Fig.~\ref{six_gamma_im_pi_diff_cut}).


\begin{figure}
	\begin{center}
		\includegraphics[width=220pt]{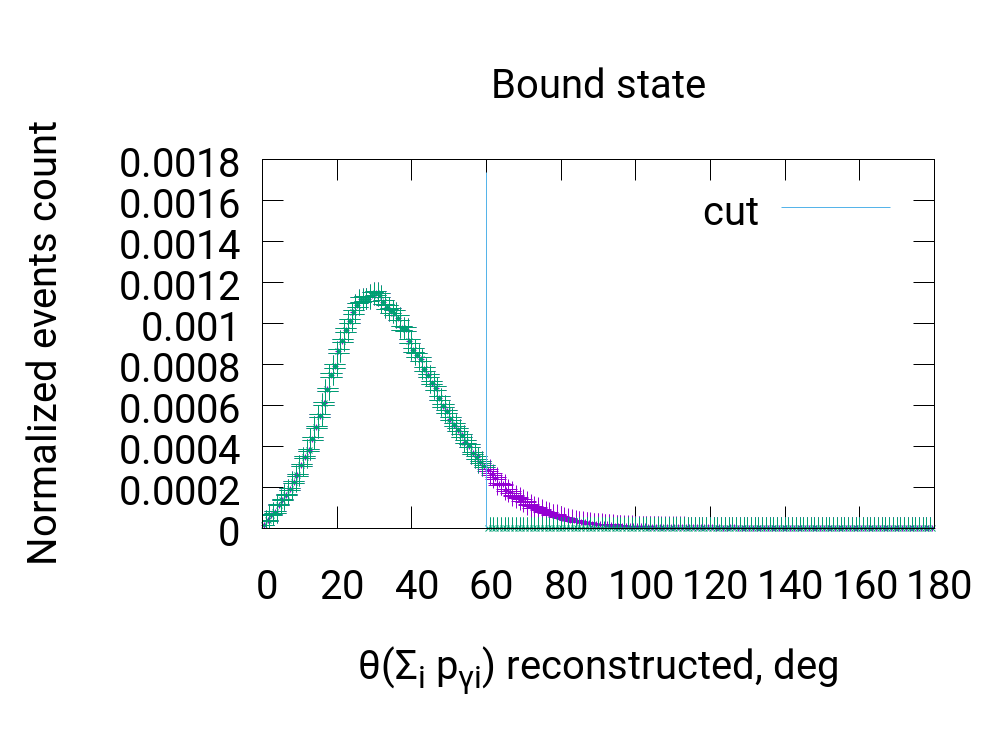}
		\includegraphics[width=220pt]{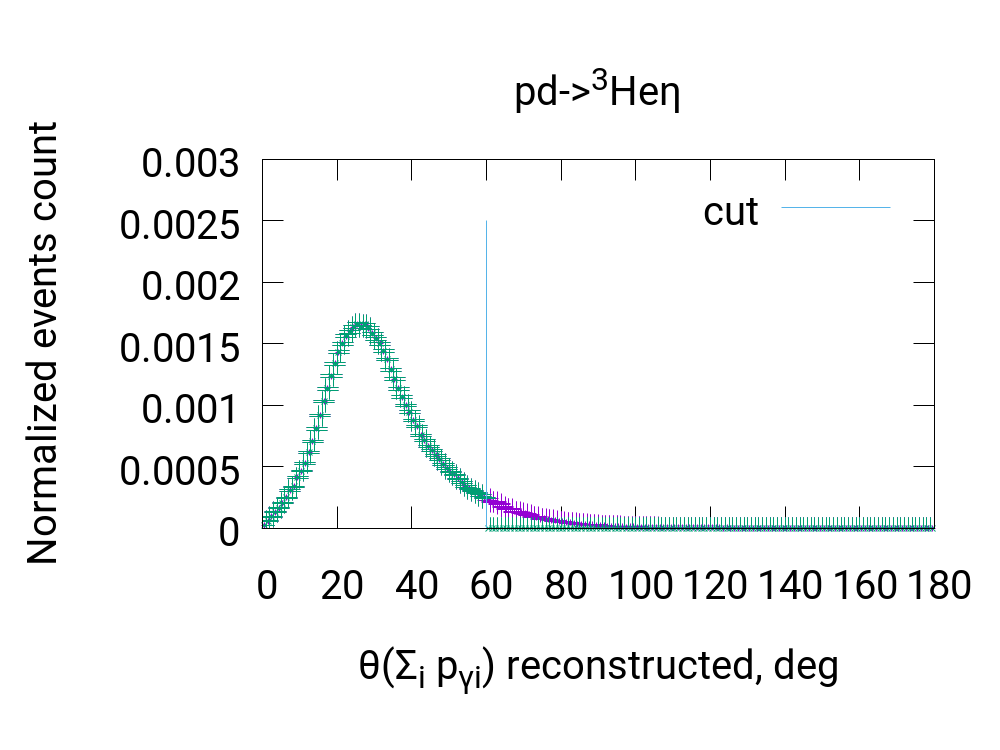}\\
		~\\~\\
		\includegraphics[width=220pt]{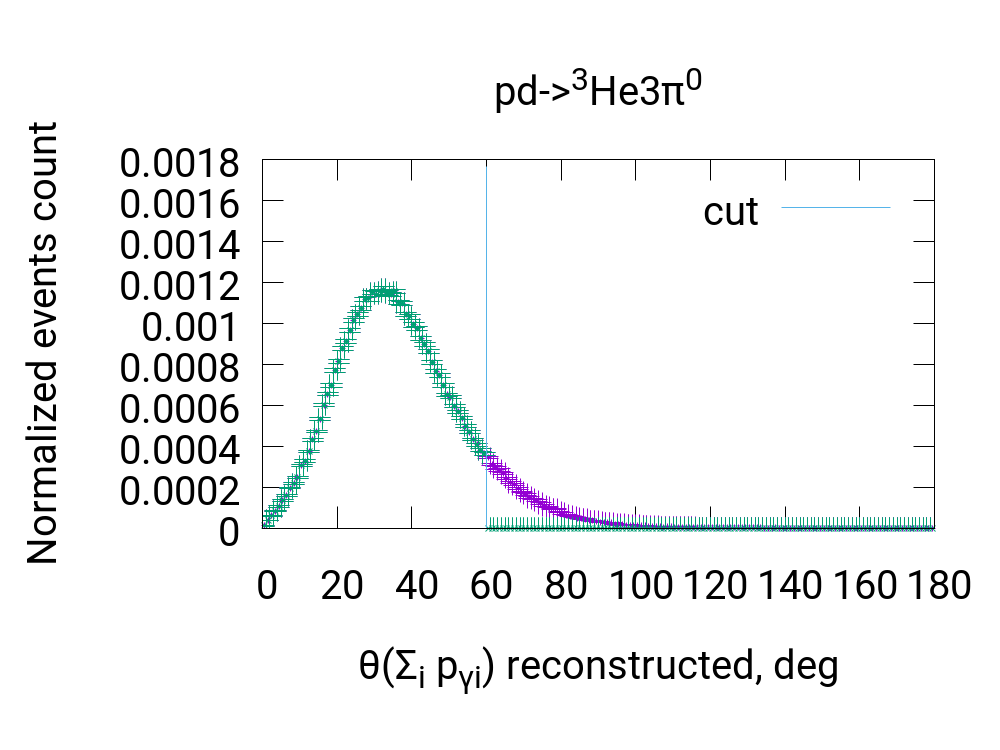}
		\includegraphics[width=220pt]{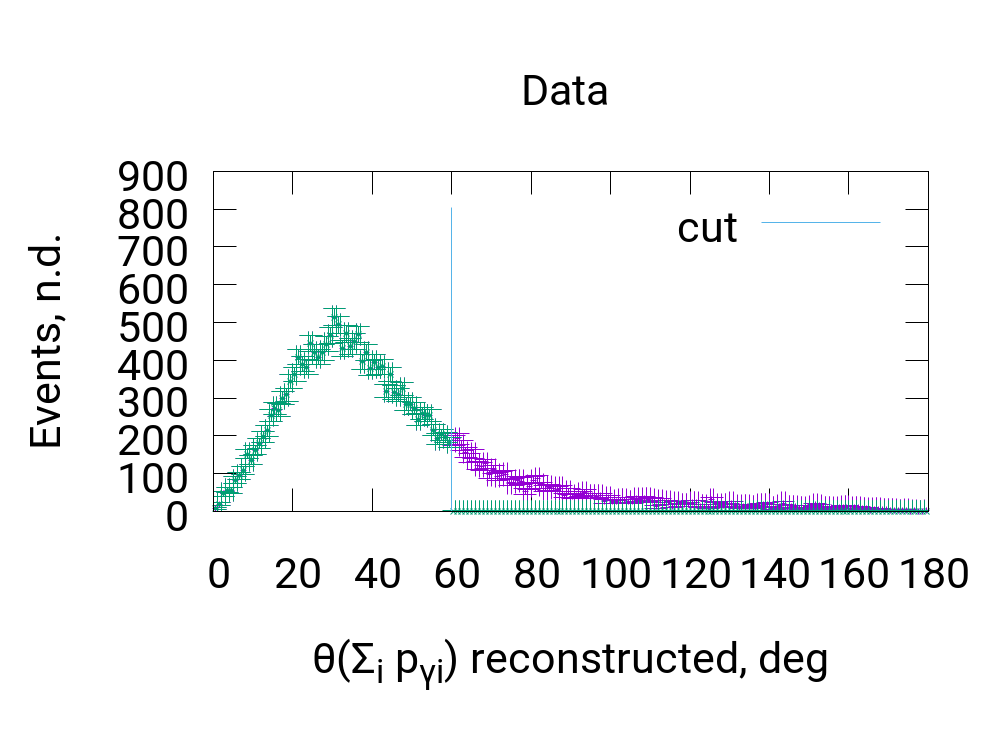}
	\end{center}
	\caption{
$\theta(\sum_i \vec{p}_{\gamma_i})$ distribution obtained in
 $pd\rightarrow^3$$He6\gamma$ reaction analysis.
Vertical line shows cut position.
As indicated in the legend above the pictures, the figure shows results of the analysis of data simulated for
$pd\rightarrow(^3$$He\eta)_{bound}\rightarrow^3$$He6\gamma$, 
$pd\rightarrow^3$$He\eta\rightarrow^3$$He6\gamma$, 
$pd\rightarrow^3$$He\pi^0\pi^0\pi^0\rightarrow^3$$He6\gamma$
and result of analysis of experimental data.
Magenta points show the distributions before applying the conditions and green points show the distributions after applying the conditions.
	}
	\label{six_gamma_eta_theta_cut}
\end{figure}

\begin{figure}
	\begin{center}
		\includegraphics[width=220pt]{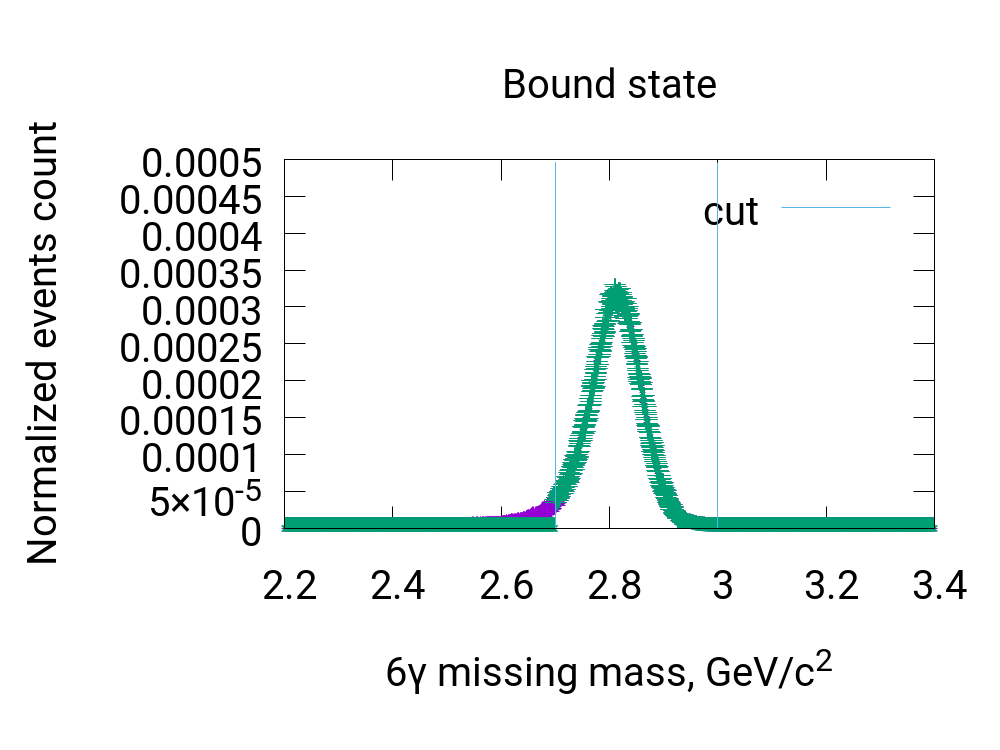}
		\includegraphics[width=220pt]{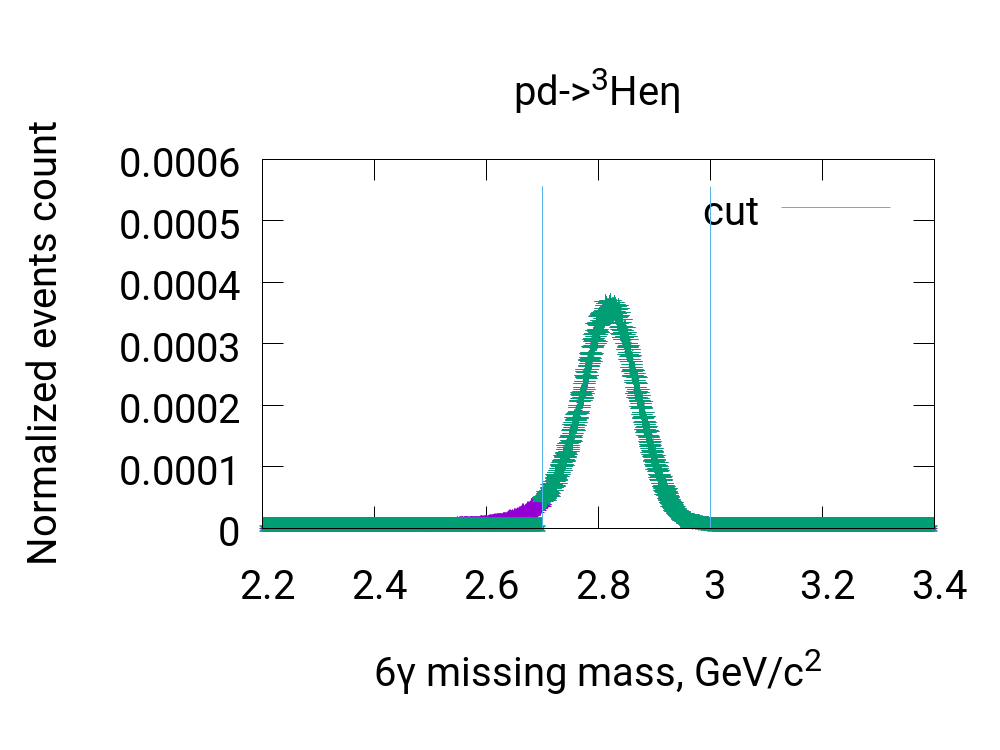}\\
		~\\~\\
		\includegraphics[width=220pt]{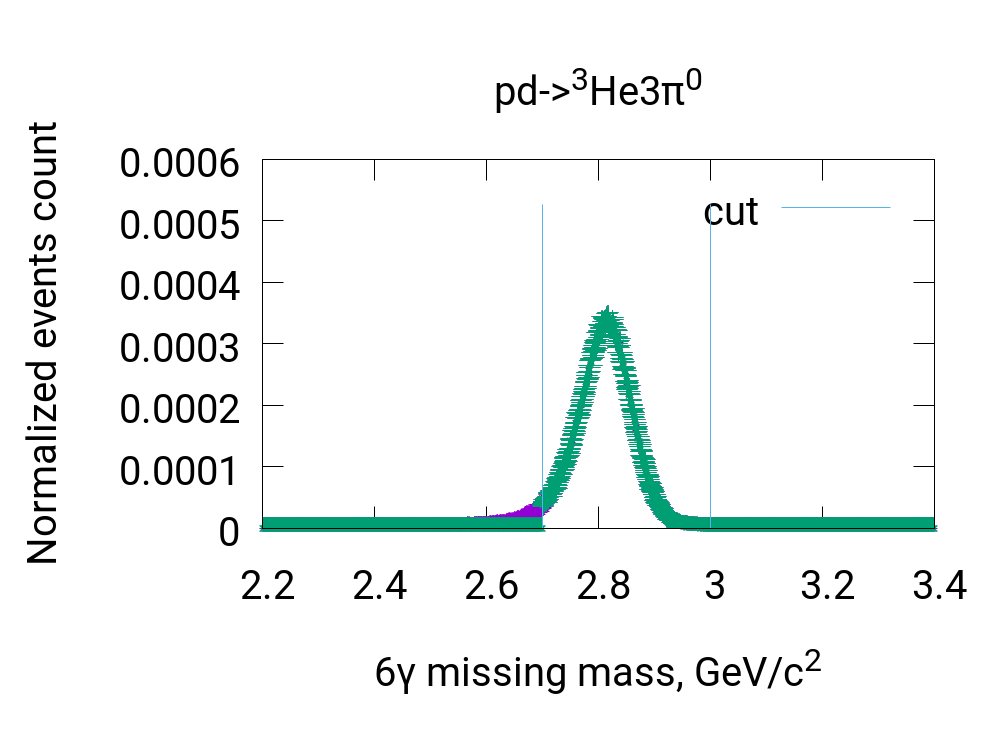}
		\includegraphics[width=220pt]{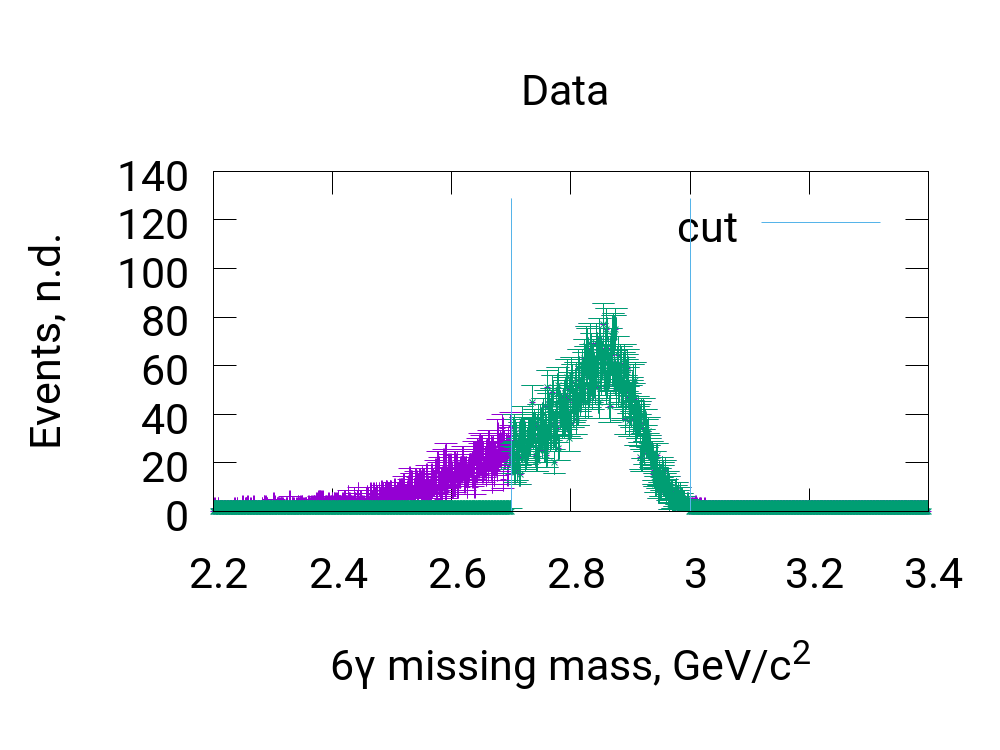}
	\end{center}
	\caption{
		$6\gamma$ missing mass distribution obtained in
		$pd\rightarrow^3$$He6\gamma$
		reaction analysis.
		Vertical lines show cuts position.
As indicated in the legend above the pictures, the figure shows results of the analysis of data simulated for
$pd\rightarrow(^3$$He\eta)_{bound}\rightarrow^3$$He6\gamma$, 
$pd\rightarrow^3$$He\eta\rightarrow^3$$He6\gamma$, 
$pd\rightarrow^3$$He\pi^0\pi^0\pi^0\rightarrow^3$$He6\gamma$
and result of analysis of experimental data.
Magenta points show the distributions before applying the conditions and green points show the distributions after applying the conditions.
	}
	\label{six_gamma_gamma_missing_mass_cut}
\end{figure}

\begin{figure}
	\begin{center}
		\includegraphics[width=220pt]{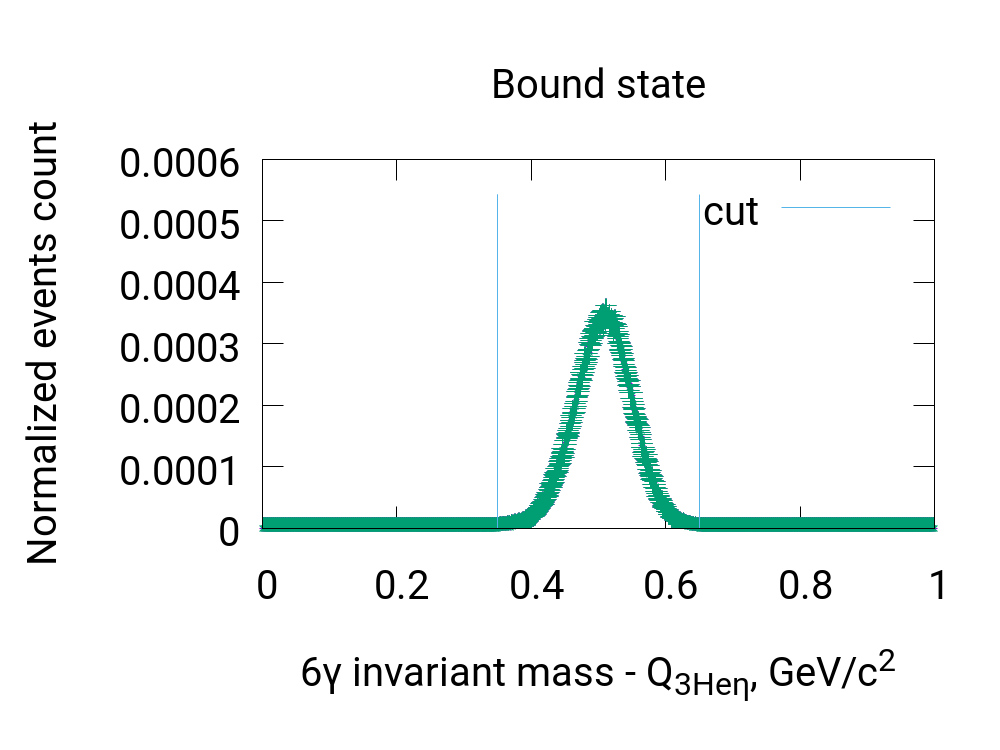}
		\includegraphics[width=220pt]{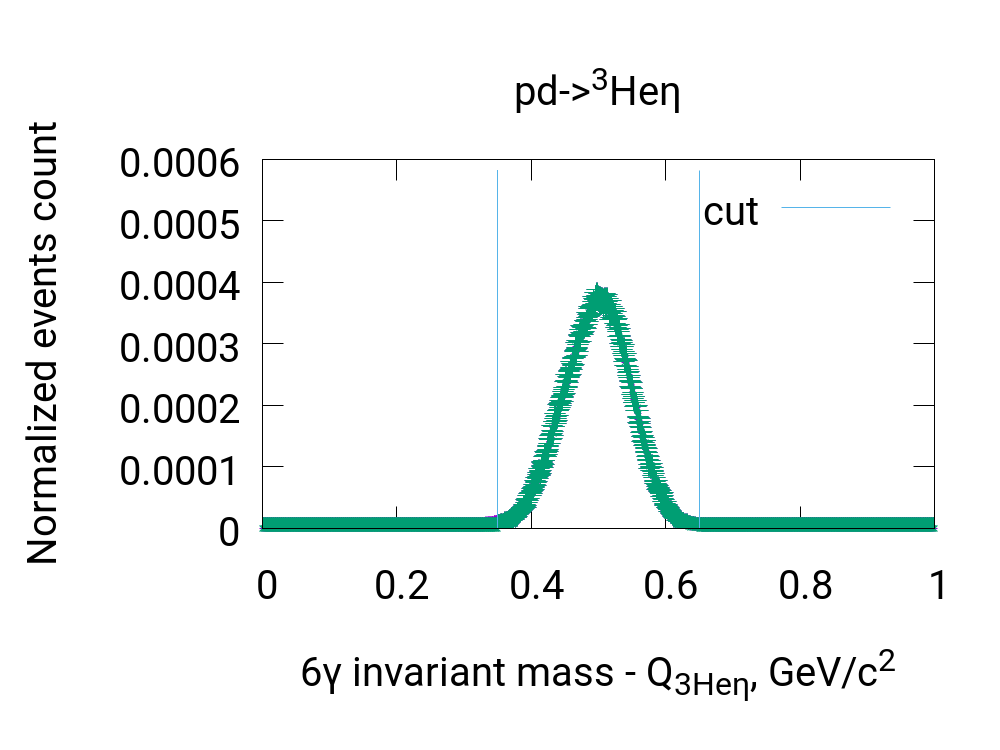}\\
		~\\~\\
		\includegraphics[width=220pt]{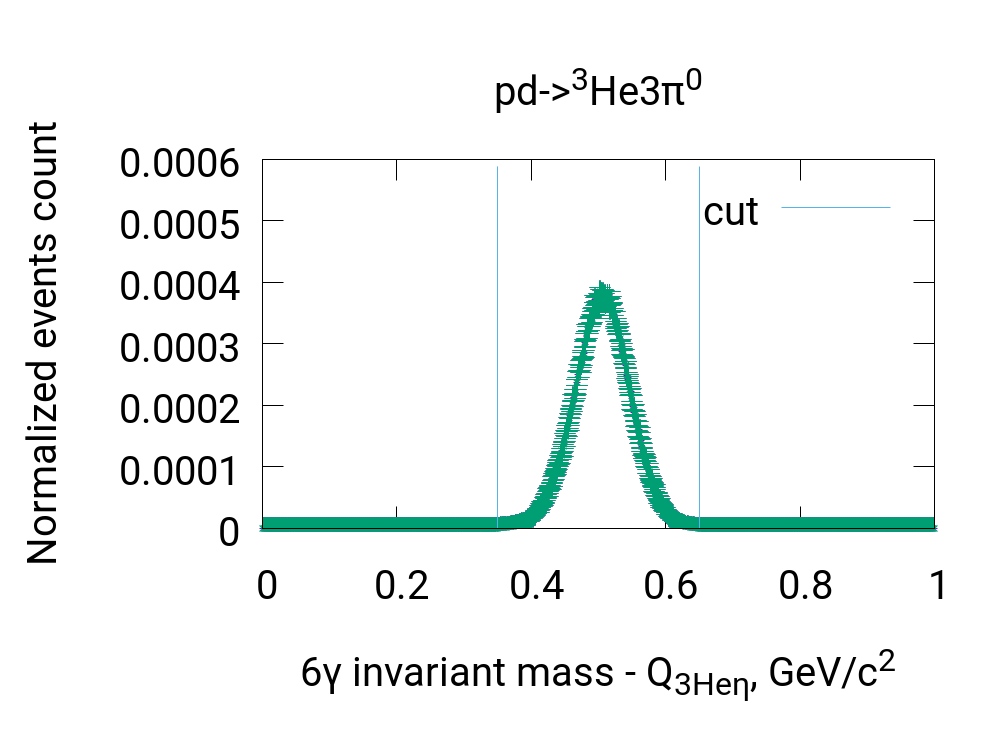}
		\includegraphics[width=220pt]{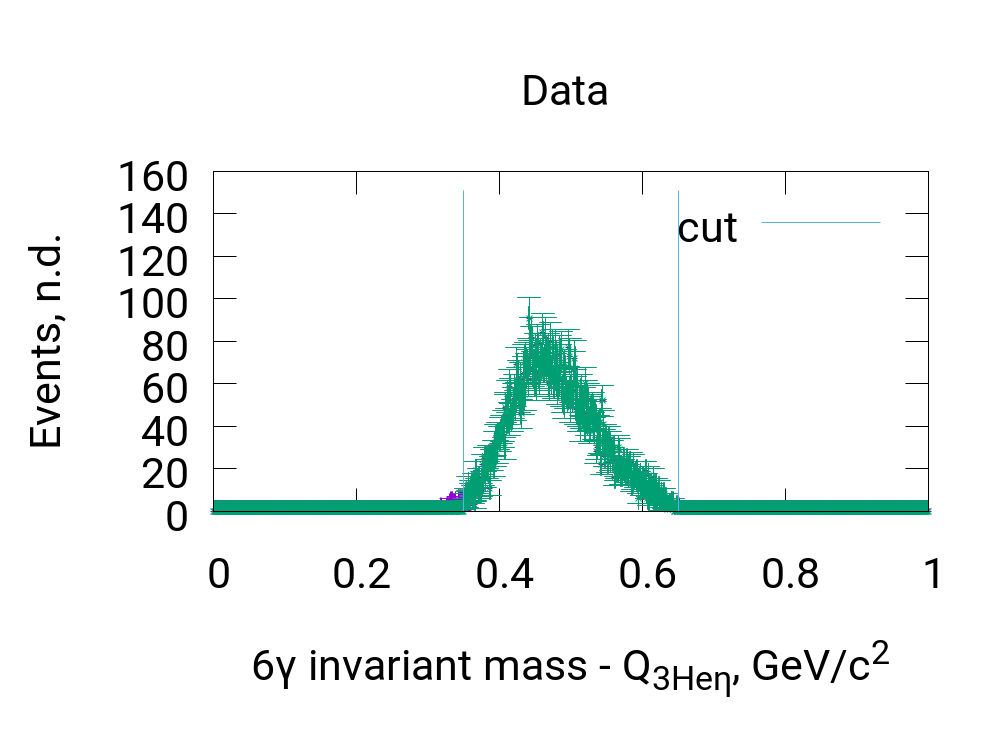}
	\end{center}
	\caption{
		The distribution of $6\gamma$ invariant mass corrected by $Q_{^3He\eta}$ obtained in
		$pd\rightarrow^3$$He6\gamma$ reaction analysis.
		Vertical lines show the cuts positions.
As indicated in the legend above the pictures, the figure shows results of the analysis of data simulated for
$pd\rightarrow(^3$$He\eta)_{bound}\rightarrow^3$$He6\gamma$, 
$pd\rightarrow^3$$He\eta\rightarrow^3$$He6\gamma$, 
$pd\rightarrow^3$$He\pi^0\pi^0\pi^0\rightarrow^3$$He6\gamma$
and result of analysis of experimental data.
Magenta points show the distributions before applying the conditions and green points show the distributions after applying the conditions.
	}
	\label{six_gamma_gamma_invariant_mass_cut}
\end{figure}

Similarly to $pd\rightarrow^3$$He2\gamma$ reaction analysis, the condition on $\theta(\sum_i\vec{p}_i)$ is applied (Fig.~\ref{six_gamma_eta_theta_cut}).
Next applied conditions are the $6\gamma$ missing mass cut (Fig.~\ref{six_gamma_gamma_missing_mass_cut}) and $6\gamma$ invariant mass cut (Fig.~\ref{six_gamma_gamma_invariant_mass_cut}).

\begin{figure}
	\begin{center}
		\includegraphics[width=300pt]{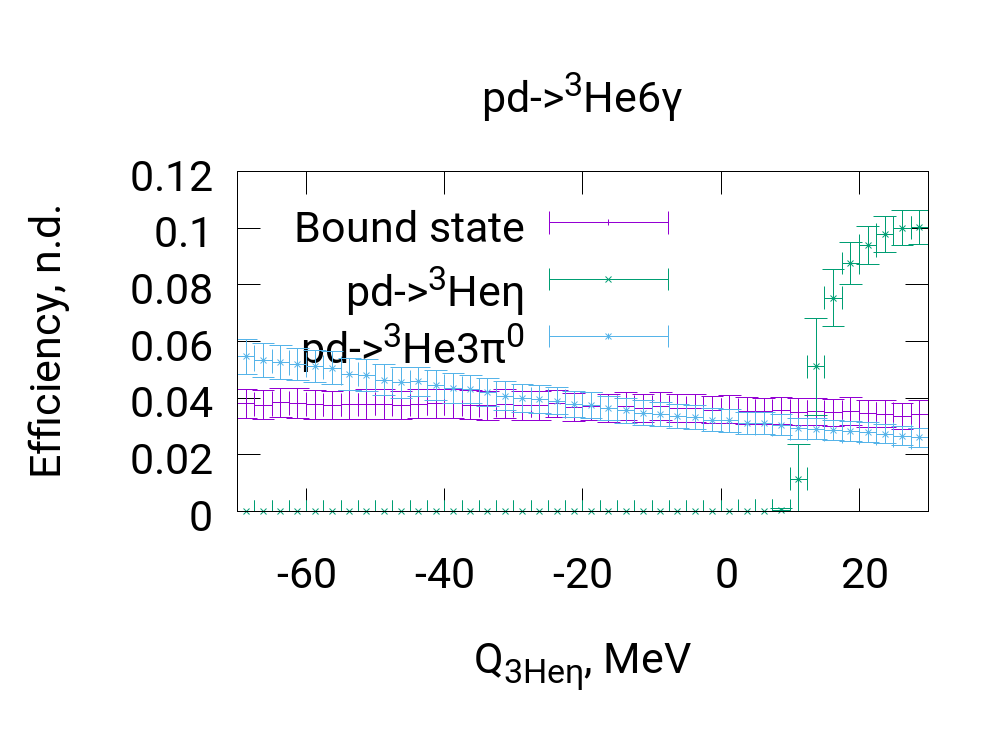}
	\end{center}
	\caption{
		Efficiency distribution for different reactions that are visible for 
		$pd\rightarrow^3$$He6\gamma$ reaction analysis algorithm.
		Systematic uncertainties are shown.
	}
	\label{six_gamma_acceptance}
\end{figure}

The $pd\rightarrow^3$$He\eta$ reaction is visible in the region of $Q_{^3He\eta}~>~10~MeV$ with the efficiency
increasing up to $10\%$ (Fig.~\ref{six_gamma_acceptance}).
The bound state registration efficiency weakly depends on beam momentum and is about $4\%$.
$pd\rightarrow^3$$He3\pi^0$ background reaction registration efficiency decreases from $6\%$ down to $2\%$ with the beam momentum increasing.

\begin{figure}
	\begin{center}
		\includegraphics[width=220pt]{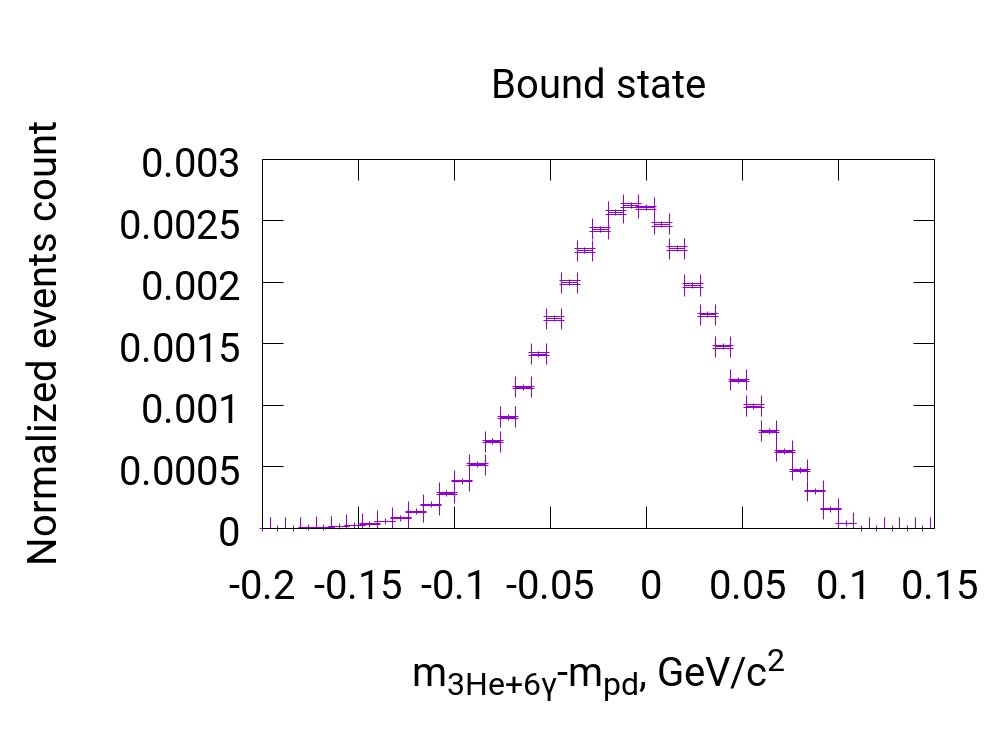}
		\includegraphics[width=220pt]{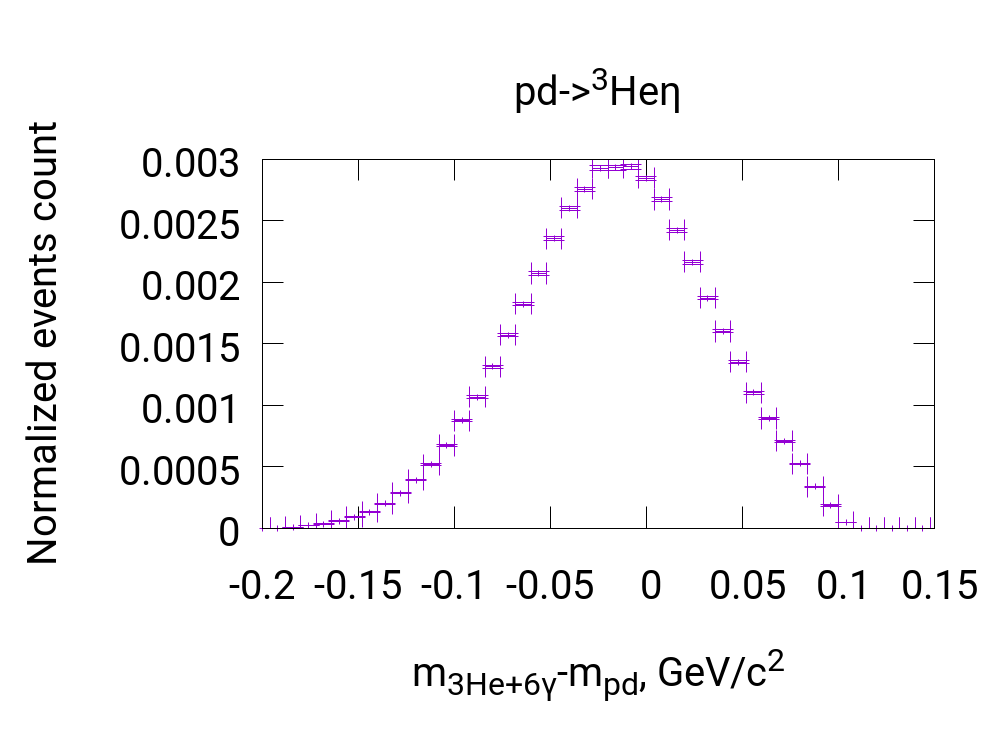}\\
		~\\~\\
		\includegraphics[width=220pt]{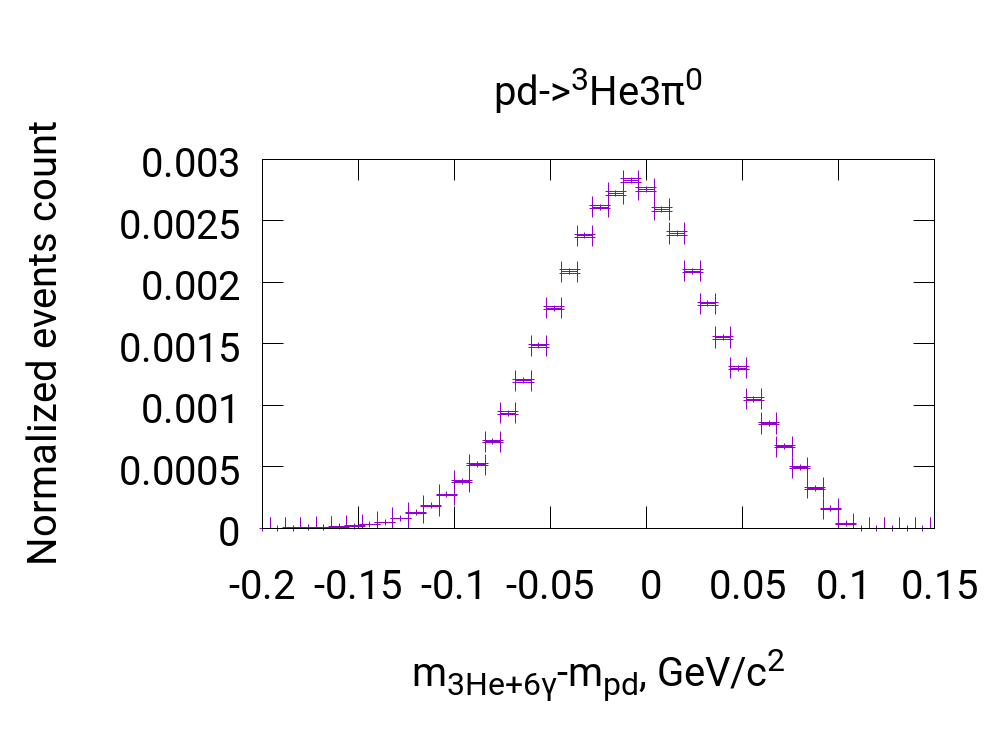}
		\includegraphics[width=220pt]{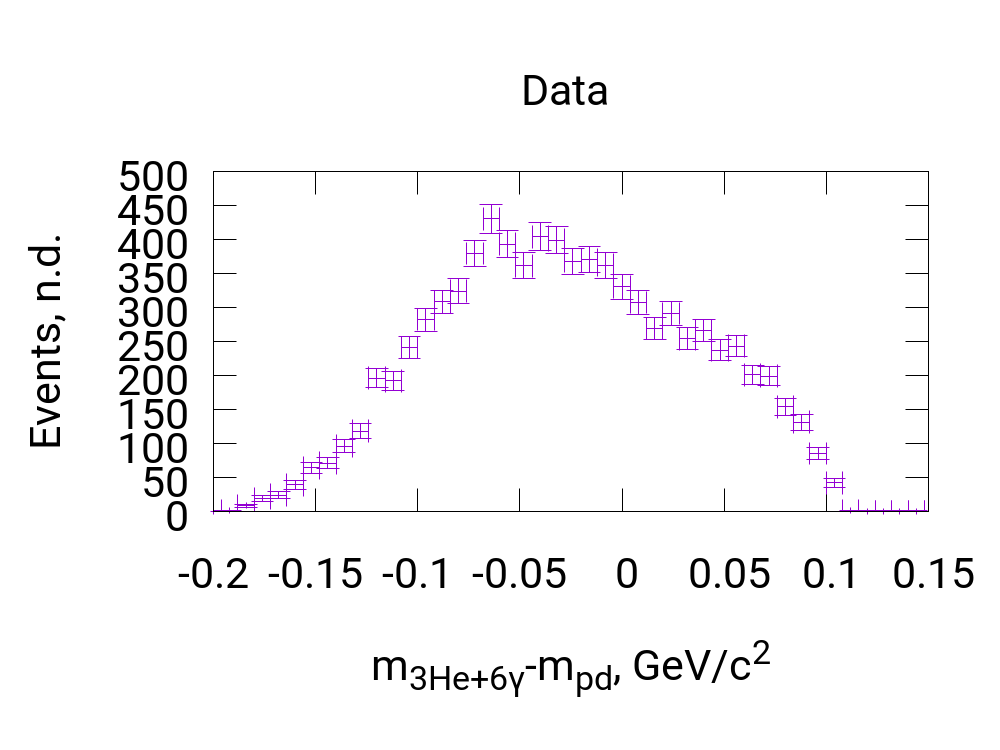}
	\end{center}
	\caption{
		Distribution of $m_{^3He6\gamma}-m_{pd}$ obtained in
		$pd\rightarrow^3$$He6\gamma$ reaction analysis.
As indicated in the legend above the pictures, the figure shows results of the analysis of data simulated for
$pd\rightarrow(^3$$He\eta)_{bound}\rightarrow^3$$He6\gamma$, 
$pd\rightarrow^3$$He\eta\rightarrow^3$$He6\gamma$, 
$pd\rightarrow^3$$He\pi^0\pi^0\pi^0\rightarrow^3$$He6\gamma$
and result of analysis of experimental data.
	}
	\label{six_gamma_total_invariant_mass_plot}
\end{figure}

\begin{figure}
	\begin{center}
		\includegraphics[width=220pt]{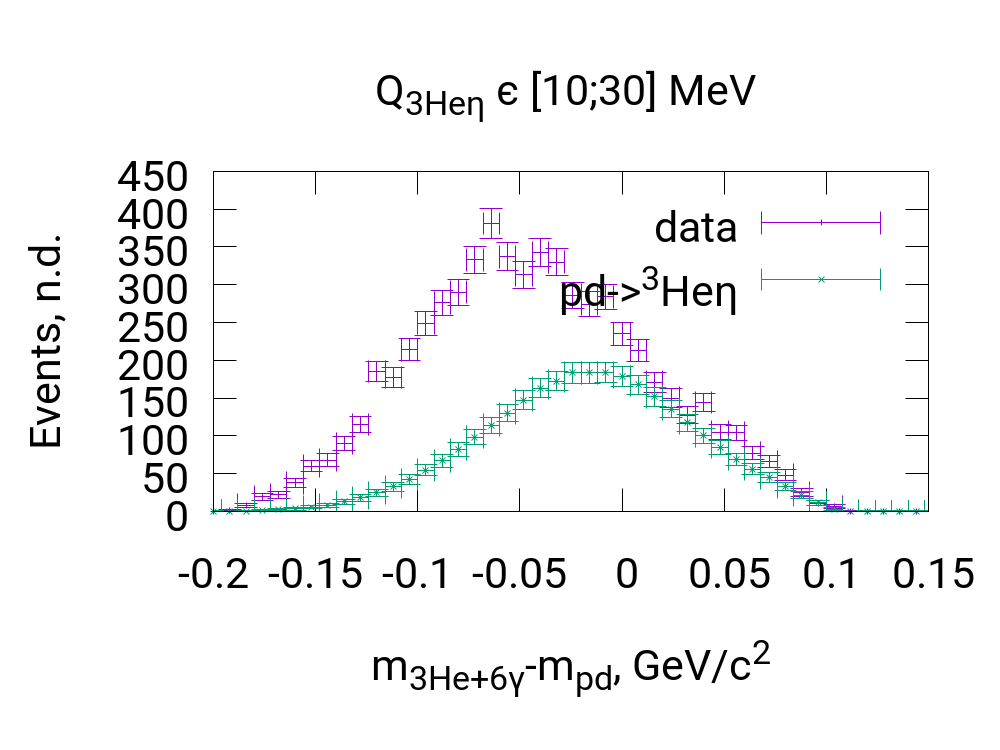}
		\includegraphics[width=220pt]{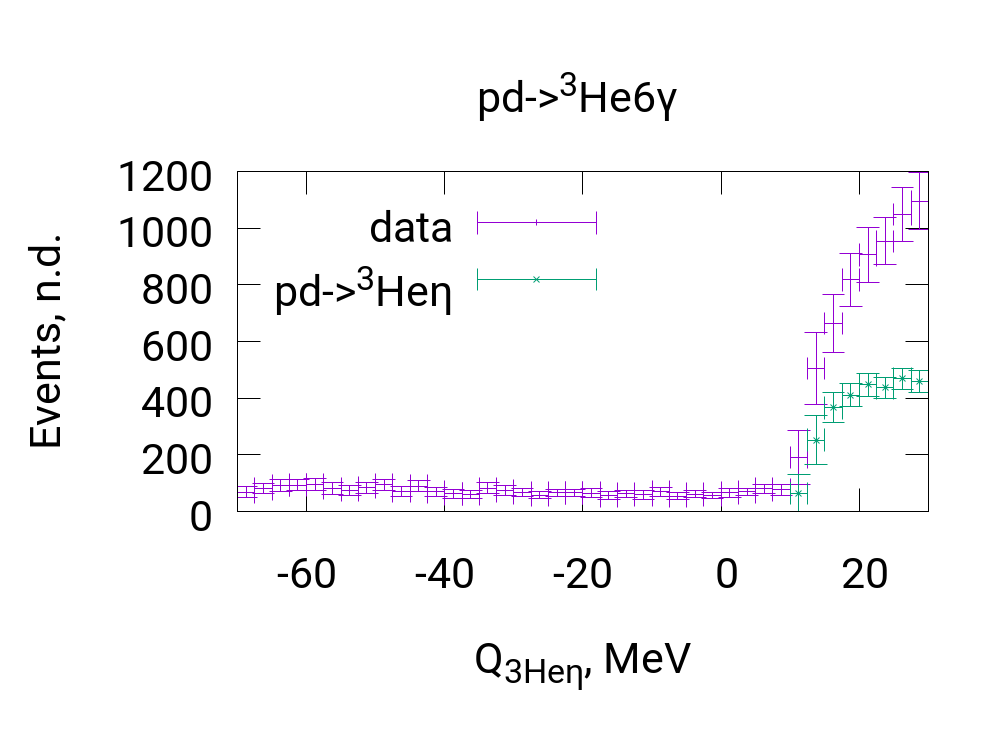}
	\end{center}
	\caption{
		Left:
		The distribution of invariant mass difference  $m_{^3He6\gamma}-m_{pd}$
		for events in excess energy region $Q_{^3He\eta}~>~10~MeV$;
		registered events count (blue),
		 and $pd\rightarrow^3$$He\eta$ events count estimation (green).
		Right:
		The dependence of events count on $Q_{^3He\eta}$ for 
		$pd\rightarrow^3$$He6\gamma$ reaction analysis;
		registered events count (blue color), 
		and
		estimation based on $pd\rightarrow^3$$He\eta$ reaction (green color).
	}
	\label{six_gamma_events}
\end{figure}

In the total invariant mass difference $m_{^3He\gamma\gamma}-m_{pd}$ distribution 
(Fig.~\ref{six_gamma_total_invariant_mass_plot}, \ref{six_gamma_events}),
some extra background with the intensity comparable to $pd\rightarrow^3$$He\eta$ is visible.
These background events are also visible in excitation curve in the excess energy region $Q_{^3He\eta}>10~MeV$ as the difference between registered events count and the estimation based on $pd\rightarrow^3$$He\eta$ reaction and luminosity (Fig.~\ref{six_gamma_events}).
The difference between registered events count and the estimation based on $pd\rightarrow^3$$He\eta$ reaction for these two distributions is in agreement.
\newpage

\section{The excitation curves}

\label{3Hegammas_discussion}
$pd\rightarrow^3$$He2\gamma$
and
$pd\rightarrow^3$$He6\gamma$
reactions have been analyzed in order to search for the signal from $^3$$He-\eta$ bound state.
The dependencies of events count on $Q_{^3He\eta}$ (sec.~\ref{Q-bins}) have been obtained for both of the reactions (Fig.~\ref{two_gamma_events}, \ref{six_gamma_events}).
For each $Q_{^3He\eta}$ interval separately, this events counts are normalized using the formula:
\begin{equation}
	N_{norm} = \frac{N_{data}~S_{trigger}}{\int Ldt~\epsilon},
\end{equation}
where $N_{data}$ is events count obtained by data analysis algorithm, $S_{trigger}$ is trigger scaling coefficient,
$\int Ldt$ is integrated luminosity (eq.~\ref{luminosity_description}) and $\epsilon$ is the efficiency.

\begin{figure}
	\begin{center}
		\includegraphics[width=220pt]{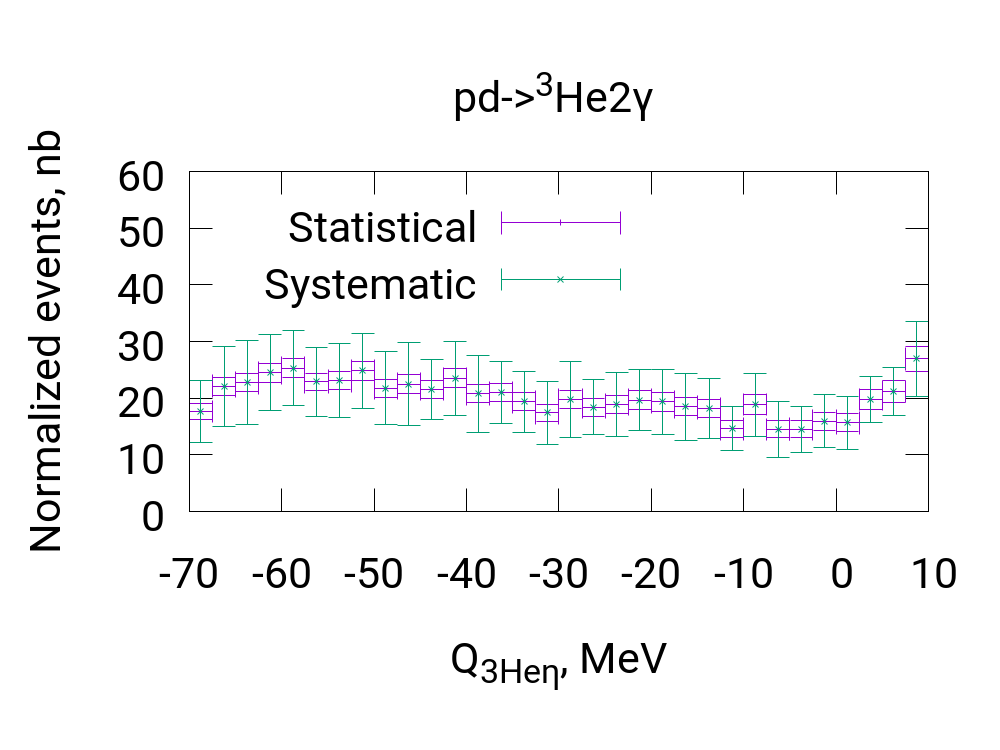}
		\includegraphics[width=220pt]{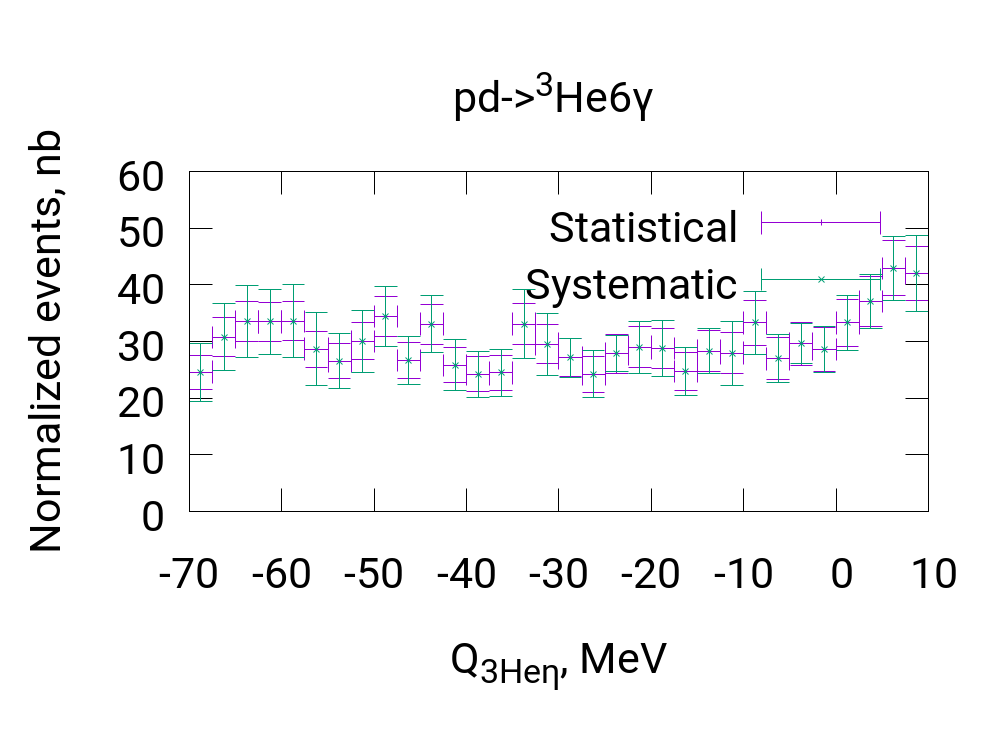}
	\end{center}
	\caption{
		The dependence of events count on $Q_{^3He\eta}$ for 
		$pd\rightarrow^3$$He2\gamma$ (left panel)
		and  $pd\rightarrow^3$$He6\gamma$ (right panel) reactions 
		normalized to cross section units.
	}
	\label{two_six_gamma_events_norm}
\end{figure}

Normalized excitation curves for excess energy region $Q_{^3He\eta}\in[-70;10]~MeV$ obtained for both reactions are shown in Fig.~\ref{two_six_gamma_events_norm}.	

\newpage
\section{The systematic uncertainties estimation}
\label{reactions_systematics}
For the excitation curves obtained in $pd\rightarrow^3$$He2\gamma$ and $pd\rightarrow^3$$He6\gamma$ analysis (Fig.~\ref{two_six_gamma_events_norm}),
the systematic error is caused by cuts positions, beam momentum correction constant, model parameters inaccuracy,
and systematic error of luminosity determination.
As far as the luminosity values obtained in $pd\rightarrow ppn_{spectator}$ reaction analysis have been used, all the parameters contributing to integrated luminosity systematic uncertainty (Table.~\ref{systematics_ppn_lum_parameters}) need to be taken into account.
The values of all analysis parameters contributing into systematic uncertainties are given in Table.~\ref{systematics_central_gamma_parameters}.

The total systematic uncertainty was calculated by the formula
\begin{equation}
	\Delta N_{syst} = \sqrt{ \sum_i (\frac{
			|N_{norm}^{P_i = P^{max}_i} - N_{norm}^{final}| + |N_{norm}^{P_i = P^{min}_i} - N_{norm}^{final}|
		}{2})^2},
	\label{events_count_systematics_formulae}
\end{equation}
where $N_{norm}$ is the events count obtained in the analysis, and $i$ index and $P_i$  have the same meaning like in eq.~\ref{luminosity_systematics_formulae}

\begin{table}[h!]
	\begin{tabular}{|p{40pt}|p{260pt}|p{50pt}|p{50pt}|}
		\hline
		Analysis&Parameter description & Value & Variation \\
		\hline
		both&\multicolumn{3}{|c|}{All parameters from Table.~\ref{systematics_ppn_lum_parameters}}\\
		\hline
		both&$\theta$ angular cut for $^3$$He$ tracks (Fig.~\ref{he3_reconstruction_kinematic_hist}, Eq.~\ref{he3_theta_cut_pos}) & $4.5^o$& $\pm0.2^o$\\
		\hline
		both&$^3$$He$ identification cut height (Fig.~\ref{he3_cut_data}, Eq.~\ref{he3_cut_height}) & $10~MeV$ & $\pm0.5~MeV$ \\
		\hline
		both&Real part of potential assumed in the model~\cite{Bound_state_model_Hirenzaki} (Fig.~\ref{bound_fermi_momentum_fig}, Table.~\ref{bound_state_potential_parameters})& $80~MeV$& $+10 MeV$, $-5~MeV$\\
		\hline
		both& $^3$$He$ missing mass cut position (Fig.~\ref{two_gamma_3He_missing_mass_cut},\ref{six_gamma_3He_missing_mass_cut}) & $510~\frac{MeV}{c^2}$ & $\pm2~\frac{MeV}{c^2}$\\
		\hline
		both& $\gamma$ energy threshold& $25~MeV$ & $\pm2~MeV$\\
		\hline
		both& Time cut  (Fig.~\ref{two_gamma_time_condition}, \ref{six_gamma_time_condition}, left panel) & $15~ns$ & $\pm2~ns$\\
		\hline
		both& Time cut  (Fig.~\ref{two_gamma_time_condition}, \ref{six_gamma_time_condition}, right panel, left border) & $0~ns$ & $\pm2~ns$\\
		\hline
		both& Time cut  (Fig.~\ref{two_gamma_time_condition}, \ref{six_gamma_time_condition}, right panel, right border) & $30~ns$ & $\pm2~ns$\\
		\hline
		both& Cut on $\theta(\sum \vec{p}_{\gamma_i})$ (Fig.~\ref{two_gamma_eta_theta_cut}, \ref{six_gamma_eta_theta_cut}) & $60^o$ & $\pm2^o$\\
		\hline
		$6\gamma$& Cut for $3\pi^0$ decay identification (Fig.~\ref{six_gamma_im_pi_diff_cut}) & $50~\frac{MeV^2}{c^4}$ & $\pm5~\frac{MeV^2}{c^4}$\\
		\hline
		both& $\gamma$ quanta missing mass cut position (Fig.~\ref{two_gamma_ggmm_cut}, \ref{six_gamma_gamma_missing_mass_cut}, left border) & $2700~\frac{MeV}{c^2}$ & $\pm10~\frac{MeV}{c^2}$\\
		\hline
		both& $\gamma$ quanta missing mass cut position (Fig.~\ref{two_gamma_ggmm_cut}, \ref{six_gamma_gamma_missing_mass_cut}, right border) & $3000~\frac{MeV}{c^2}$ & $\pm10~\frac{MeV}{c^2}$\\
		\hline
		$2\gamma$& Cut on $\gamma-\gamma$ invariant  mass corrected by $Q_{^3He\eta}$ (Fig.~\ref{two_gamma_ggim_cut}, left border) & $450~\frac{MeV}{c^2}$ & $\pm10~\frac{MeV}{c^2}$\\
		\hline
		$2\gamma$& Cut on $\gamma-\gamma$ invariant mass corrected by $Q_{^3He\eta}$ (Fig.~\ref{two_gamma_ggim_cut}, right border) & $650~\frac{MeV}{c^2}$ & $\pm10~\frac{MeV}{c^2}$\\
		\hline
		$6\gamma$& Cut on $\gamma$ quanta invariant  mass corrected by $Q_{^3He\eta}$ (Fig.~\ref{six_gamma_gamma_invariant_mass_cut}, left border) & $350~\frac{MeV}{c^2}$ & $\pm10~\frac{MeV}{c^2}$\\
		\hline
		$6\gamma$& Cut on $\gamma$ quanta invariant mass corrected by $Q_{^3He\eta}$ (Fig.~\ref{six_gamma_gamma_invariant_mass_cut}, right border) & $650~\frac{MeV}{c^2}$ & $\pm10~\frac{MeV}{c^2}$\\
		\hline
	\end{tabular}
	\caption{The list of parameters contributing into systematic error for $pd\rightarrow^3$$He2\gamma$ and $pd\rightarrow^3$$He6\gamma$ reactions analysis.}
	\label{systematics_central_gamma_parameters}
\end{table}
\chapter{Results and interpretation}
\section{Upper limit for the $\eta-^3$$He$  bound state production cross section}
\label{upper_limits}
\begin{figure}[h!]
	\begin{center}
		\includegraphics[width=220pt]{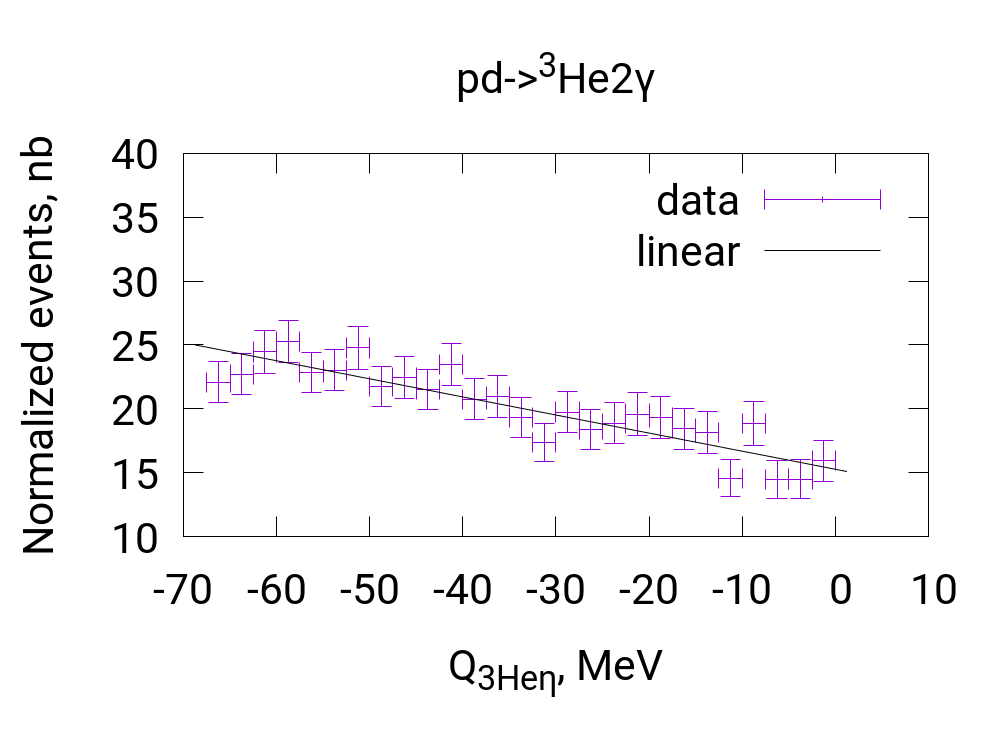}
		\includegraphics[width=220pt]{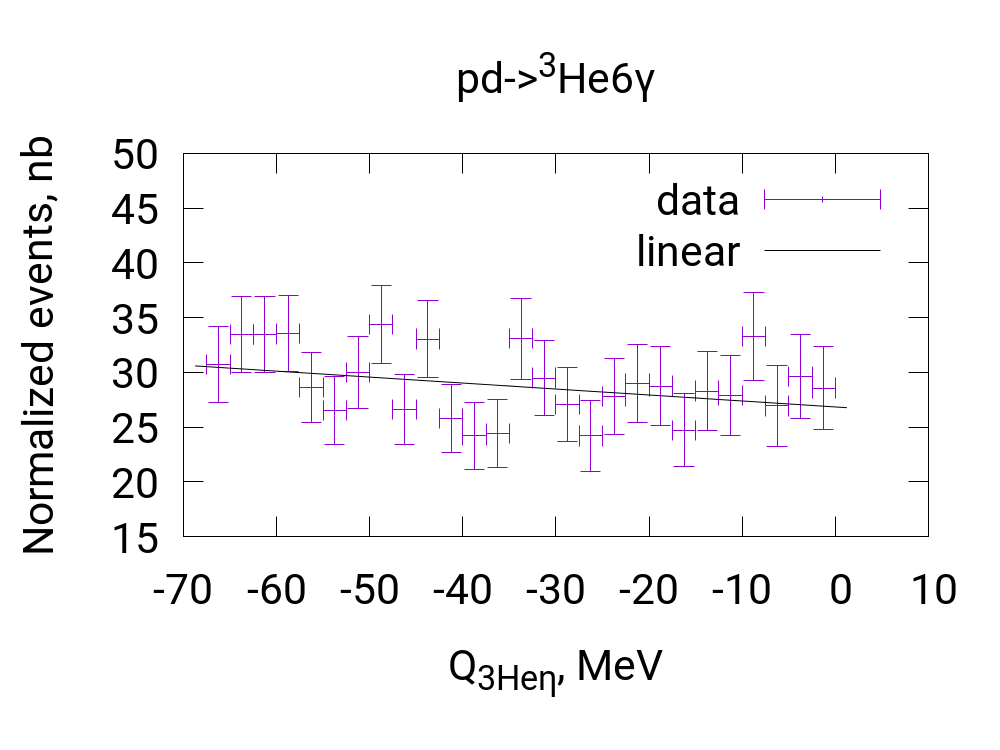}
	\end{center}
	\caption{
		Excitation curves fit by independent linear background.
	}
	\label{upper_limit_fit_linear}
\end{figure}

\begin{figure}[h!]
	\begin{center}
		\includegraphics[width=250pt]{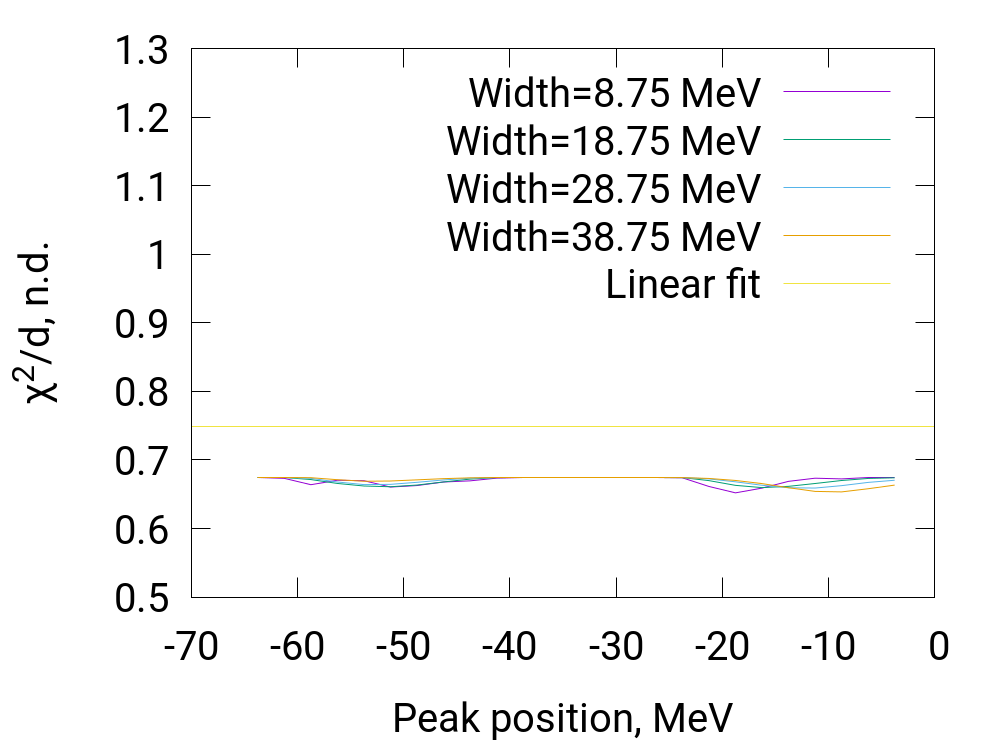}
	\end{center}
	\caption{
		The $\chi^2$ per degree of freedom values obtained from excitation curves fits for different bound state widths and peak positions (Fig.~%
		\ref{upper_limit_fit1}, \ref{upper_limit_fit2}, \ref{upper_limit_fit3}, \ref{upper_limit_fit4}%
		).
		The $\chi^2/d$ value for linear fit (Fig.~\ref{upper_limit_fit_linear}) is shown by horizontal line.
	}
	\label{upper_limit_chisq_figure}
\end{figure}

During the analysis of data collected in current experiment, the excitation curves for $pd\rightarrow^3$$He2\gamma$ and $pd\rightarrow^3$$He6\gamma$ reactions have been obtained (Fig.~\ref{two_six_gamma_events_norm}) in order to search for the $^3$$He\eta$ bound state.
If the bound state is observed in current experiment, Breit-Wigner shaped peaks would be present in both excitation curves in the same position.
Nevertheless, the shape of the curves can be well described with linear function fit resulting in the $\chi^2$ per degree of freedom less than 1 (Fig.~\ref{upper_limit_fit_linear}, \ref{upper_limit_chisq_figure}), that means good description of the experiment without assuming the $\eta$-mesic $^3$$He$.

Therefore, we can only determine the upper limit for $^3$$He\eta$ bound state formation cross section.
Though there are theoretical works resulting in that $\eta$ meson embedded in nuclear matter can change its properties~\cite{bound_theory8,bound_theory9},  in current analysis, an assumption that the branching ratio for $\eta\rightarrow2\gamma$ and $\eta\rightarrow3\pi^0$ decay channels
(Table~\ref{eta_decay_modes_table})
for bound $\eta$ meson remains the same was made:
\begin{equation}
	P_{\eta\rightarrow2\gamma} = 0.3941\pm0.0020,~
	P_{\eta\rightarrow3\pi^0} = 0.3268\pm0.0023.
	\label{branching_ratio}
\end{equation}

The production cross section of the hypothetical bound state is assumed to have a following energy dependence:
\begin{equation}
	\sigma_b(Q_{^3He\eta},B_s,\Gamma,\sigma) 
	= \sigma~\frac{\Gamma^2/4}{(Q_{^3He\eta}-B_s)^2+\Gamma^2/4},
	\label{Breight_Wigner_formulae}
\end{equation}
where $B_s$ is the binding energy, $\Gamma$ is the bound state width, and $\sigma$ is the amplitude.

\begin{figure}
	\begin{center}
		\includegraphics[width=220pt]{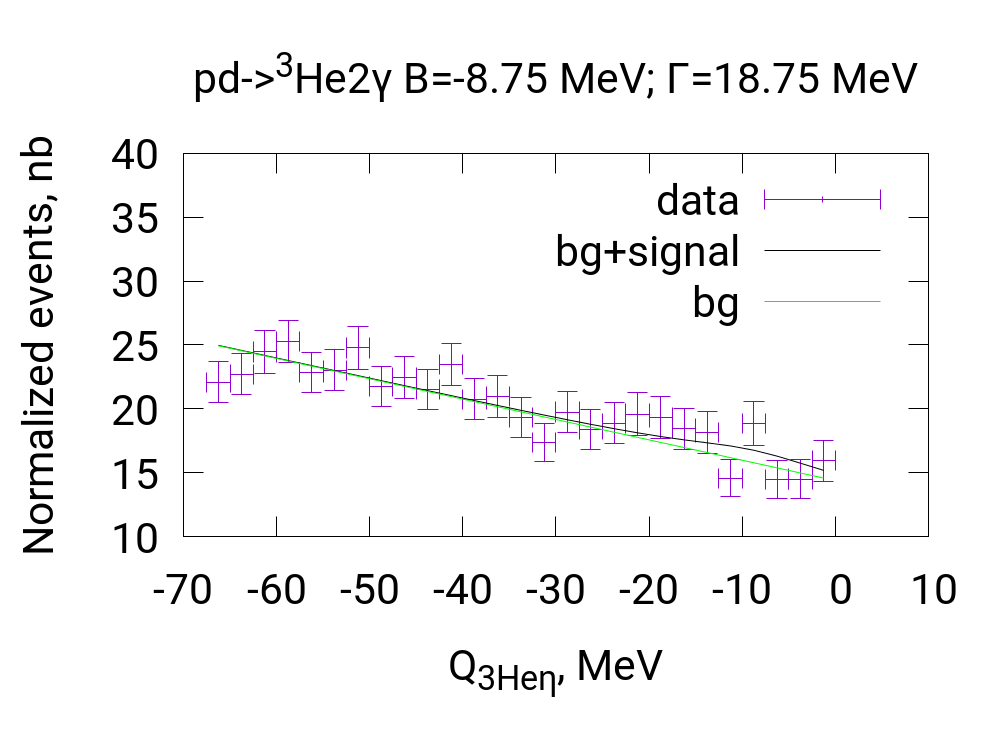}
		\includegraphics[width=220pt]{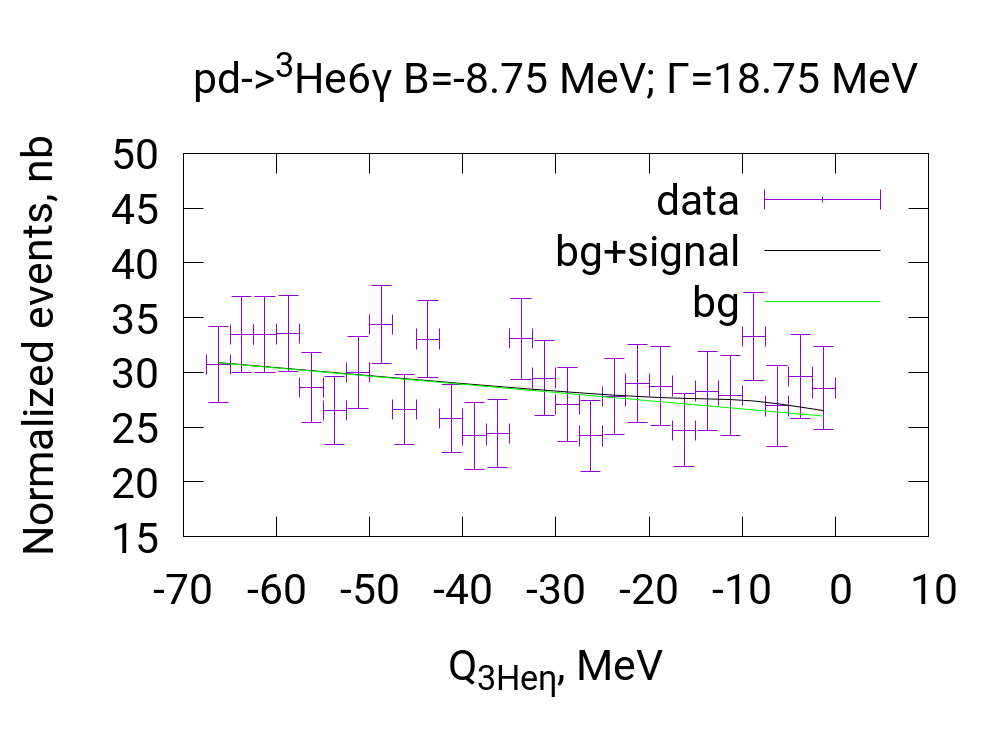}\\
		\includegraphics[width=220pt]{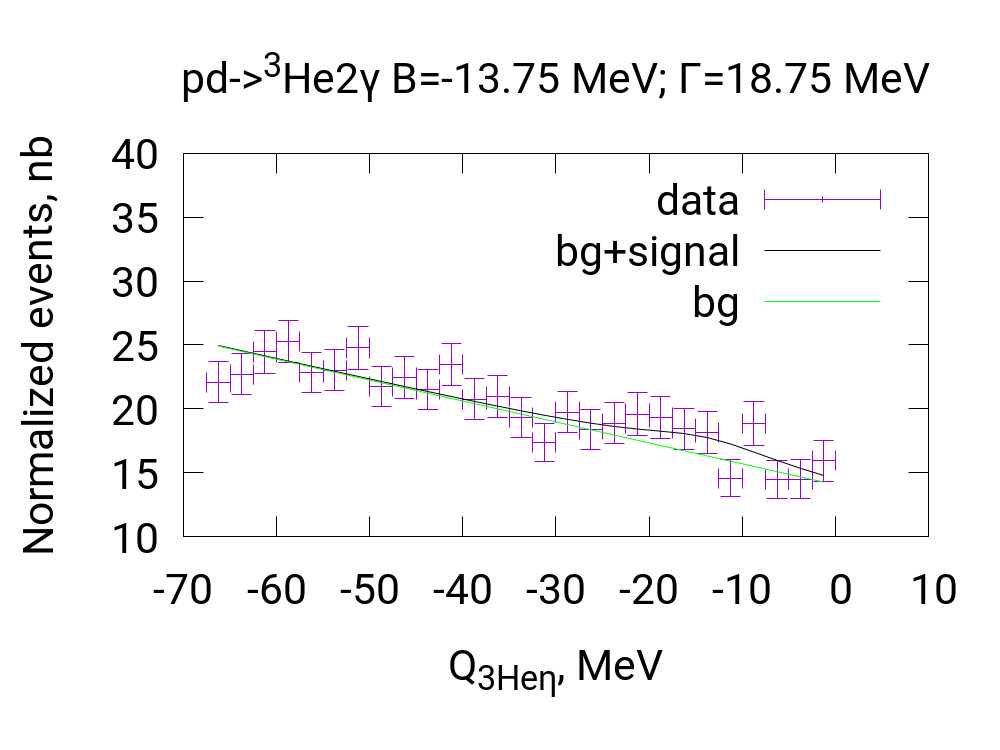}
		\includegraphics[width=220pt]{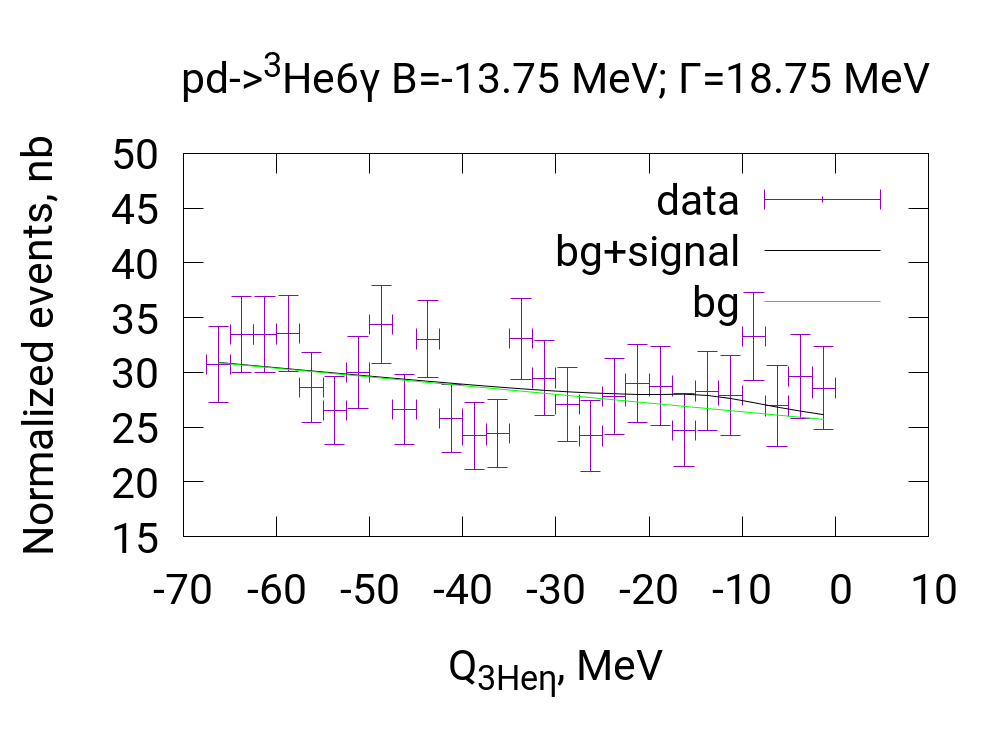}\\
		\includegraphics[width=220pt]{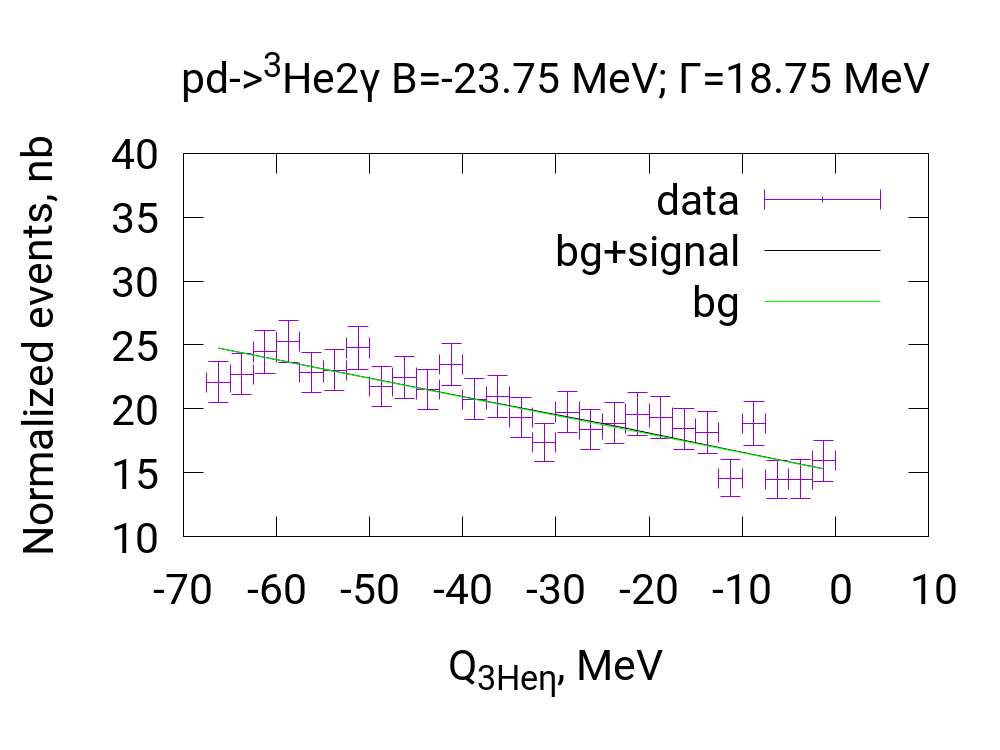}
		\includegraphics[width=220pt]{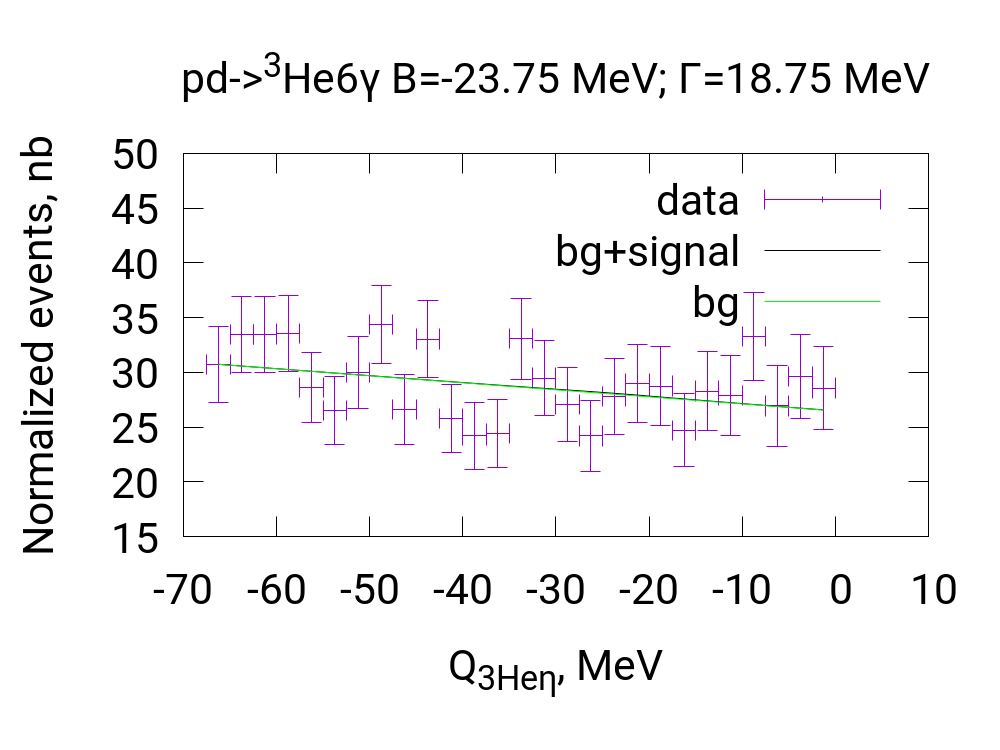}
	\end{center}
	\caption{
		Excitation curves fit using independent linear background shapes for the reactions and dependent hypothetical bound state contribution taking the branching ratio into account.
		Different bound state parameters were used
		as it is shown in the legend above the plots.
		Blue points show the experimental points taken into account,
		black line shows the fit result, and the green line shows the background function.
	}
	\label{upper_limit_fit1}
\end{figure}

\begin{figure}
	\begin{center}
		\includegraphics[width=220pt]{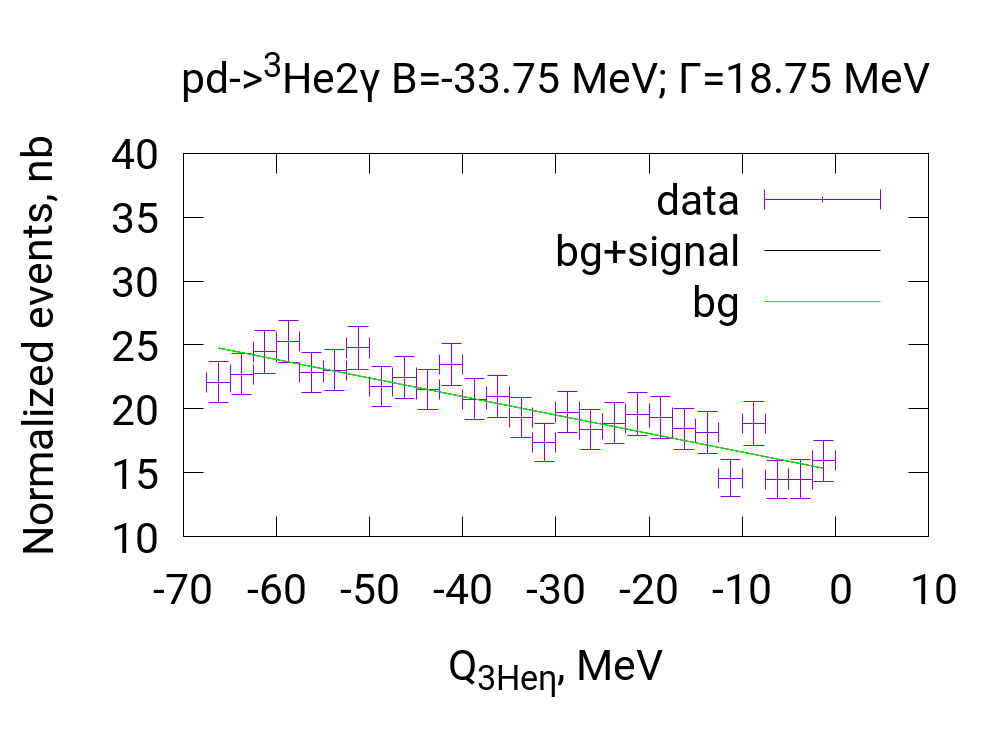}
		\includegraphics[width=220pt]{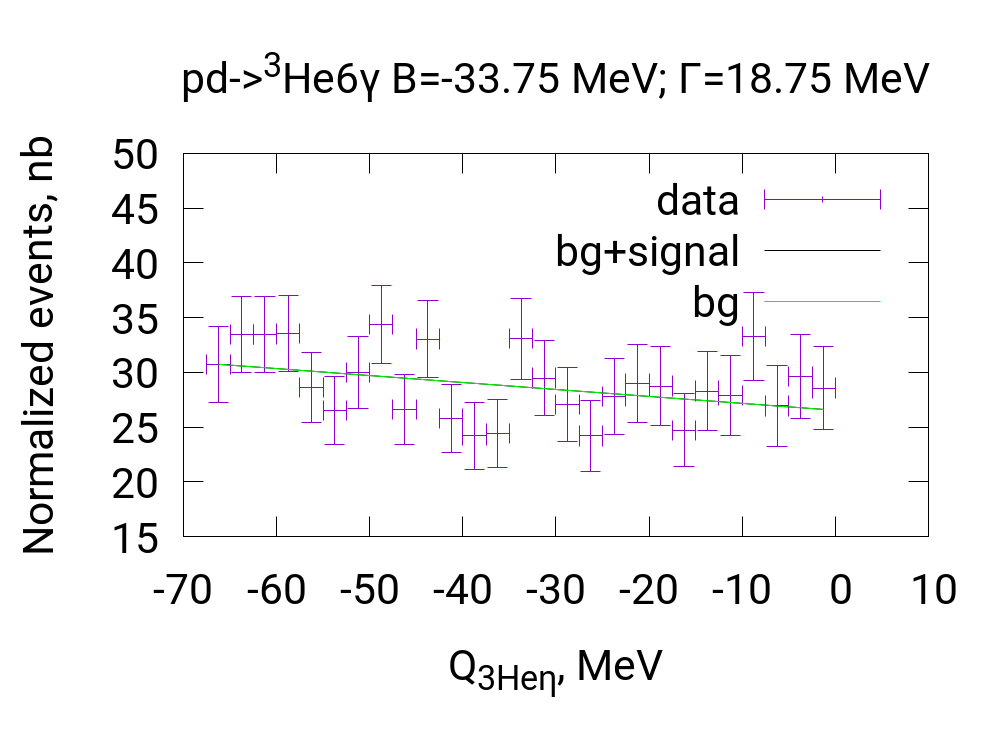}\\
		\includegraphics[width=220pt]{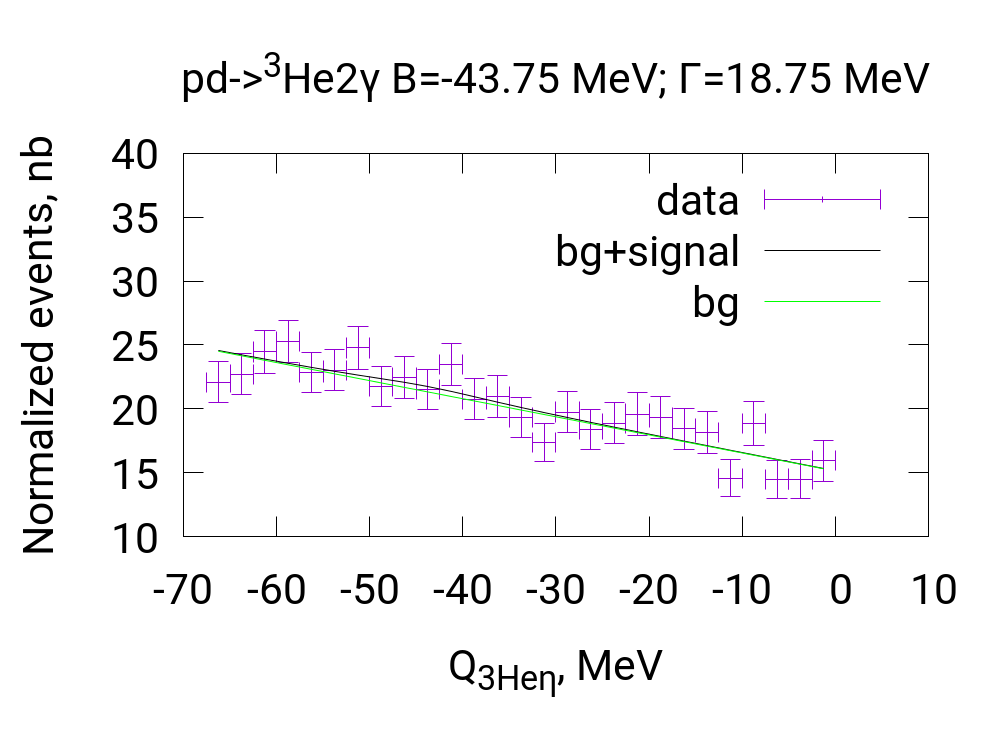}
		\includegraphics[width=220pt]{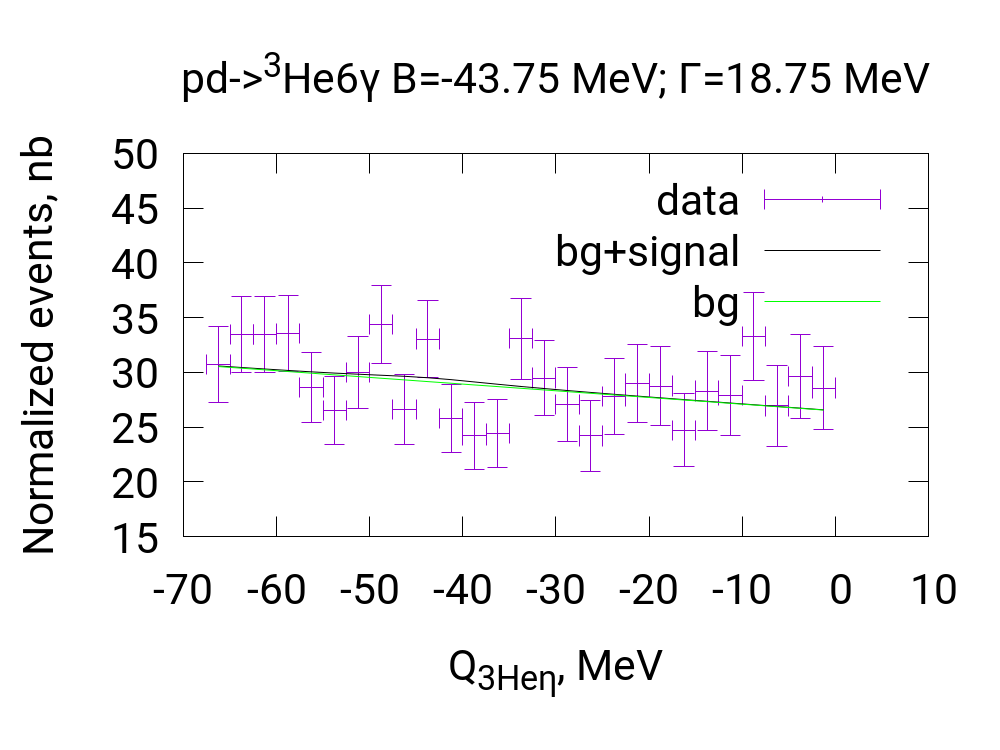}\\
		\includegraphics[width=220pt]{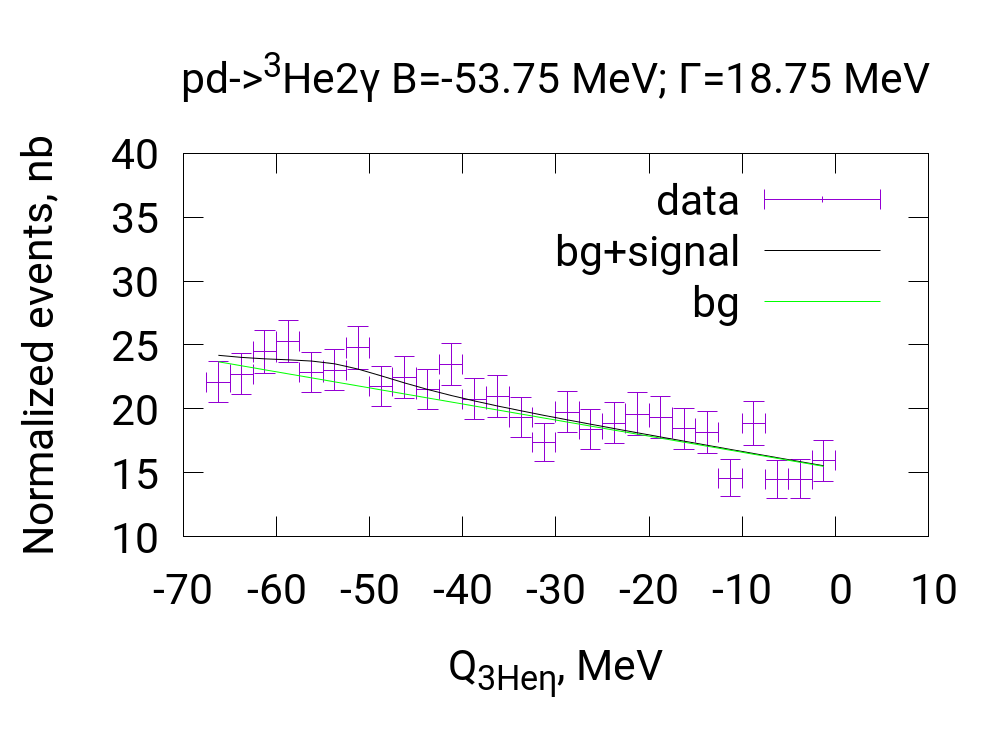}
		\includegraphics[width=220pt]{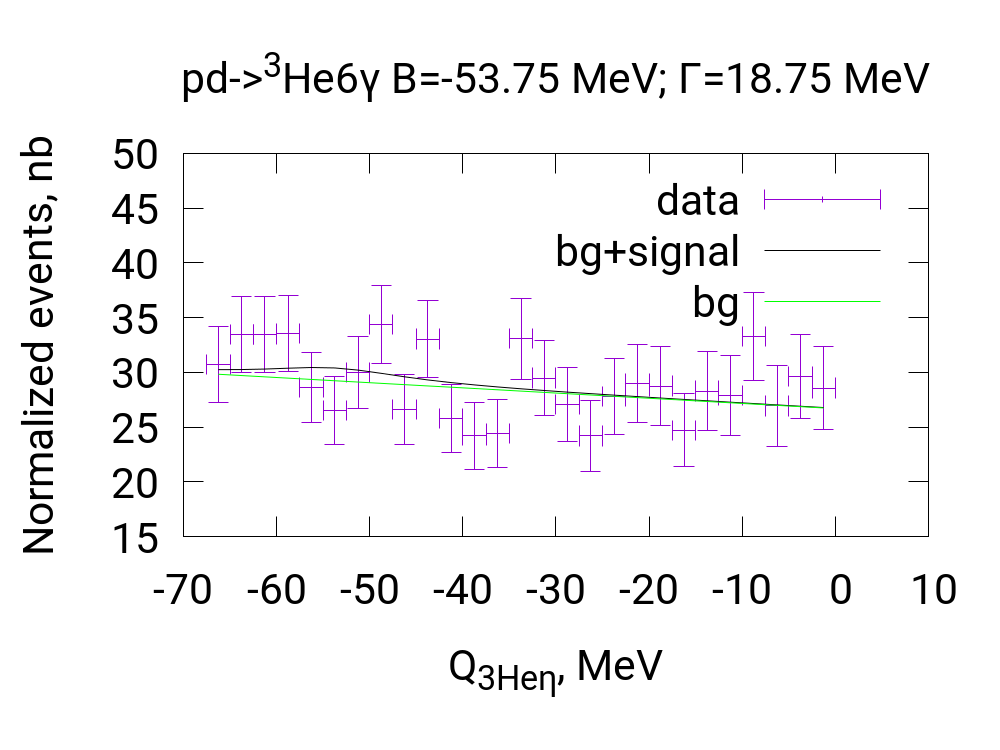}
	\end{center}
	\caption{
		Excitation curves fit using independent linear background shapes for the reactions and dependent hypothetical bound state contribution taking the branching ratio into account.
		Different bound state parameters were used
		as it is shown in the legend above the plots.
		Blue points show the experimental points taken into account,
		black line shows the fit result, and the green line shows the background function.
	}
	\label{upper_limit_fit2}
\end{figure}

\begin{figure}
	\begin{center}
		\includegraphics[width=220pt]{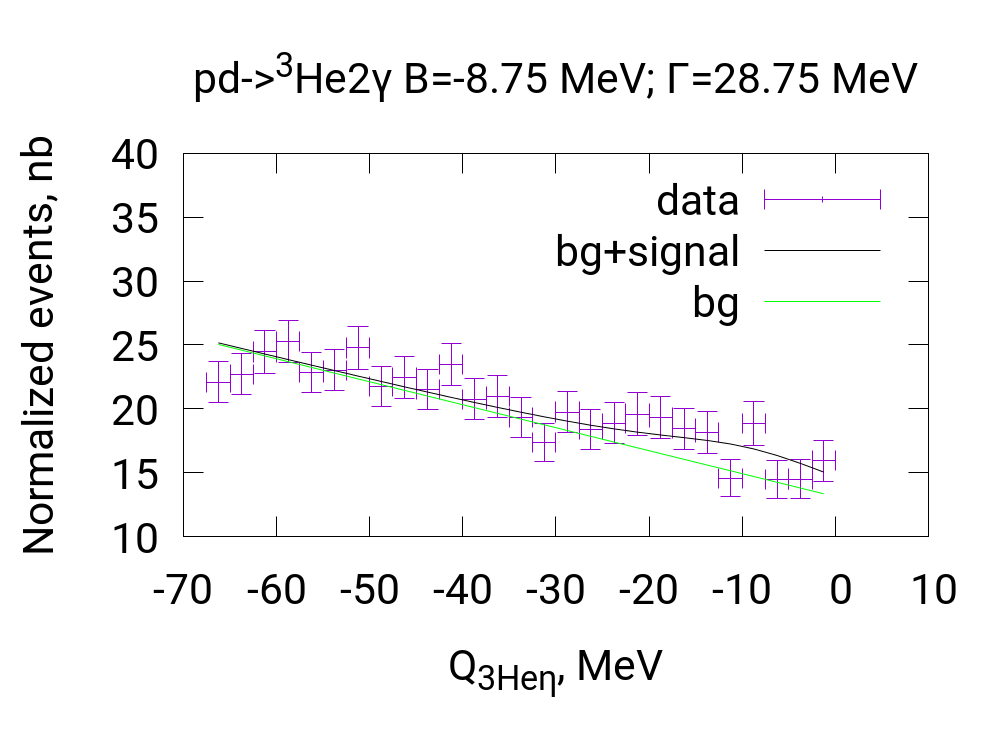}
		\includegraphics[width=220pt]{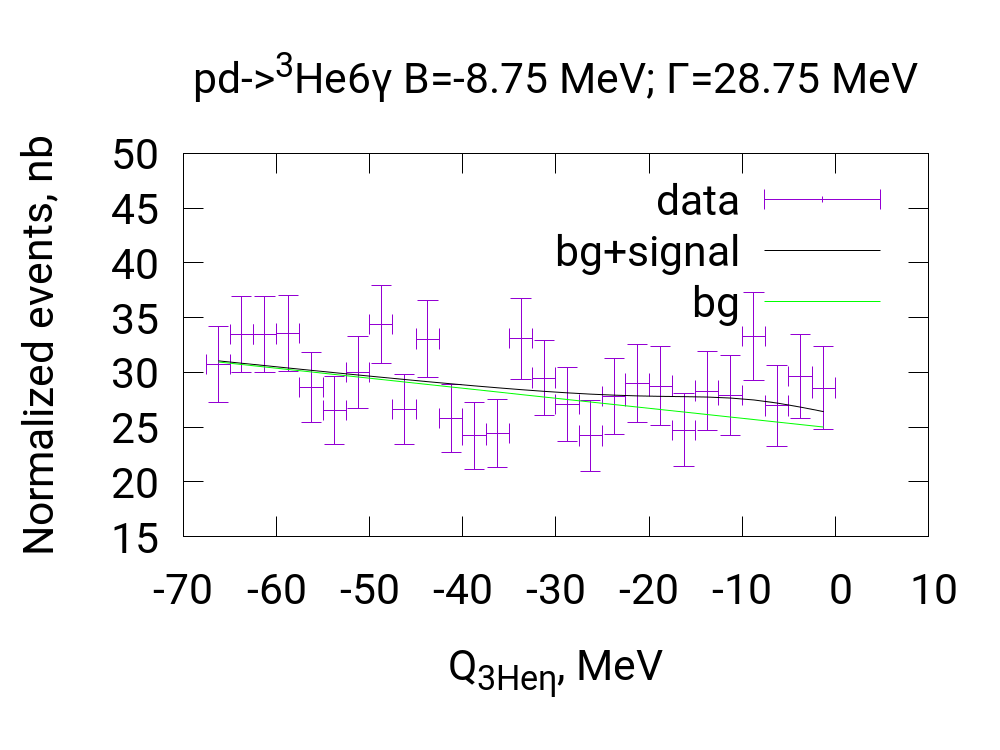}\\
		\includegraphics[width=220pt]{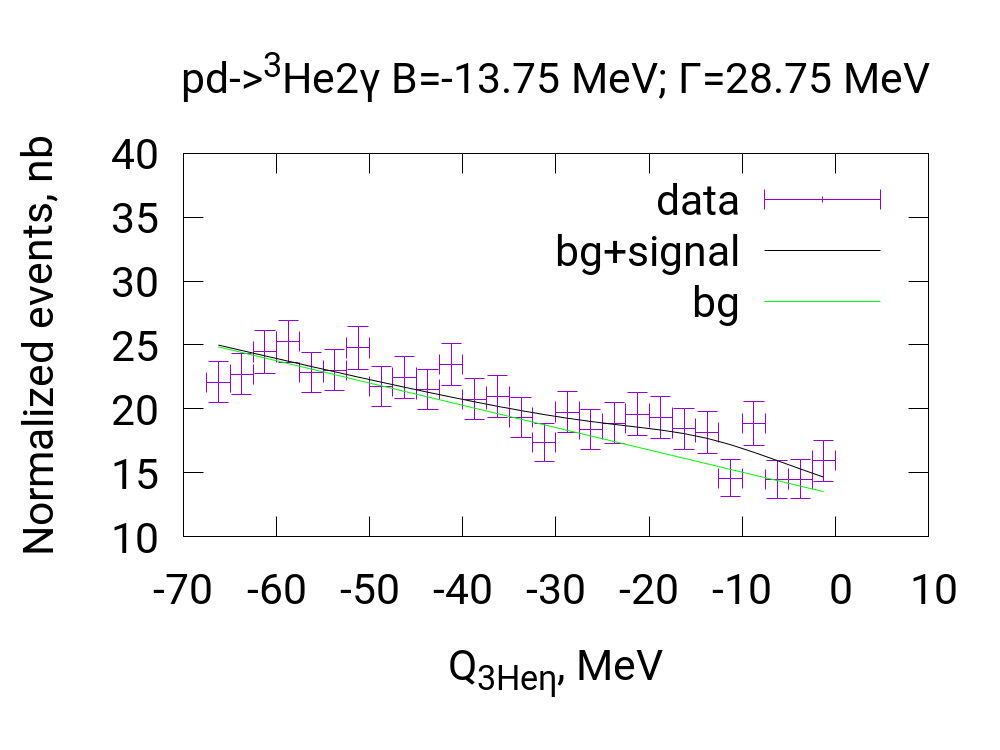}
		\includegraphics[width=220pt]{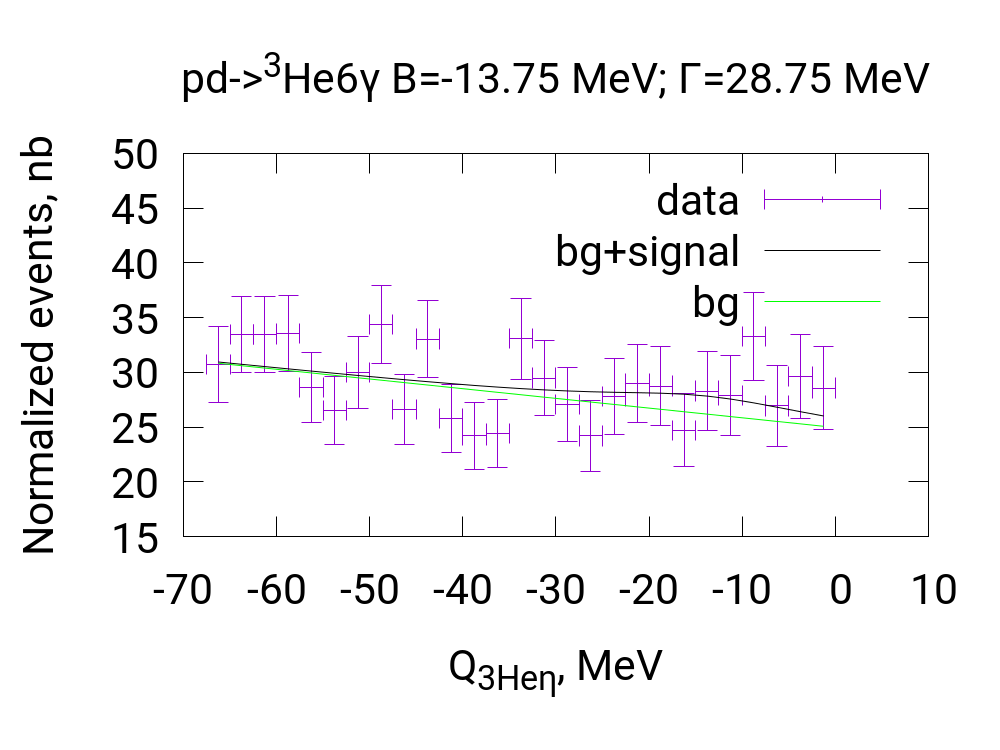}\\
		\includegraphics[width=220pt]{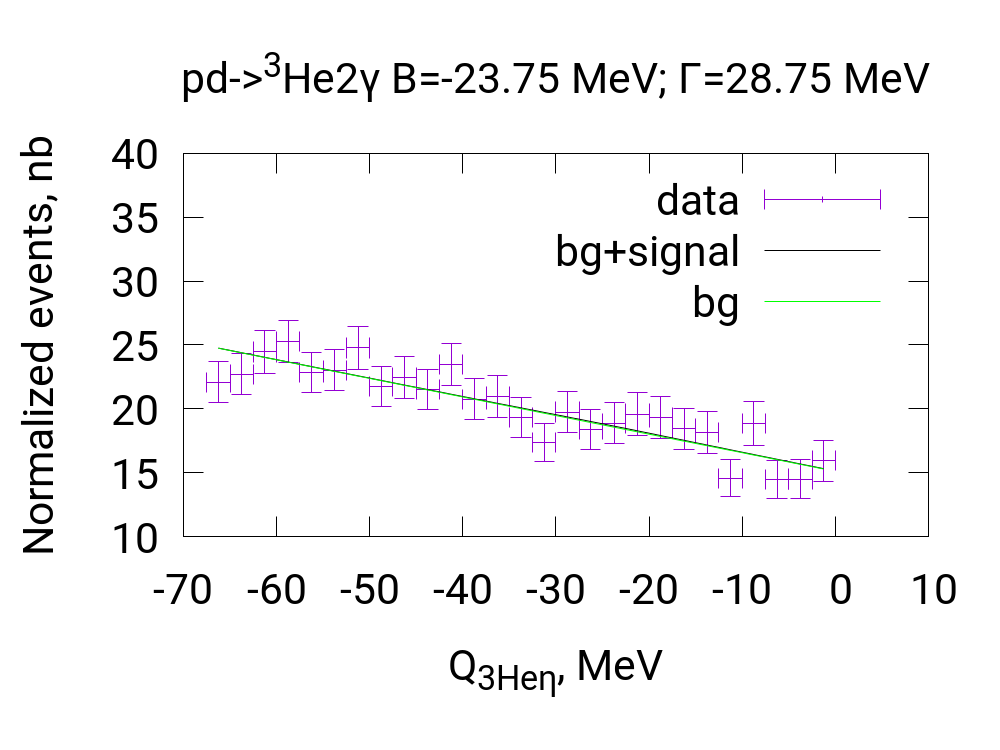}
		\includegraphics[width=220pt]{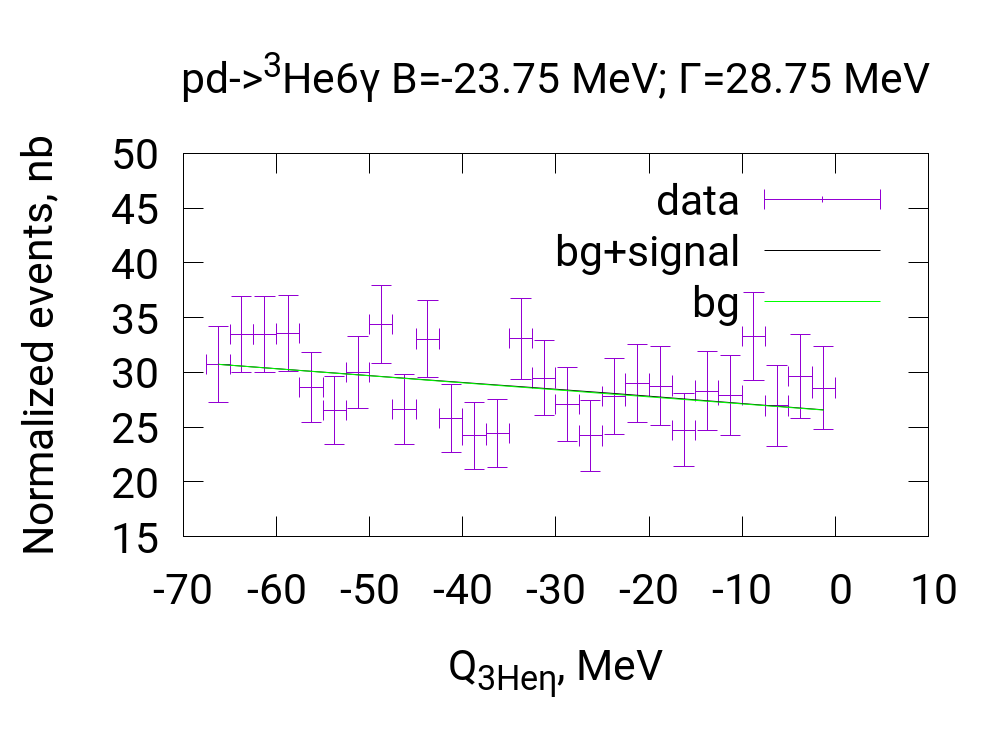}
	\end{center}
	\caption{
		Excitation curves fit using independent linear background shapes for the reactions and dependent hypothetical bound state contribution taking the branching ratio into account.
		Different bound state parameters were used
		as it is shown in the legend above the plots.
		Blue points show the experimental points taken into account,
		black line shows the fit result, and the green line shows the background function.
	}
	\label{upper_limit_fit3}
\end{figure}

\begin{figure}
	\begin{center}
		\includegraphics[width=220pt]{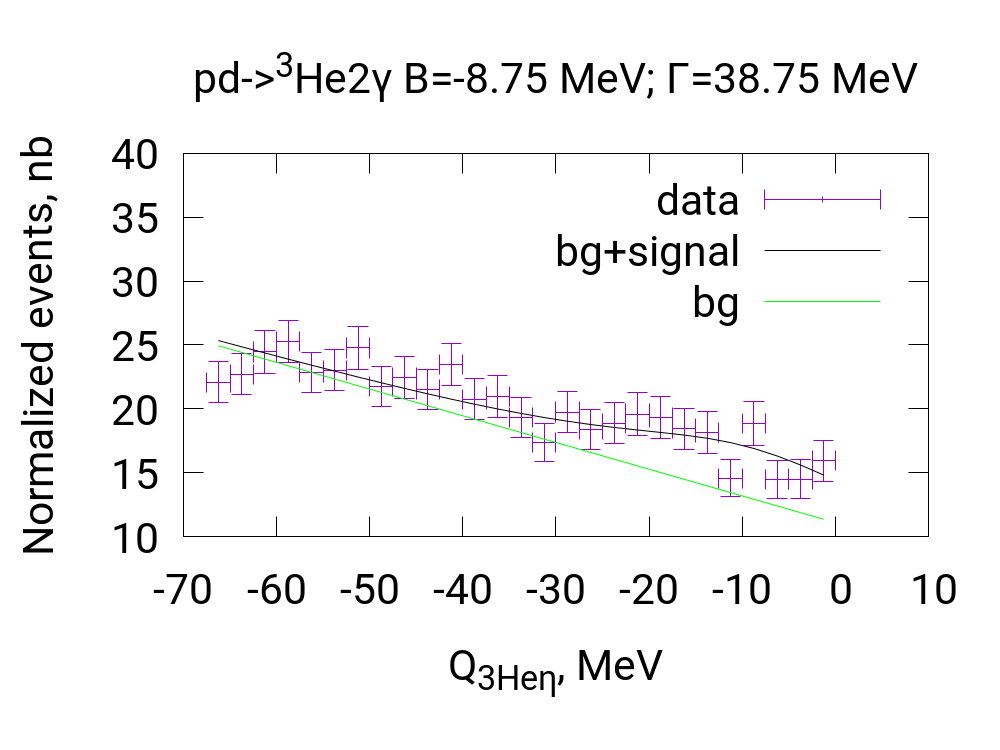}
		\includegraphics[width=220pt]{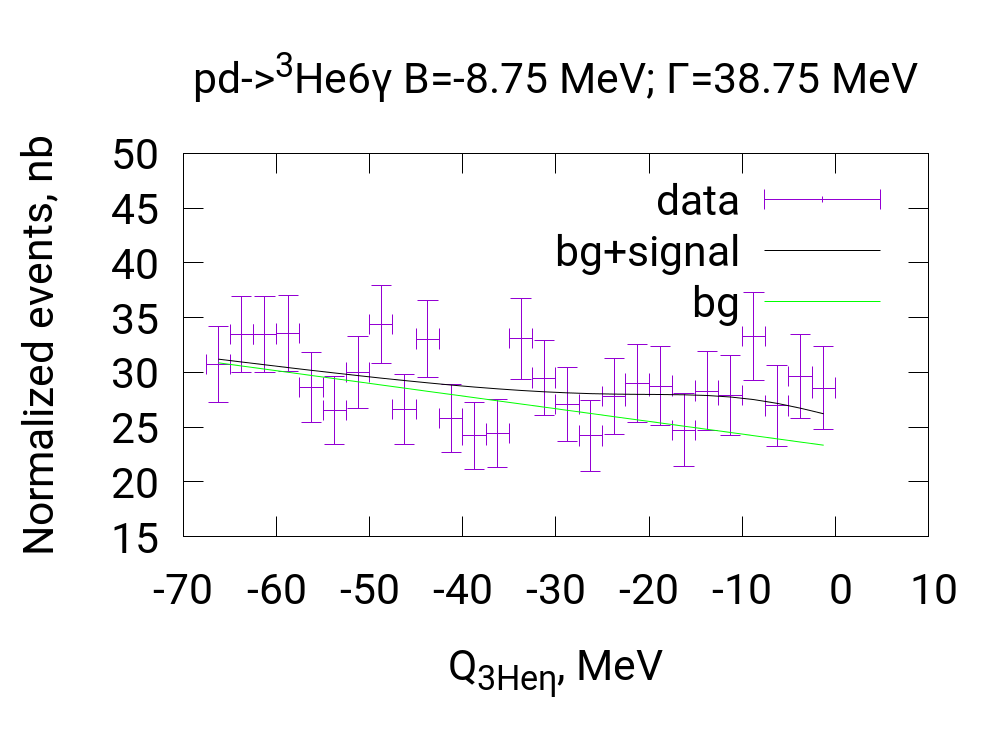}\\
		\includegraphics[width=220pt]{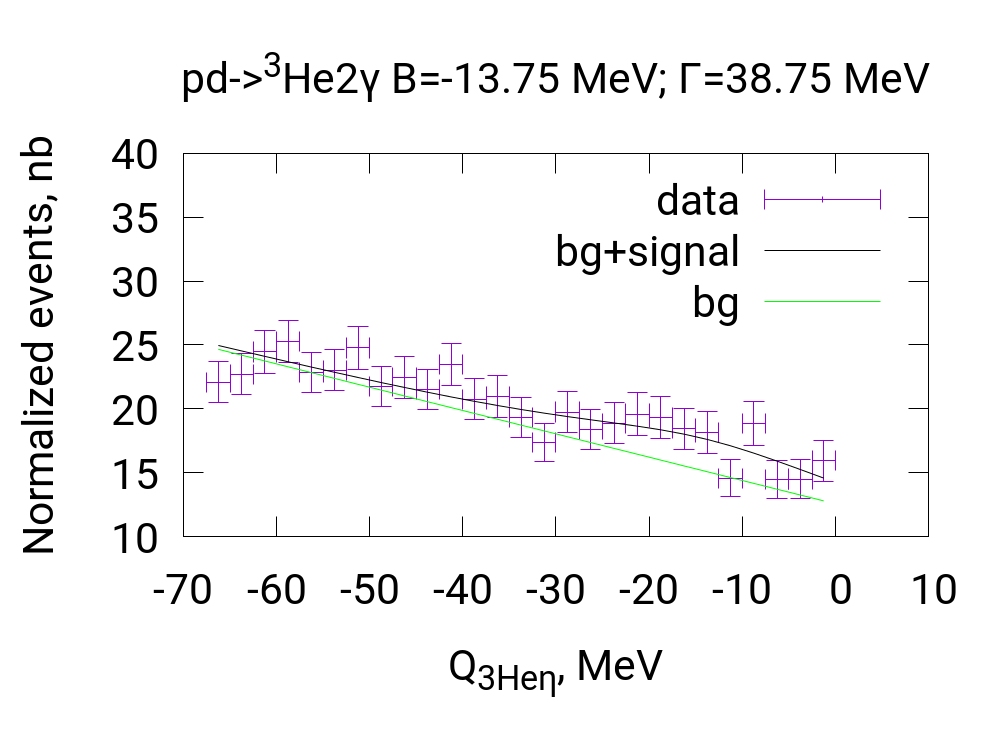}
		\includegraphics[width=220pt]{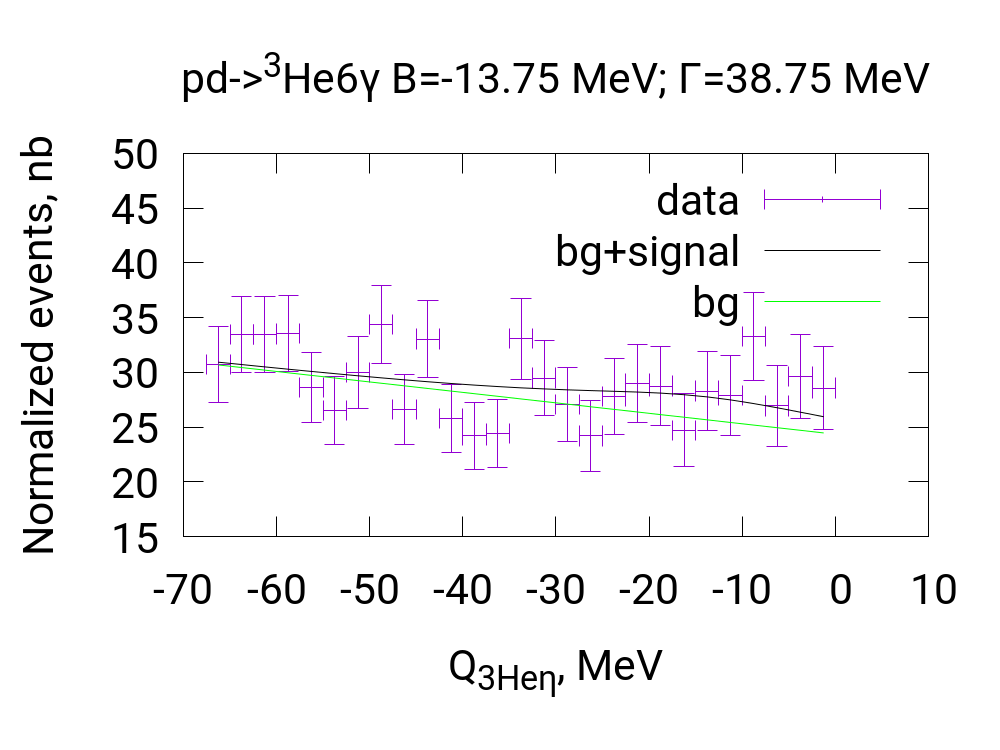}\\
		\includegraphics[width=220pt]{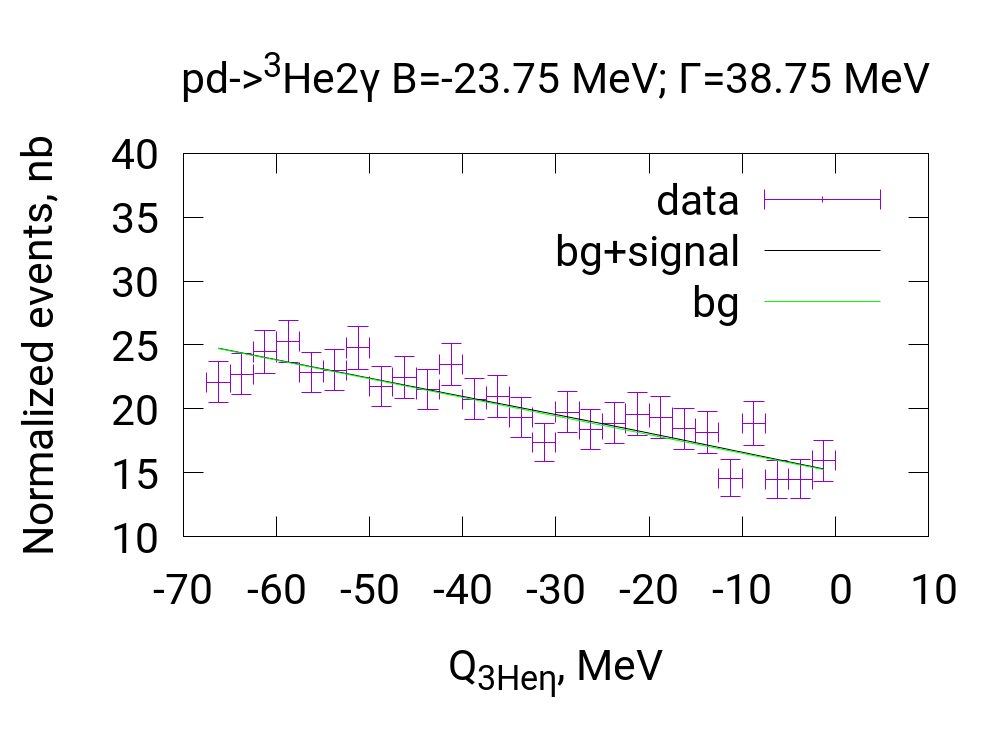}
		\includegraphics[width=220pt]{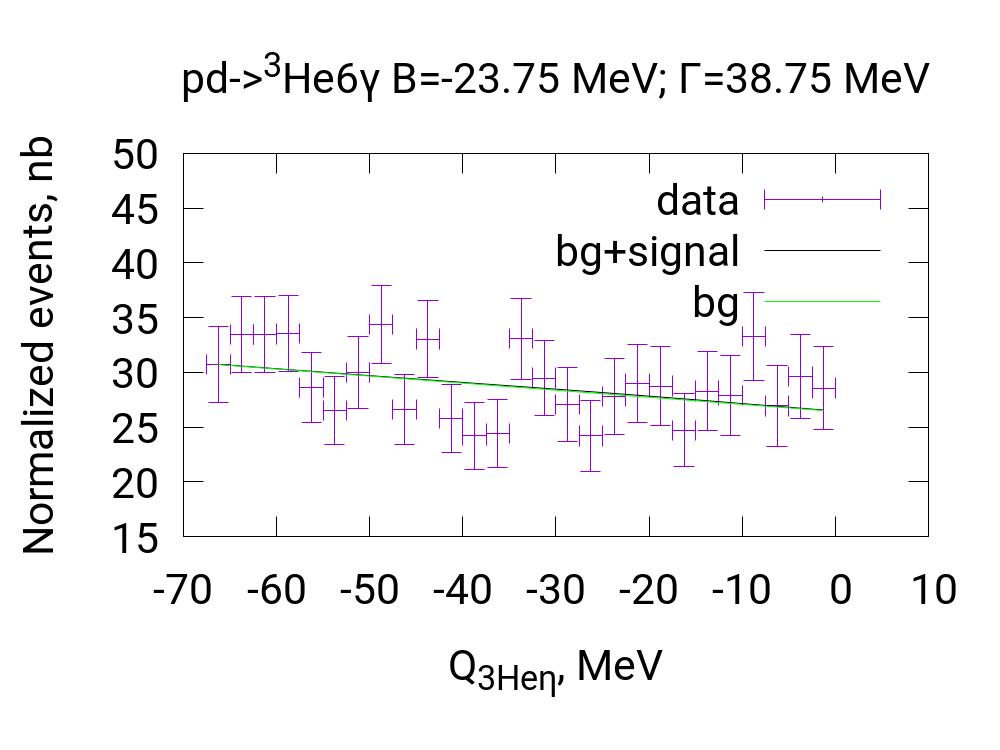}
	\end{center}
	\caption{
		Excitation curves fit using independent linear background shapes for the reactions and dependent hypothetical bound state contribution taking the branching ratio into account.
		Different bound state parameters were used
		as it is shown in the legend above the plots.
		Blue points show the experimental points taken into account,
		black line shows the fit result, and the green line shows the background function.
	}
	\label{upper_limit_fit4}
\end{figure}

For different assumed $B_s$ and $\Gamma$ values, the excitation curves for 
$pd\rightarrow^3$$He2\gamma$ and $pd\rightarrow^3$$He6\gamma$ reactions
(Fig~\ref{two_six_gamma_events_norm}) have been simultaneously fitted by the combinations of Breit-Wigner and linear functions:
\begin{equation}
	\rho^{fit}_{^3He2\gamma}(Q_{^3He\eta}) = P_{\eta\rightarrow2\gamma} \sigma_b + p_1 Q_{^3He\eta} + p_2,
	\label{excitation_curves_fit_function1}
\end{equation}
\begin{equation}
\rho^{fit}_{^3He6\gamma}(Q_{^3He\eta}) = P_{\eta\rightarrow3\pi^0} \sigma_b + p_3 Q_{^3He\eta} + p_4,
\label{excitation_curves_fit_function2}
\end{equation}
where $\sigma$, $p_1$, $p_2$, $p_3$, and $p_4$ are the fitted parameters (Fig%
~\ref{upper_limit_fit1}, \ref{upper_limit_fit2}, \ref{upper_limit_fit3}, \ref{upper_limit_fit4}%
).

The $\chi^2$ value is defined by the formula:
\begin{equation}
\chi^2 (\sigma,p_1,p_2,p_3,p_4) = \sum_i {
	(\frac{
		\rho^{measured}_{^3He2\gamma,i}-\rho^{fit}_{^3He2\gamma}(Q_{i})
	}{
		\Delta\rho^{measured}_{^3He2\gamma,i}
	})^2
}+	\sum_i {
	(\frac{
		\rho^{measured}_{^3He6\gamma,i}-\rho^{fit}_{^3He6\gamma}(Q_{i})
	}{
		\Delta\rho^{measured}_{^3He6\gamma,i}
	})^2
},
\label{excitation_curves_fit_chisq}
\end{equation}
where the $i$ index denotes sum over all $Q_{^3He\eta}$ bins (Sec.~\ref{Q-bins}) used in current analysis,
$Q_i$ means $Q_{^3He\eta}$ value corresponding to $i$th bin center,
$\rho^{measured}_{^3He2\gamma,i}$ and $\rho^{measured}_{^3He6\gamma,i}$ normalized events counts measured for $i$th bin,
$\Delta\rho^{measured}_{^3He2\gamma,i}$ and $\Delta\rho^{measured}_{^3He6\gamma,i}$ are the statistical uncertainties obtained for $i$th bin,
and $\rho^{fit}_{^3He2\gamma}(Q_{i})$ and $\rho^{fit}_{^3He6\gamma}(Q_{i})$ are the fitting functions from Eq.~\ref{excitation_curves_fit_function1} and \ref{excitation_curves_fit_function2} (Fig.~\ref{upper_limit_chisq_figure}).
The fit algorithm minimizes $\chi^2$ value varying the fit parameters $\sigma$, $p_1$, $p_2$, $p_3$, and $p_4$.

If one assumes that $\chi^2$ distribution is parabolic near its minimum:
\begin{equation}
	\chi^2 (\sigma) = (\frac{\sigma-\sigma_{fit}}{\Delta \sigma})^2 + 	\chi^2 (\sigma_{fit}),
\end{equation}
where $\sigma_{fit}$ denotes the fit parameter value corresponding to the minimum position,
and defines the statistical fit parameter uncertainty as parameter change $\Delta\sigma$ increasing $\chi^2$ by 1,
the following formula for the statistical uncertainty can be deduced:
\begin{equation}
\Delta \sigma^{stat}_{fit} = \sqrt{\frac{2}{\delta^2\chi^2/\delta\sigma^2}},
\label{fit_parameter_uncertainty}
\end{equation}
where 
$\delta^2\chi^2/\delta\sigma^2$ is the second $\chi^2$ derivative by this parameter, and
$\Delta\sigma^{stat}_{fit}$ is the fit parameter's uncertainty.
The statistical uncertainties for $p_1$, $p_2$, $p_3$, and $p_4$ fit parameters can be obtained by the same formula.
\begin{figure}[h!]
	\begin{center}
		\includegraphics[width=300pt]{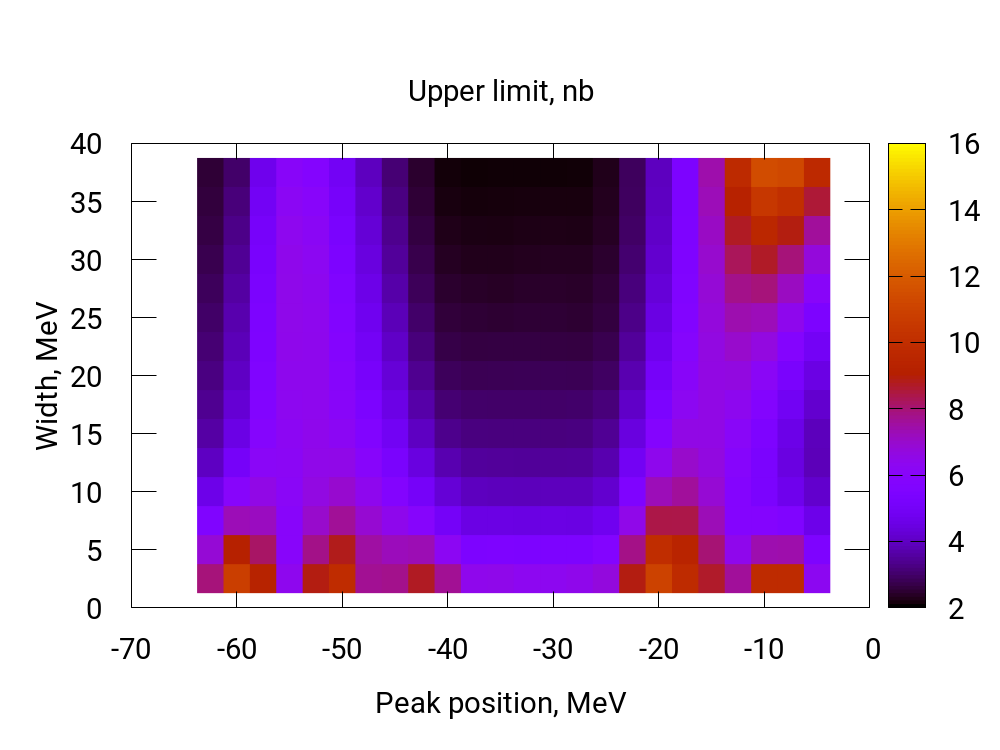}
	\end{center}
	\caption{
		The upper limit values obtained 
		based on excitation curves fit assuming different bound state parameters.
		The estimation is made with using the statistical uncertainties corresponding to $90\%$ confidence level.
	}
	\label{upper_limit_2d}
\end{figure}


\begin{figure}[h!]
	\begin{center}
		\includegraphics[width=200pt]{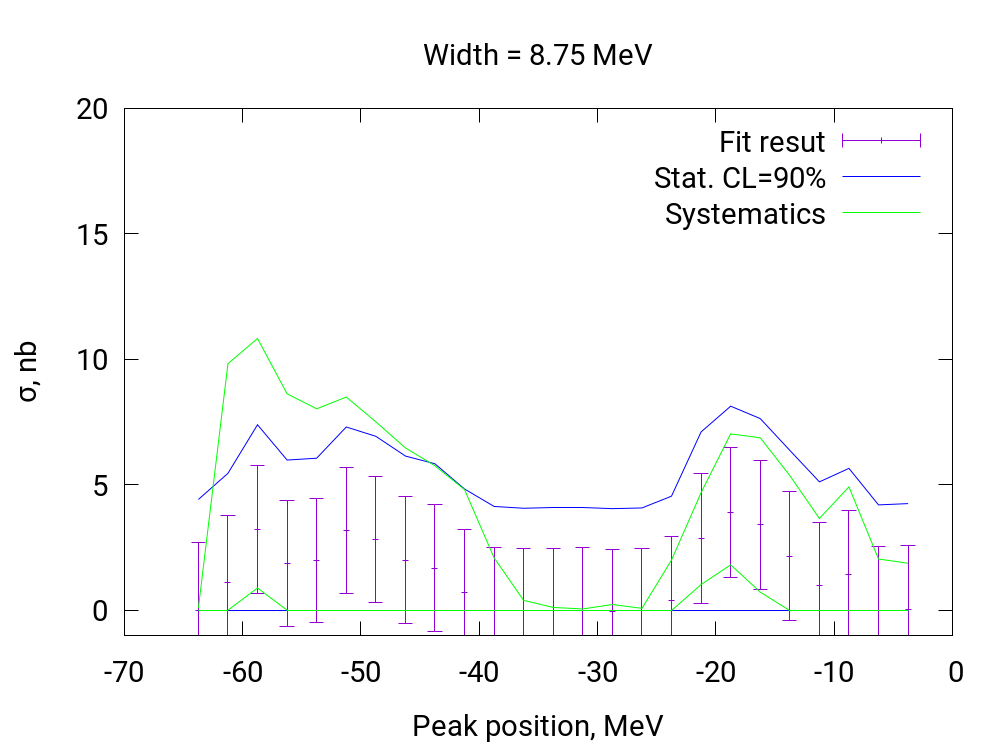}
		\includegraphics[width=200pt]{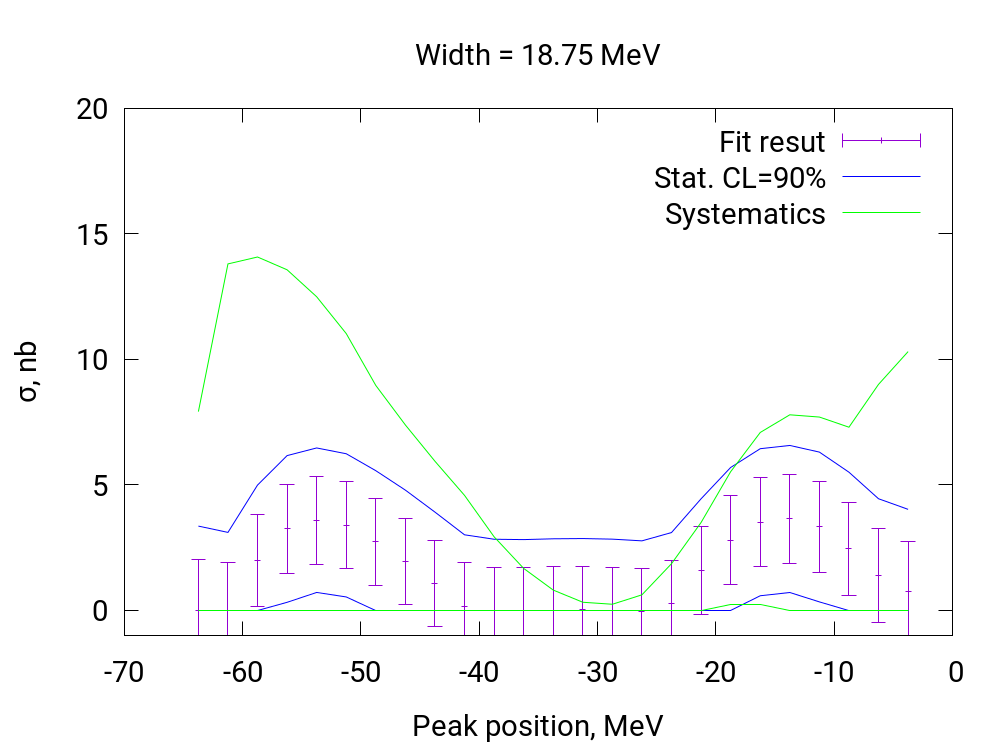}\\
		\includegraphics[width=200pt]{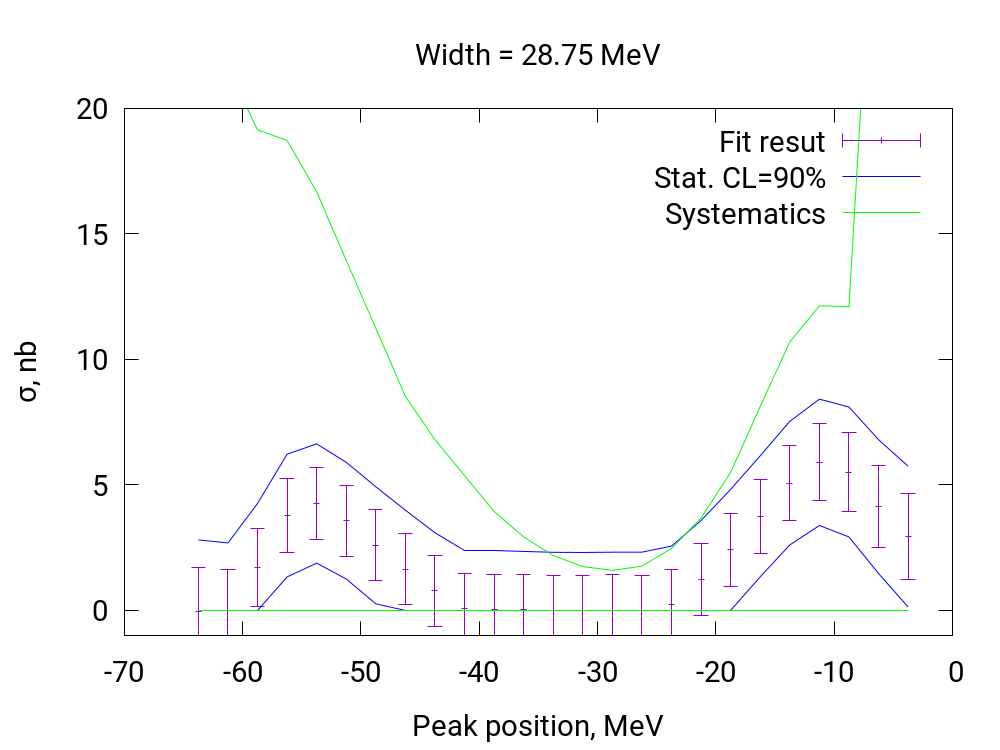}
		\includegraphics[width=200pt]{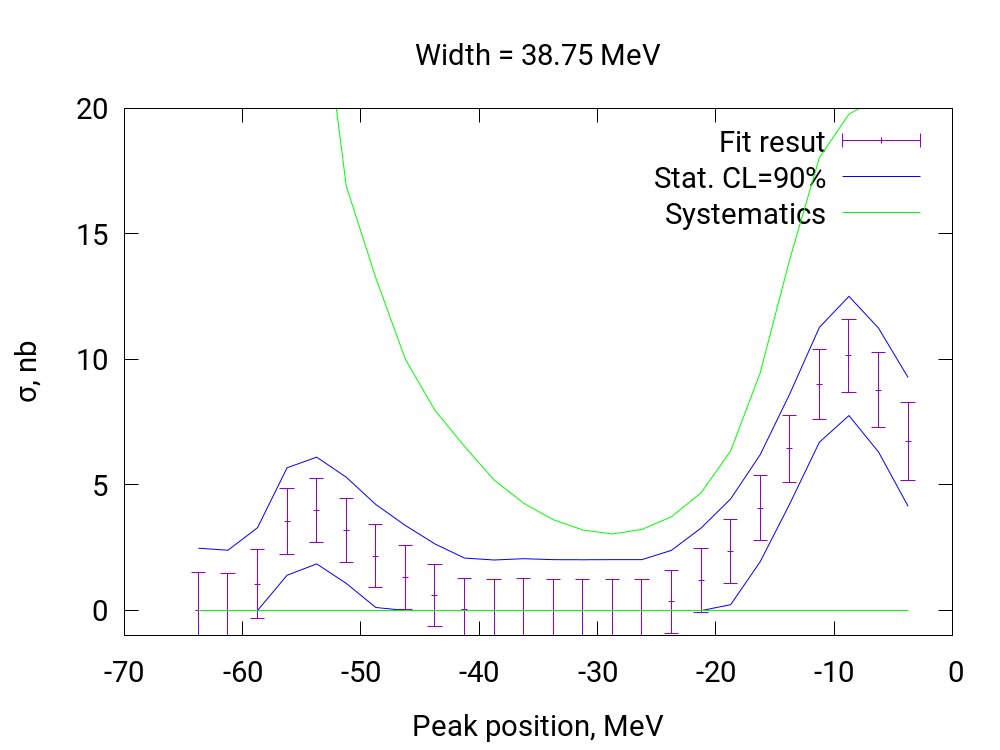}
	\end{center}
	\caption{
		The results of fit performed in order to obtain the upper limit for the bound state forming cross section.
		The error bars show amplitude values obtained by fitting algorithm with statistical uncertainties.
		The range of possible bound state forming cross section obtained based on statistical uncertainty corresponding to $90\%$ confidence level is shown by blue lines.
		The range of possible bound state forming cross section obtained based on systematic uncertainty is shown by green lines.
	}
	\label{upper_limit_1d}
\end{figure}

As far as the bound state peak is not observed, the upper limit for the bound state forming cross section is determined as fit parameter uncertainty:
\begin{equation}
\sigma_{upper}^{CL=90\%} (B_s,\Gamma)~
=~ \sigma_{fit} + \Delta \sigma^{CL=90\%}_{fit}
= \sigma_{fit} + k \Delta\sigma^{stat}_{fit},
\label{upper_limit_estimation_statistical_formulae}
\end{equation}
where $k$ is the statistical factor equal to $1.64485$ corresponding to $90\%$ confidence level.
The result is shown in Fig.~\ref{upper_limit_2d},~\ref{upper_limit_1d}.

There is statistically significant difference between the fit parameter and zero that is observed if large bound state widths are assumed (Fig.~\ref{upper_limit_1d}).
However, this difference cannot be interpreted as bound state observation in current experiment because linear fit describes the excitation curves obtained in current experiment satisfactorily from statistical point of view.
Furthermore, the interval of possible cross section values determined based on systematic uncertainty contains zero for all range of assumed hypothetical bound state parameters.
All these arguments do not allow to claim that this statistically significant difference can be interpreted as $\eta$-mesic $^3$$He$ observation.

\newpage
\section{The systematic uncertainties estimation}

The parameters contributing into upper limit estimation systematic uncertainty are all parameters contributing the systematics of luminosity estimation, $pd\rightarrow^3$$He2\gamma$, and $pd\rightarrow^3$$He6\gamma$ reaction analysis.
All these parameters are listed in Table.~\ref{systematics_ppn_lum_parameters} and \ref{systematics_central_gamma_parameters}.

\begin{figure}[h!]
	\begin{center}
		\includegraphics[width=220pt]{UpperLimit--87-287Fit1.png}
		\includegraphics[width=220pt]{UpperLimit--87-287Fit2.png}\\
		\includegraphics[width=220pt]{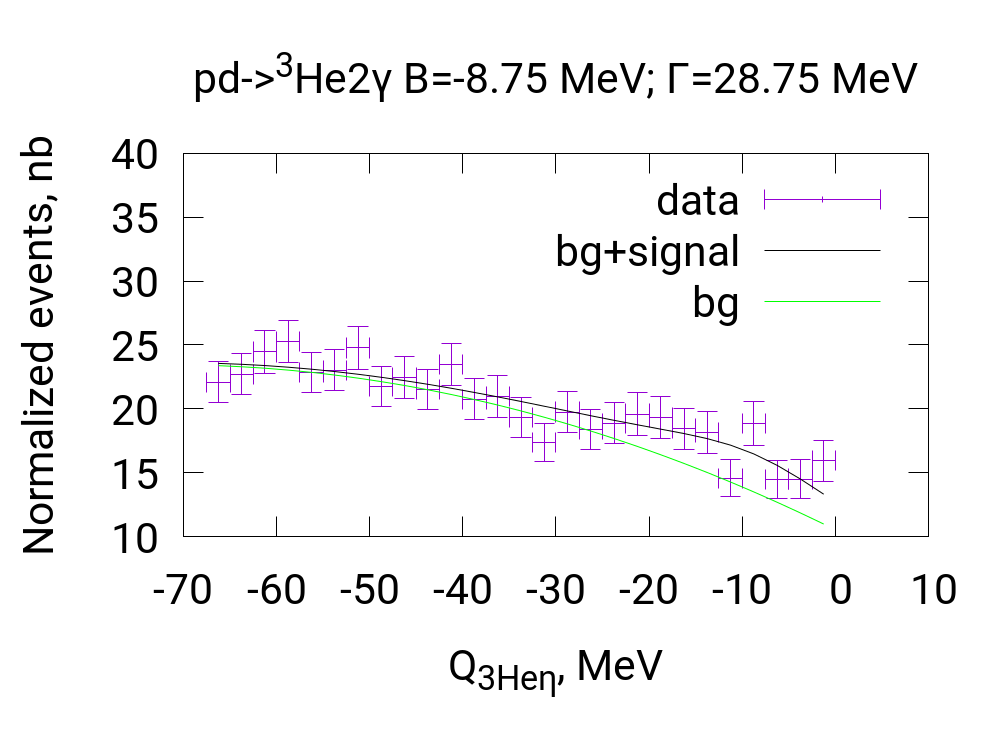}
		\includegraphics[width=220pt]{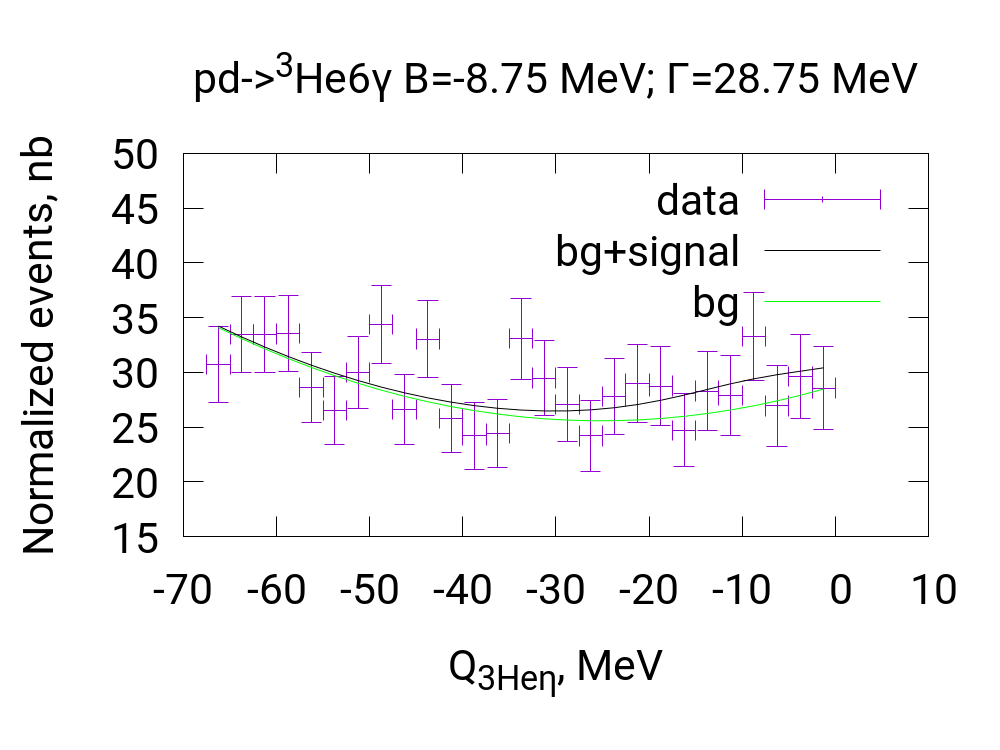}
	\end{center}
	\caption{
		Excitation curves fit using linear (upper panels) and quadratic (lower panels) background shapes for the hypothetical bound state parameters resulting in the largest upper limit for the cross section (Fig.~\ref{upper_limit_2d}).
		The fit with quadratic background was performed in order to estimate the systematic uncertainty (Eq.~\ref{upper_systematic_quadratic_eq}).
		The hypothetical bound state parameters are denoted in the legend above the plots.
		Blue points show the experimental points taken into account,
		black line shows the fit, and the green line shows the background function.
	}
	\label{upper_limit_fit_quad_syst}
\end{figure}

One more source of systematic error is connected with background fit function.
The analysis was performed using linear fit for background.
For systematic uncertainty estimation, fit with quadratic background function was used as well (Fig.~\ref{upper_limit_fit_quad_syst}).

The contribution to systematic
originating from the choice of the function to fit the background was calculated by the formula:
\begin{equation}
	\Delta \sigma^{(1)\pm}_{fit} = {| \sigma^{quadratic}_{fit} - \sigma^{linear}_{fit}|},
	\label{upper_systematic_quadratic_eq}
\end{equation}
where $\sigma_{fit}$ means the cross section value obtained from fit and the upper index denotes the background function chosen for particular fit. 
The index $+$ or $-$ is selected depending on if the fit value is increased or decreased by the background polynomial power change.
The component with opposite sign is assumed to be equal to zero.

\begin{figure}
	\begin{center}
		\includegraphics[width=220pt]{UpperLimit--87-287Fit1.png}
		\includegraphics[width=220pt]{UpperLimit--87-287Fit2.png}\\
		\includegraphics[width=220pt]{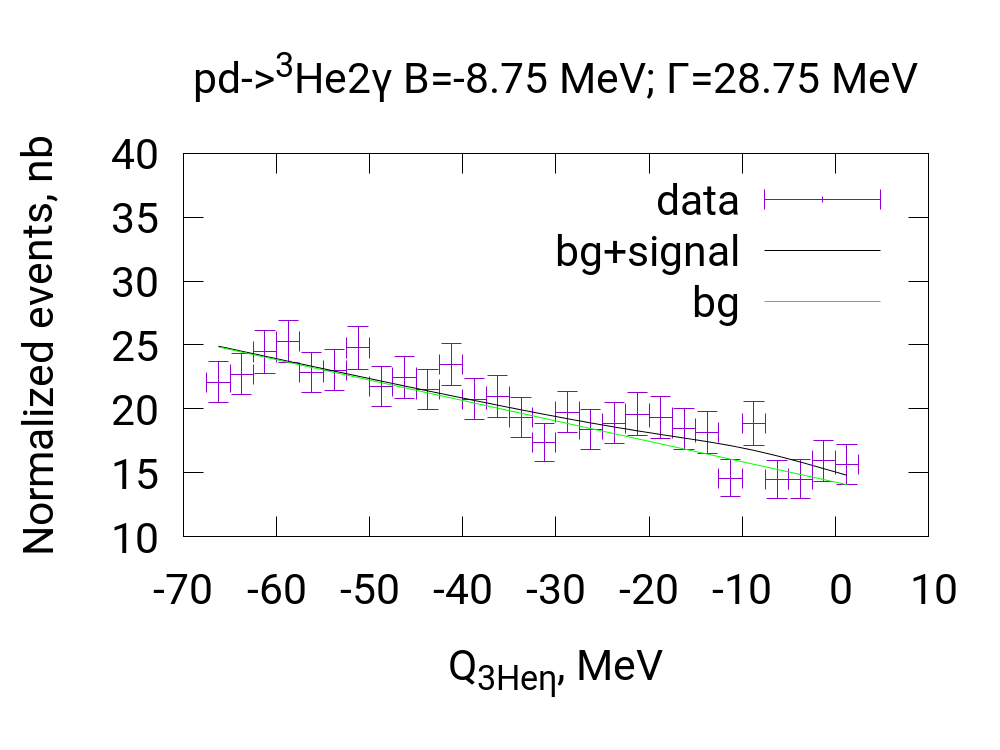}
		\includegraphics[width=220pt]{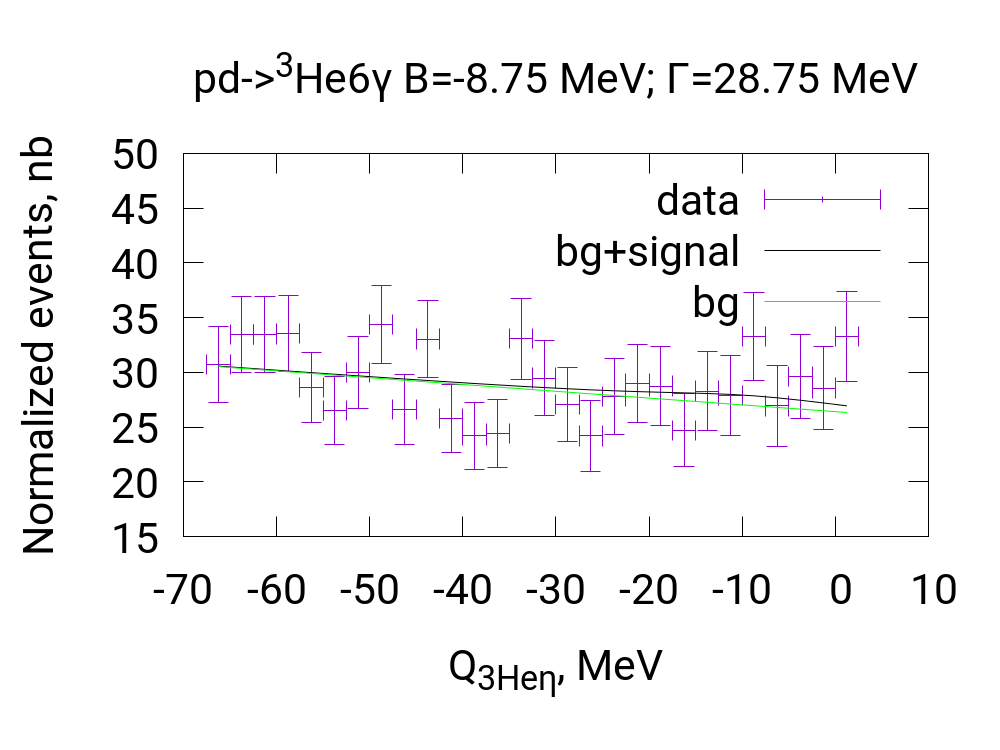}\\
		\includegraphics[width=220pt]{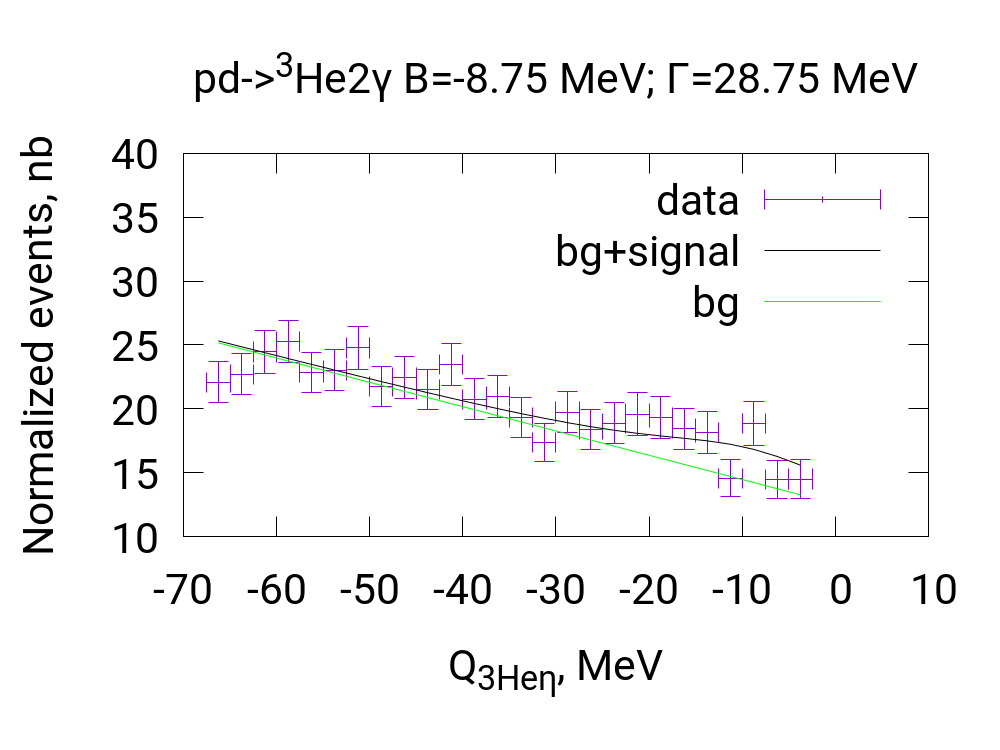}
		\includegraphics[width=220pt]{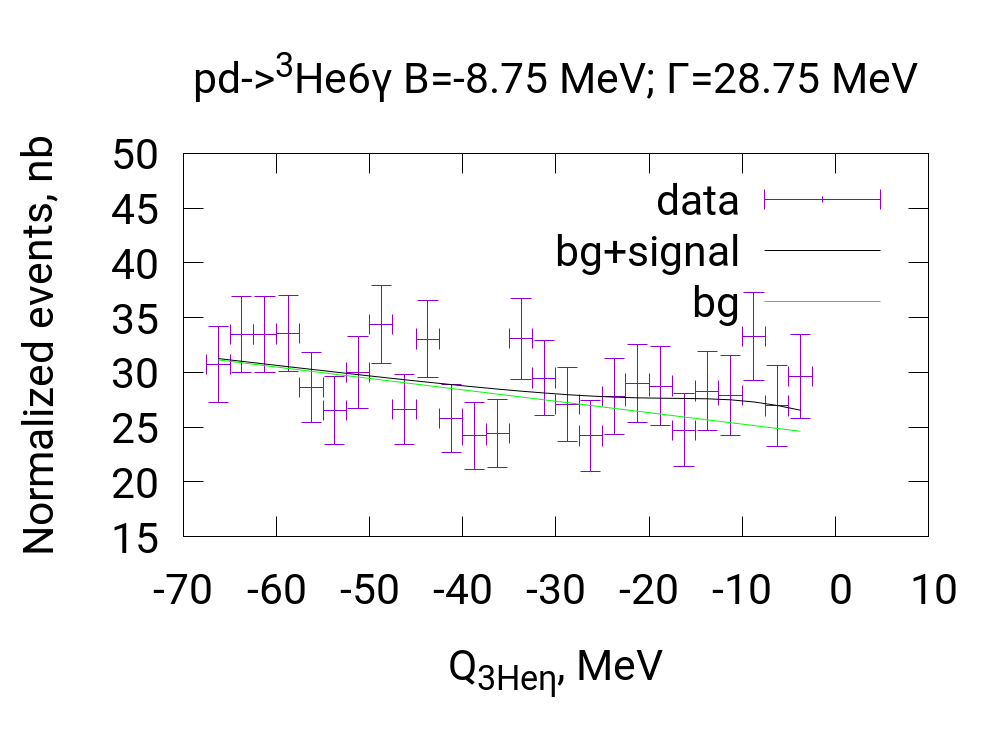}
	\end{center}
	\caption{
		Excitation curves fit using normal fit range (upper panels) and changed ones in order to determine the systematic uncertainty (Eq.~\ref{upper_systematic_ranges_eq1}~\ref{upper_systematic_ranges_eq2}).
		The hypothetical bound state parameters resulting in the largest upper limit for the cross section (Fig.~\ref{upper_limit_2d}) are used for this figure.
		These parameters are denoted in the legend above the plots.
		Blue points show the experimental points taken into account,
		black line shows the fit, and the green line shows the background function.
	}
	\label{upper_limit_fit_move_syst}
\end{figure}

One more contribution into the systematic error was calculated by varying the fit ranges by one bin (Fig.~\ref{upper_limit_fit_move_syst}):
\begin{equation}
\Delta \sigma^{(2)\pm}_{fit} =|\sigma^{left\pm}_{fit}-\sigma^{final}_{fit}|,
\label{upper_systematic_ranges_eq1}
\end{equation}
\begin{equation}
\Delta \sigma^{(3)\pm}_{fit} =
|\sigma^{right\pm}_{fit}-\sigma^{final}_{fit}|,
\label{upper_systematic_ranges_eq2}
\end{equation}
where the indices $left+$ and $right+$ denote changes of left or right fit range respectively by one bin that increases the fit result.
$left-$ and $right-$ indices denote the fit range changes that decrease the fit result.
If both changes of some fit range result in the same fitted parameter change direction they are included to one systematic uncertainty component.

The total systematic uncertainty is calculated by the formula
\begin{equation}
	\Delta\sigma^{syst\pm}_{fit} = \sqrt{ 
		\sum_i |\sigma^{P_i = P^{\pm}_i}_{fit}-\sigma^{final}_{fit}|^2
	+ \sum_{k=1}^{3}(\Delta \sigma^{(k)\pm}_{fit})^2 
},
	\label{upper_limit_systematics_formulae}
\end{equation}
where $i$ index, $P_i$, and $final$ have the same meaning as in eq.~\ref{luminosity_systematics_formulae},~\ref{events_count_systematics_formulae},
 $\Delta\sigma^{(k)}_{fit}$ denotes the uncertainties estimated in Eq.~\ref{upper_systematic_quadratic_eq}-\ref{upper_systematic_ranges_eq2},
 $P^{+}_i$ denotes the change of $i$th parameter that increases the fit result,
 and $P^{-}_i$ denotes the parameter change that decreases the result.
 If both changes of $P_i$ result in the same fitted parameter change direction they are included to one systematic uncertainty component.

According to Ref.~\cite{systematic_uncertainties_link}, not all parameters were taken into account but only the ones fulfilling the condition described below.

\begin{figure}[h!]
	\begin{center}
		\includegraphics[width=220pt]{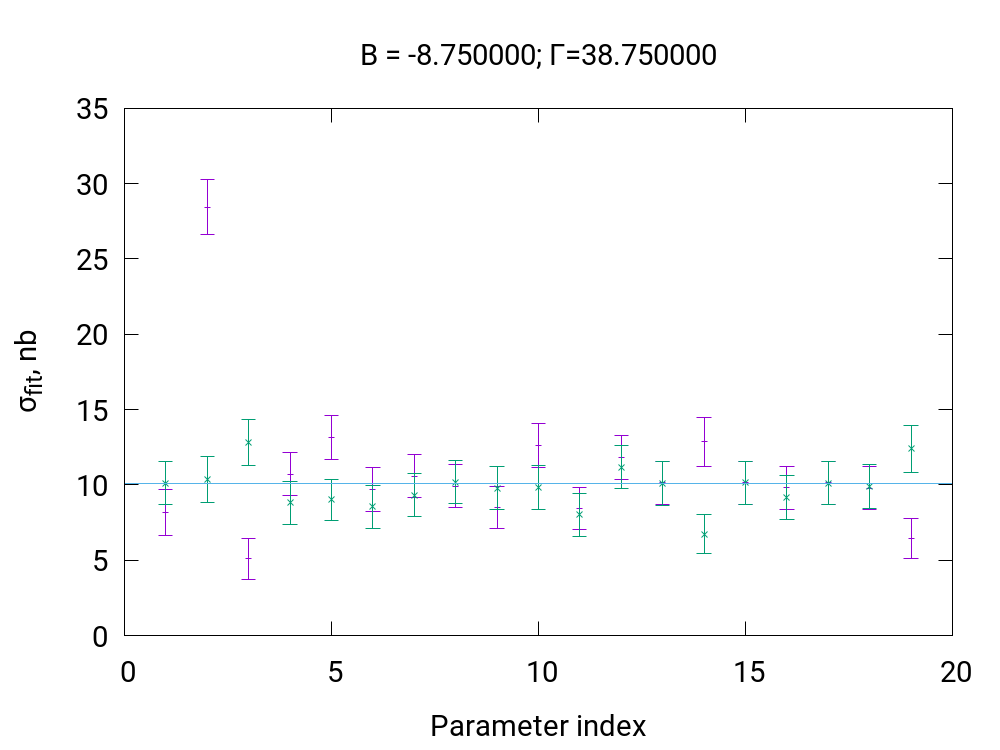}
		\includegraphics[width=220pt]{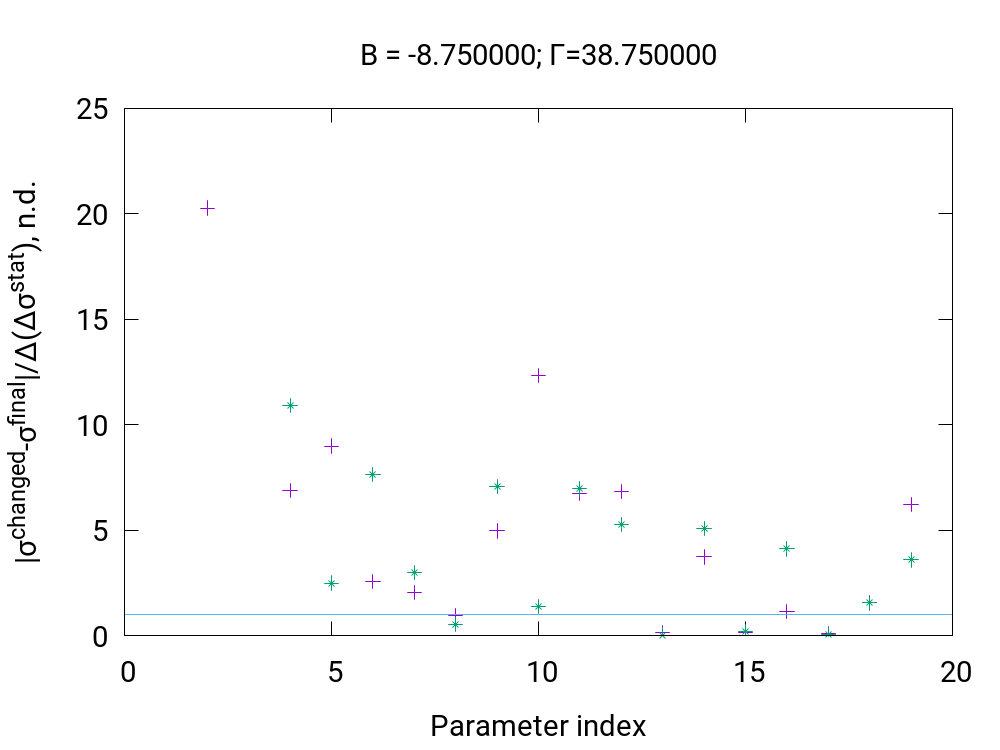}
	\end{center}
	\caption{
Left panel: The changes of fit result for particular values of  assumed bound state parameters caused by varying the parameters taken into account in systematic uncertainty calculations. 
The error bars denote the statistical uncertainties.
The horizontal line shown the $\sigma_{fit}^{final}$ value.
Right panel: The magnitude from Eq.~\ref{systematic_parameter_condition} denoting the condition applied to the parameters taken into account in systematic uncertainty calculation.
The horizontal line shows the value of $1$ used in the condition.
The parameter indices used in this figure are explained in Table.~\ref{upper_limit_systematic_details_table}.
	}
	\label{upper_limit_systematic_details_figure}
\end{figure}

For each parameter change, the following magnitude is calculated (Fig.~\ref{upper_limit_systematic_details_figure}):
\begin{equation}
	\Delta(\Delta\sigma^{stat}_{fit})^{\pm} = \sqrt{|(\Delta\sigma^{final}_{fit})^2-(\Delta\sigma^{changed\pm}_{fit})^2|},
\end{equation}
where $\Delta\sigma^{final}_{fit}$ and $\Delta\sigma^{changed\pm}_{fit}$ denote the statistical uncertainties for the final result and changed parameter value respectively.
The meaning of $+$ or $-$ indices is the same as in the equations above.
The statistical uncertainties are calculated by Eq.~\ref{fit_parameter_uncertainty}.

When the parameter changes fulfill the condition:
\begin{equation}
	\frac{|\sigma^{changed\pm}_{fit}-\sigma^{final}_{fit}|}{\Delta(\Delta\sigma^{stat}_{fit})^{\pm}}<1,
	\label{systematic_parameter_condition}
\end{equation}
it is excluded from Eq.~\ref{upper_limit_systematics_formulae}.

For each hypothetical bound state parameters,
two values of uncertainties were calculated: $\Delta\sigma_{fit}^{syst+}$ and $\Delta\sigma_{fit}^{syst-}$.
These values allow to assume the range of possible bound state forming cross section values:
\begin{equation}
	\sigma^{final}_{fit}-\Delta\sigma_{fit}^{syst-} < \sigma < \sigma^{final}_{fit}+\Delta\sigma_{fit}^{syst+}.
\end{equation}

This range is shown in Fig.~\ref{upper_limit_1d} by green lines.

\begin{table}[h!]
	\begin{tabular}{|p{50pt}|p{400pt}|}
\hline
		Index&Parameter description \\
\hline
		1& Background polynomial power (Eq.~\ref{upper_limit_fit_quad_syst}) \\
\hline
		2& Left excitation curve fit range (Eq.~\ref{upper_systematic_ranges_eq1}) \\
\hline
		3& Right excitation curve fit range (Eq.~\ref{upper_systematic_ranges_eq2}) \\
\hline
		4& Beam momentum correction constant (Eq.~\ref{beam_momentum_correction_constant}) \\
\hline
		5& $^3$$He$ identification cut height (Fig.~\ref{he3_cut_data}, Eq.~\ref{he3_cut_height}) \\
\hline
		6&  $\gamma$ energy threshold \\
\hline
		7& Time cut  (Fig.~\ref{two_gamma_time_condition}, \ref{six_gamma_time_condition}, left panel)  \\
\hline
		8& Time cut  (Fig.~\ref{two_gamma_time_condition}, \ref{six_gamma_time_condition}, right panel, left border)  \\
\hline
		9& Time cut  (Fig.~\ref{two_gamma_time_condition}, \ref{six_gamma_time_condition}, right panel, right border)  \\
\hline
		10 &Cut on $\theta(\sum \vec{p}_{\gamma_i})$ (Fig.~\ref{two_gamma_eta_theta_cut}, \ref{six_gamma_eta_theta_cut}) \\
\hline
		11 & $^3$$He$ missing mass cut position (Fig.~\ref{two_gamma_3He_missing_mass_cut}, \ref{six_gamma_3He_missing_mass_cut})  \\
\hline
		12 &  $\gamma$ quanta missing mass cut position (Fig.~\ref{two_gamma_ggmm_cut}, \ref{six_gamma_gamma_missing_mass_cut}, left border) \\
\hline
		13 &$\gamma$ quanta missing mass cut position (Fig.~\ref{two_gamma_ggmm_cut}, \ref{six_gamma_gamma_missing_mass_cut}, right border) \\
\hline
		14 & Cut on $\gamma\gamma$ invariant  mass corrected by $Q_{^3He\eta}$ (Fig.~\ref{two_gamma_ggim_cut}, left border) \\
\hline
		15 &Cut on $\gamma\gamma$ invariant mass corrected by $Q_{^3He\eta}$ (Fig.~\ref{two_gamma_ggim_cut}, right border) \\
\hline
		16 & Cut for $3\pi^0$ decay identification (Fig.~\ref{six_gamma_im_pi_diff_cut})\\
\hline
		17 &  Cut on 6 $\gamma$ quanta invariant  mass corrected by $Q_{^3He\eta}$ (Fig.~\ref{six_gamma_gamma_invariant_mass_cut}, left border) \\
\hline
		18 & Cut on $\gamma$ quanta invariant mass corrected by $Q_{^3He\eta}$ (Fig.~\ref{six_gamma_gamma_invariant_mass_cut}, right border) \\
\hline
		19 & $\theta$ angular cut for $^3$$He$ tracks (Fig.~\ref{he3_reconstruction_kinematic_hist}, Eq.~\ref{he3_theta_cut_pos}) \\
\hline
	\end{tabular}
\caption{
		The list of parameters contributing into systematic error.
		The indices given in the left column correspond to the horizontal coordinates in Fig.~\ref{upper_limit_systematic_details_figure}.
}
\label{upper_limit_systematic_details_table}
\end{table}
\chapter{Conclusions and outlook}
This dissertation describes the search for $^3$$He-\eta$ bound state via the study of $pd\rightarrow^3$$He2\gamma$ and $pd\rightarrow^3$$He6\gamma$ reactions.
This is the first experiment considering the reaction mechanism with direct decay of the bound $\eta$ meson.
The dissertation contains detailed description of the experiment and the data analysis.

The measurement was performed in 2014 at COSY accelerator in Jülich (Germany) using WASA-at-COSY detection system.
Ramped beam technique was used that allowed to reduce systematic uncertainties.
The beam momentum was varied in the range $[1.426;1.635]~\frac{MeV}{c}$ corresponding to $[-70;30]~MeV$ range for excess energy of $^3$$He-\eta$ system.

The performed analysis allows to determine the upper limit for $^3$$He-\eta$ bound state formation and decay cross sections.
The conditions applied for the events selection are based on the bound state forming and decay Monte Carlo simulation.
$\eta\rightarrow2\gamma$ and $\eta\rightarrow3\pi^0$ decay channels of bound $\eta$ meson are taken into account.

Total integrated luminosity was obtained based on 
$pd\rightarrow^3$$He\eta$ reaction in the range of $Q_{^3He\eta}~\in~[10;30]~MeV$ 
and quasielastic proton-proton scattering in the range of $Q_{^3He\eta}~\in~[-70;30]~MeV$.
The results are consistent within systematic errors.
The excess energy dependence of luminosity
obtained based on quasifree proton-proton scattering
was used for
$pd\rightarrow^3$$He2\gamma$ and $pd\rightarrow^3$$He6\gamma$
excitation curves normalization.

The analysis of the obtained excitation functions for the  $pd\rightarrow^3$$He\eta$ indicates slightly the signal from the bound state for $\Gamma>20~MeV$ and $B\in[0;15]~MeV$.  
However, the observed indication is in the range of the systematic error. 
Therefore, the final conclusion of this thesis is that no narrow structure that could be interpreted as $\eta$-mesic nucleus was observed in both excitation curves.

The obtained excitation curves do not reveal resonance-like structures that could be interpreted as $^3$$He\eta$ bound state with binding energy less than $60~MeV$ and width less than $40~MeV$.
The fit of the excitation curves with combined independent linear functions and Breit-Wigner distribution multiplied by branching ratios of corresponding $\eta$ decay channels allows to determine the upper limit for $pd\rightarrow(^3$$He\eta)_{bound}$ process 
assuming that the $\eta$ decay channels branching ratio is not influenced by coupling with $^3$$He$ nucleus.
The upper limit for the cross section of the bound state forming is varying between $2$ and $15~nb$ depending on the bound state parameters.
This is the first result obtained for the direct decay of bound $\eta$ meson.
The upper limit is essentially lower than the limit of $70~nb$ for $pd\rightarrow(^3$$He\eta)_{bound}\rightarrow^3$$He\pi^0$  reaction obtained by COSY-11 Collaboration~\cite{light_he3_old_upperlimit}.
\appendix\newpage

\addcontentsline{toc}{chapter}{Appendix}
\chapter{$\eta$ meson properties}
\label{eta_properties}
$\eta$ meson was discovered in 1961~\cite{eta_meson_discovered} and was intensively studied experimentally and theoretically.
According to the Particle Data Group (PDG) the $\eta$ mass is equal to $547.862~\pm~0.017~MeV/c^2$ and the full width is $1.31~\pm~0.05~keV$~\cite{eta_decay_modes_new}.
According to $\eta$ meson's spin $J=0$ and odd parity $P=-1$, it is a pseudoscalar meson.
$\eta$ is a neutral meson with zero isospin and even charge parity and G-parity.

The most intensive $\eta$ meson decay channels are shown in the Table.~\ref{eta_decay_modes_table}.
The decay probabilities are given according to the PDG~\cite{eta_decay_modes_new}.
Many $\eta$ decay channels are energetically possible but forbidden according to the conservation of C, P, and CP symmetry and are used for these symmetries test~\cite{particle_masses,eta_decay_modes_new}.
\begin{table}
	\begin{center}
		\begin{tabular}{|p{100pt}|p{100pt}|}
			\hline
			Decay modes & Probability \\
			\hline
			Neutral modes    & $(71.91\pm0.34)\%$\\
			$2\gamma$     & $(39.41\pm0.20)\%$\\
			$3\pi^0$         & $(32.68\pm0.23)\%$\\
			\hline
			Charged modes               & $(28.10\pm0.34)\%$\\
			$\pi^{+}\pi^{-}\pi^0$     & $(22.92\pm0.28)\%$\\
			$\pi^{+}\pi^{-}\gamma$ & $(4.22\pm0.08)\%$\\
			\hline
		\end{tabular}
	\end{center}
	\caption{
		The most intensive $\eta$ meson decay modes~\cite{eta_decay_modes_new}.
	}
	\label{eta_decay_modes_table}
\end{table}

\begin{figure}
	\begin{center}
		\includegraphics[width=220pt]{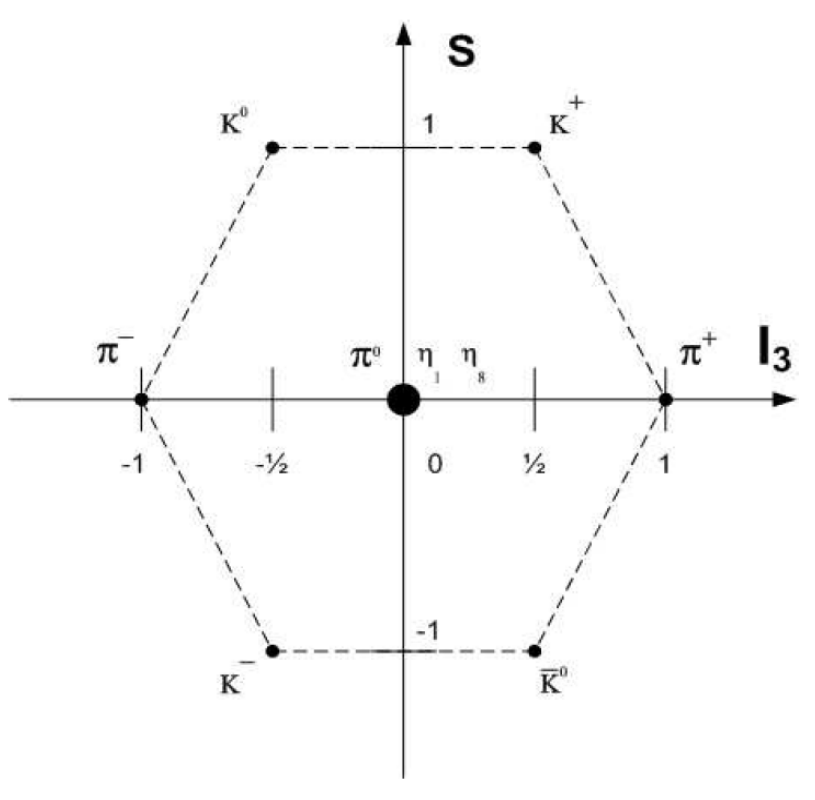}
	\end{center}
	\caption{The nonet of pseudoscalar mesons. The axes show isospin third component ($I_3$) and the strangeness ($S$). The picture is taken from Ref.~\cite{particle_book2}.}
	\label{nonet}
\end{figure}
In frame of quark model, the $\eta$ meson is classified as lightest pseudoscalar mesons SU(3)-flavor nonet component~\cite{particle_book,particle_book2} (Fig.~\ref{nonet}).
The $\eta_1$ and $\eta_8$ are the flavor singlet and the flavor octet states respectively:
\begin{equation}
\Ket{\eta_1} = \frac{1}{\sqrt{3}}(d\tilde{d}+u\tilde{u}+s\tilde{s}),
\end{equation}
\begin{equation}
\Ket{\eta_8} = \frac{1}{\sqrt{6}}(d\tilde{d}+u\tilde{u}-2s\tilde{s}).
\end{equation}
These clean states are not observed but their superpositions:
\begin{equation}
\Ket{\eta} = \cos\theta \Ket{\eta_8} - \sin\theta\Ket{\eta_1}\label{eta_mixing},
\end{equation}
\begin{equation}
\Ket{\eta'} = \sin\theta \Ket{\eta_8} + \cos\theta\Ket{\eta_1}\label{eta_prim_mixing},
\end{equation}
where $\theta$ is the mixing angle equal to about $-15^o$~\cite{eta_mixing_angle1,eta_mixing_angle2}.
Relatively small mixing angle value allows to treat $\eta$ meson as $\eta_8$ with a small admixture of $\eta_1$ component.
\label{eta_is_influenced_when_embeded_in_nuclear_matter}
There are presumptions that $\eta_1$ component can mix with pure gluonic states that can influence the properties of $\eta$ meson embedded in nuclear matter~\cite{bound_theory8,bound_theory9}.

\label{eta_nucleon}
Short $\eta$ meson lifetime does not allow to create $\eta$ beams.
Thus, one cannot measure data about the elastic scattering that makes the theoretical investigations more complicated.

In the low energy region, the dominating $\eta$-nucleon interaction mechanism is the excitation of $S_{11}$ resonance $N^{*}(1535)$.
Its mass is $1535~MeV/c^2$ and the width is of about $150~MeV$~\cite{eta_decay_modes_new,particle_masses}.
This resonance is strongly coupled to s-wave $\pi-N$ and $\eta-N$ channels~\cite{Nstar_coupled_with_channels1,Nstar_coupled_with_channels2}.
It also decays via $\eta N$, $\pi N$, $\Delta\pi$, $\gamma N$, $\pi\pi N$ channels.

\chapter{Bound states and resonances in the scattering theory}
\label{scattering_theory}
The detailed review of bound state description in frame of scattering theory can be found in Ref.~\cite{krzemien_phd}.

Applying the perturbative approach, the scattering theory determines the scattering operator transforming initial state wave function into the final state:
\begin{equation}
\Ket{\Psi_{final}} = \hat{S} \Ket{\Psi_{initial}}.
\label{scattering_operator_eq}
\end{equation}
This operator is analytic, unitary, time reversal symmetric, and Lorentz invariant~\cite{scattering_theory_book1,scattering_theory_book2}.
These properties allow to represent it as scattering matrix expanding the wave functions in a basis of orthonormal states:
\begin{equation}
\Ket{\Psi_{final}} = \sum_{f,i} \Ket{f} S_{fi}\Bra{i}\Ket{\Psi_{initial}},
\label{sczttering_matrix_eq}
\end{equation}
where $S_{fi}$ is the scattering matrix element defined as $\Bra{f}\hat{S}\Ket{i}$.

For spinless particles, in angular momentum basis, the scattering matrix is diagonal and depends on relative momentum $p$:
\begin{equation}
S_{ll'}(p) = S_{l}(p) \delta(l-l'),
\end{equation}
where $l$ denotes the angular momentum of the system, $\delta(l-l')$ is the Kronecker delta, and $S_{l}$ is defined by the following formula:
\begin{equation}
S_l(p) = e^{2 i \delta_l(p)} ,
\end{equation}
where $\delta_l(p)$ is the phase shift. Thus, in the unit system where $\hbar=1$, the scattering amplitude can be defined as~\cite{scattering_theory_book2}:
\begin{equation}
f_l = \frac{S_l - 1}{2ip} = \frac{1}{p~cot(\delta_l) - ip}.
\label{scattering_amplitude_definition}
\end{equation}

For $l=0$, the phase shift can be approximated by the formula~\cite{scattering_theory_book2}:
\begin{equation}
p~cot(\delta_0) = -\frac{1}{a} + \frac{r_0 p^2}{2},
\label{scattering_length_definition}
\end{equation}
where $a$ is the scattering length and $r_0$ is the effective scattering potential range.
There are different conventions about the sign before the first term in Eq.~\ref{scattering_length_definition} defining the scattering length.
In this thesis, the convention with minus sign before the first term of this expression is used.

In case of small momentum one can neglect the second term of Eq.~\ref{scattering_length_definition} and write the scattering amplitude as:
\begin{equation}
f_0 = \frac{1}{-\frac{1}{a} -ip} = \frac{a}{-1 - iap}.
\label{scattering_amplitude_from_scattering_length}
\end{equation}

When inelastic channels are also open, the scattering length appears as a complex magnitude.
Because of the scattering matrix unitarity, the imaginary part of scattering length is positive.
The bound state existing requires~\cite{scattering_length_bound_state_condition} a condition:
\begin{equation}
|\Re a| > \Im a,
\label{scattering_length_bound_state}
\end{equation}
where the symbols $\Re$ and $\Im$ mean the real and imaginary part of a complex number, correspondingly.
This is necessary condition but it is not enough to conclude that the bound state exists.

One of powerful techniques in scattering theory is generalizing the scattering matrix in the complex momentum plane~\cite{scattering_theory_book2}.
In this approach the matrix elements $S_l(p)$ are treated as analytic complex momentum functions.
The analytical properties of scattering matrix elements depend on interaction potential, especially on its asymptotic behavior.
If the potential exponentially decreases in $r\rightarrow\infty$ and is an analytic function for complex argument $\Re r>0$, then the scattering matrix is analytic at the whole complex plain except maybe a finite number of points (poles)~\cite{scattering_theory_book2}. 
If the interaction potential is a real function (no absorption channels) then the poles can be either on the imaginary axis ($\Re p=0$) 
or on the lower half plane ($\Im p<0$)~\cite{scattering_with_absorptive_interaction}.
Poles laying at $\Im p>0$ correspond to bound states
and poles laying at $\Im p<0$ correspond to resonances.
However, poles located far from real axis ($\Im p=0$) and not leading to observable resonances can exist.

The simplest form of the scattering matrix 	with a pole corresponding to a bound state is~\cite{scattering_theory_book2}:
\begin{equation}
S_0 = \frac{-p - i p_0}{ p - i p_0},
\label{simplest_scattering_matrix}
\end{equation}
where $i p_0$ is the pole position ($p_0>0$).
For this scattering matrix, the scattering amplitude is the following~\cite{scattering_theory_book2}:
\begin{equation}
f_0 = \frac{1}{-p_0 - ip}.
\label{simplest_scattering_amplitude}
\end{equation}

The similarity of Eq.~\ref{scattering_amplitude_from_scattering_length} and \ref{simplest_scattering_amplitude} points at the connection between the scattering length and the pole position:
\begin{equation}
p_0 = \frac{1}{a}.
\label{scattering_param_conn}
\end{equation}

In case when the scattering length is much larger than the potential range, one can determine the binding energy~\cite{scattering_theory_book2}:
\begin{equation}
- E_{bound} = \frac{p^2_0}{2m} = \frac{1}{2ma^2},
\label{binding_energy_from_the_scattering_length}
\end{equation}
where $m$ is the reduced mass.

\begin{figure}[h!]
	\begin{center}
		\includegraphics[width=250pt]{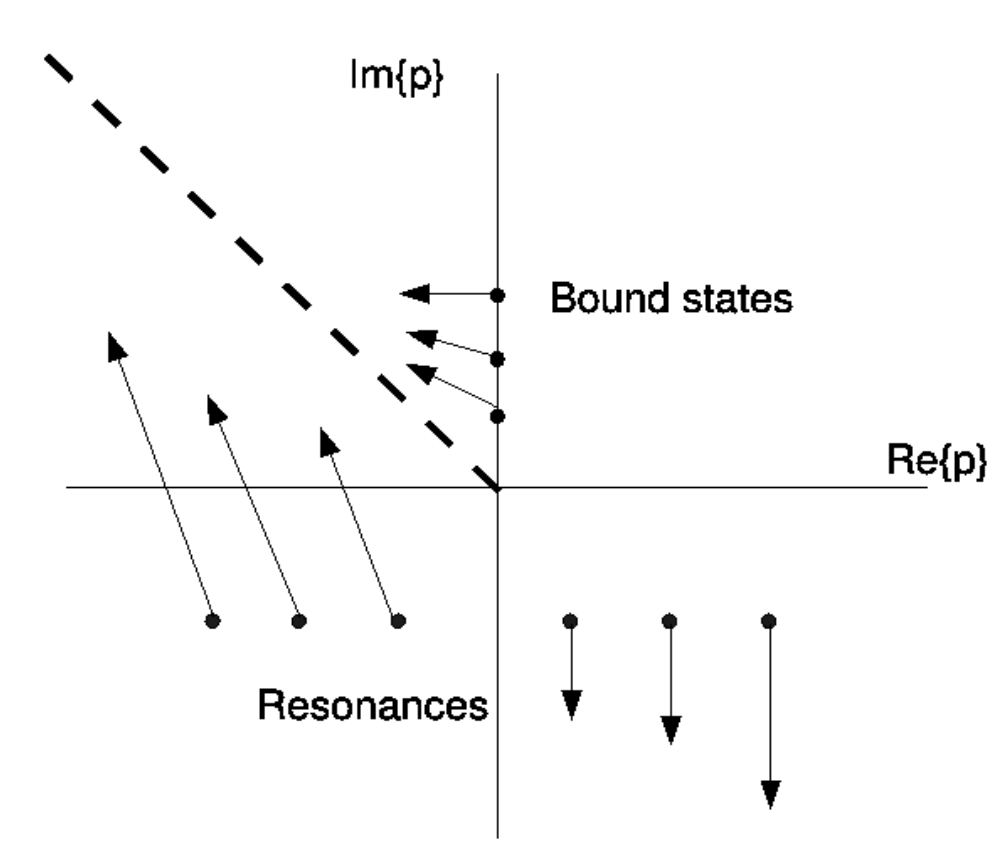}
	\end{center}
	\caption{The scheme showing the scattering matrix poles movement when increasing the imaginary part of interaction potential. The picture was taken from~\cite{krzemien_phd}.}
	\label{poles_movement}
\end{figure}

In case when the interaction potential contains an imaginary part (inelastic channels are also taken into account), the situation is more complicated.
The poles positions are shifted~\cite{scattering_with_absorptive_interaction,scattering_bound_state_fig} like it is shown on Fig.~\ref{poles_movement}.

The scattering length can be measured in experiments studying low momentum elastic scattering. 
It can also be determined on the basis of final state interaction (FSI) between produced particles.
Bound states or resonances can be observed as a change in $S(p)$ phase or as a peak in the cross section as a function of energy.
\chapter*{List of Abbreviations}
\addcontentsline{toc}{chapter}{List of Abbreviations}
\begin{itemize}
\item \textbf{FSI} Final State Interaction
\item \textbf{QCD} Quantum Chromodynamics
\item \textbf{QMC} quark-meson-coupling
\item \textbf{COSY} Cooler Synchrotron
\item \textbf{CD} Central Detector
\item \textbf{WASA} Wide Angle Shower Aparatus
\item \textbf{MDC} Mini Drift Chamber
\item \textbf{PSB} Plastic Scintillator Barrel
\item \textbf{SEC} Scintillation Electromagnetic Calorimeter
\item \textbf{FD} Forward Detector
\item \textbf{FWC} Forward Window Counter
\item \textbf{FPC} Forward Proportional Chamber
\item \textbf{FTH} Forward Trigger Hodoscope
\item \textbf{FRH} Forward Range Hodoscope
\item \textbf{DAQ} Data Acquisition
\item \textbf{FPGA} Field-Programmable Gate Array
\item \textbf{QDC} Charge-to-Digital Converter
\item \textbf{TDC} Time-to-Digital Converter
\item \textbf{FIFO} First In First Out

\end{itemize}
\chapter*{Acknowledgements}
\addcontentsline{toc}{chapter}{Acknowledgements}

First of all, I would like to thank WASA-at-COSY Collaboration members for organizing the experiment that became the topic of my thesis.
I am especially thankful to my thesis advisor prof.~dr~hab.~Paweł Moskal and co-advisor dr.~Magdalena~Skurzok.
Also, I would like to mention dr.~Wojciech~Krzemień, prof.~Volker~Hejny and prof.~Magnus~Volke who have also consulted me much in data analysis.

I want to thank professors from Jagiellonian University who provided interesting courses that I attended in frame of the PhD studies program: prof.~dr~hab.~Józef~Spałek, prof.~dr~hab.~Jarosław~Koperski, prof.~dr~hab.~Tomasz~Kozik, prof.~dr~hab.~Jacek~Kołodziej, and dr~hab.~Paweł~Korecki.
Also I want to thank Janusz~Konarski for help in teaching practicing that also was a part of the studies.

I am thankful as well to people who taught me in Taras Shevchenko University National in Kyiv for the knowledge they have transferred to me.
Among them, I want to mention dr.~Larysa~Holinka-Bezshyjko, dr.~Oleh~Bezshyjko, prof.~Volodymyr~Pliuyko, dr.~Yuri~Onyshchuk, prof.~Ihor~Kadenko,
and \fbox{dr.~Yuriy~Pavlenko} who worked in Kyiv Institute for Nuclear Research and was my diploma thesis advisor.

I am especially thankful to my wife Iryna~Rundel for her support that made preparing of this thesis possible.

\bibliographystyle{unsrt}
\end{document}